\documentclass[twocolappendix,numberedappendix]{emulateapj}
\usepackage{natbib}
\usepackage{apjfonts}
\usepackage{placeins}
\usepackage{multirow}
\usepackage{amsmath}
\usepackage{amssymb}
\usepackage{epsfig}
\usepackage{graphicx}
\usepackage{color}
\newcommand{\kms}{\mbox{km s$^{-1}$}}

\newcommand{\Msun}{\mbox{$M_{\odot}$}}

\newcommand{\hii}{\ion{H}{2}}
\newcommand{\co}{\mbox{CO$(1\rightarrow0)$}}
\newcommand{\cotwo}{\mbox{CO$(2\rightarrow1)$}}

\newcommand{\degrees}{\arcdeg}
\newcommand{\avg}[1]{\left< #1 \right>}

\shorttitle{Molecular clouds in spiral galaxies}
\shortauthors{Rebolledo et al.}

\begin{document}
\submitted{Accepted for publication in The Astrophysical Journal, June 14, 2015}

\title{Scaling relations of the properties for CO resolved structures in nearby spiral galaxies}

\author{David Rebolledo$^{1,2,3}$, Tony Wong$^{2}$, Rui Xue$^{4}$, Adam Leroy$^{5,6}$, Jin Koda$^{7}$ and Jennifer Donovan Meyer$^{5}$}

\affil{$^{1}$Sydney Institute for Astronomy, School of Physics, The University of Sydney, NSW 2006, Australia; davidr@physics.usyd.edu.au\\
$^{2}$Astronomy Department, University of Illinois, Urbana, IL 61801, USA\\
$^{3}$School of Physics, University of New South Wales, Sydney, NSW 2052, Australia\\
$^{4}$Department of Physics, Purdue University, 525 Northwestern Avenue, West Lafayette, IN 47907, USA\\
$^{5}$National Radio Astronomy Observatory, 520 Edgemont Road, Charlottesville, VA 22903, USA\\
$^{6}$Department of Astronomy, The Ohio State University, 140 West 18th Avenue, Columbus, OH 43210, USA\\
$^{7}$Department of Physics and Astronomy, Stony Brook University, Stony Brook, NY 11794-3800, USA}

\begin{abstract}

Complementing the observations on the eastern part of galaxy NGC 6946 presented in a previous work, we report high spatial resolution observations of Giant Molecular Clouds (GMCs) in the nearby spiral galaxies M101 and NGC 628 obtained with the Combined Array for Research in Millimeter-wave Astronomy (CARMA).  We observed $\co$ over regions with active star formation extending from 2 kpc to 15 kpc galactocentric radius.  Higher resolution observations of $\cotwo$ toward the brightest regions observed in $\co$ have allowed us to resolve some of the largest GMCs.  We have recovered short-spacing $u$-$v$ components by using single dish observations from the Nobeyama 45m and IRAM 30m telescopes.  Using the automated CPROPS algorithm we identified 112 CO complexes in the $\co$ maps and 144 GMCs in the $\cotwo$ maps.  Using a Bayesian fitting approach, we generate scaling relations for the sizes, line widths, and virial masses of the structures identified in this work.  We do not find evidence for a tight power law relation between size and line width, although the limited dynamic range in cloud size remains a clear issue in our analysis.  Additionally, we use a Bayesian approach to analyze the scaling relation between the star formation and molecular gas surface density, known as the Kennicutt-Schmidt relation.  When we perform our analysis using the boundaries of the structures identified by CPROPS, we find that the distribution of slopes are broadly distributed, mainly due to the limited dynamic range of our measured $\Sigma_\mathrm{H2}$.  In the case of the $\co$ complexes, the slope distributions are most consistent with super-linear relations, although sub-linear relations cannot be excluded for NGC 628 and NGC 6946.  The GMCs from higher resolution $\cotwo$ maps follow a similar behavior, but with larger scatter.   As a complementary study, we use the Bayesian approach to analyze the Kennicutt-Schmidt relation for a uniform grid covering the areas surveyed, and with $\Sigma_\mathrm{H2}$ non-detections included in the analysis.  The distributions of slopes is consistent with sub-linear relations for NGC 6946 and NGC 628, but is less constrained for M101.  This picture is preserved after a 24$\mu$m background component is subtracted from the $\Sigma_\mathrm{SFR}$ measurements.  On-arm regions tend to have higher star formation rates than inter-arm regions.  Similar to what we find in our study of the eastern part of NGC 6946, in M101 we find regions where the star formation efficiency (SFE) shows marked peaks at specific galoctocentric radii.  On the other hand, the distribution of SFE in NGC 628 is more contiguous.  We hypothesize that differences in the distribution of SFE may be indicative of different processes driving the spiral structure.

\end{abstract}


\section{INTRODUCTION}

The formation of stars is the final stage of many physical processes acting on the complex structure of the ISM at different temporal and spatial scales.  As a consequence, a complete understanding of the processes responsible for how stars form out of the ISM remains a challenging task.  However, it is observed that most of the massive star formation in the Milky Way (MW) is associated with GMCs, and the corrrelation between star formation and molecular mass has been observed to hold up to much larger sizes, even $\sim 1$ kpc (\citealt{2013AJ....146...19L}; \citealt{2008AJ....136.2846B}, \citealt{2011ApJ...730L..13B}).  Therefore, studying GMCs in different environments seems to be a crucial tool to connect the processes that trigger and regulate the star formation at large scales with those playing an important role for star formation within the clouds, allowing us to establish a unified picture of the star formation process across the full range of sizes observed.

Due to observational limitations, most of our understanding of the physical processes involved in the creation of molecular clouds, their subsequent fragmentation into smaller and denser structures such as clumps and cores, and finally the birth of clusters and individual stars, has come from the detailed exploration of individual star-forming regions in the Milky Way.  On the other hand, extragalactic studies have focused mainly on the integrated star formation over regions commonly larger than 100 pc, or even over the entire galactic disk.  Resolving structures close to the scales where GMCs form ($\sim$100 pc) in nearby galaxies is still observationally challenging.  Thus, high resolution CO surveys of galaxies (\citealt{1999ApJS..124..403S}; \citealt{2003ApJS..145..259H}) have focused primarily on the central regions.  Single-dish observations (\citealt{2007PASJ...59..117K}; \citealt{2009AJ....137.4670L}) sometimes offer a complete mapping of the disk, but the resulting resolutions have been inadequate to resolve GMCs.  A complete census of GMCs generally requires a dedicated observing program over the disk of a single galaxy, such as has been conducted for M33 (\citealt{2003ApJS..149..343E}; \citealt{2010ApJ...722.1699S}), the Magellanic Clouds (NANTEN surveys,  \citealt{2008ApJS..178...56F}), and M51 (\citealt{2012ApJ...761...41K}; \citealt{2013ApJ...779...42S}).  

In the past few years, numerical simulations have suggested that the formation and evolution of molecular clouds may depend on the dynamical process that drives the spiral arm structure (\citealt{2006MNRAS.367..873D}; \citealt{2008MNRAS.385.1893D}; \citealt{2011ApJ...735....1W}).  In the case of grand-design spiral galaxies, massive molecular clouds are thought to be formed by the passing of spiral density waves, inducing the formation of spurs downstream of the spiral arms due to the interaction between the ISM and the stellar spiral arm (\citealt{2002ApJ...570..132K}; \citealt{2004MNRAS.349..270W}).  Thus, an offset between the spiral shock and the potential minimum is expected (e.\ g.\ M51, \citealt{2013ApJ...763...94L}).  On the other hand, in galaxies with a multi-arm spiral structure, simulations find that the gaseous spiral structure coincides with the potential minimum over several crossing times.  In this case, the collision or merging between spiral arms leads to the formation of the most massive structures, increasing the star formation in those overdense regions.  Therefore, high resolution observations of the molecular gas in spiral galaxies with different morphologies can help disentangle the real processes involved in the formation of dense gas, and the subsequent formation of stars.

\subsection{Molecular Cloud Scaling Relations }\label{intr-scaling}
Scaling relations between properties of resolved structures have been useful tools to decipher the physical conditions present in the interstellar medium (ISM).  Studies conducted in the MW showed that resolved properties such as velocity dispersion, luminosity and size of GMCs were correlated, following empirical relations known as Larson's laws (\citealt{1981MNRAS.194..809L}; \citealt{1987ApJ...319..730S}).  The scaling relation between the line width and the size, commonly referred as the first Larson's law, has been used to investigate the character of the turbulence in the ISM.  Nevertheless, the dominant source of turbulence in the ISM, in particular whether it is internally or externally driven, remains controversial (see \citealt{2007ARA&A..45..565M}).  The second and third Larson's laws dictate that the GMC's are roughly self-gravitating and have approximately similar surface densities respectively.  Although Larson's relations have been shown to hold across a wide range of extragalactic environments (\citealt{2007prpl.conf...81B}, \citealt{2008ApJ...686..948B}), recent galactic (\citealt{2009ApJ...699.1092H}) and extragalactic studies (\citealt{2011ApJS..197...16W}, \citealt{2013ApJ...779...46H}, \citealt{2014ApJ...784....3C}) have questioned the universality of those relations.  They usually find a weak correlation between the size and line width of the identified structures, while inferring genuine differences in their physical properties such as velocity dispersion, CO peak brightness and mass surface densities.  

Although Larson's laws originally relate properties such as size, line width and luminosity, an extra relation can be built assuming virialization of the clouds.  Virial masses derived from the size and line width can be compared to the luminosity of the GMCs, thus defining a scaling factor that relates the intensity of the CO emission with the H$_2$ column density, the $X_\mathrm{CO}$ factor.  This conversion factor is thought to depend on the metallicity and the local radiation field acting on the regions probed by the CO observations (\citealt{2011ApJ...737...12L}; \citealt{2013ApJ...777....5S}; \citealt{2013ARA&A..51..207B}).  The current generation of radio facilities has provided CO maps of galaxies at a sufficient spatial resolution to resolve individual GMCs.  Thus, the $X_\mathrm{CO}$ factor can be inferred using the virial mass method.  The average values thus obtained for the conversion factor in nearby spiral galaxies are within a factor of two of the MW value (\citealt{2012ApJ...744...42D}; \citealt{2013ApJ...772..107D}; \citealt{2012ApJ...757..155R}).  However, the $X_\mathrm{CO}$ factor can be estimated using other methods.  For example, by assuming a constant H$_2$ depletion time in dwarf galaxies, \citet{2012AJ....143..138S} find $X_\mathrm{CO}$ factor values more than one order of magnitude higher than spiral galaxies with solar metallicity.  By assuming a power-law relation between the star formation rate and the molecular gas surface density (see Section \ref{intr-sfr-recipes}), \citet{2013ApJ...764..117B} measure the dependence of the $X_\mathrm{CO}$ factor with the galactocentric radius in NGC 628.  They find that the $X_\mathrm{CO}$ factor increases with radius, which added to the observed metallicity gradient across the disk, implies that lower metallicity regions coincides with larger values of $X_\mathrm{CO}$.  Alternatively, optically thin dust emission can be used as an independent tracer of the gas (\citealt{2011ApJ...737...12L}, \citealt{2013ApJ...777....5S}).  The surface density of the gas $\Sigma_\mathrm{gas}$ is estimated from the dust surface density $\Sigma_\mathrm{dust}$ through $\Sigma_\mathrm{gas}=\delta_\mathrm{gtd} \Sigma_\mathrm{dust}$, where $\delta_\mathrm{gtd}$ is the gas to dust ratio.  To constrain $\delta_\mathrm{gtd}$, $\Sigma_\mathrm{gas}$ is independently obtained from the $\Sigma_\mathrm{H2}=X_\mathrm{CO} I_\mathrm{CO}$ and $\Sigma_\mathrm{HI}$ maps.  For a set of galaxies in the Local Group, \citet{2011ApJ...737...12L} solved simultaneously for $X_\mathrm{CO}$ and $\delta_\mathrm{gtd}$ over regions with size scales $\sim$1 kpc$^2$.  For galaxies with higher metallicity, they found $X_\mathrm{CO} \sim 1-4.5 \times 10^{20}\ \mathrm{cm}^{-2} (\mathrm{K}\ \kms)^{-1}$, similar to MW values.  On the other hand, for systems with lower metallicity, they found larger $X_\mathrm{CO}$ by a factor of 10 compared to the conversion factor obtained for high metallicity regions, similar to the conclusion obtained by  \citet{2012AJ....143..138S}.  Additionally, \citet{2013ApJ...777....5S} found a depression in $X_\mathrm{CO}$ in the central part of the disk of some galaxies in their sample. 

\begin{figure*}
\centering
\begin{tabular}{c}
\epsfig{file=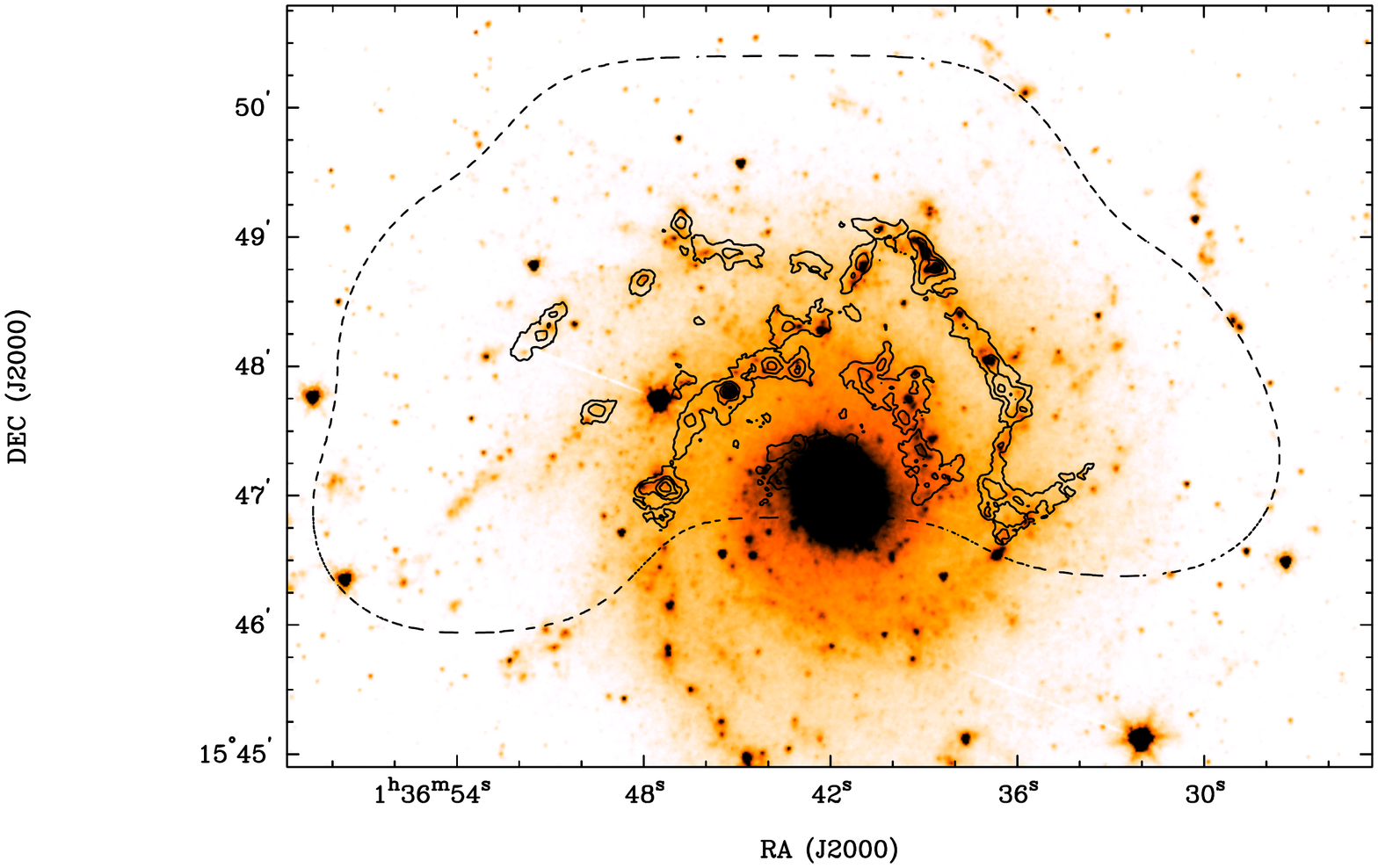,width=0.5\linewidth,angle=-90}
\end{tabular}
\caption{$\co$ integrated intensity contours overlaid on 3.6 $\mu$m map from SINGS for the northern arms of NGC 628.  Contours begin at 3.56 K $\kms$ and are spaced by 2$\times$3.56 K $\kms$.  The noise at the center of the map is 1.5 K km s$^{-1}$. The dashed black line illustrates the region where full gain correction was made.}
\label{figure_ngc628N}
\end{figure*}

\subsection{Star Formation Relations}\label{intr-sfr-recipes}
Another relation used to study the properties of the ISM is the empirical power-law relation that connects the star formation surface density ($\Sigma_\mathrm{SFR}$) and the gas surface density ($\Sigma_\mathrm{gas}$) available to form stars, i.\ e., the relation given by $\Sigma_\mathrm{SFR} \propto \Sigma_\mathrm{gas}^{\alpha}$ (\citealt{2012ARA&A..50..531K}, and references therein).  This relationship, known as the Kennicutt-Schmidt law (abbreviated as K-S henceforth), has been the subject of several works that have used observations of the ISM in both the Milky Way ( \citealt{2009ApJS..181..321E}; \citealt{2010ApJ...723.1019H}; \citealt{2011ApJ...739...84G}) and nearby galaxies (\citealt{2008AJ....136.2846B}; \citealt{2011ApJ...735...63L}; \citealt{2011ApJ...730...72R}; \citealt{2013AJ....146...19L}; \citealt{2013ApJ...772L..13M}).  Nevertheless, due to the difficulties in measuring the molecular gas content and star formation rates across a wide dynamic range, the intrinsic form of the K-S law has remained elusive.  The different slopes reported in previous studies mostly fall into two groups: those that report linear relations ($\alpha \sim$ 1), and those that find a super-linear relations ($\alpha >$ 1) in the K-S law.  Several explanations have been proposed to explain the wide range of slopes ($\sim$ 1-3) found in extragalactic studies, e.\ g., biases in the tracer used to estimate the star formation rate or the molecular gas, biases in the fitting procedure used to find the power-law, variations in the fraction of dense gas, and the size scales over which averaging is performed (\citealt{2011ApJ...735...63L}; \citealt{2012ApJ...752...98C}; \citealt{2013ApJ...772L..13M}).  More recently, it has been proposed that diffuse emission in the maps used to estimate the star formation rates (FUV, 24$\mu$m and H$\alpha$) may be the reason for the divergent results.  \citet{2011ApJ...735...63L}, for example, were able to reconcile the linear relation found by \citet{2008AJ....136.2846B} with the super-linear relation found by \citet{2007ApJ...671..333K} in M51a by subtracting or preserving a diffuse background component in the star formation tracers.  

Although the correlation between gas and star formation rate (SFR) surface densities has been observed to be tight when averaged over substantial areas of galaxies, recent high resolution observations of molecular gas in nearby galaxies have revealed a poorer correlation between these two quantities (\citealt{2010ApJ...722.1699S}; \citealt{2012ApJ...757..155R}, hereafter Paper I). These recent studies concluded that the evolution of individual regions is the main source of the increased scatter observed at small scales.  Similarly, \citet{2012ApJ...752...98C} find that the stochastic sampling of the molecular cloud mass function determines the slope and the scatter of their simulated K-S relation.  They find that at scales below $\sim$ 1 kpc the K-S relation is super-linear, while at scales $\sim$ 1-2 kpc the relation becomes linear as the cloud mass function is fully sampled. 

\begin{table}
\caption{Properties of sample galaxies.}
\centering
\begin{tabular}{lccc}
\hline\hline
Property  &  NGC 6946  &  NGC 628  &  M101 \\
Morph. \tablenotemark{a} & SABcd & SA(s)c  & SAB(rs)cd  \\ 
R.A. (J2000)$^\mathrm{a}$ & 20:34:52.3  &  01:36:41.7  &  14:03:12.5   \\ 
Decl. (J2000$^\mathrm{a}$ & 60:09:14 & 15:47:01 &  54:20:56  \\
Distance (Mpc) & 5.5$^\mathrm{b}$  & 7.3$^\mathrm{d}$  & 7.4$^\mathrm{e}$  \\   
Incl. (\degrees)$^\mathrm{c}$ & 33   &  7   &  18   \\  
P.A. (\degrees)$^\mathrm{c}$ & 243  & 20  & 39  \\
\hline
\multicolumn{2}{l}{{\bf Notes.}} \\
\multicolumn{4}{l}{$^\mathrm{a}$ NASA/IPAC Extragalactic Database (NED).}\\
\multicolumn{2}{l}{$^\mathrm{b}$ \citet{1988JBAA...98..316T}.}\\
\multicolumn{2}{l}{$^\mathrm{c}$  \citet{2008AJ....136.2563W}.} \\
\multicolumn{2}{l}{$^\mathrm{d}$  \citet{1996A&AS..119..499S}.}\\
\multicolumn{2}{l}{$^\mathrm{e}$  \citet{2000ApJS..128..431F}.}
\end{tabular}
\label{galaxy-prop}
\end{table}

\begin{table}
\caption{CARMA data properties.}
\centering
\begin{tabular}{lccc}
\hline
\hline
Property  &  NGC 6946 \tablenotemark{a}   & NGC 628   &  M101 \\
\hline
\multicolumn{4}{c}{$\co$ maps}  \\ 
\hline
Sensitivity (K) & 0.40  & 0.42  & 0.37   \\ 
Velocity Resolution ($\kms$) & 2.5 & 2.5   & 2.5    \\
Angular Resolution (") & 5.2 $\times$ 5.0  & 4.5 $\times$ 4.4   & 4.4 $\times$ 4.2     \\
Linear Resolution (pc) \tablenotemark{b} & 136  & 157  & 154    \\ 
\hline
\multicolumn{4}{c}{$\cotwo$ maps}  \\ 
\hline
Sensitivity (K) & 0.53  &  0.31 &  0.29  \\ 
Velocity Resolution ($\kms$) & 2.5 & 2.5   & 2.5    \\
Angular Resolution (")  & 2.5 $\times$ 2.0 &  2.0 $\times$ 1.8   &  2.0 $\times$ 1.9   \\
Linear Resolution (pc) \tablenotemark{b} & 60   &  67  & 70    \\ 
\hline
\multicolumn{2}{l}{{\bf Notes.}} \\
\multicolumn{4}{l}{$^\mathrm{a}$ Properties of NGC 6946 as in Paper I.}\\
\multicolumn{2}{l}{$^\mathrm{b}$ Linear resolution is calculated from $\sqrt{b_\mathrm{min} \times b_\mathrm{maj}}$.}\\
\end{tabular}
\label{data-prop}
\end{table}

We have recently presented a study of the properties of the molecular gas in the eastern part of the nearby galaxy NGC 6946 (Paper I). We found that the properties of the clouds follow relations similar to those found previously for extragalactic clouds.  In particular, the trends exhibited by the $\co$ complexes presented in that study were consistent with those found in the center of NGC 6946 by \citet{2012ApJ...744...42D} (hereafter DM12), despite the differences in resolution and cloud identification algorithms; the exception is the set of GMCs located within 400 pc of the galactic center with large velocity dispersions ($>$ 10 $\kms$).  Additionally, our virial mass-CO luminosity relation is consistent with the relation found by DM12 for the central part of NGC 6946 and is consistent with our choice of the CO to H$_2$ conversion factor $X_\mathrm{CO}$=$2\times10^{20}\mathrm{cm}^{-2}(\mathrm{K}\ \kms)^{-1}$.  

\begin{figure*}
\centering
\begin{tabular}{c}
\epsfig{file=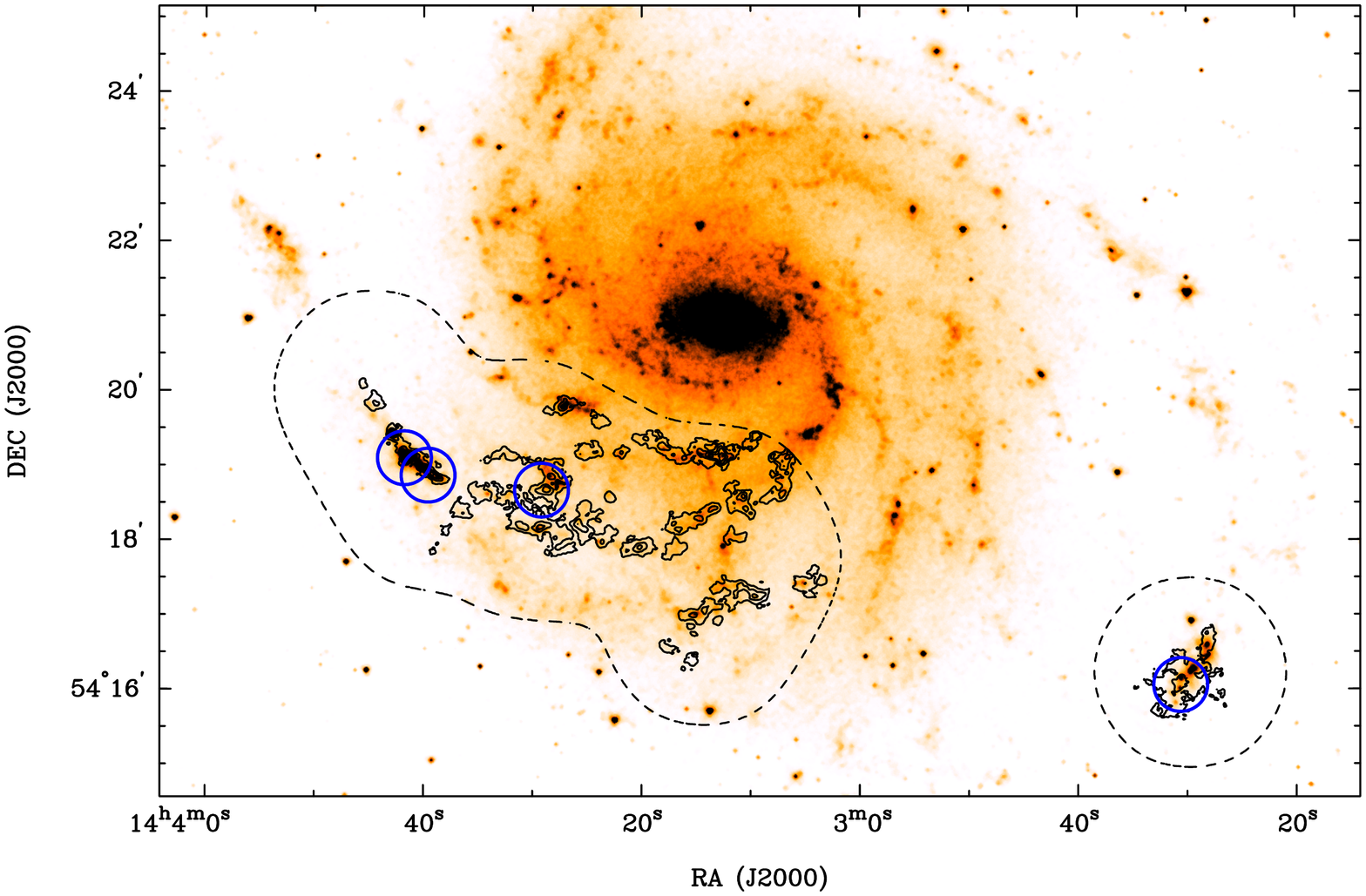,width=0.5\linewidth,angle=-90}
\end{tabular}
\caption{$\co$ integrated intensity contours overlaid on 3.6 $\mu$m map from SINGS for the regions observed in M101.  Contours begin at 5.1 K $\kms$ and are spaced by 2$\times$5.1 K $\kms$.  The noise at the center of the maps is 2.5 K km s$^{-1}$.  The blue circles indicates the regions with enhanced star formation reported in Section \ref{rad-prop}.}
\label{figure_m101S}
\end{figure*}

In this paper, we compare the properties of molecular structures found in NGC 6946 with those located in sub-regions of two other nearby spiral galaxies: M101 and NGC 628.  We have continued our observational strategy previously used for NGC 6946 (Paper I) which consists of starting with large-scale, lower resolution $\co$ maps, and following up with higher resolution $\cotwo$ imaging on smaller areas.  We select regions in the disks with active star formation, and near the transition radius from the H$_2$-dominated inner galactic region to the HI-dominated outer region. As with the galaxy NGC 6946, these two galaxies represent excellent targets because of their proximity, high CO surface brightness, low inclination, and the availability of high-quality datasets at different wavelengths, including HI THINGS maps (\citealt{2008AJ....136.2563W}), GALEX UV imaging (\citealt{2007ApJS..173..185G}) and multiband Spitzer imaging from SINGS (\citealt{2003PASP..115..928K}).  

We present our study as follows:  in Section \ref{obs} we describe our observations of NGC 6946, M101 and NGC 628 using CARMA, and we describe the archival data at several wavelengths that we include in our analysis.  In Section \ref{cprops} we summarize the technique used to identify GMCs and to measure their physical properties.  In Section \ref{cl-prop} we present some statistics of the measured properties, and we introduce the Bayesian inference method used to study the different ISM scaling relations found in our sample of galaxies.  In Section \ref{discuss} we discuss the implications of our results, and we investigate whether the star formation properties of the clouds differ between on-arm and inter-arm regions or between galaxies.  Section \ref{caveats} discusses the limitations of our analysis.  In Section \ref{summary} we summarize the work presented in this paper.

\begin{figure*}
\centering
\begin{tabular}{c}
\epsfig{file=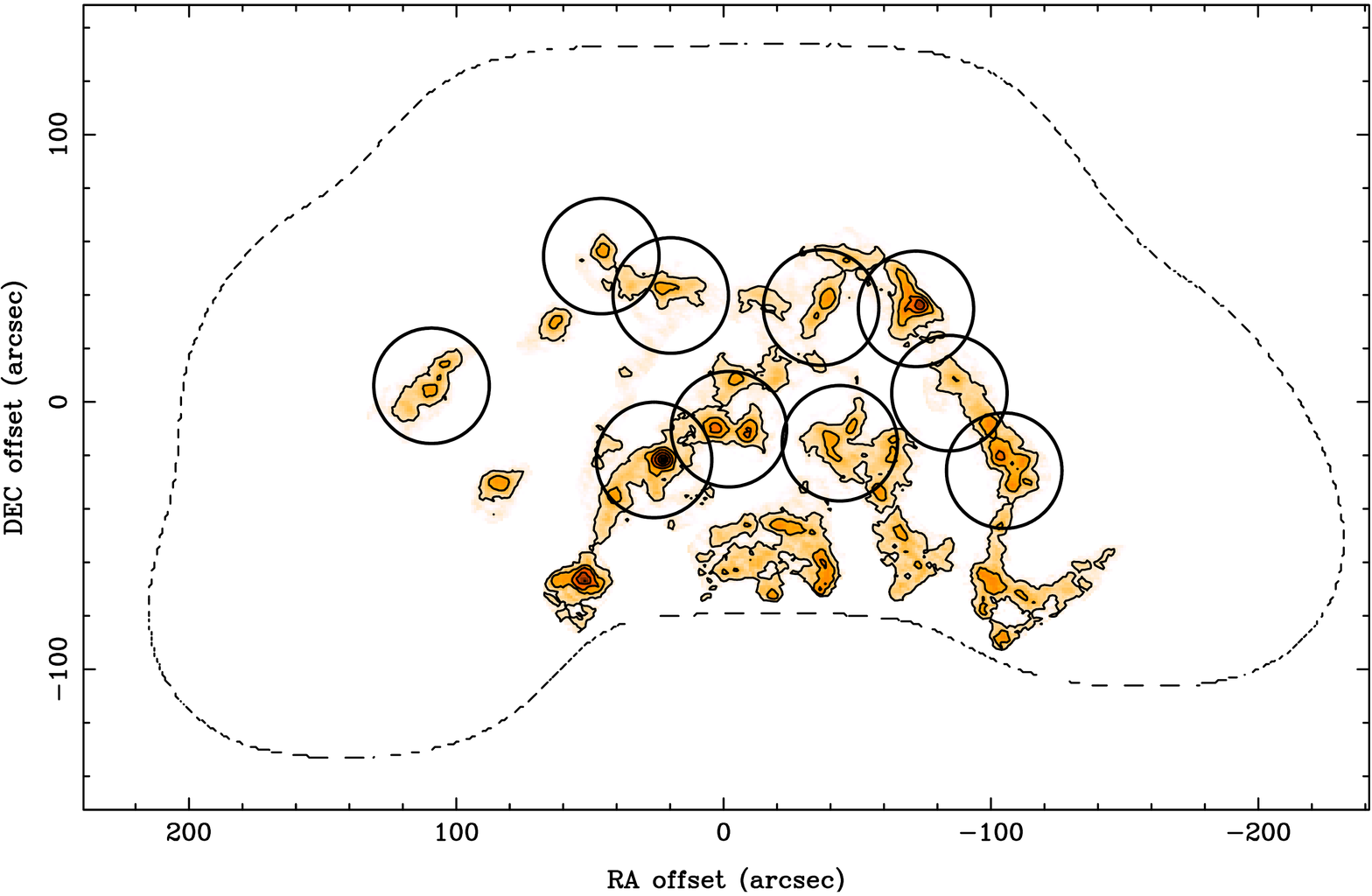,width=0.4\linewidth,angle=-90}
\end{tabular}
\caption{$\co$ integrated intensity map of the region observed in NGC 628.  Contours are as in Figure \ref{figure_ngc628N}.  Black circles indicate the regions where we performed $\cotwo$ observations.}
\label{figure_ngc628}
\end{figure*}

\begin{figure*}
\centering
\begin{tabular}{c}
\epsfig{file=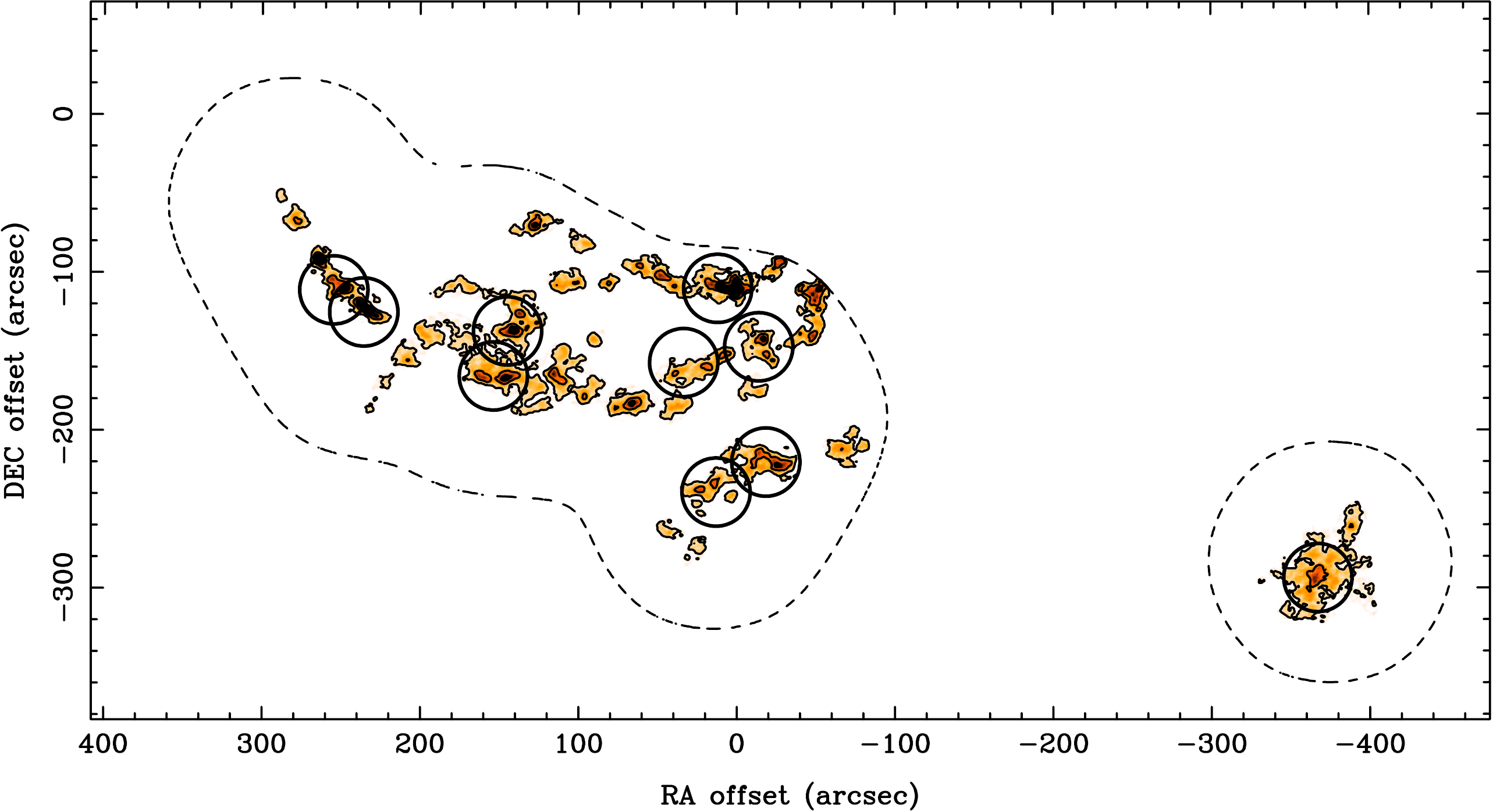,width=0.4\linewidth,angle=-90}
\end{tabular}
\caption{$\co$ integrated intensity map of the regions observed in M101.  Contours are as in Figure \ref{figure_m101S}.  Black circles indicate the regions where we performed $\cotwo$ observations.}
\label{figure_m101}
\end{figure*}

\section{DATA}\label{obs}
\subsection{CARMA observations}\label{carma}

\subsubsection{3 mm}

We performed high spatial resolution observations of $\co$ for the selected regions in NGC 6946, NGC 628 and M101 from July of 2009 to September 2012.  We used the Combined Array for Research in Millimeter-wave Astronomy (CARMA) in E, D and C array configurations, which have baselines of  8.5-66 meters, 11-148 meters, and 30-350 meters respectively. In this section we describe the details of the NGC 628 and M101 maps, whereas we refer the reader to Paper I for the observation strategy and details of NGC 6946 maps.  Table \ref{galaxy-prop} shows the basic parameters of NGC 6946, NGC 628 and M101.  

For NGC 628, we observed $\co$ in a 53-point mosaic area, which covers 5.5$\times$2.7 arcmin$^2$, corresponding to a physical scale of 11.7$\times$5.7 kpc$^2$ at a distance of 7.3 Mpc (calculated from the luminosity of the three brightest blue supergiant stars, \citealt{1996A&AS..119..499S}). The correlator was set to have the $\co$ line in the upper side band (USB).  We have placed three overlapping 62 MHz bands to cover the line.  One band was centered in the $\co$ line frequency, while the other two were placed with an offset of 15 MHz from the rest frequency, achieving a total velocity coverage of 240 $\kms$ and a channel width of 0.4 $\kms$.  A 500 MHz wide-band was placed outside the spectral windows for calibration.  We observed every pointing for 30 seconds, yielding a 26.5-minute observation time per cycle.  At the beginning of the track we observed 3C454.3 and 3C84 as the passband calibrators, and at the end of every cycle, we observed 0108+015 as the gain calibrator.  Calibration, imaging and deconvolution were performed using standard procedures of the MIRIAD software package. The clean $\co$ map has a resolution of $4\farcs50\times4\farcs37$, a $\sigma_\mathrm{rms}$ of 0.423 K, a channel size of 2.5 $\kms$, and a pixel size of 1$\arcsec$. Figure \ref{figure_ngc628N} shows the $\co$ map of the region observed (Section \ref{moments} explains our approach to generate moment maps) overlaid on a 3.6 $\mu$m image from SINGS (\citealt{2003PASP..115..928K}).

In the case of M101, we observed $\co$ in two different areas of the southern region.  The main area was covered with a 53-point mosaic, corresponding to a physical scale of 13.7$\times$5 kpc$^2$ at a distance of 7.4 Mpc (calculated from the luminosity of the Cepheids, \citealt{2000ApJS..128..431F}). Additionally, we have observed a small region in the western region of the disk with a 7-point mosaic, corresponding to a physical scale of 1 kpc$^2$. The correlator setup is the same as for the NGC 628 observations.   We have used 1153+495 as the gain calibrator, and 3C279 and 3C273 as passband calibrators.  Every pointing was observed for 40 seconds, but in this case the mosaic was separated into two cycles of 30 pointings, yielding a 20-minute observation time per cycle. The clean $\co$ map has a resolution of $4\farcs37\times4\farcs16$, a $\sigma_\mathrm{rms}$ of 0.370 K, a channel size of 2.5 $\kms$, and a pixel size of 1$\arcsec$. Figure \ref{figure_m101S} shows the $\co$ map of the regions observed overlaid on a on a 3.6 $\mu$m image from SINGS.  Table \ref{data-prop} summarizes the basic properties of our data cubes.

\subsubsection{1 mm}\label{1mm_data}
\begin{figure*}
\centering
\begin{tabular}{ccc}
\epsfig{file=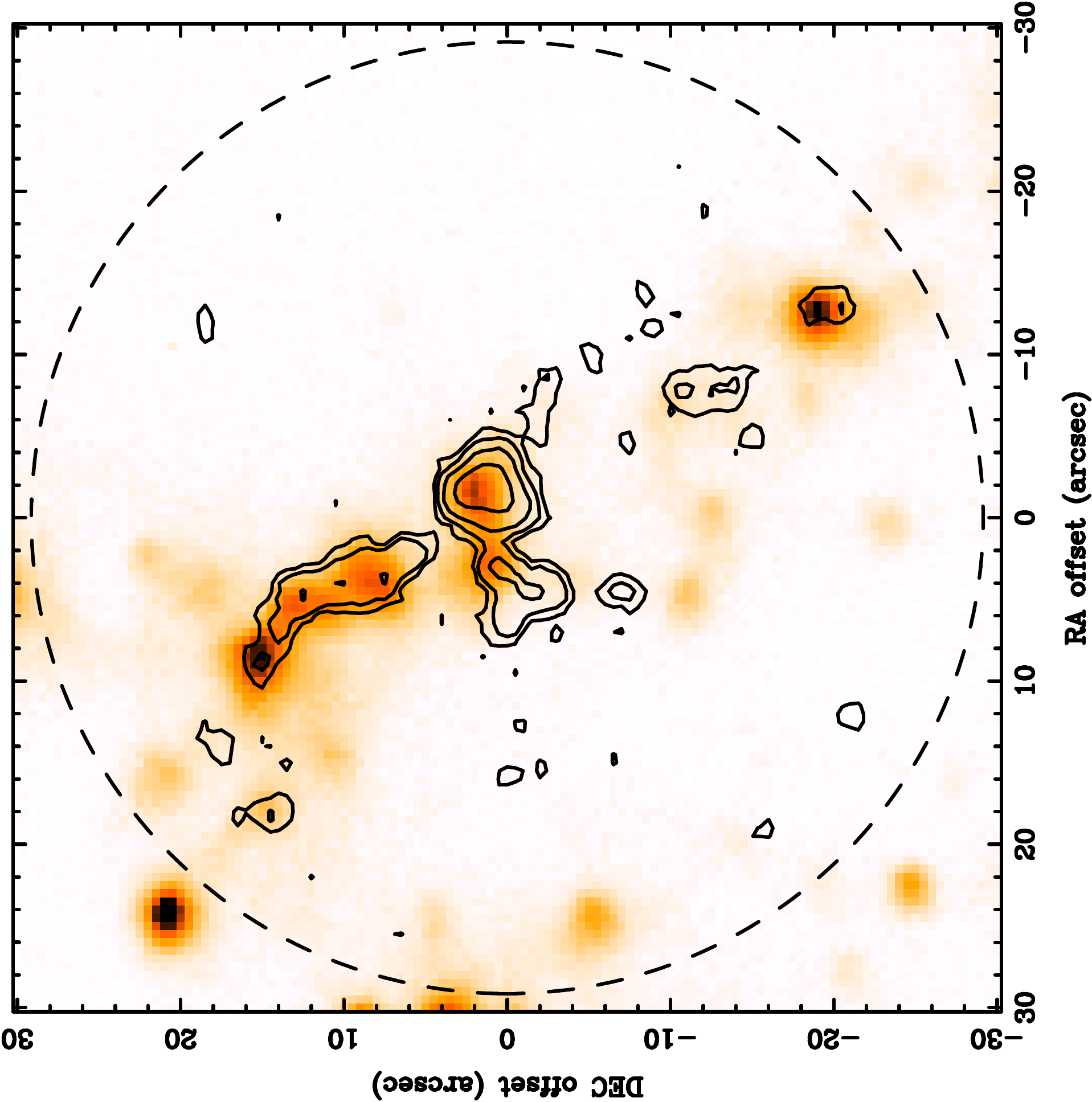,width=0.27\linewidth,angle=-90}
\epsfig{file=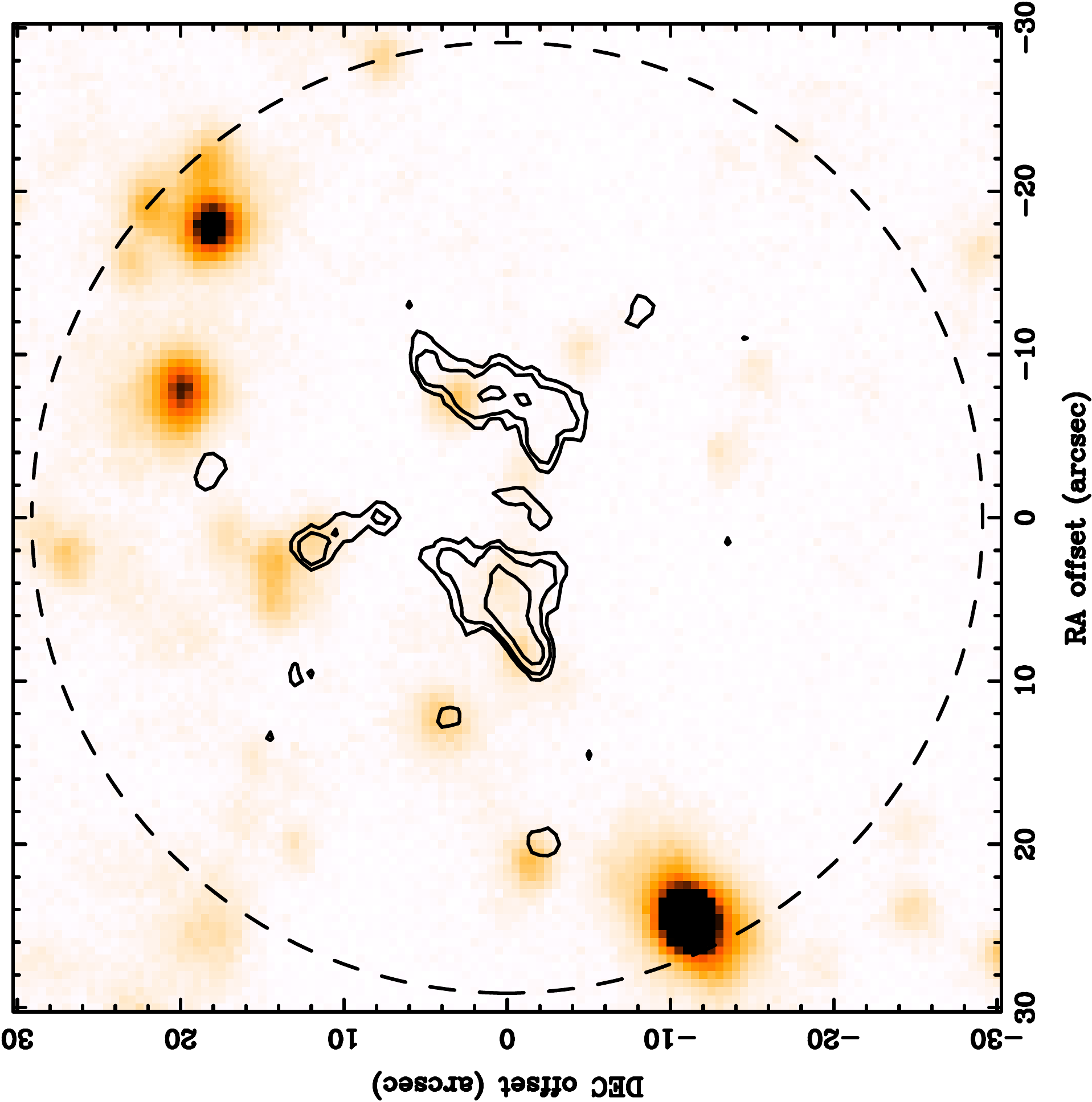,width=0.27\linewidth,angle=-90}
\epsfig{file=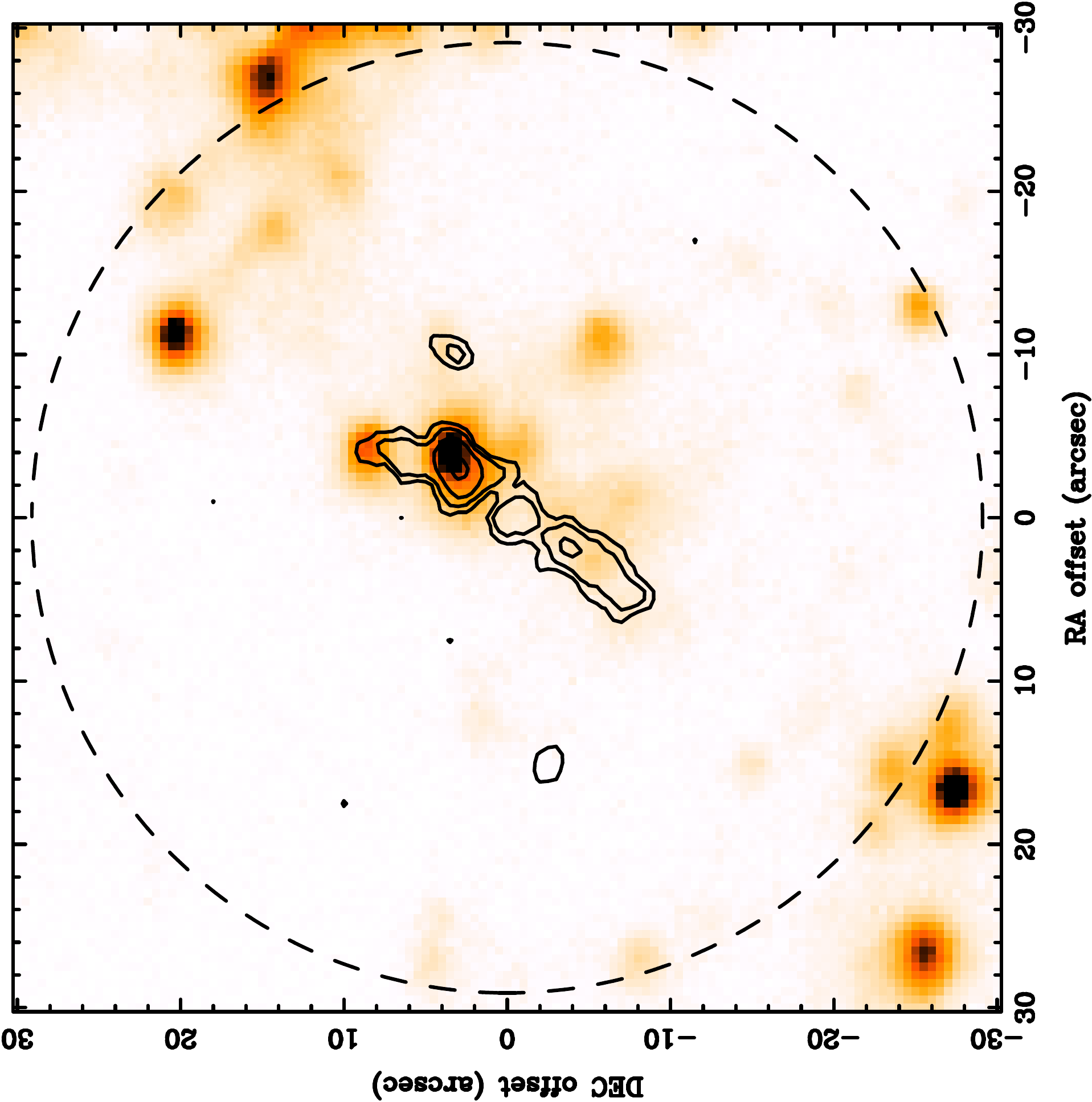,width=0.27\linewidth,angle=-90} \\
\epsfig{file=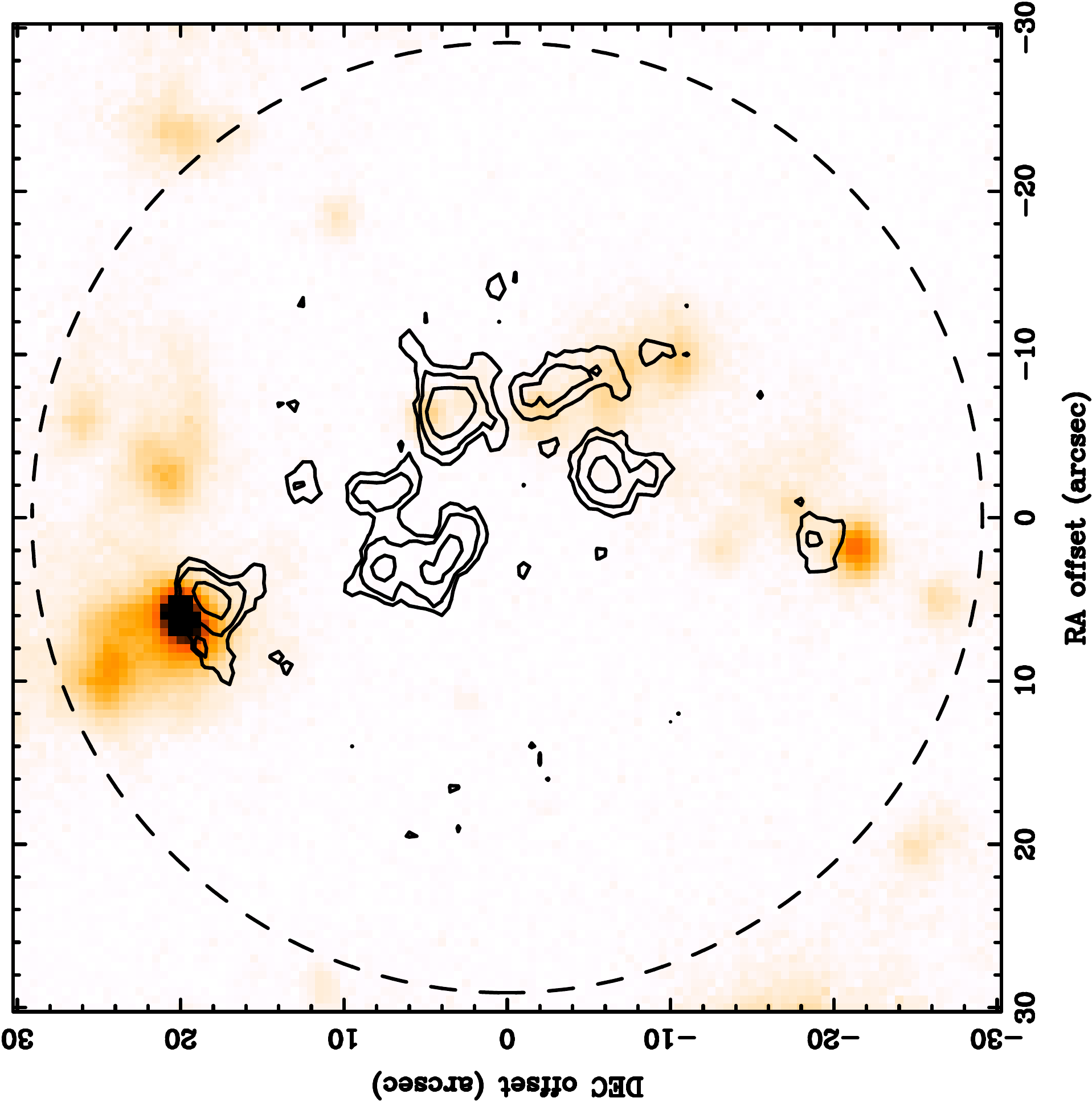,width=0.27\linewidth,angle=-90}
\epsfig{file=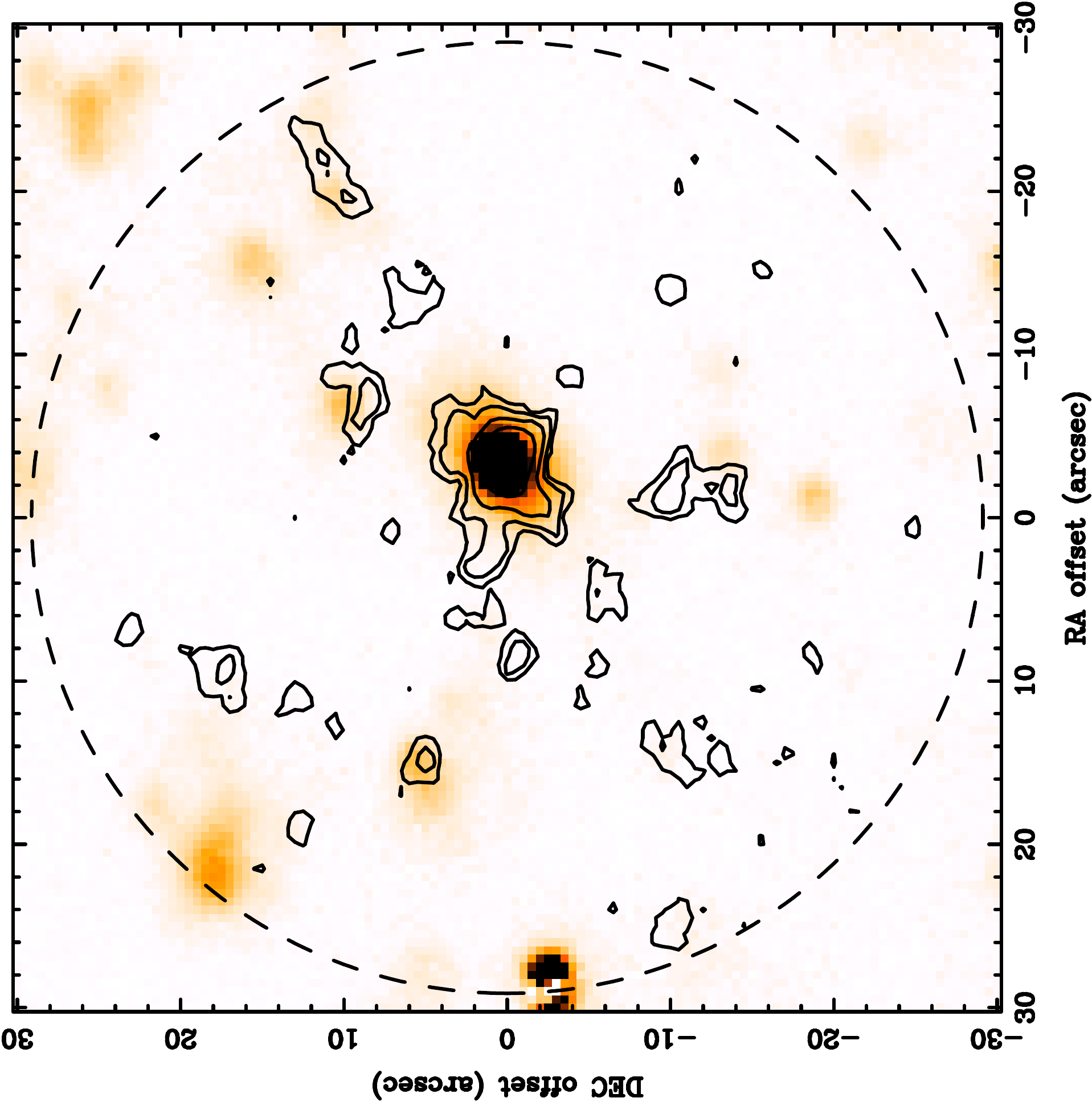,width=0.27\linewidth,angle=-90}
\epsfig{file=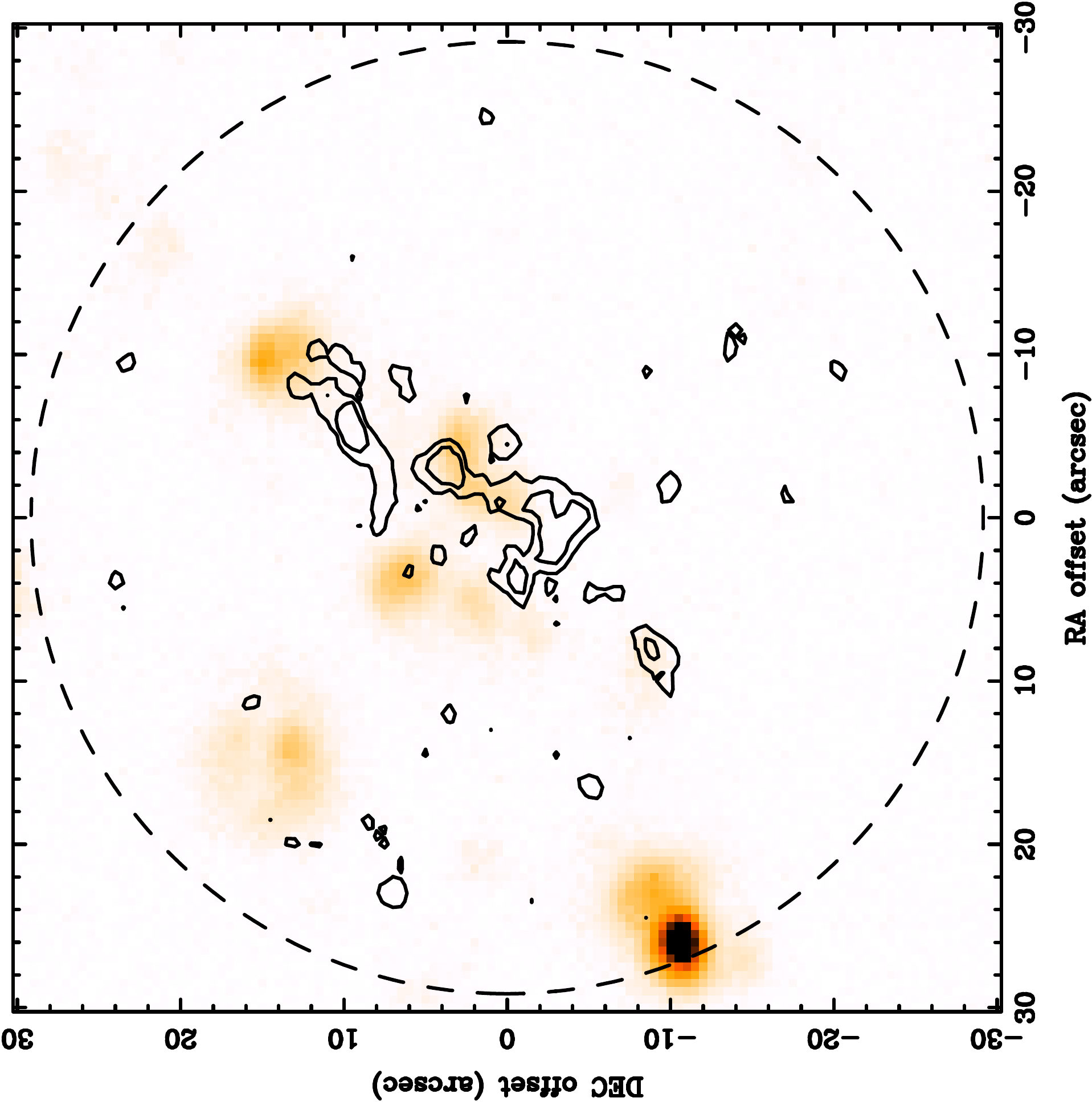,width=0.27\linewidth,angle=-90 } \\
\epsfig{file=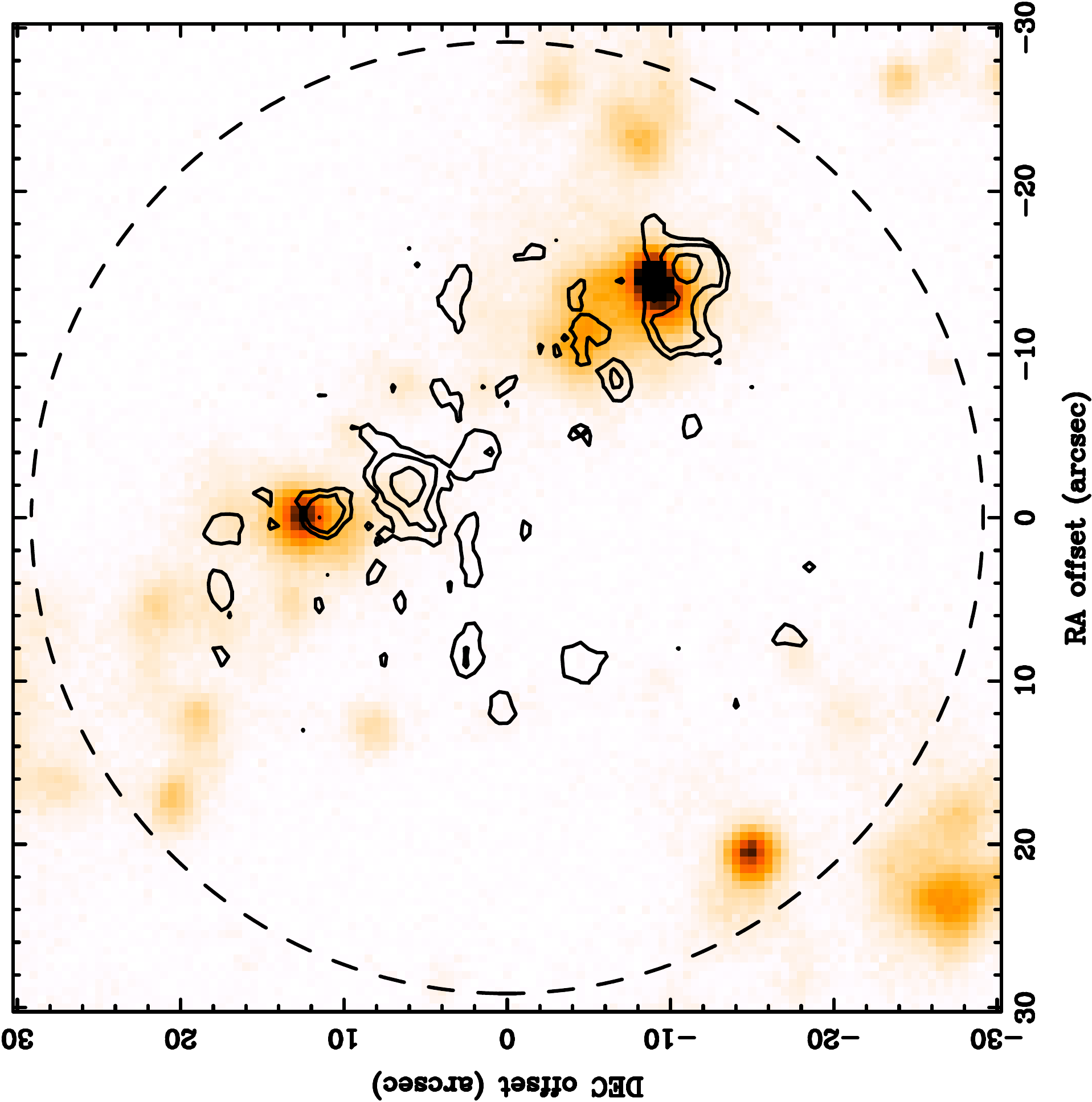,width=0.27\linewidth,angle=-90}
\epsfig{file=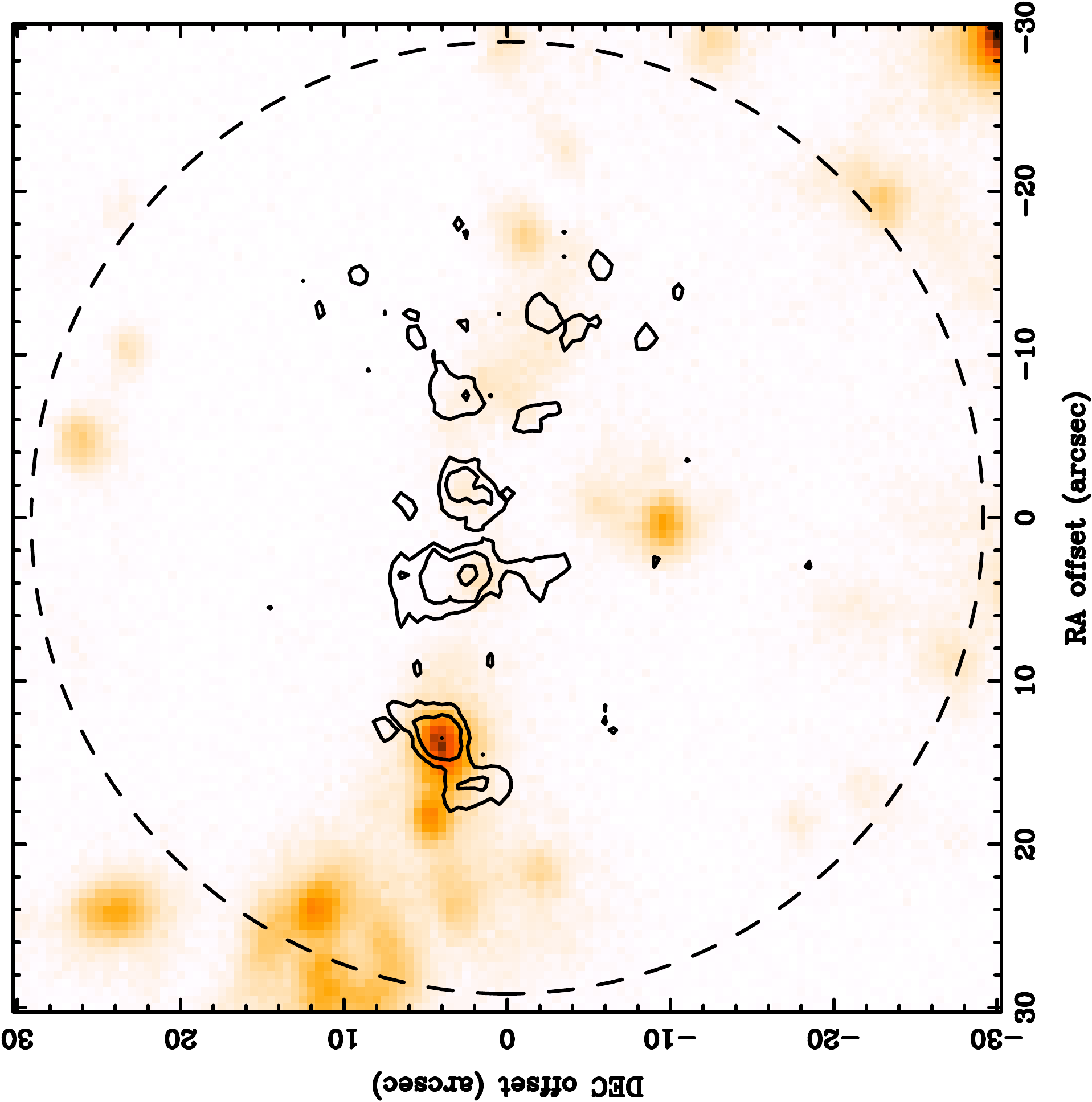,width=0.27\linewidth,angle=-90}
\epsfig{file=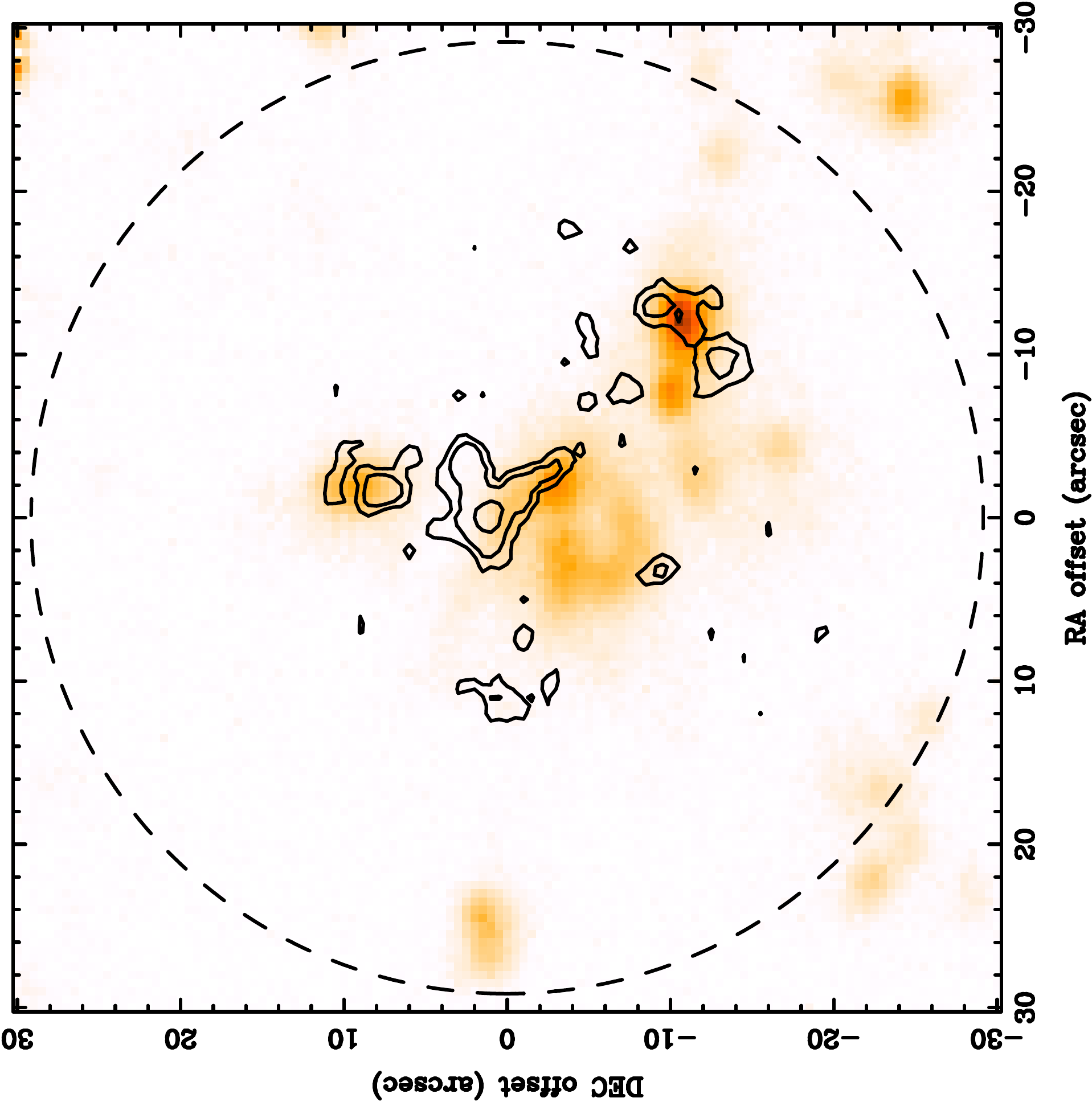,width=0.27\linewidth,angle=-90} \\
\epsfig{file=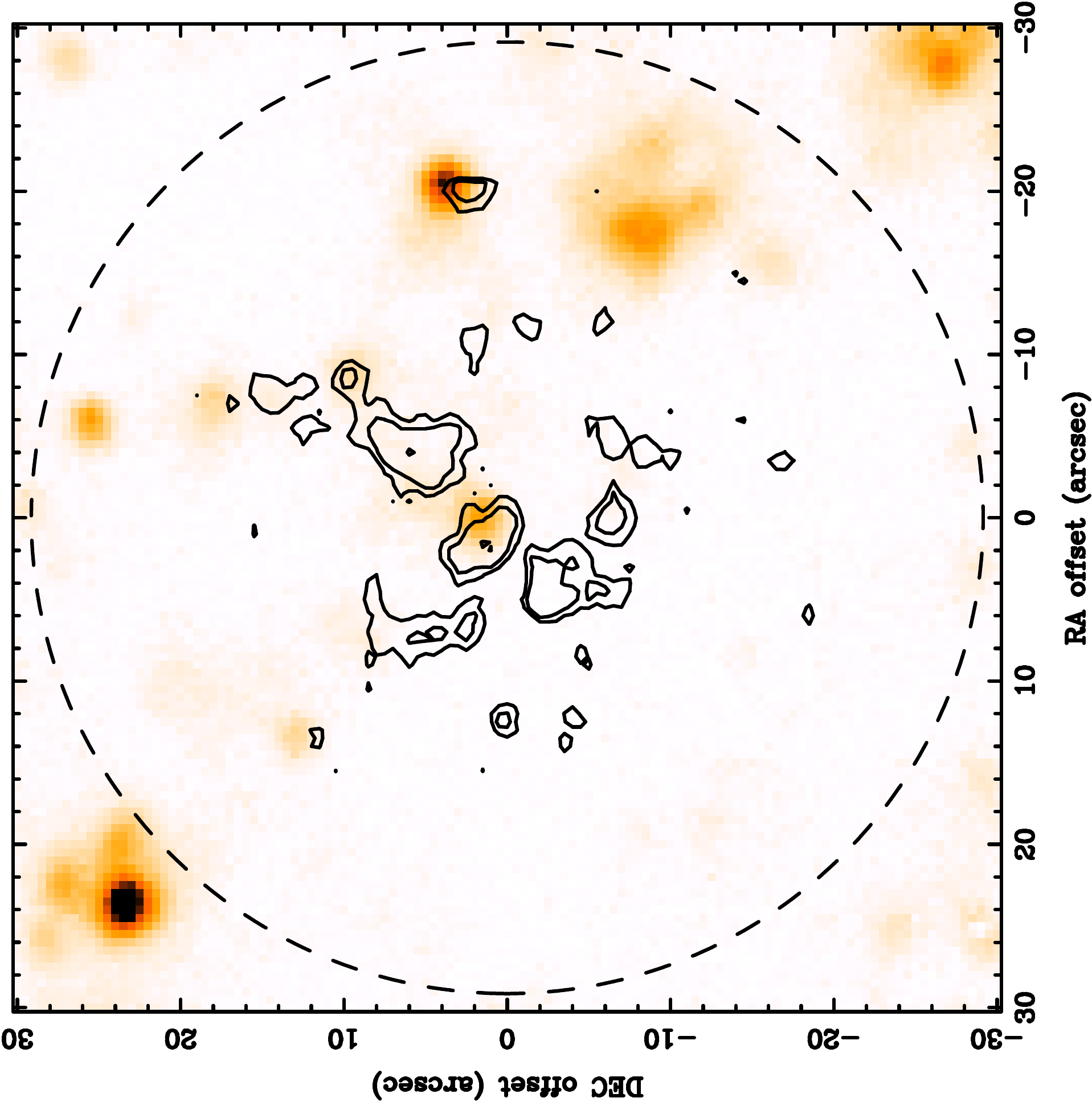,width=0.27\linewidth,angle=-90}
\end{tabular}
\caption{$\cotwo$ integrated intensity contours overlaid on the H$\alpha$ image for all the regions targeted in the northern part of NGC 628.  Contours begin at 2.5$\sigma$ and are spaced by 2$\sigma$.  The noise at the center of the map is $\sim$ 2 K km s$^{-1}$.}
\label{fig_co21maps_628}
\end{figure*}

\begin{figure*}
\centering
\begin{tabular}{ccc}
\epsfig{file=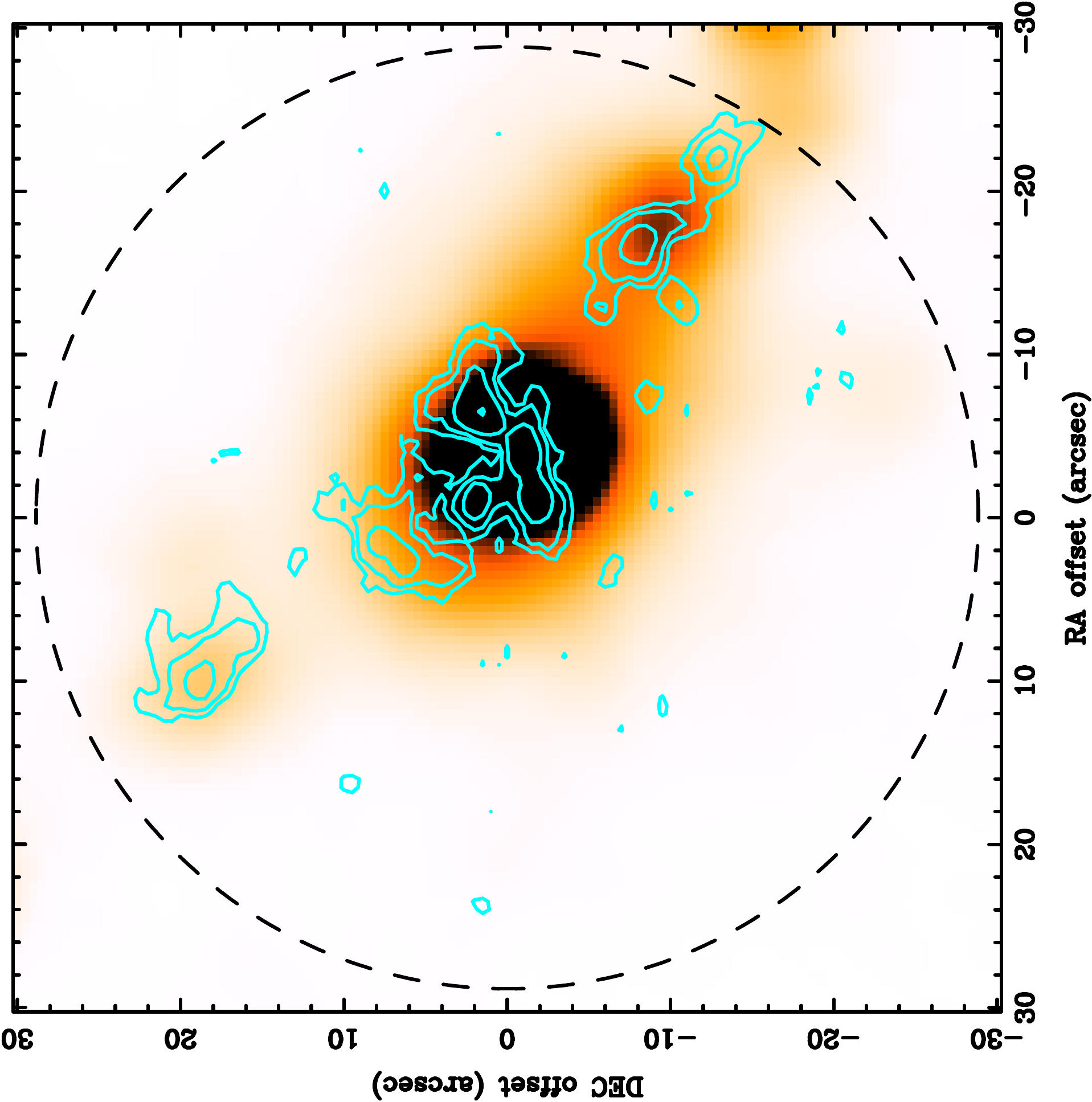,width=0.27\linewidth,angle=-90}
\epsfig{file=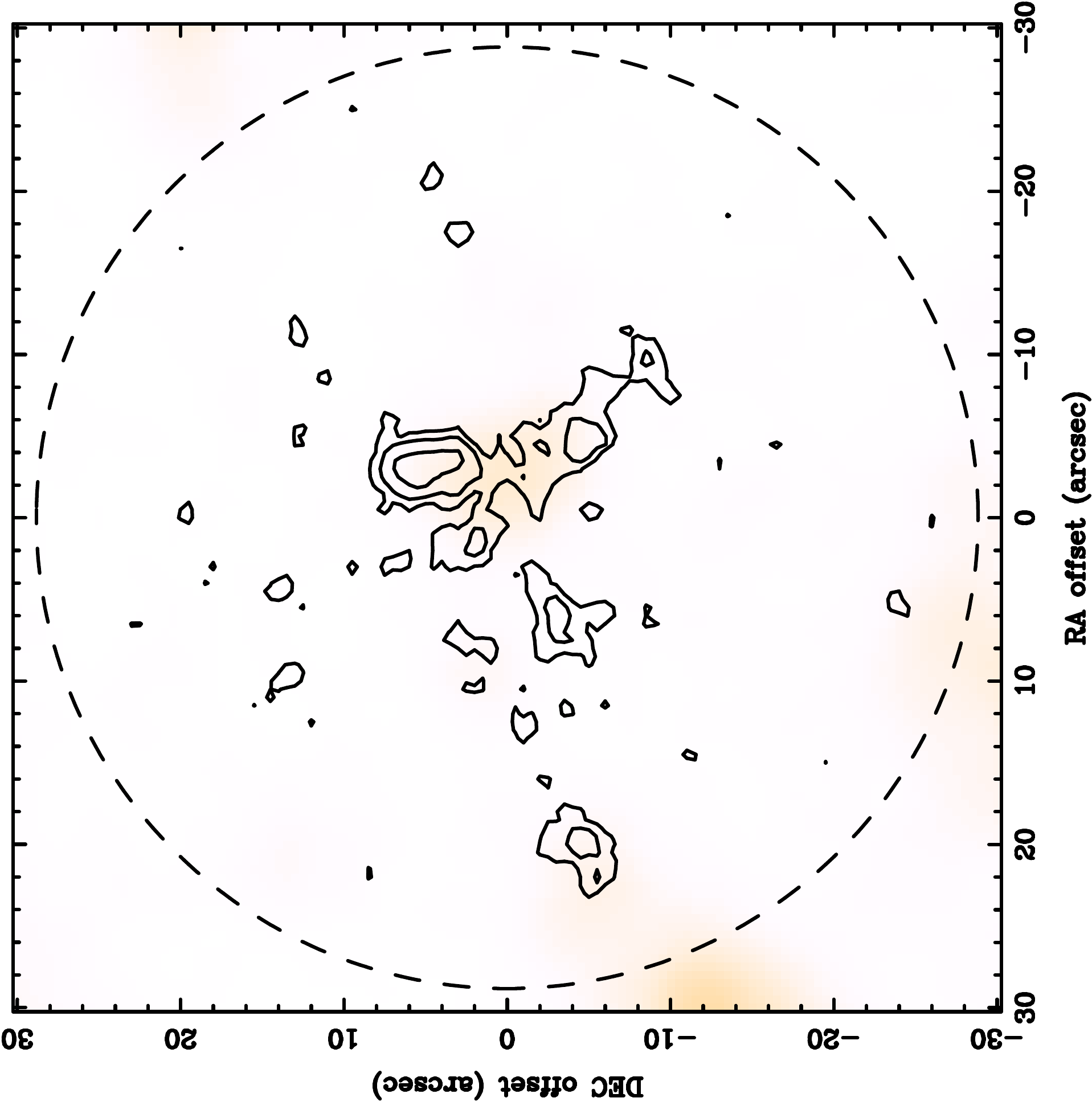,width=0.27\linewidth,angle=-90}
\epsfig{file=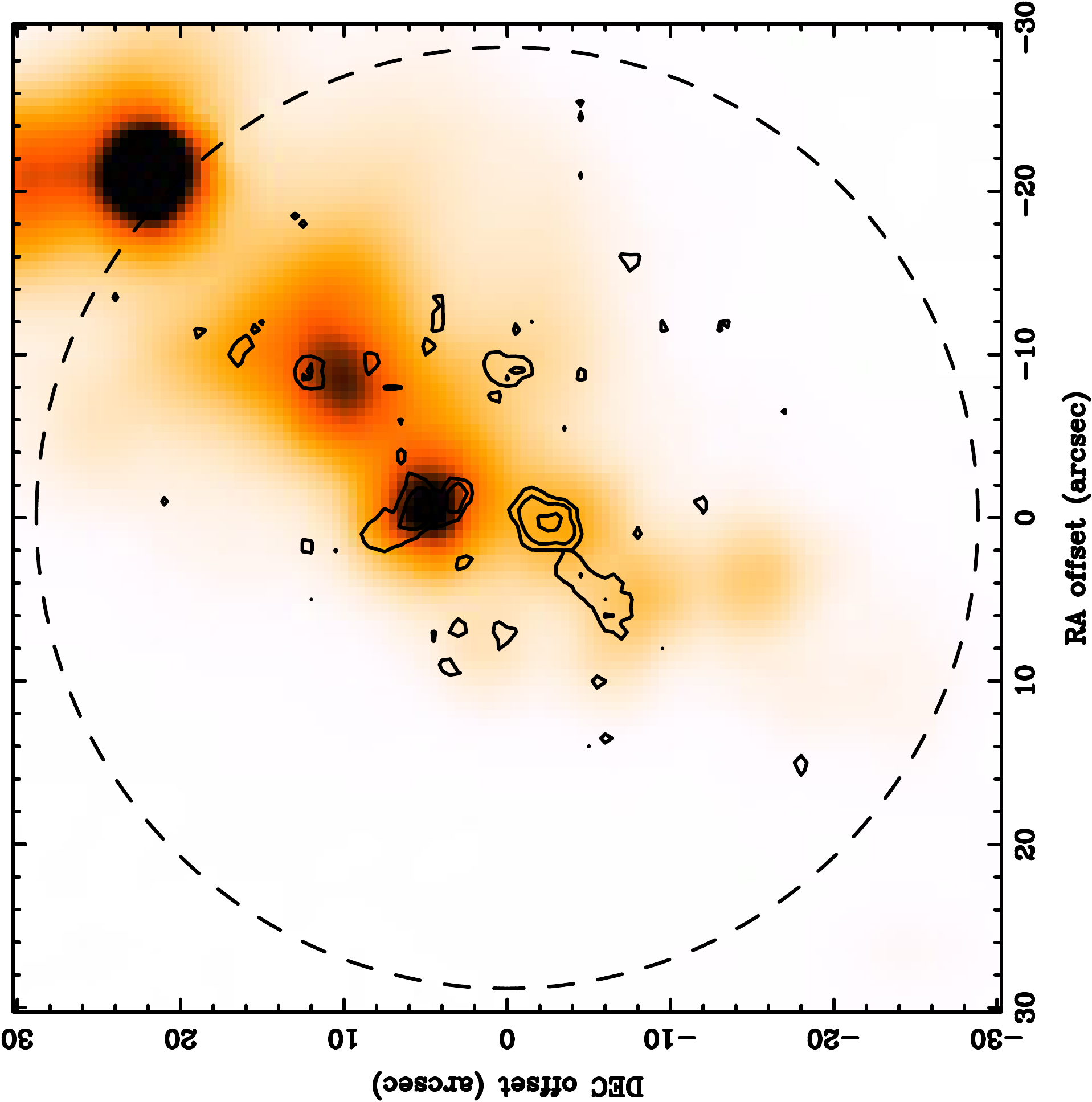,width=0.27\linewidth,angle=-90} \\
\epsfig{file=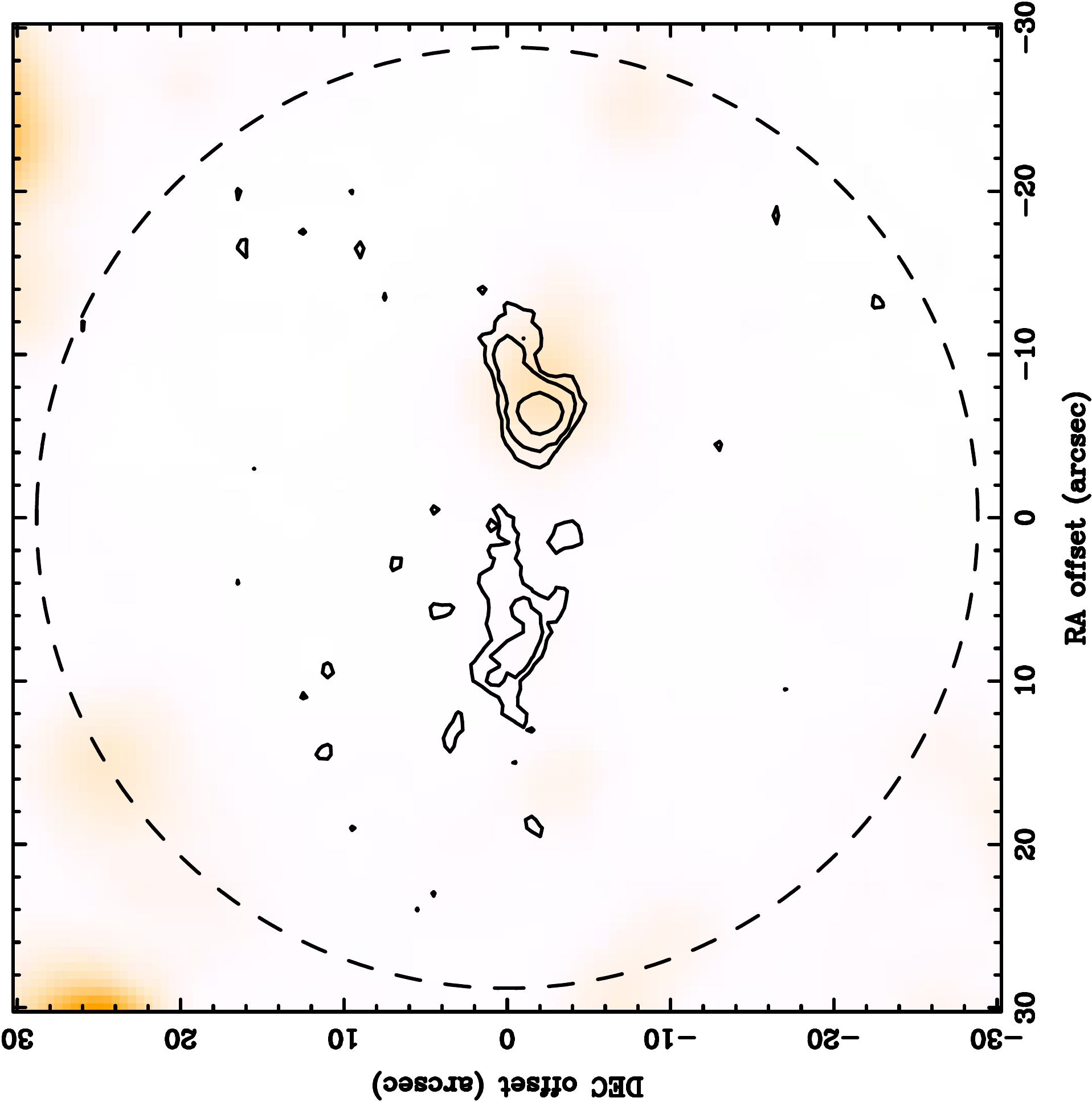,width=0.27\linewidth,angle=-90}
\epsfig{file=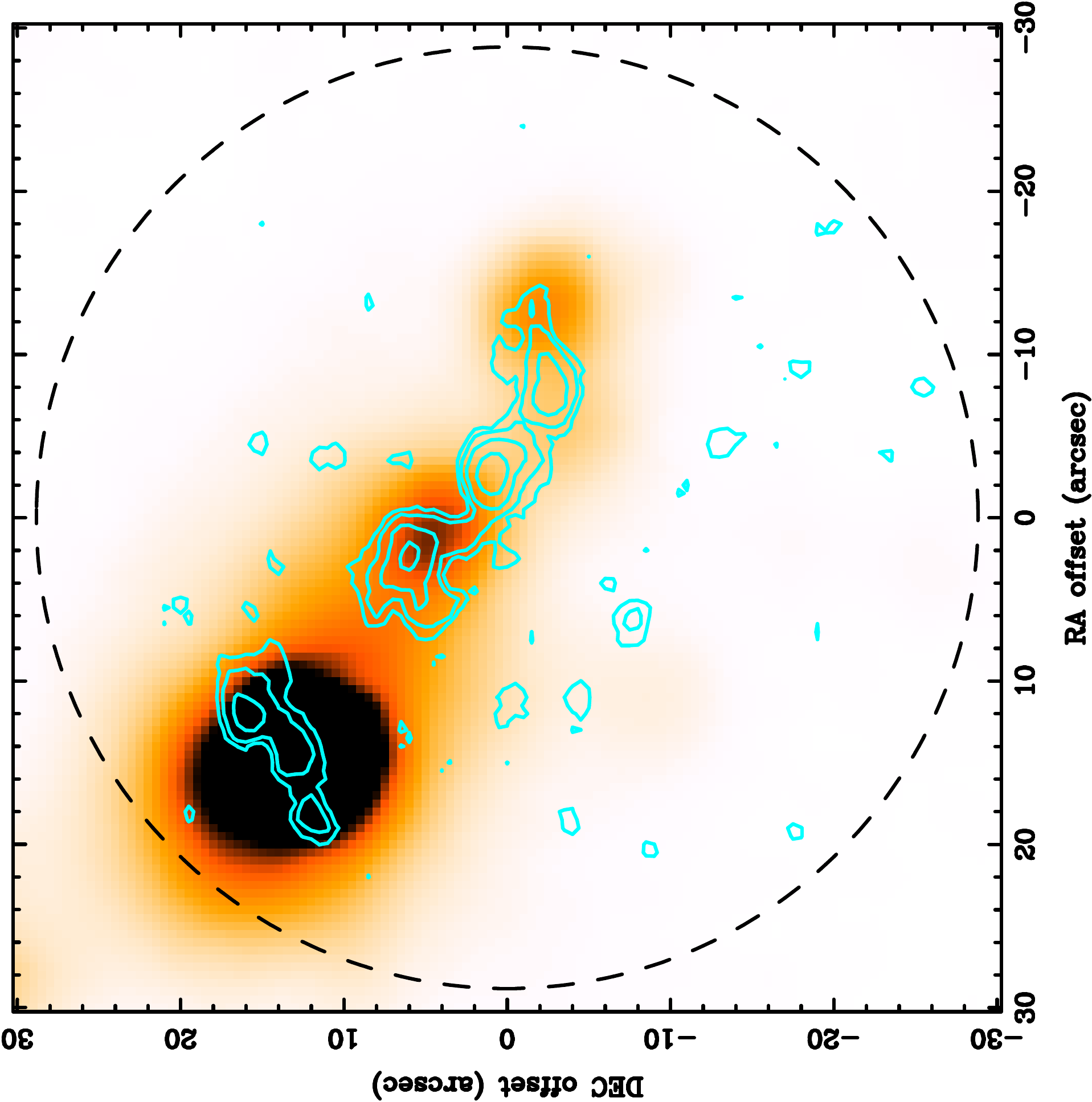,width=0.27\linewidth,angle=-90}
\epsfig{file=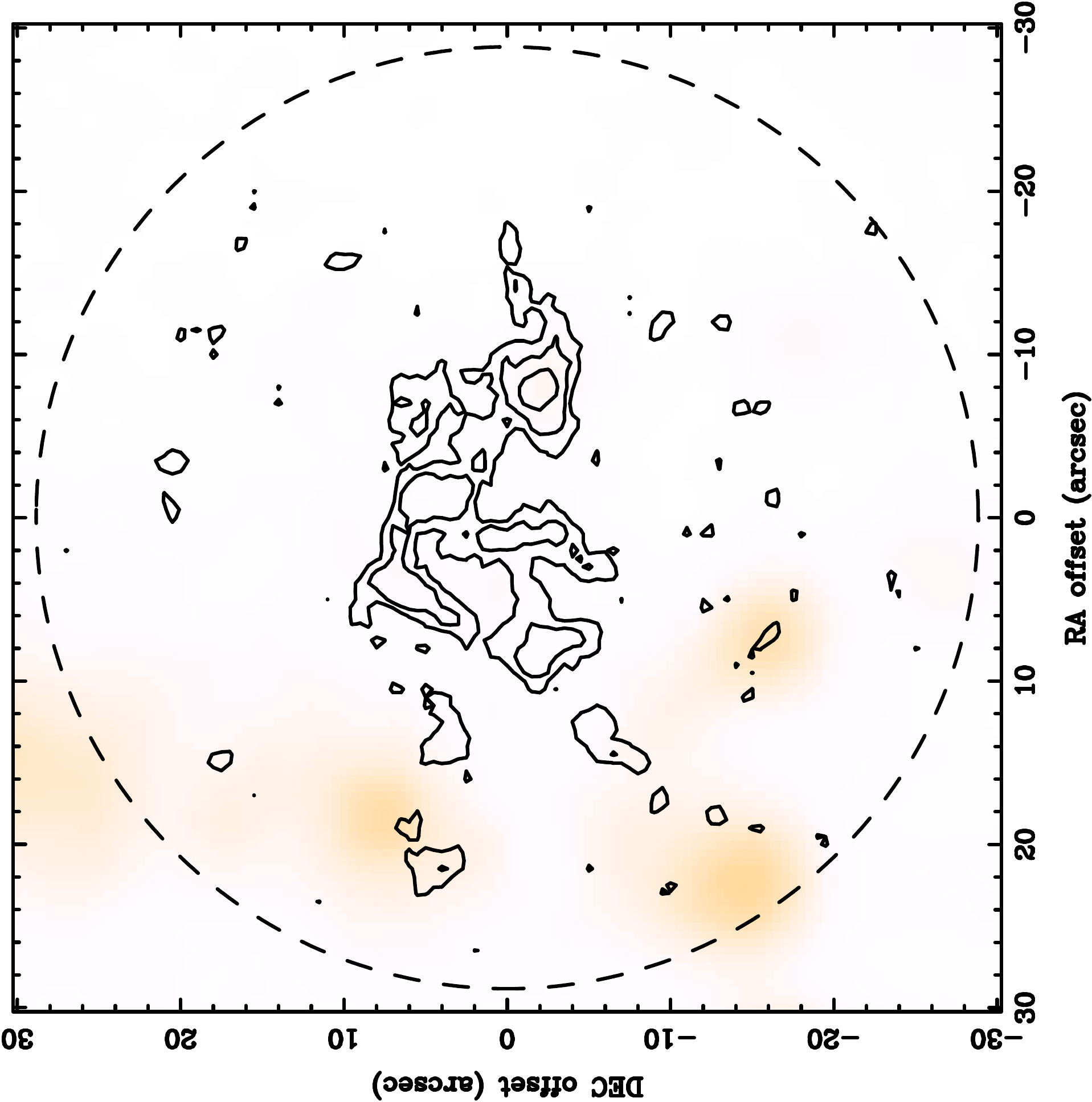,width=0.27\linewidth,angle=-90 } \\
\epsfig{file=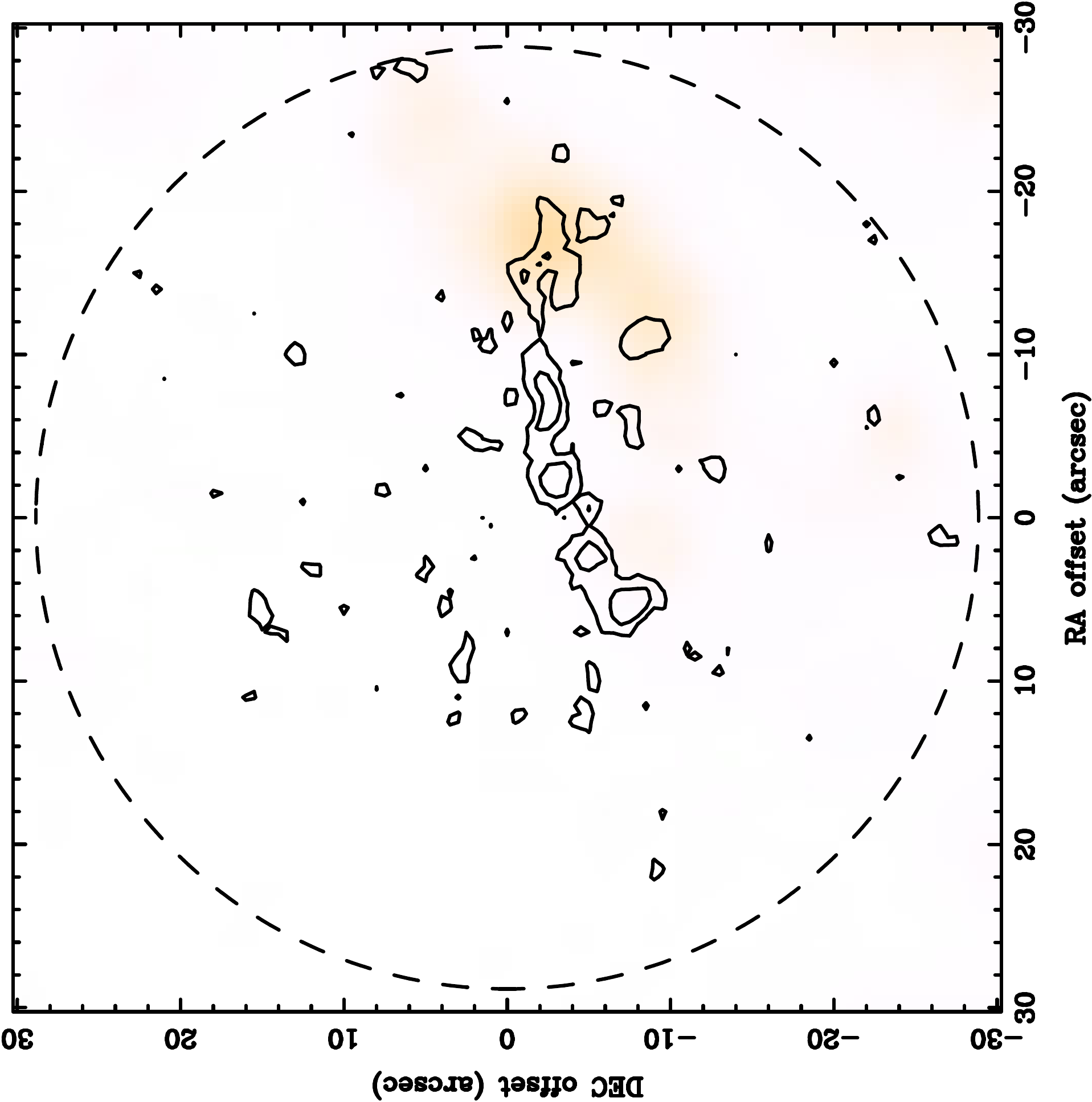,width=0.27\linewidth,angle=-90}
\epsfig{file=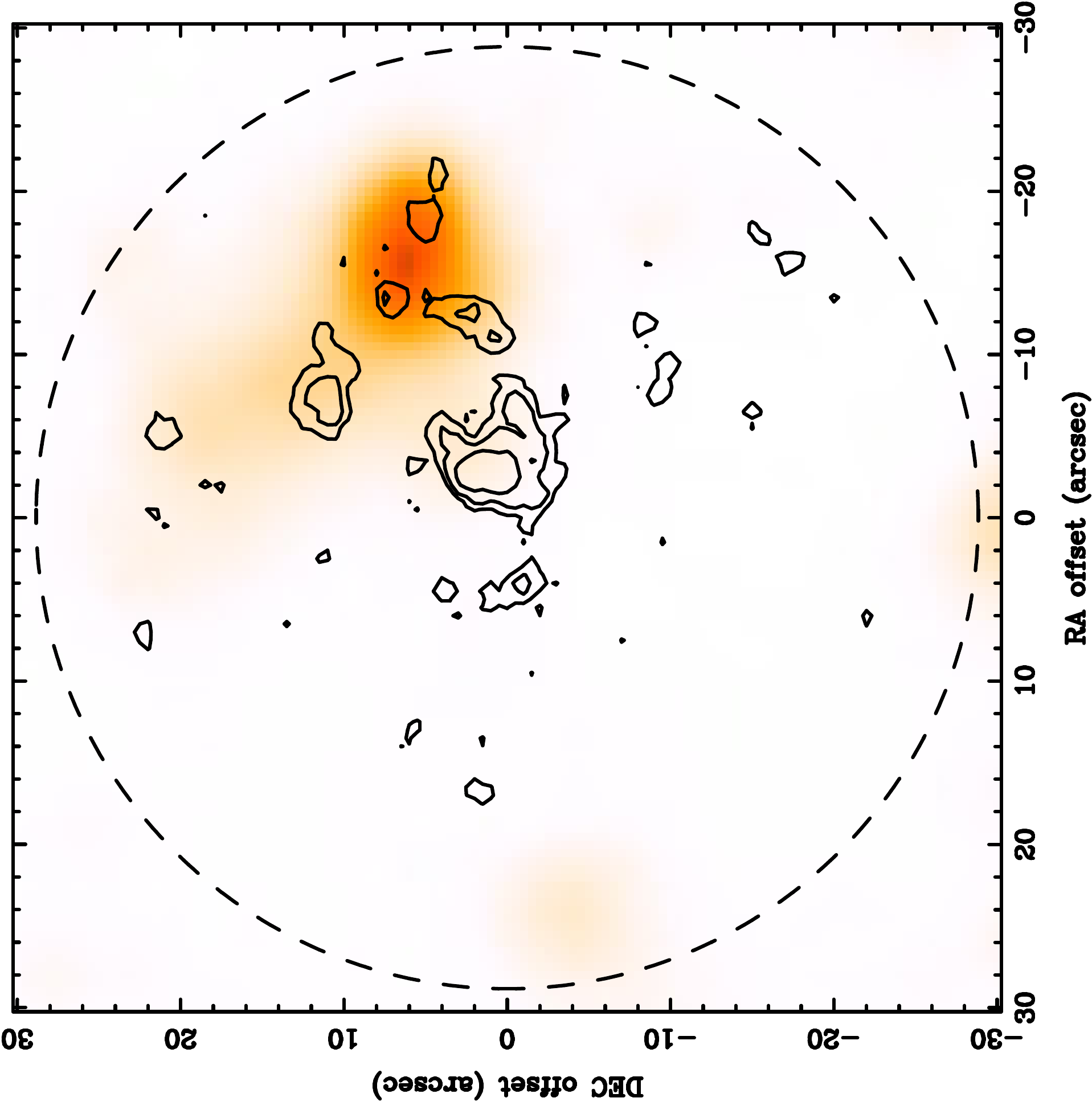,width=0.27\linewidth,angle=-90}
\epsfig{file=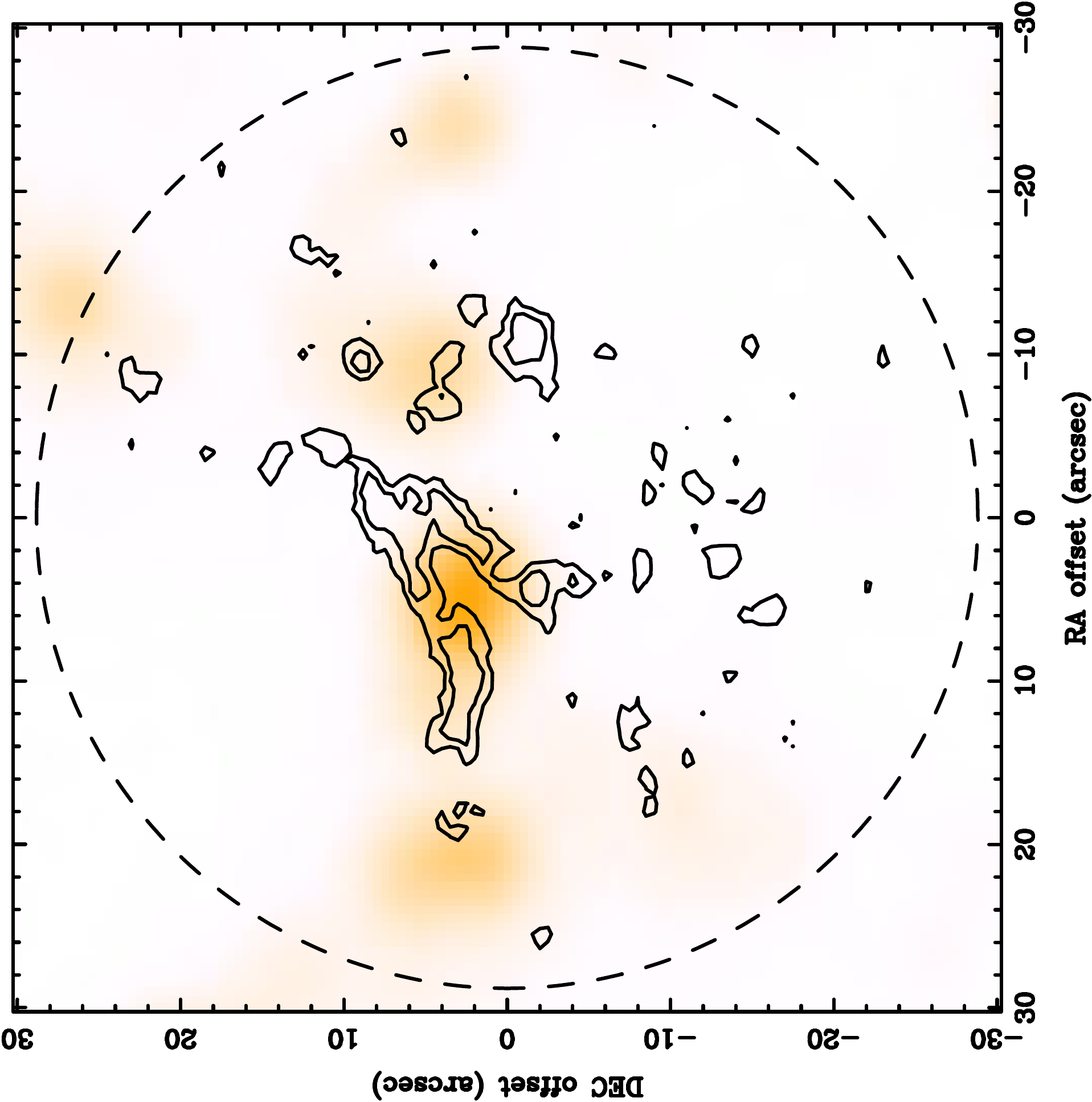,width=0.27\linewidth,angle=-90} \\
\epsfig{file=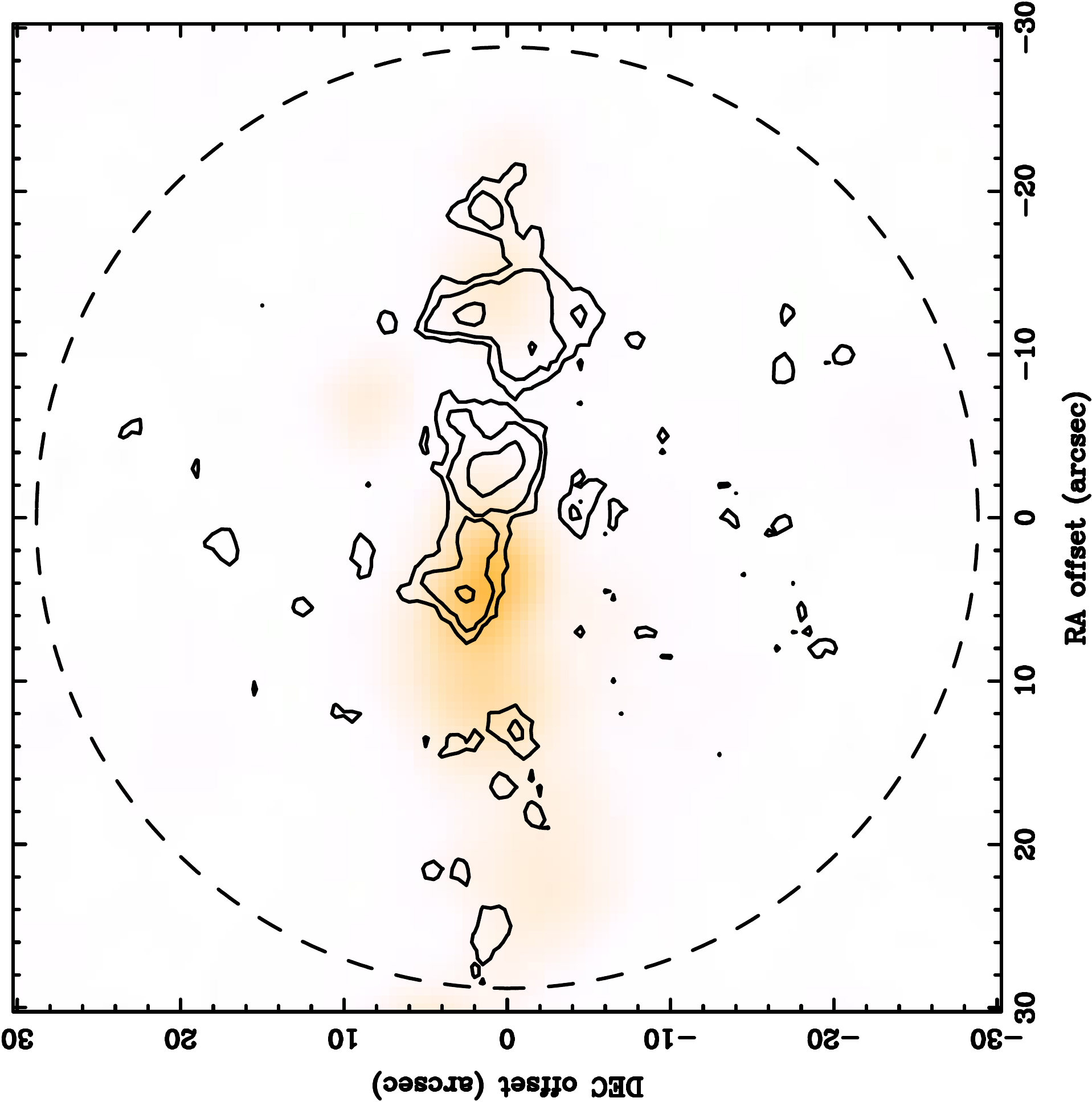,width=0.27\linewidth,angle=-90}
\end{tabular}
\caption{$\cotwo$ integrated intensity contours overlaid on the H$\alpha$ image for all the regions targeted in the southern arm of M101.  Contours begin at 2.5$\sigma$ and are spaced by 2$\sigma$.  The noise at the center of the map is $\sim$ 2 K km s$^{-1}$.}
\label{fig_co21maps_m101}
\end{figure*}

In order to resolve the largest GMCs in the galaxies, we performed high resolution observations in D array towards some CO complexes using the $\cotwo$ line. We selected 10 regions in each galaxy based on the highest $\co$ integrated intensity in the area covered by 3 mm observations.  Figures \ref{figure_ngc628} and \ref{figure_m101} illustrate the regions targeted with $\cotwo$ observations in NGC 628 and M101 respectively. The observations were taken between October and November 2011.  For both galaxies, we set the correlator to have a 125 MHz band centered on the rest frequency of the $\cotwo$ line in the LSB, and seven 500 MHz wide bands to observe the continuum at 1 mm.  This correlator configuration yields a velocity coverage of 160 $\kms$ and a channel width of 0.5 $\kms$.  Every pointing was observed for 90 seconds, which yields a total cycle time of 15 minutes.  For NGC 628 we observed the gain calibrator 0108+015 after every cycle, and 3C84 was used as passband calibrator, while for M101 we observed 1642+689 and 3C279 as gain and passband calibrators respectively.  The clean $\cotwo$ maps for NGC 628 have a 2.5 $\kms$ velocity resolution, and an angular resolution of 2\farcs03$\times$1\farcs80 corresponding to $\sim$ 70 pc.  The rms noise of the clean maps is 0.312 K.  For M101, the clean maps have a 2.5 $\kms$ velocity resolution, and a resolution of 2\farcs01$\times$1\farcs87 corresponding to $\sim$ 72 pc.  The rms noise of the clean maps is 0.287 K.  The pixel size of the maps is 0.5$\arcsec$.  The $\cotwo$ maps are illustrated in Figures \ref{fig_co21maps_628} and \ref{fig_co21maps_m101} for NGC 628 and M101 respectively.

Through this paper, we convert the $\cotwo$ maps to $\co$ assuming a fixed $I_\mathrm{CO}(2 \rightarrow 1)$ to $I_\mathrm{CO}(1 \rightarrow 0)$ ratio of 1 across the regions observed, in order to be consistent with the work presented in Paper I.  Although previous studies have used fixed values of this quantity (e.\ g., 0.8, \citealt{2008AJ....136.2782L}, henceforth L08), this ratio has been observed to vary from 0.6 to 1.0 in different parts of the disk of nearby spiral galaxies (e.\ g., HERACLES, \citealt{2009AJ....137.4670L}; M 51, \citealt{2012ApJ...761...41K}).  

\subsection{Single dish maps}\label{single-dish}

\subsubsection{NRO 45m $\co$ map}\label{nro45}
Extended flux is recovered by merging our $\co$ CARMA data with single dish maps from the Nobeyama 45-meter single dish telescope (NRO 45m).  The galaxy NGC 628 is part of the CARMA and NObeyama Nearby-galaxies (CANON) CO(1-0) Survey project, which combines observations from the CARMA and NRO 45m telescopes to resolve GMCs in disks of nearby galaxies (Koda et al., in preparation).  The $\co$ map of galaxy NGC 628 was obtained using the Beam Array Receiver System (BEARS) instrument.  The FWHM of the 45 meter dish is 15$\arcsec$ at the $\co$ rest frequency which is degraded to 19$\farcs$7 after regridding.  The rms noise of the $\co$ single dish map is 0.06 K (0.3 Jy beam$^{-1}$) in a channel width of 2.54 $\kms$, which corresponds to $\sigma(\Sigma_\mathrm{H_2}) \sim 7\ \Msun\ \mathrm{pc}^{-2}$ assuming a line width of 20 $\kms$ and $X_\mathrm{CO}=2\times10^{20}\ \mathrm{cm}^{-2}(\mathrm{K}\ \kms)^{-1}$.

For the main mosaic observed in M101, we use NRO 45m observations from \citet{2007PASJ...59..117K} to recover the extended flux, and we refer the reader to that paper for the observation details. The rms noise in the maps is 0.07 K, and the FWHM of the 45 meter dish is 15$\arcsec$ before regridding at the $\co$ rest frequency. The velocity resolution of the map is 5 $\kms$.  On the other hand, since the area covered by the \citet{2007PASJ...59..117K} observations does not overlap with the small region located in the western arm of M101 covered with a 7-pointing mosaic, we have used a $\cotwo$ single dish map from the HERACLES project (see \ref{heracles}) to combine with our CARMA map instead, assuming a line ratio of 1 as noted in Section \ref{1mm_data}.

\subsubsection{Heracles $\cotwo$ map}\label{heracles}
Following the same approach we adopted in Paper I, we have combined our $\cotwo$ CARMA data with single dish maps from the HERACLES project (\citealt{2009AJ....137.4670L}).  HERACLES observed the $\cotwo$ line toward the full optical disk of 18 nearby galaxies with the HERA receiver at the IRAM 30m telescope, producing maps with $13\arcsec$ and 2.6 $\kms$ of angular and velocity resolution respectively.  The $\sigma_\mathrm{rms}$ in the data cube is $\sim$ 20 mK, which yields an integrated intensity map with a noise level of $\sigma(\Sigma_\mathrm{H_2}) \sim 2\ \Msun\ \mathrm{pc}^{-2}$ assuming a velocity width of 20 $\kms$ and  $X_\mathrm{CO}=2\times10^{20}\ \mathrm{cm}^{-2}(\mathrm{K}\ \kms)^{-1}$.

\subsection{CARMA-Single dish merging procedure}\label{merging}

\subsubsection{Combination of $\co$ data}
We have used the approach introduced by \citet{2011ApJS..193...19K} to combine our $\co$ maps from CARMA with the single dish maps from NRO 45m single dish telescope.  A detailed description of the procedure can be found in \citet{2011ApJS..193...19K}, but here we summarize the basic concepts.  The method consists in converting the NRO 45m map from image space to the $uv$ plane, i. e., it generates the visibilities of the single dish map.  Then, the NRO 45m visibilities are combined with the CARMA visibilities, and the merged $uv$ samples are imaged together.  Following the suggestion from \citet{2011ApJS..193...19K}, we have excluded single-dish visibilities beyond a given threshold in the baseline length.  This threshold in chosen based on the the distribution of the sensitivity of the NRO 45m along the $uv$ plane, and the flux recovery from the combined NRO 45m + CARMA map relative to the single dish flux calculated over regions of significant emission.  For NGC 6946 we have cut at 7 $k \lambda$, for NGC 628 we cut at 10 $k \lambda$, and for M101 we cut at 6 $k \lambda$.

The NRO 45m $\co$ map of M101 does not cover the small region located in the western region of the outer disk.  In this case, we have used the IRAM 30m $\cotwo$ map to recover the extended flux using the MIRIAD task {\it immerge} (see Section \ref{immerge} for details).

\subsubsection{Combination of $\cotwo$ data}\label{immerge}
In the case of  $\cotwo$ maps, we have used the MIRIAD task {\it immerge} to combine our CARMA maps with the single dish maps from the IRAM 30m.  {\it Immerge} linearly merges a low resolution image with a high resolution image, which is equivalent to include short and large spacings in the $uv$ plane into the imaging.  In order to prevent this procedure to be affected by regions of low signal to noise ratio, we have excluded edge velocity channels and we have masked the region where the sensitivity of the CARMA map falls below half of the maximum.  As we did in Paper I, we have allowed {\it immerge} to solve for the factor that put the low resolution image on the same flux scale as the high resolution image.  In our combination procedure, we obtained values close to 1.05, expected for data sets that have correct calibrated flux scales.  Using regions of significant emission identified by CPROPS (see Section \ref{cprops} for details) for both the CARMA observations, and the CARMA + IRAM 30m combined map, we found that CARMA+IRAM 30m images have typically 20\% more flux inside the region of significant $\cotwo$ emission. 

\subsubsection{Generation of integrated intensity maps}\label{moments}
We use CO spectral cubes and corresponding signal masks to produce integrated intensity maps.  To generate a signal mask, we first degraded the data cube in angular and velocity resolutions by a factor of 3 and 4, respectively.  Then we identify continuous regions above 3$\times \sigma_{\rm smo}$ having at least two pixels, where $\sigma_{\rm smo}$ is the rms intensity of the smoothed cube.  We further expanded each region to include any adjacent pixels above 2$\times \sigma_{\rm smo}$ and used the resulting regions to construct the signal mask.  This signal mask was later used to exclude noise pixels in the cube, and thus generate the integrated intensity images.  The associated uncertainty map for each image was derived from the error propagation through pixels in the signal mask.  Smoothing can enhance the signal-to-noise ratio for extended structures in the position-position-velocity cube, and the dilated mask approach, similar to the one used in CPROPS (see Section \ref{ident} for more details), was designed to help identify extended low brightness emission contiguous with significant emission.

\subsection{GALEX FUV data}
FUV data for NGC 628 and M101 are available from the {\it GALEX} Nearby Galaxies Survey (NGS; \citealt{2007ApJS..173..185G}).  The FUV band covers the wavelength range 1350-1750 \AA, and the maps have an angular resolution of 5\farcs6.  We applied the same procedure followed in Paper I to generate a sky background subtracted FUV image.  We used the extinction map of \citet{1998ApJ...500..525S} to correct the FUV maps for Galactic extinction.  The FUV extinction was estimated by using $A_\mathrm{FUV}=8.24\times\  E(B-V)$ from \citet{2007ApJS..173..293W}.

\subsection{SINGS data}

\subsubsection{24 $\mu$m map}
In order to estimate the amount of star formation obscured by dust, we use the 24 $\mu$m data obtained with the {\it Spitzer} MIPS instrument provided by SINGS (\citealt{2003PASP..115..928K}) for NGC 628, and the {\it Spitzer} Local Volume Legacy (LVL, \citealt{2009ApJ...703..517D}) for M101.  The angular resolution of MIPS is $5\farcs7$ (FWHM).  We used the same approach we did in Paper I to estimate the background emission.  We perform a median spatial filtering to remove bright spots and other features over a region away from the galaxy, and then measure the mean flux over the filtered region.

\subsubsection{H$\alpha$ map}\label{halpha-sfr}
Given the limited resolution offered by 24 $\mu$m ($\sim 6\arcsec$), we need an alternative star formation tracer with a spatial resolution similar to that provided by our $\cotwo$ maps ($\sim 2\arcsec$).  In the case of NGC 628, an H$\alpha$ image is available in the SINGS fourth data release as part of the SINGS ancillary data program.  The angular resolution of this H$\alpha$ map is $\sim 2 \arcsec$, a good match to the resolution of our $\cotwo$ maps.  We used the stellar continuum-subtracted H$\alpha$ image, and we convert to flux units using the SINGS data release documentation.  In order to remove the [N II] contribution, we have assumed ratios of [N II]$\lambda$6584/H$\alpha$=0.5 and [N II]$\lambda$6548/[N II]$\lambda$6584=0.335 (\citealt{2005ApJ...633..871C}).  In the case of M101, we have used a H$\alpha$ map observed with the Burrell-Schmidt telescope at Kitt Peak National Observatory (KPNO) (\citealt{2001ApJ...559..878H}).  The map was calibrated by comparing the luminosities of specific \hii\ regions to published values (\citealt{1998PhDT........16G}).

\section{Cloud Properties}\label{cprops}

\subsection{Identification}\label{ident}
We follow the same procedure used in Paper I to identify cloud structures, and we refer the reader to that paper for details. The methodology is based on the cloud properties algorithm (CPROPS) described in \citet{2006PASP..118..590R}.  CPROPS identifies clouds by isolating regions of significant emission that are both spatially and kinematically connected.  More specifically, regions of significant emission are defined by the pixels with values greater than a threshold of $n_\mathrm{th}\times \sigma_\mathrm{rms}$ across two consecutive velocity channels.  These regions of significant emission are further extended to adjacent regions with emission greater than $n_\mathrm{edge} \times \sigma_\mathrm{rms}$ in two adjacent velocity channels.  In this study we have used $n_\mathrm{th}=4$ and $n_\mathrm{edge}=2$ for both $\co$ and $\cotwo$ maps.  We keep the same nomenclature we used in Paper I for structures identified in our maps, referring to structures identified in $\co$ maps as ``molecular complexes'', while the structures identified in $\cotwo$ will be called ``clouds''. 

The rms size $\sigma_{r}$ of each identified structure is determined by the geometric mean of the second moments of emission along the major and minor axes, and the velocity dispersion $\sigma_{v}$ is provided by the second moment of the emission along the velocity axis.  The total flux of the structure is calculated by adding the flux in each pixel associated to the structure.  A pixel with coordinates {\it $x_i, y_i$} , velocity {\it $v_i$} and temperature {\it $T_i$} is associated to a specific structure if its position is bordered by the temperature isosurface defined by $T_\mathrm{edge}$, where $T_\mathrm{edge}$ is the lowest brightness temperature uniquely associated to the structure.  We reduce the sensitivity bias by measuring these properties at many contour levels inside the boundary of the identified region and linearly extrapolating the size and the velocity width to the value where the brightness temperature of the cloud is equal to 0 K.  On the other hand, the extrapolation applied to the flux is quadratic.  The effect of finite spatial and spectral resolution is corrected by subtracting in quadrature the beam size and the spectral channel profile, approximated by a gaussian, from the extrapolated measurements of size and velocity width respectively (see Paper I for details).  The structures identified by CPROPS with properties corrected by sensitivity and resolution bias are referred as {\it resolved} structures henceforth in this paper.

Following the definition of \citet{1987ApJ...319..730S} (henceforth S87), the cloud size is defined as $R=1.91\sigma_r$.  The CO luminosity $L_\mathrm{CO}$ is given by

\begin{equation}\label{Lco-equ}
\frac{L_\mathrm{CO}}{\mathrm{K\ \kms\ pc^{2}}}=\frac{F_\mathrm{CO}}{\mathrm{K\ \kms\ arcsec^2}}\left(\frac{D}{\mathrm{pc}}\right)^{2}\left(\frac{\pi}{180 \times 3600}\right)^{2}
\end{equation}

\noindent where $D$ is the distance to the galaxy in parsecs.  The luminosity-based mass is obtained from $L_{\mathrm{CO}}$ by using  

\begin{equation}\label{Mco-equ}
\frac{M_\mathrm{lum}}{M_\odot}=4.4 \frac{L_\mathrm{CO}}{\mathrm{K\ \kms\ pc^{2}}} X'_\mathrm{CO},
\end{equation}

\noindent where $X'_\mathrm{CO}$ is the assumed CO-to-H$_2$ conversion factor in units of $2\times10^{20}\mathrm{cm}^{-2}(\mathrm{K}\ \kms)^{-1}$, and includes a factor of 1.36 to account for the presence of helium.  Several observational (\citealt{2006MNRAS.371.1865B}; \citealt{2011ApJ...737...12L}; \citealt{2013ApJ...777....5S}) and theoretical (\citealt{2012ApJ...747..124F}; \citealt{2012MNRAS.421.3127N}) works have studied the dependence of the CO-to-H$_2$ conversion factor on local ISM properties such as metallicity or radiation field and spatial scale.  Different approaches to estimating the CO-to-H$_2$ conversion factor in normal galaxies consistently have found values within a factor of 2 of that found in the MW ($\sim 2\times10^{20}\ \mathrm{cm}^{-2}(\mathrm{K}\ \kms)^{-1})$.  Thus, in this study we assume a fixed value of $X'_\mathrm{CO}=1$ in our mass calculations.

Following the approach in Paper I, we calculate the virial mass for spheroidal clouds (see Appendix A of \citealt{1992ApJ...395..140B}).  Ignoring the magnetic energy and external pressure terms in the virial equilibrium equation, the virial mass is given by

\begin{equation}\label{Mvir-equ}
\frac{M_\mathrm{vir}}{M_\odot}=1040\left(\frac{\sigma_{v}}{\kms}\right)^{2}\left(\frac{R}{\mathrm{pc}}\right)\frac{1}{a_2},
\end{equation}

\noindent where $R$ is the radius perpendicular to the axis of symmetry of the cloud.  Equation \eqref{Mvir-equ} corresponds to the expression for the virial mass assuming a spherically symmetric geometry except for a shape-dependent correction factor $a_2$.  The $a_2$ factor is given by $\arcsin{\epsilon}/\epsilon$ if the shape of the cloud is oblate, where $\epsilon=(\sqrt{1-y^{2}})$ is the eccentricity of the cloud, and $y=Z/R$ is the ratio between the size along and perpendicular to the axis of symmetry.  On the other hand, If the cloud has a prolate shape, then $a_2=\mathrm{arcsinh}\ \epsilon/\epsilon$.  As we did in Paper I, in the present work we assume that the strucutures are fully described by the sky-projected major and minor axis, the axis of symmetry is given by the major axis, and that the structures are prolate.  We found that the mean of $a_2$ for $\co$ complexes was 0.8 with a rms of 0.1 for M101, and a mean of 0.8 and rms of 0.3 for NGC 628.  On the other hand, the mean of $a_2$ for $\cotwo$ clouds was 0.8, with a rms of 0.1.

We use the bootstrapping method implemented in the CPROPS package to estimate the uncertainties in the moment measurements, and this is the only source of uncertainties that we include in our analysis.

\subsection{Mass surface density} 
We estimate the mass surface density ($\Sigma_\mathrm{H2}$) by dividing the luminosity-based mass by the corresponding area covered by the identified structure, i.\ e., $\Sigma_\mathrm{H2}=M_\mathrm{lum}/(\pi R^2)$.  Following Paper I, in order to estimate $\Sigma_\mathrm{H2}$ we have used the non-extrapolated and non-deconvolved values of the luminosities and areas determined at level $T_\mathrm{edge}$, as we intend to compare molecular gas surface density with the SFR tracers (24$\mu$m, H$\alpha$, FUV) within the boundaries of the structures identified by the CPROPS mask.

\subsection{Star Formation surface density}\label{sfr-mol}

\subsubsection{SFR in $\co$ complexes}
Following Paper I, we estimate the star formation rate surface density in $\co$ complexes using {\it GALEX} FUV and {\it Spitzer} 24 $\mu$m maps (see L08 for details).  FUV traces the photospheric emission of O and B stars, measuring the star formation not obscured by dust over time scales of 10-100 Myr (\citealt{2005ApJ...633..871C}).  In turn, obscured star formation is traced by the 24 $\mu$m emission from small dust grains heated by UV photons radiated from young stars.  The 24 $\mu$m emission traces star formation over time scales of $\sim$10 Myr (\citealt{2005ApJ...633..871C}; \citealt{2007ApJ...666..870C}).  L08 found that the total star formation surface density $\Sigma_\mathrm{SFR}$ can be estimated by:

\begin{equation}\label{sfr}
\frac{\Sigma_\mathrm{SFR}}{ \Msun \mathrm{yr}^{-1}\ \mathrm{kpc}^{-2}}=(8.1\times 10^{-2} I_\mathrm{FUV} + 3.2\times 10^{-3} I_\mathrm{24\mu m}) \cos i 
\end{equation}

\noindent where $I_\mathrm{FUV}$ and $I_{24\mu m}$ are in units of MJy/sr, and $i$ is the inclination.  Equation \ref{sfr} assumes continuous star formation and a fully sampled Initial Mass Function (IMF) over the regions being probed.  We discuss the caveats of these assumptions at the spatial resolution of our maps in Section \ref{caveats}.

In general, the amount of SF traced by $\mathrm{24\mu m}$ dominates over that traced by the $\mathrm{FUV}$ component in the the boundaries of the complexes in the three galaxies.  The fraction of SF from $\mathrm{24\mu m}$ is 0.85 $\pm$ 0.08, 0.82 $\pm$ 0.06 and 0.81 $\pm$ 0.08 for NGC 6946, NGC 628, and M101 respectively.

\begin{figure}
\centering
\begin{tabular}{l}
\epsfig{file=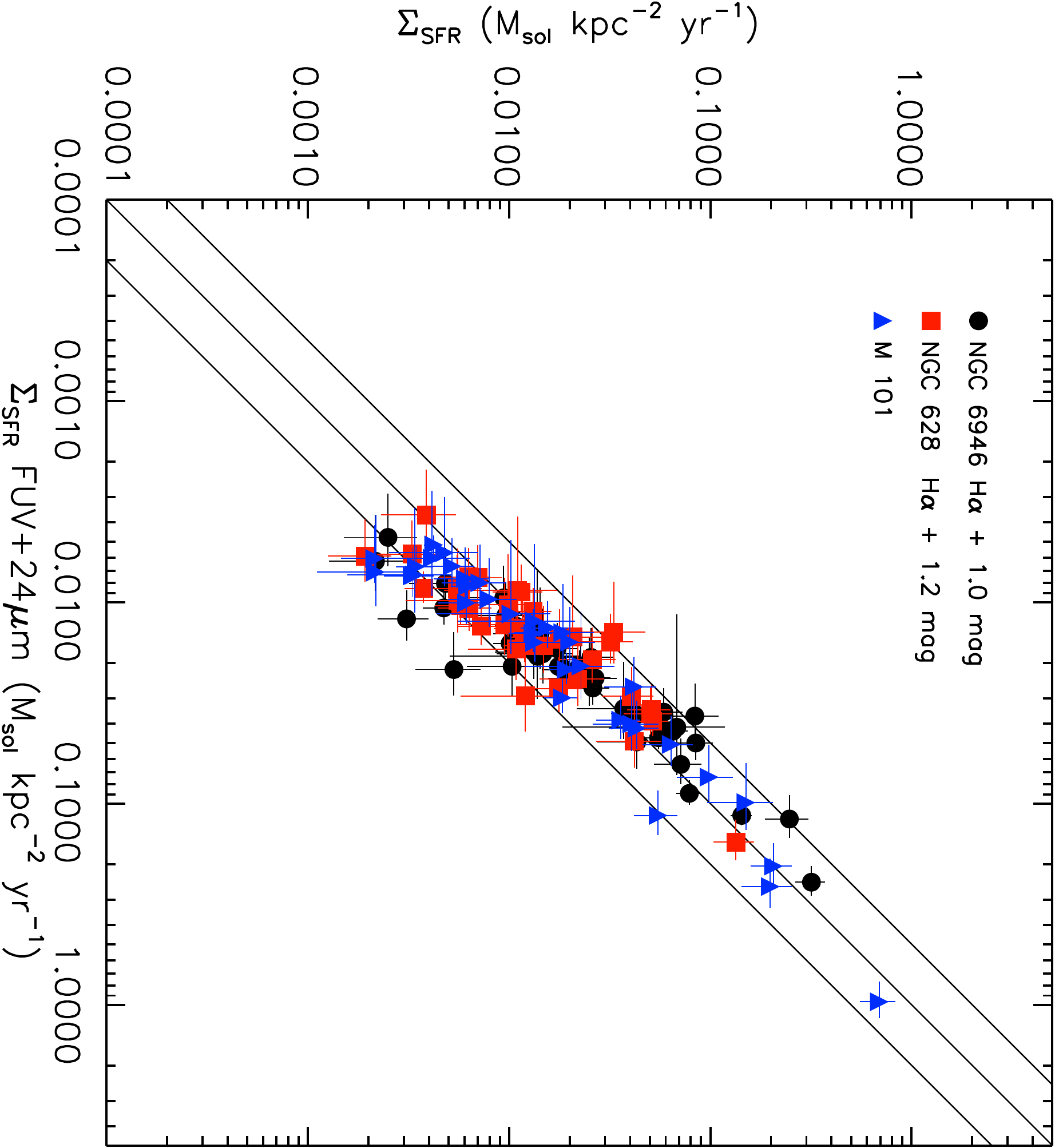,width=0.8\linewidth,angle=90}
\end{tabular}
\caption{$\Sigma_\mathrm{SFR}$ estimates using H$\alpha$ for $\co$ complexes as a function of $\Sigma_\mathrm{SFR}$ predicted using a combination of FUV + 24$\mu$m, following L08.  Black dots represent the complexes identified in NGC 6946, while red squares correspond to $\co$ complexes detected in NGC 628 and blue triangles represent complexes identified in M101.  Solid black lines show slopes of 2, 1 and 0.5.  We use A$_\mathrm{H\alpha}=1.0$ and $1.2$ magnitude extinction for NGC 6946 and NGC 626 respectively.  For M101, we have applied a calibration factor to convert from H$\alpha$ flux map to SFR using FUV + 24$\mu$m as comparison.}
\label{fig_ha_fuv}
\end{figure}

\subsubsection{SFR in \cotwo\ clouds}\label{sfr-co21}
We use H$\alpha$ images to estimate the amount of star formation in $\cotwo$ clouds.  The H$\alpha$ images offer a finer resolution that allows us to directly compare the star-forming regions to the structures identified in $\cotwo$ maps.  As H$\alpha$ emission can be strongly affected by dust extinction, we have corrected the H$\alpha$ maps to account for the obscured star formation.  According to \citet{2007ApJ...666..870C}, the total star formation can be estimated by

\begin{equation}\label{sfr_ha}
\frac{\mathrm{SFR}_\mathrm{tot}}{ \Msun\ \mathrm{yr}^{-1}}=(5.3\times 10^{-42} L_\mathrm{H\alpha})\times 10^{A_{\mathrm{H}\alpha}/2.5}
\end{equation}

\noindent where $L_\mathrm{H\alpha}$ is in units of ergs s$^{-1}$, and $A_{\mathrm{H}\alpha}$ is the H$\alpha$ extinction.

Due to the lack of a calibrated H$\alpha$ image for M101 in the region we targeted, for this galaxy we have used the following relation to estimate the SFR

\begin{equation}\label{sfr_ha_m101}
\frac{\mathrm{SFR}_\mathrm{tot}}{ \Msun\ \mathrm{yr}^{-1}}=(F_\mathrm{H\alpha}\times f_\mathrm{FUV+24\mu m}) 
\end{equation}

\noindent where $F_\mathrm{H\alpha}$ is the total H$\alpha$ flux in the region considered, and $f_\mathrm{FUV+24\mu m}$ is a calibration factor to convert from the H$\alpha$ map units to SFR rate using the FUV+24$\mu$m map as a reference for the total star formation rate.    

If we assume that the SFR values estimated using FUV+ $24\mu$m are correct, we can estimate the H$\alpha$ extinction $A_{\mathrm{H}\alpha}$ for NGC 628 and the calibration factor $f_\mathrm{FUV+24\mu m}$ for M101 by comparing the star formation rate calculated using Equations \eqref{sfr_ha} and \eqref{sfr_ha_m101} to the values derived from Equation \eqref{sfr}.  In both cases, the comparison was performed by eye.  Figure \ref{fig_ha_fuv} shows the comparison between the different methods to estimate the SFR.  For NGC 628, we found that a mean value of $A_{\mathrm{H}\alpha}=1.2$ applied to Equation \eqref{sfr_ha} roughly recovers the SFR traced by FUV+ $24\mu$m.  


By assuming a constant value of $A_{\mathrm{H}\alpha}$ for NGC 6946 and NGC 628, and a fixed calibration factor for M101, we may be introducing some scatter in our SFR measurements for $\cotwo$ clouds.  According to Figure \ref{fig_ha_fuv}, there is a scatter of a factor of $\sim$ 2 about the identity line between $\Sigma_\mathrm{SFR}$ from H$\alpha$ and the $\Sigma_\mathrm{SFR}$ from FUV+24$\mu$ m.  This scatter is similar to that shown in Figure 23 of L08.  However, we notice that the $\Sigma_\mathrm{SFR}$ from FUV+24$\mu$ m consistently overestimates the $\Sigma_\mathrm{SFR}$ from H$\alpha$ for $\Sigma_\mathrm{SFR} \lesssim 0.01\ \Msun \mathrm{yr}^{-1}\ \mathrm{kpc}^{-2}$.  We further discuss this discrepancy in the context of a cirrus component in 24$\mu$m in Appendix \ref{24um_cirr}.

\begin{table}
\caption{Mean values of resolved structure properties.\label{table-mean-props}}
\centering
\begin{tabular}{lccc}
\hline\hline
Property  &   NGC 6946   &  NGC 628  &  M101 \\
\hline
\multicolumn{4}{c}{$\co$ complexes}  \\ 
\hline
No. identified structures & 43  & 34 & 35 \\
No. resolved structures\tablenotemark{a}  &   33 &  19 & 17 \\
$\avg{R}$(pc) & 180 $\pm$ 9 & 142 $\pm$ 12 & 162 $\pm$ 12 \\
$\avg{\sigma_v} ($\kms$)$ & 6.7 $\pm$ 0.3 & 4.8 $\pm$ 0.4 &  5.7 $\pm$ 0.4  \\
$\avg{L_\mathrm{CO}} (10^5\ \mathrm{K}\ \kms \mathrm{pc}^{2}$) & 20 $\pm$ 1  & 11 $\pm$ 1 & 15 $\pm$ 1   \\
$\avg{M_\mathrm{vir}} (10^{6}\ \Msun)$ & 11 $\pm$ 2 & 5 $\pm$ 1& 7 $\pm$ 1   \\
$\avg{\Sigma_\mathrm{H2}} (\Msun\ \mathrm{pc}^{-2})$ & 44  $\pm$ 2 & 36 $\pm$ 2 &  49 $\pm$  3  \\
$\avg{\Sigma_\mathrm{SFR}} (10^{-3} \Msun\ \mathrm{yr}^{-1}\ \mathrm{kpc}^{-2})$ & 36 $\pm$ 2 & 21 $\pm$ 2  & 62 $\pm$ 7   \\
\hline
\multicolumn{4}{c}{$\cotwo$ clouds}  \\ 
\hline
No. identified structures & 64  & 28 & 52 \\
No. resolved structures\tablenotemark{a}  &  34 &  25 & 38  \\
$\avg{R}$(pc) & 65 $\pm$ 4 & 98 $\pm$ 6 & 96 $\pm$ 6  \\
$\avg{\sigma_v} ($\kms$) $ & 5.7 $\pm$  0.3 & 5.1 $\pm$ 0.3 & 5.4 $\pm$ 0.3  \\
$\avg{L_\mathrm{CO}} (10^4\ \mathrm{K}\ \kms \mathrm{pc}^{2}) $ & 40 $\pm$ 2 & 59 $\pm$ 3 & 46 $\pm$ 3   \\
$\avg{M_\mathrm{vir}} (10^{5}\ \Msun) $ & 24 $\pm$ 1 & 33 $\pm$ 4 & 37 $\pm$ 5  \\
$\avg{\Sigma_\mathrm{H2}} (\Msun\ \mathrm{pc}^{-2})$ & 70 $\pm$ 3 & 52 $\pm$ 3 & 37 $\pm$ 2  \\
$\avg{\Sigma_\mathrm{SFR}} (10^{-3} \Msun\ \mathrm{yr}^{-1}\ \mathrm{kpc}^{-2})$ & 90 $\pm$ 6 &  62  $\pm$ 5 & 134  $\pm$ 15   \\
\hline
\multicolumn{4}{l}{{\bf Notes.}} \\
\multicolumn{4}{l}{$^\mathrm{a}$ Structures with properties corrected by sensitivity and resolution bias.}
\end{tabular}
\label{cprops-prop}
\end{table}

\section{Results}\label{cl-prop}

\subsection{Statistics of identified structures}\label{cl-prop-corr}



The number of identified structures in each galaxy by CPROPS is shown in Table \ref{table-mean-props}, along with the number of structures with measured values of $R$ and $\sigma_v$ after correction by sensitivity and resolution bias is applied (i.\ e., resolved structures).  Also, Table \ref{table-mean-props} shows the mean values of the properties for resolved structures.  The $\co$ complexes have resolved radii between 23 pc and 426 pc, and velocity dispersions between 2.6 $\kms$ and 13.1 $\kms$.  The luminosity $L_\mathrm{CO}$ for complexes has values between 1.3 $\times 10^5$ $\mathrm{K}\ \kms\ \mathrm{pc}^{2}$ and 108 $\times 10^5$ $\mathrm{K}\ \kms\ \mathrm{pc}^{2}$, and virial masses from $4.7 \times 10^{5}\ \Msun$ to 766 $\times 10^{5}\ \Msun$.

On the other hand $\cotwo$ clouds have resolved radii between 25 pc to 171 pc, and velocity dispersions between 1.2 $\kms$ to 9.6 $\kms$.  $L_\mathrm{CO}$ for clouds has values between 2.7 $\times 10^4$ $\mathrm{K}\ \kms\ \mathrm{pc}^{2}$ and 280 $\times 10^4$ $\mathrm{K}\ \kms\ \mathrm{pc}^{2}$, and virial masses from $7.6 \times 10^{4}\ \Msun$ to 1810 $\times 10^{4}\ \Msun$.

At scales sampled by the $\co$ observations ($\sim$ 150-200 pc), the mean surface density for each galaxy, $\avg{\Sigma_\mathrm{H2}}$, is in the range of $\sim 35-50\ \Msun\ \mathrm{pc}^{-2}$.  In contrast, the higher resolution $\cotwo$ observations reveal some differences in the mean values.  NGC 6946 shows the highest mean of $\sim 70\ \Msun\ \mathrm{pc}^{-2}$ for the $\cotwo$ clouds followed by NGC 628 with a mean of $\sim 50\ \Msun\ \mathrm{pc}^{-2}$, and finally M101 shows the smallest mean surface density $\sim 40\ \Msun\ \mathrm{pc}^{-2}$.  This difference in the $\avg{\Sigma_\mathrm{H2}}$ of NGC 6946 basically reflects the fact that the overall CO brightness of NGC 6946 is higher than the other two galaxies, M101 and NGC 628.

Overall, the $\Sigma_\mathrm{SFR}$ values for the regions observed in the three galaxies cover similar ranges, from $\log(\frac{\Sigma_\mathrm{SFR}}{\Msun\ \mathrm{yr}^{-1}\ \mathrm{kpc}^{-2}}) \sim$ -2.7 to 0 for both complexes and clouds.  We identify some complexes in NGC 6946 and M101 with $\Sigma_\mathrm{SFR}> 0.1\ \Msun\ \mathrm{yr}^{-1}\ \mathrm{kpc}^{-2}$.  These complexes are located in on-arm regions, where local star formation activity is higher compared to other areas surveyed over the three galaxies.  


At higher resolution, clouds follow a behavior similar to the complexes, with clouds located in specific regions of the spiral arms exhibiting active star formation ($> 0.1\ \Msun\ \mathrm{yr}^{-1}\ \mathrm{kpc}^{-2}$) constituting the high end of the $\Sigma_\mathrm{SFR}$ values.


\begin{table*}
\caption{Bayesian regression parameters for $\co$ complex scaling relations.\label{table-bayes}}
\centering
\begin{tabular}{ccccccccc}
\hline\hline
Relation & $\alpha$ & 90\% HDI & & $A$ &  90\% HDI & & $\sigma$ &  90\% HDI \\
\hline
& & & & \multicolumn{1}{c}{NGC 6946} &  &  & &\\ 
 \hline
$\sigma_{v}-R $& 0.23 & [-0.13, 0.67] & & 0.79  & [0.61, 0.88] & & 0.09  & [0.06, 0.14]   \\
$M_\mathrm{vir}-L_\mathrm{CO}$ & 1.15 & [ 0.84, 1.40] & & 6.56 & [6.38, 6.71] & & 0.16 & [0.09, 0.25] \\
\hline
& & & & \multicolumn{1}{c}{NGC 628}  & & & & \\
\hline
$\sigma_{v}-R $& 0.05  & [-0.35, 0.50]  & & 0.70 & [0.57, 0.78] & & 0.06 & [0.01, 0.12] \\
$M_\mathrm{vir}-L_\mathrm{CO}$ & 1.17 & [0.64, 1.63] & & 6.42 & [6.21, 6.54]& & 0.09  & [0.03, 0.25]  \\
 \hline
 & & & & \multicolumn{1}{c}{M101}  & & & & \\
 \hline
$\sigma_{v}-R $& 0.19 & [-0.20, 0.61] & & 0.70 & [0.57, 0.81] & & 0.08 & [0.03, 0.17]   \\
$M_\mathrm{vir}-L_\mathrm{CO}$ & 1.53 & [1.16, 2.18] & & 6.17 & [5.86, 6.35]& & 0.11 & [0.04, 0.32]  \\
\hline
\end{tabular}
\end{table*}

\subsection{Scaling relations}\label{scal-rel}

\subsubsection{Bayesian inference method}

We have used a Bayesian inference method to find the best fit relations between the properties of the structures.  This approach differs from the bisector linear regression method implemented in Paper I for NGC 6946.  Bayesian fitting methods have been successfully employed in several astrophysical analyses, such as the derivation of the extinction law in the Perseus molecular cloud complex (\citealt{2013MNRAS.428.1606F}), Type Ia supernova light curve inference (\citealt{2011ApJ...731..120M}), the size-line width relation in the dense ISM of the Central Molecular Zone (\citealt{2012MNRAS.425..720S}), and the extragalactic Kennicutt-Schmidt relation (\citealt{2013MNRAS.430..288S}).  We refer the reader to \citet{2007ApJ...665.1489K} for details of the Bayesian regression fit method, and we limit our description here to the basic concepts.  The Bayesian approach estimates the joint posterior probability distribution of the regression parameters given the observed data, and draws the error in each measured quantity from some a priori defined distribution which should reflect the uncertainties in the measurements.  Bayes' theorem yields the posterior distribution of the parameters $\theta$ given the observed data $(x,y)$ as 

\begin{equation}\label{eq-baye}
p(\theta | x,y) \propto p(x,y | \theta )p(\theta), 
\end{equation}

\noindent where $p(\theta)$ is the prior distribution of the parameters, and $p(x,y | \theta)$ is the probability of the data given the parameters $\theta$.  Although a direct derivation of the posterior probability distribution is computationally expensive, the Markov chain Monte Carlo (MCMC) routine allows one to sample the probability distribution of the fit parameters through random draws in parameter space.  This yields a histogram of the marginal probability distribution, allowing the estimation of the median and error for each parameter. In our analysis, we have used the {\it Gibbs sampling} method for generating random draws from the posterior distribution (\citealt{2007ApJ...665.1489K}).  As this method accounts for the uncertainties of the dependent and independent variables of the fit, it provides more realistic uncertainty estimates for each parameter of the regression.  

\begin{figure*}
\centering
\begin{tabular}{cc}
\epsfig{file=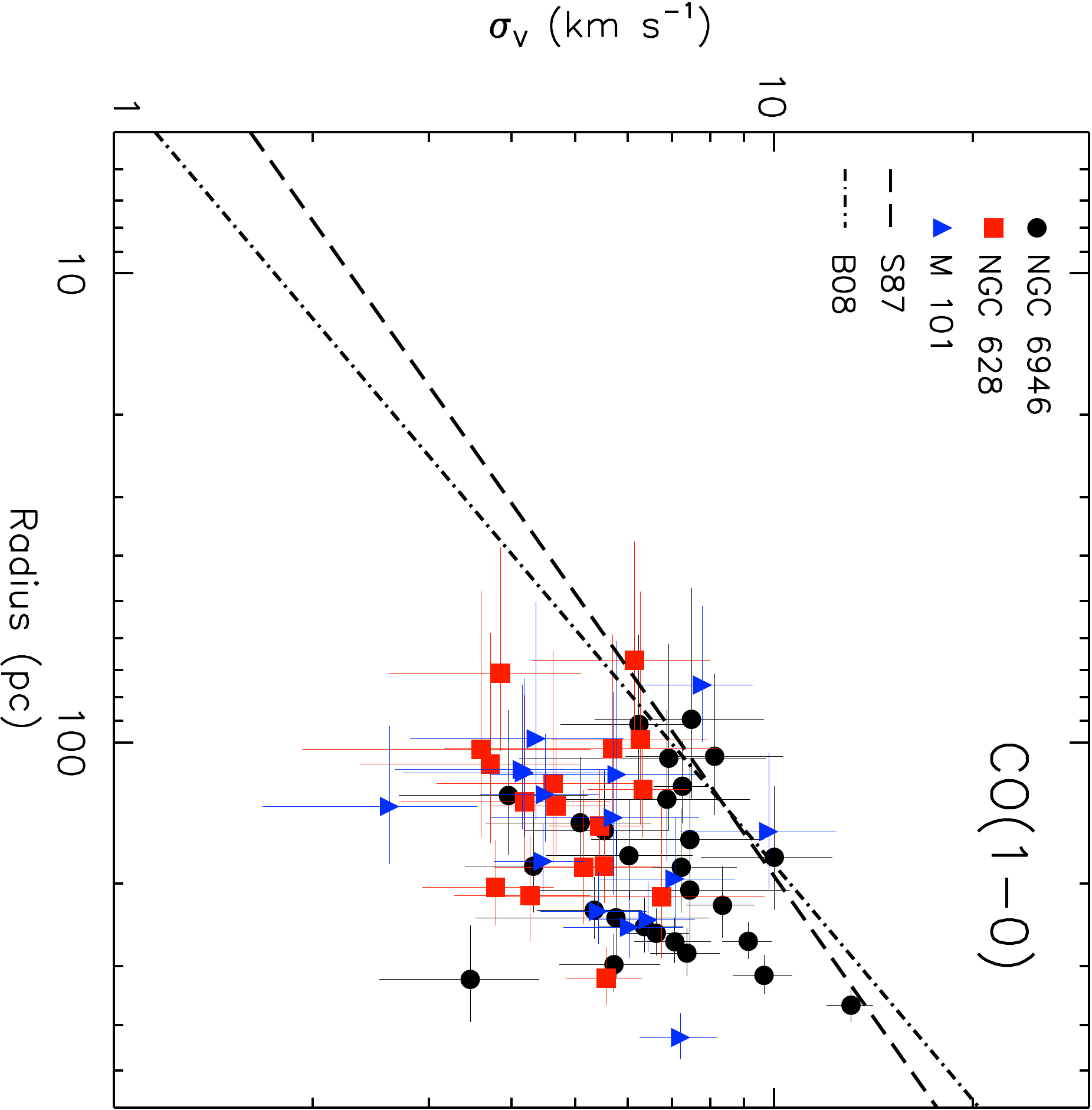,width=0.3\linewidth,angle=90}
\epsfig{file=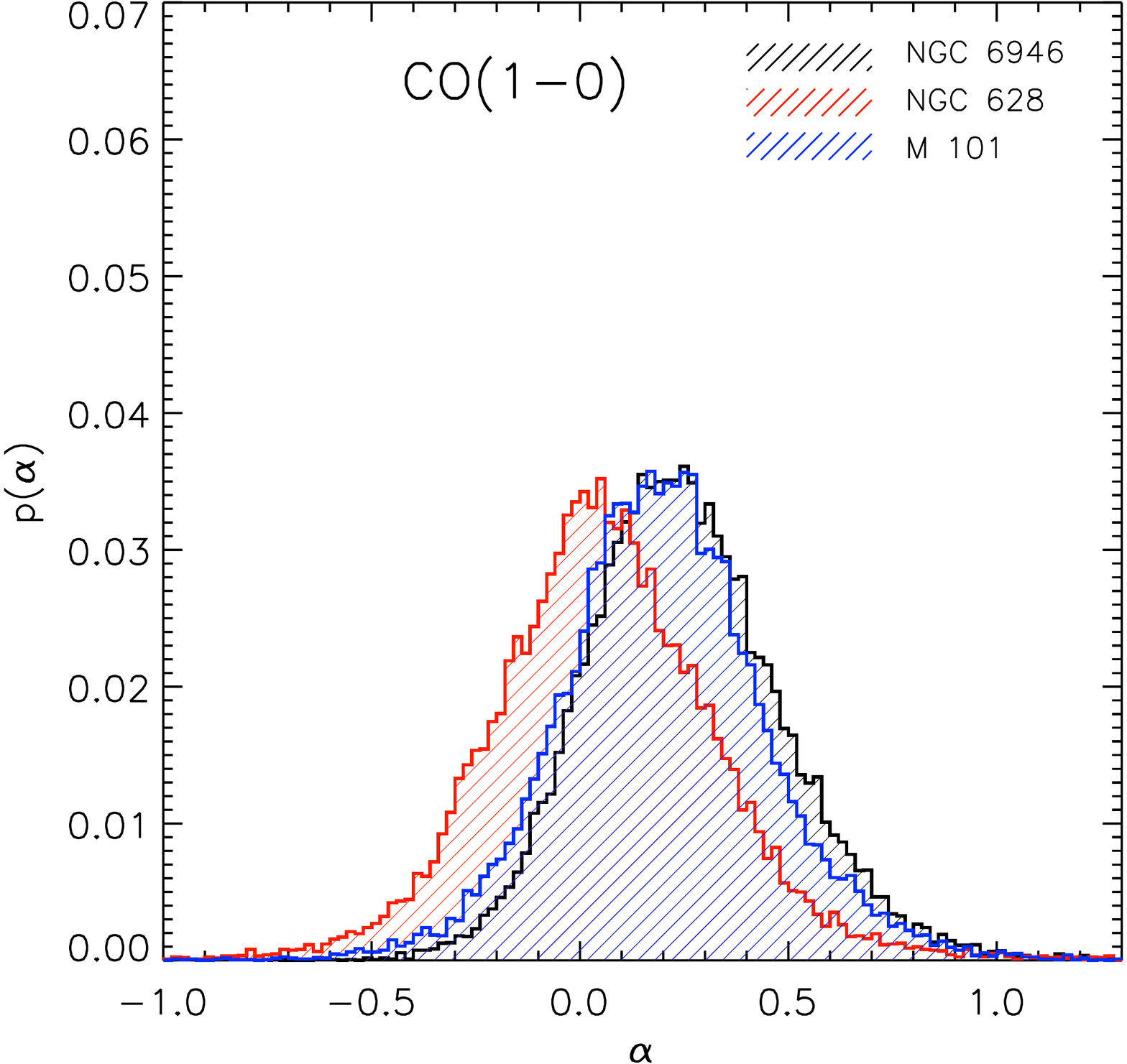,width=0.3\linewidth,angle=90}\\
\epsfig{file=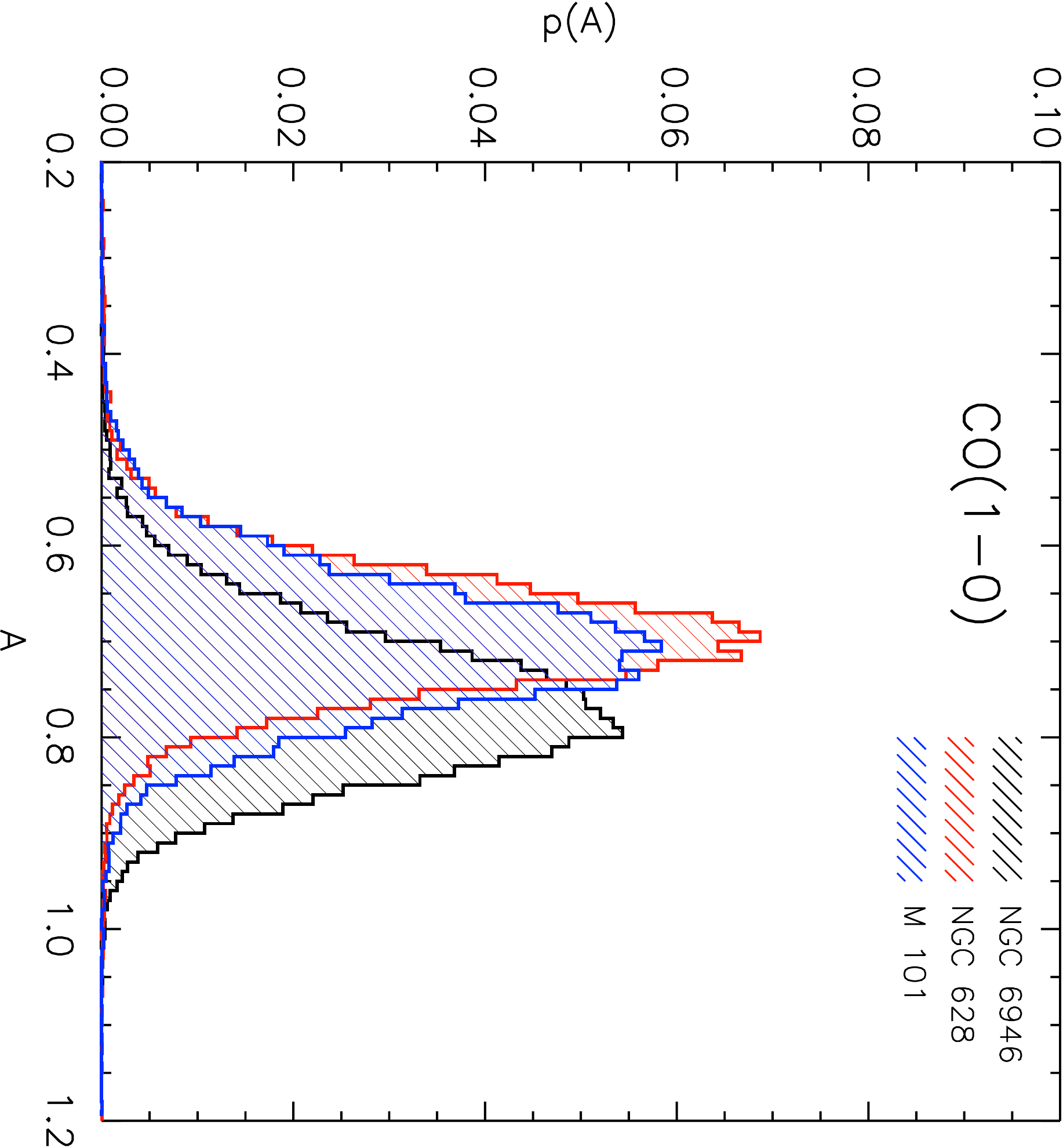,width=0.3\linewidth,angle=90}
\epsfig{file=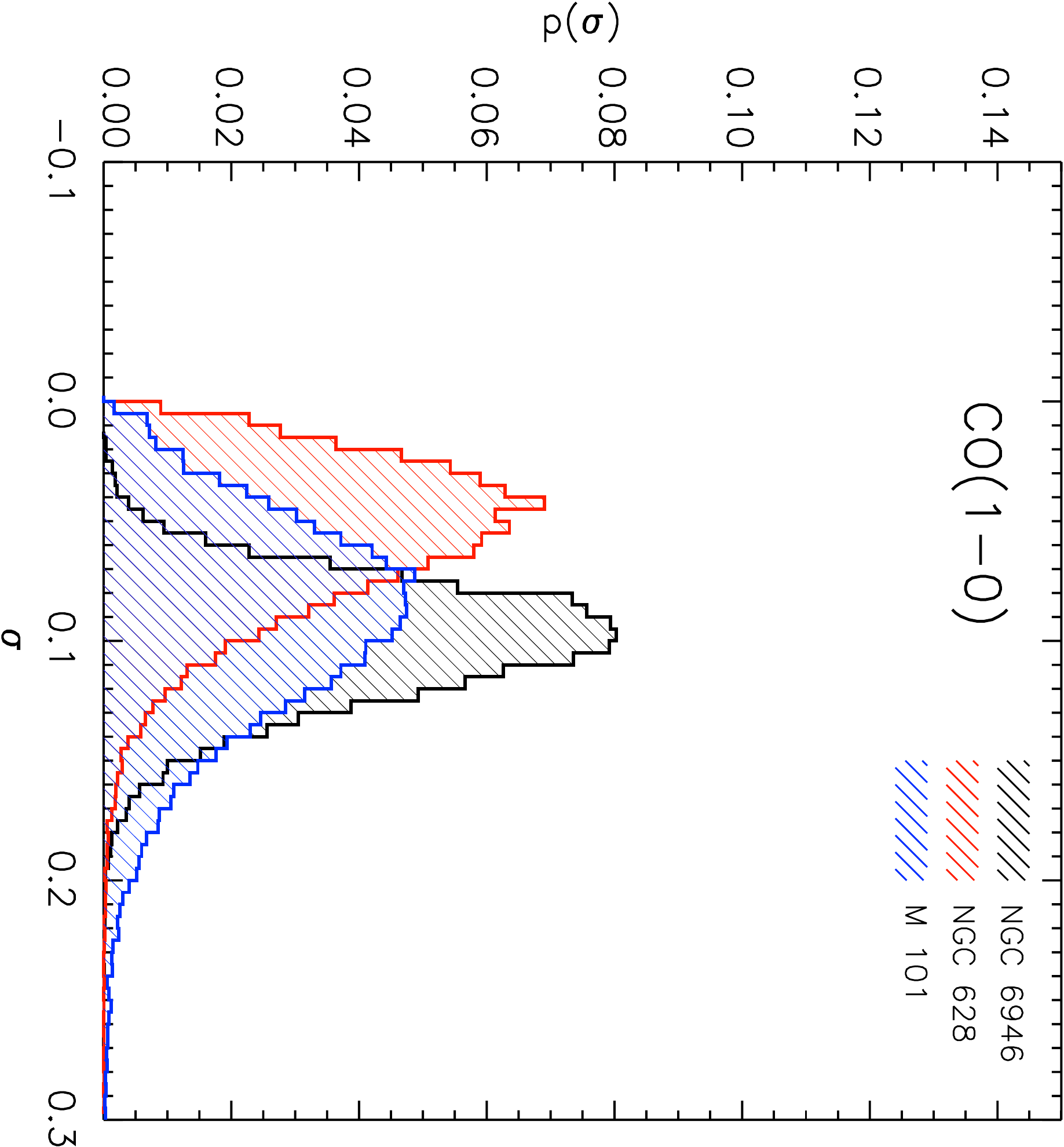,width=0.3\linewidth,angle=90}
\end{tabular}
\caption{Size-line width relation for the $\co$ complexes found in this study (top-left).  Black dots represent the complexes identified in NGC 6946, while red squares correspond to complexes detected in NGC 628 and blue triangles represent complexes identified in M101.  The dashed line illustrates the relation found by S87, and the dashed dotted line represents the fit found by B08.  The marginal distribution of the slope $\alpha$ from the Bayesian regression analysis is shown in the top-right panel, the intercept coefficient $A$ is shown in the bottom-left panel and the dispersion $\sigma$ is shown in the bottom-right panel, with $\mathrm{log}(\frac{\sigma_v}{\mathrm{km\ s^{-1}}})=A+\alpha \mathrm{log}(\frac{R}{100\ \mathrm{pc}})+ \epsilon_\mathrm{scat}$.}
\label{fig_sizew_10}
\end{figure*}

Through this paper, we model the scaling relations using the standard power law form given by $y=a x^{\alpha}$.  Following the prescription introduced by \citet{2007ApJ...665.1489K}, the fitting is performed in the log space, 

\begin{equation}\label{model-fitting}
\log{(y)}=A + \alpha \log{(x)} + \epsilon_\mathrm{scat}, 
\end{equation}

\noindent where $A=\log{(a)}$, and $\epsilon_\mathrm{scat}$ is the scatter about the regression line.  $\epsilon_\mathrm{scat}$ is assumed to have mean of 0 and dispersion of $\sigma$.  Thus, the three parameters involved in the fitting method are $A, \alpha$ and $\sigma$.  For each scaling relation, we have run $2 \times 10^{4}$ random draws to sample the probability distribution of the fit parameters.  The Bayesian regression fit method used in this paper (\citealt{2007ApJ...665.1489K}) allows us to include the non-detections in the $y$-axis only.  However, in Section \ref{k-s_grid_h2_nondetect} we will introduce a new approach to include $\Sigma_\mathrm{H2}$ non-detections in the context of the K-S relation.

\subsubsection{Size-Line width relation}\label{size-width}
 
Figure \ref{fig_sizew_10} shows the scatter plot of the line width and size of $\co$ emitting complexes on a logarithmic scale.  For comparison, we include the relations found by S87 and B08 in the figure.  Because we are interested in the variations of the regression fit parameters at size scales representative of the values observed in complexes and clouds, we have normalized the size variable by an appropriate value (100 pc).  Thus, for the size-line width relation, we have expressed the sizes in units of 100 pc, so the linear regression is performed over the relation

\begin{table*}
\caption{Bayesian regression parameters for $\cotwo$ clouds scaling relations.\label{table-bayes21}}
\centering
\begin{tabular}{ccccccccc}
\hline\hline
Relation & $\alpha$ & 90\% HDI & & $A$ &  90\% HDI & & $\sigma$ &  90\% HDI  \\
\hline
 & & & & \multicolumn{1}{c}{NGC 6946} &  & & & \\ 
 \hline
$\sigma_{v}-R $& 0.27 & [-0.11, 0.67] & & 0.83  & [0.75, 0.87] & & 0.09  & [0.06, 0.14]   \\
$M_\mathrm{vir}-L_\mathrm{CO}$ & 1.17 & [ 0.94, 1.46] & & 6.64 & [6.53, 6.74] & & 0.19 & [0.11, 0.31] \\
\hline
 & & & & \multicolumn{1}{c}{NGC 628}  & & & & \\
\hline
$\sigma_{v}-R $& -0.17  & [-0.77, 0.47]  & & 0.73 & [0.67, 0.76] & & 0.06 & [0.01, 0.10] \\
$M_\mathrm{vir}-L_\mathrm{CO}$ & 0.65 & [0.34, 1.06] & & 6.66 & [6.55, 6.74]& & 0.12  & [0.04, 0.22] \\
 \hline
 & & & & \multicolumn{1}{c}{M101}  & & & & \\
 \hline
$\sigma_{v}-R $& 0.37 & [-1.48, 2.78] & & 0.76 & [0.67, 0.80]& & 0.08 & [0.04, 0.12]    \\
$M_\mathrm{vir}-L_\mathrm{CO}$ & 1.39 & [1.03, 1.72] & & 6.88 & [6.76, 6.97]& & 0.13 & [0.05, 0.23] \\
\hline
\end{tabular}
\end{table*}

\begin{figure*}
\centering
\begin{tabular}{cc}
\epsfig{file=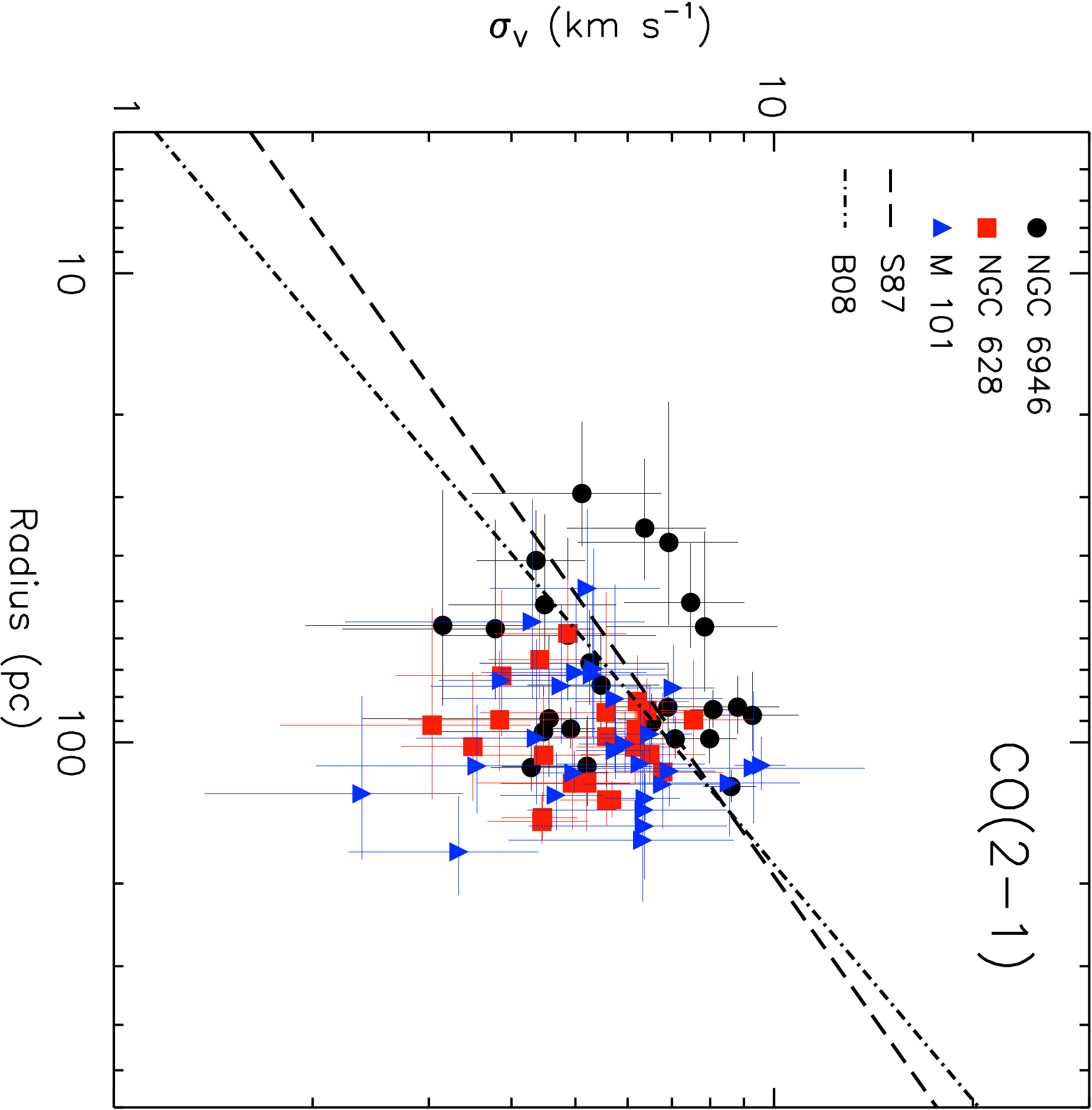,width=0.3\linewidth,angle=90}
\epsfig{file=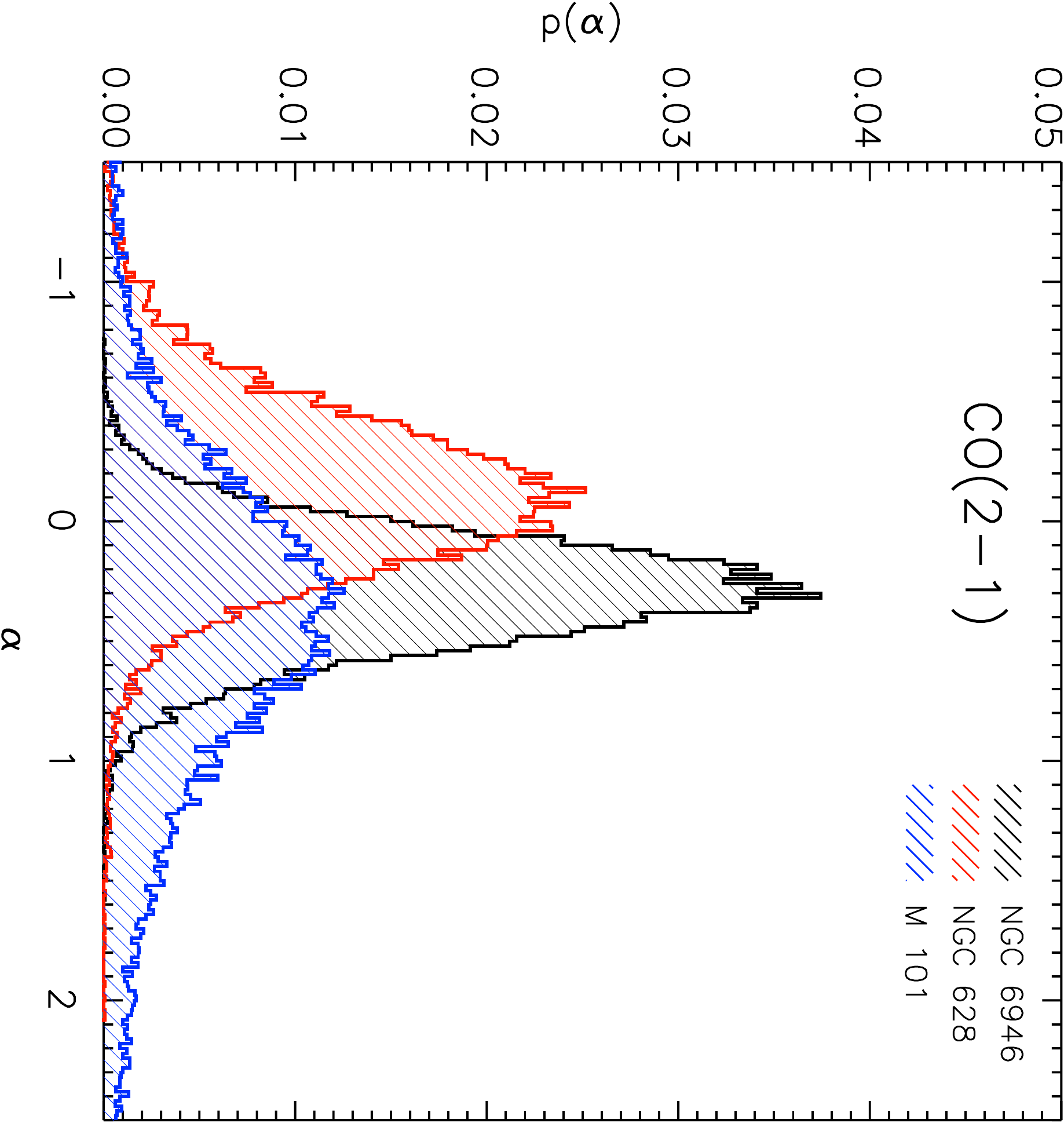,width=0.3\linewidth,angle=90}\\
\epsfig{file=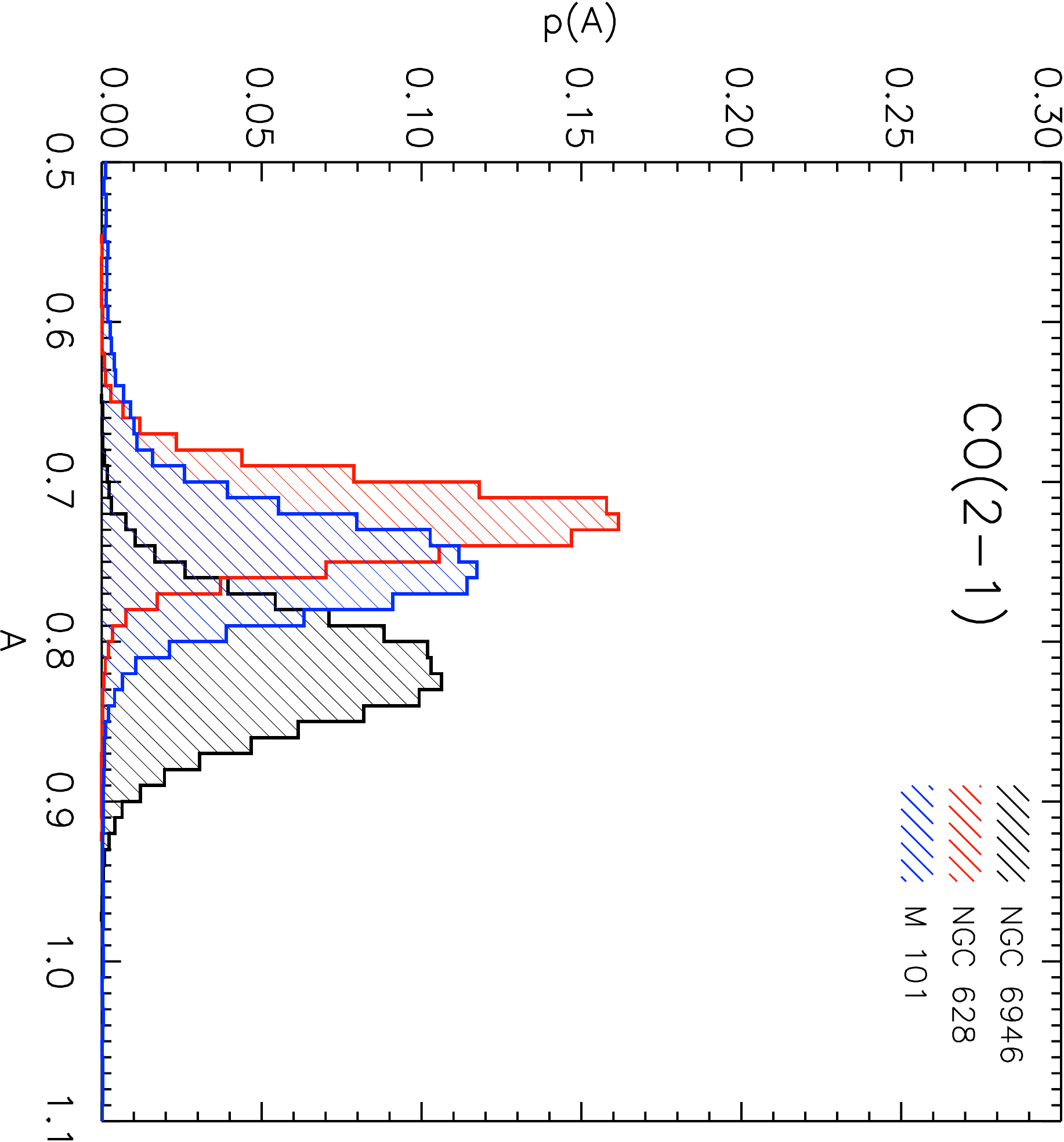,width=0.3\linewidth,angle=90}
\epsfig{file=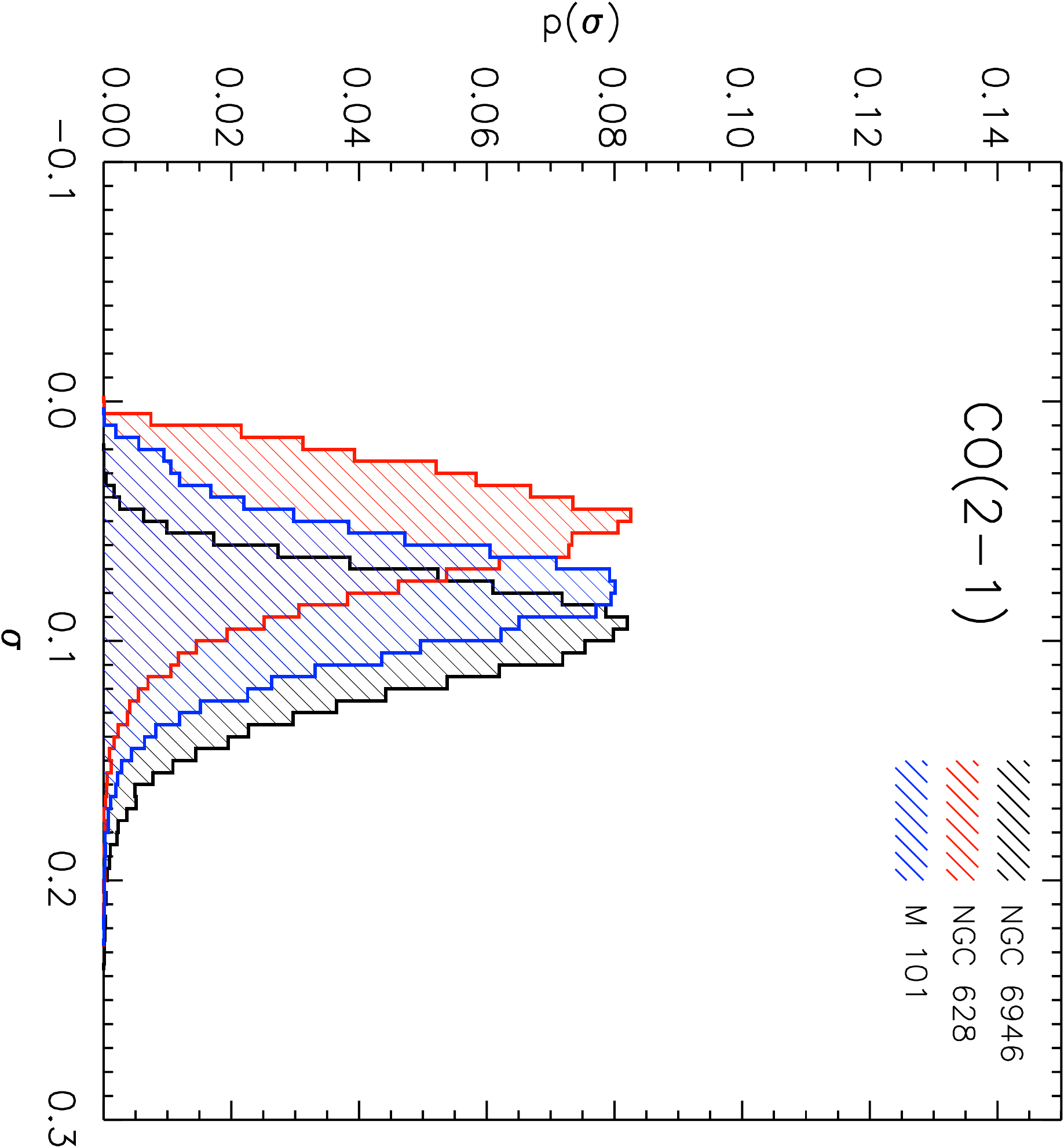,width=0.3\linewidth,angle=90}
\end{tabular}
\caption{Size-line width relation for the $\cotwo$ clouds found in this study (top-left).  Symbols and lines are the same as Figure \ref{fig_sizew_10}.  The panels showing the marginal distribution of the slope $\alpha$, the intercept coefficient $A$ and the dispersion of the scatter of the relation $\sigma$ from the Bayesian regression analysis are placed in the same way as Figure \ref{fig_sizew_10}.}
\label{fig_sizew_21}
\end{figure*}

\begin{equation}\label{size_line10}
\mathrm{log}\left(\frac{\sigma_v}{\mathrm{km\ s^{-1}}}\right)=A+\alpha\ \mathrm{log}\left(\frac{R}{100\ \mathrm{pc}}\right)+ \epsilon_\mathrm{scat}.
\end{equation}

The distributions of the regression fit parameters using the Bayesian inference approach are shown in Figure \ref{fig_sizew_10}.  The distributions of the slope $\alpha$, the y-axis intercept coefficient $A$, and the dispersion of the scatter $\sigma$ for the size-line width relations show overall consistent behaviours for the three galaxies in our sample.  The peak of the slope distribution is $< 0.5$ for the three galaxies, and the wide range of the HDI (High Density Interval, defined as the interval that encloses the 90\% of the probability distribution) for the slopes (HDI $\sim$ 0.8) indicates the $\alpha$ parameter is poorly constrained, i.\ e., we do not find a strong correlation between size and line width for the complexes.  For NGC 628, the complexes show line widths smaller than 7 $\kms$ for sizes that span a range of 50 to 300 pc.  Among the three galaxies in our sample, NGC 628 shows the smallest inclination (7$\degrees$), which suggests that the CO velocity dispersion observed in this galaxy is less affected by cloud blending or superposition of several clouds along the line of sight.  The intrinsic scatter about the linear relation is similar for the three galaxies, although the distribution of $\sigma$ is shifted to smaller values for NGC 628.  The values of the peaks and the HDI provided by the Bayesian inference method for the slope $\alpha$, the intercept coefficient $A$, and the dispersion $\sigma$ for complexes are shown in Table \ref{table-bayes}.

\begin{figure*}
\centering
\begin{tabular}{ccc}
\epsfig{file=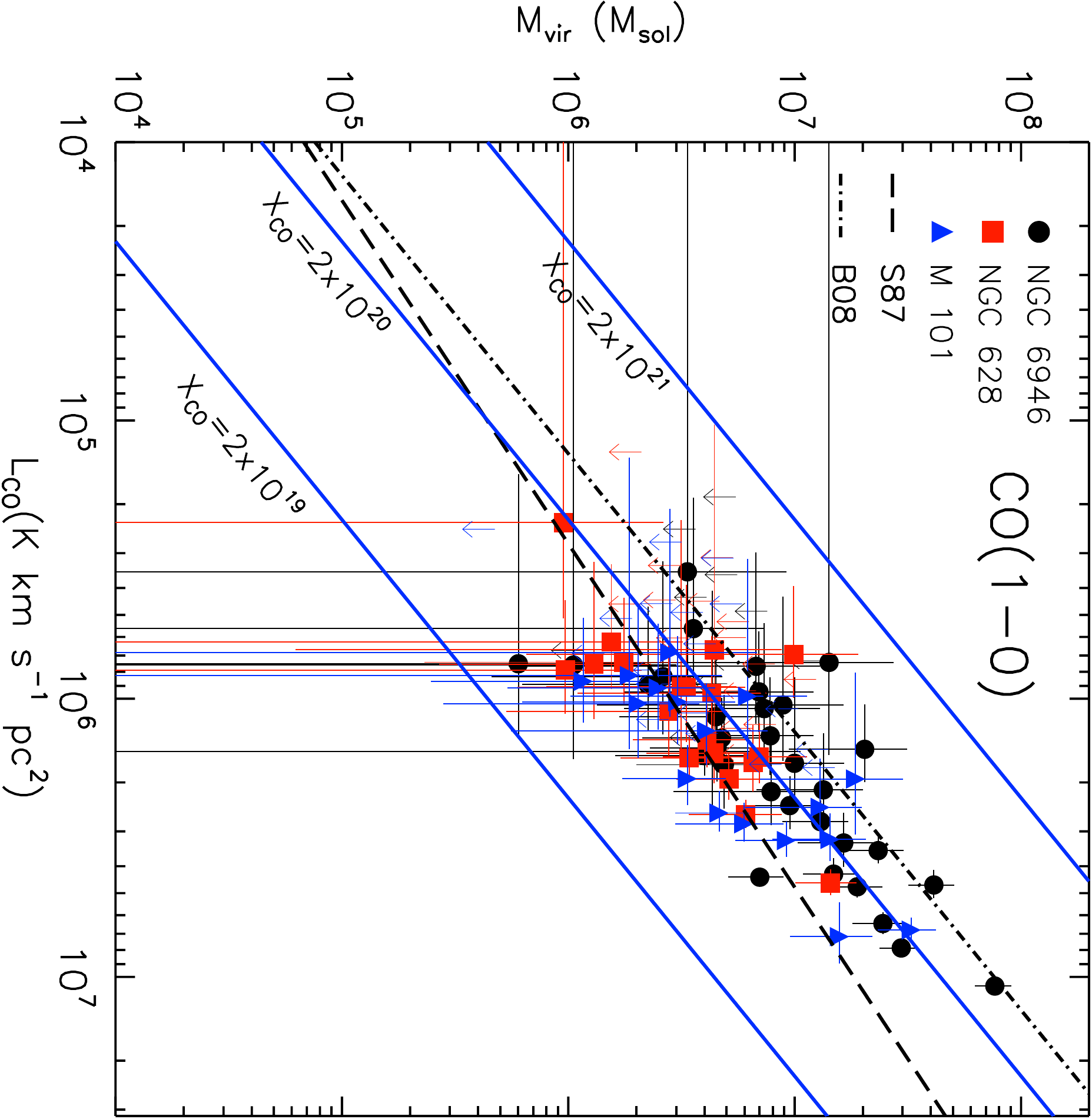,width=0.3\linewidth,angle=90}
\epsfig{file=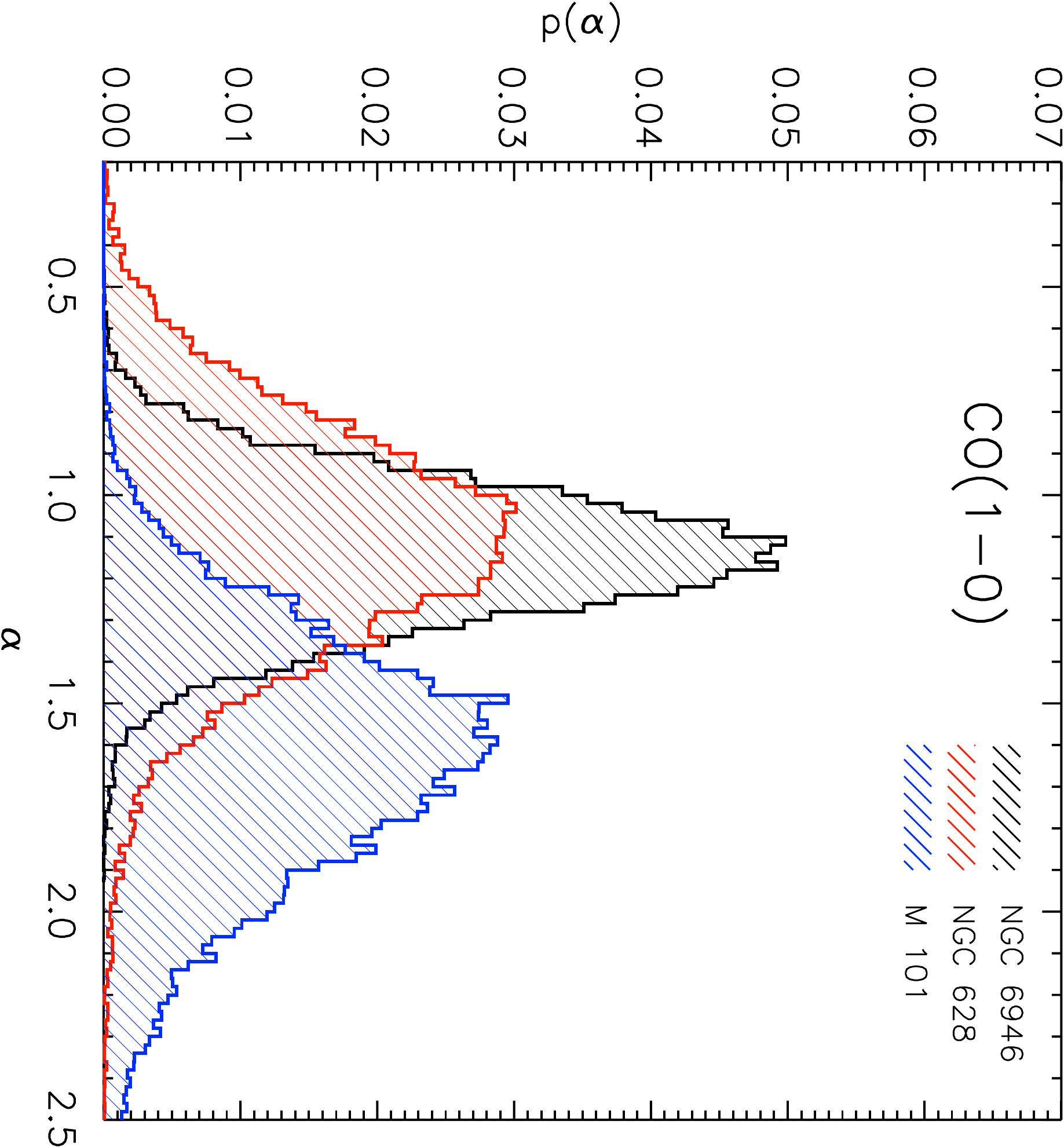,width=0.3\linewidth,angle=90}\\
\epsfig{file=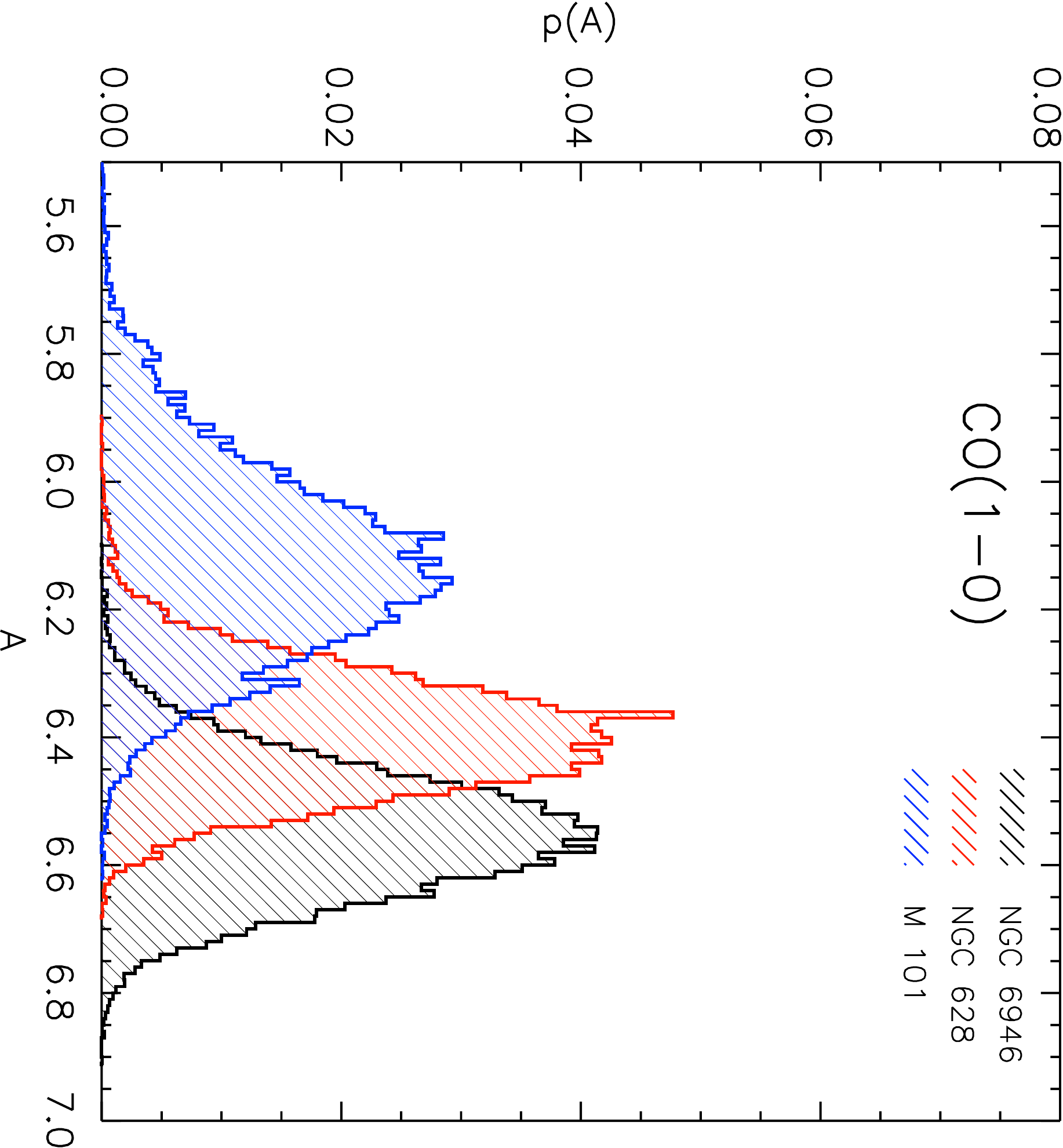,width=0.3\linewidth,angle=90}
\epsfig{file=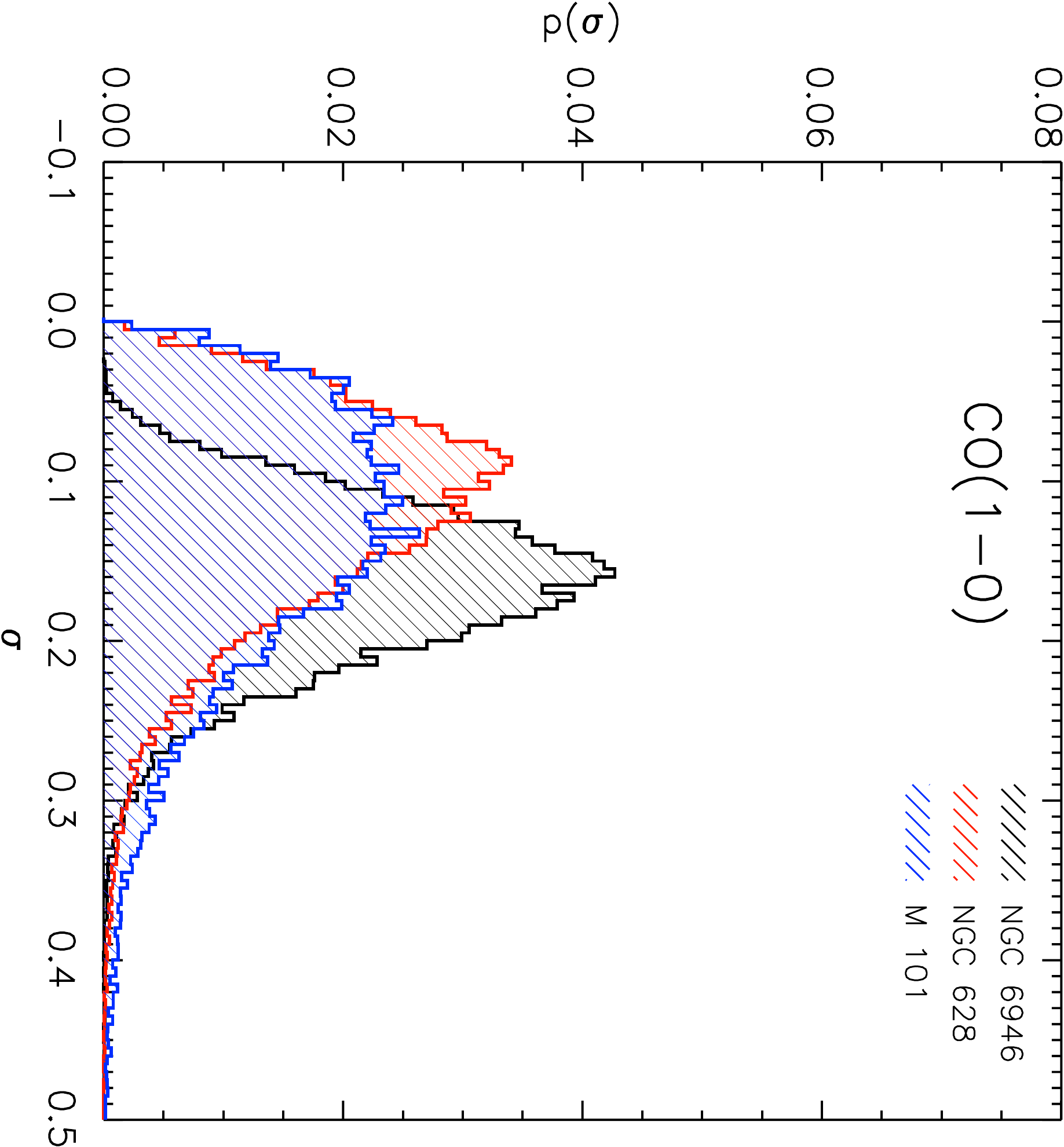,width=0.3\linewidth,angle=90}
\end{tabular}
\caption{The Virial mass-Luminosity relation for the complexes found in this study (top-left).  Solid blue lines show different $X_\mathrm{CO}$ values.  Down arrows illustrates upper limits of unresolved clouds.  The marginal distribution of the slope $\alpha$ from the Bayesian regression analysis is shown in the top-right panel, the intercept coefficient $A$ is shown in the bottom-left panel and the dispersion $\sigma$ is shown in the bottom-right panel, with $\mathrm{log}(\frac{M_\mathrm{vir}}{M_{\odot}})=A+\alpha \mathrm{log}(\frac{L_\mathrm{CO}}{ \mathrm{10^6\ K\ km\ s^{-1} pc^{2}}})+ \epsilon_\mathrm{scat}$.}
\label{fig_mvir_10}
\end{figure*}

Figure \ref{fig_sizew_21} shows the $\sigma_v-R$ plot for the $\cotwo$ clouds found in our sample of galaxies, and the values for the linear fit using Equation \ref{size_line10} are shown in Table \ref{table-bayes21}.  Similarly to the parameter distribution for complexes shown in Figure \ref{fig_sizew_10}, for clouds we observe that the distribution of the three parameters overlaps substantially for the three galaxies in our sample.  Although the slope distribution peaks at similar values in both NGC 6946  and M101 ($\sim 0.32$), the HDI for $p(\alpha)$ for M101 ($\sim$ 2.8) is wider than NGC 6946 ($\sim$ 0.8).  The intercept coefficient $A$ is larger for NGC 6946 clouds than for the M101 counterparts, indicating a larger mean velocity dispersion for clouds with similar sizes located in the NGC 6946.  Similarly to the behaviour observed at larger size scales in complexes, NGC 628 shows a nearly flat distribution of the size-line width relation ($\alpha_\mathrm{peak} \sim -0.1$ ) for clouds.  The line widths are smaller than 7 $\kms$, and generally smaller than the velocity dispersions observed in clouds with similar sizes located in the other two galaxies.

\begin{figure*}
\centering
\begin{tabular}{ccc}
\epsfig{file=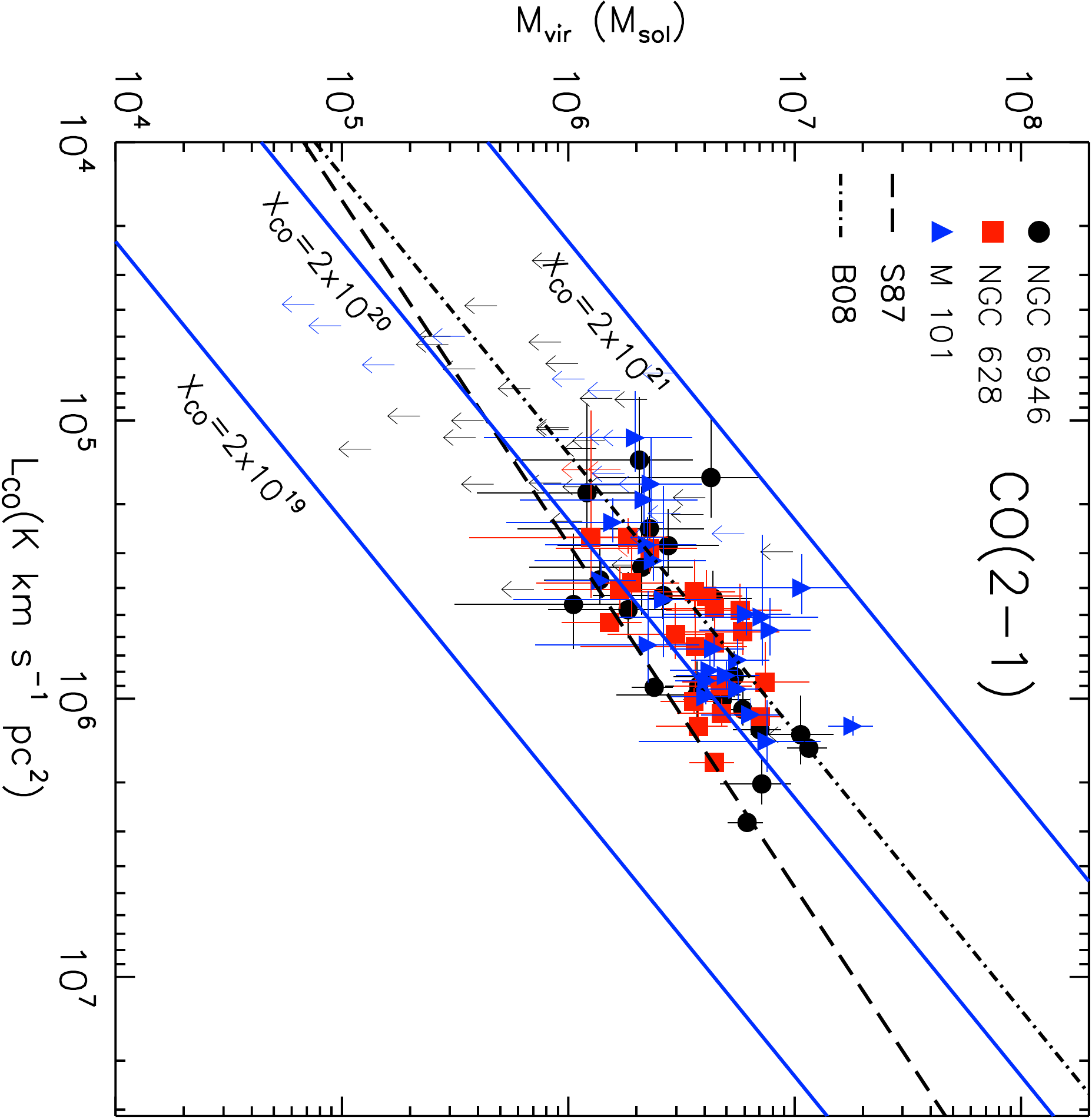,width=0.3\linewidth,angle=90}
\epsfig{file=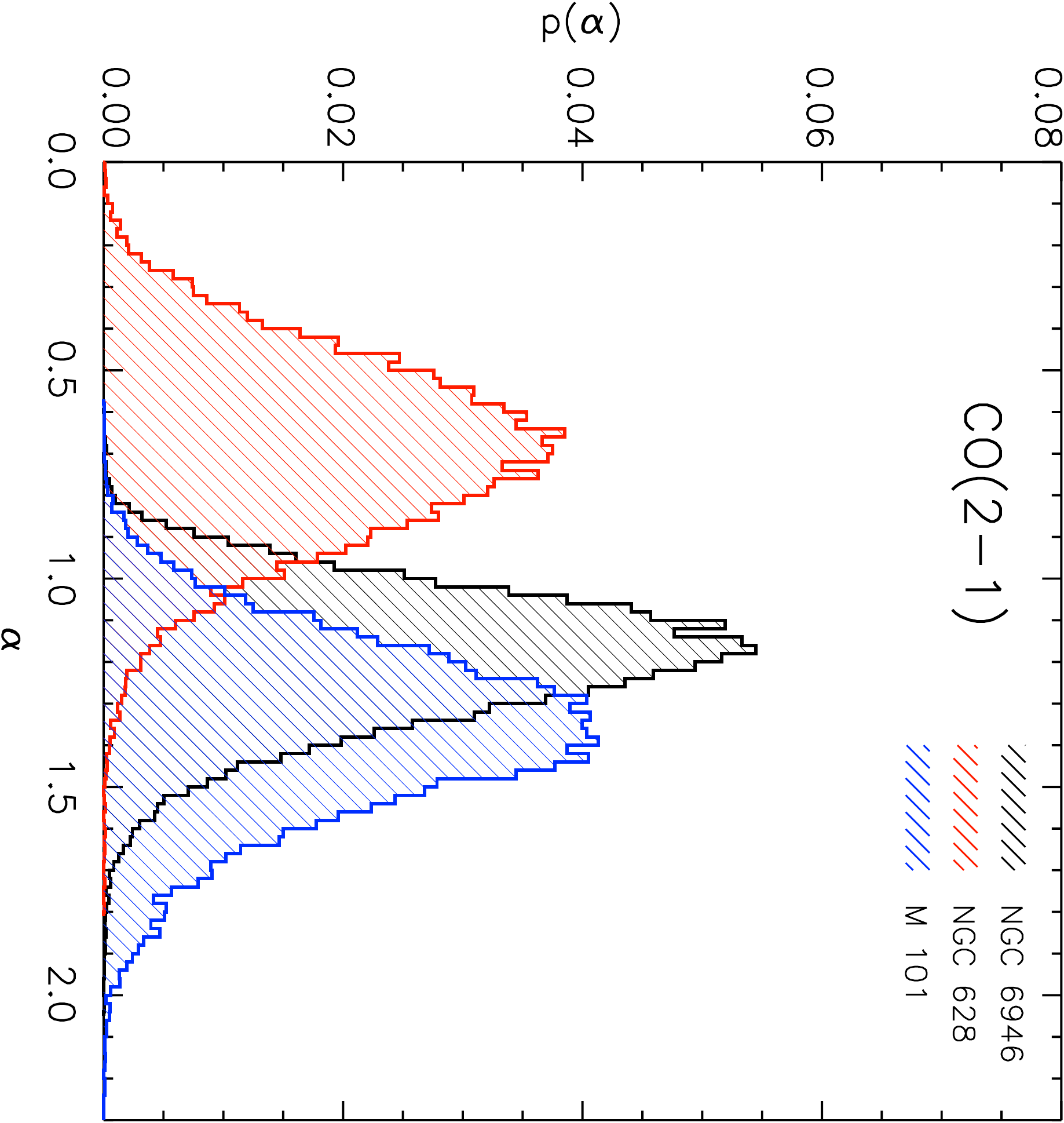,width=0.3\linewidth,angle=90}\\
\epsfig{file=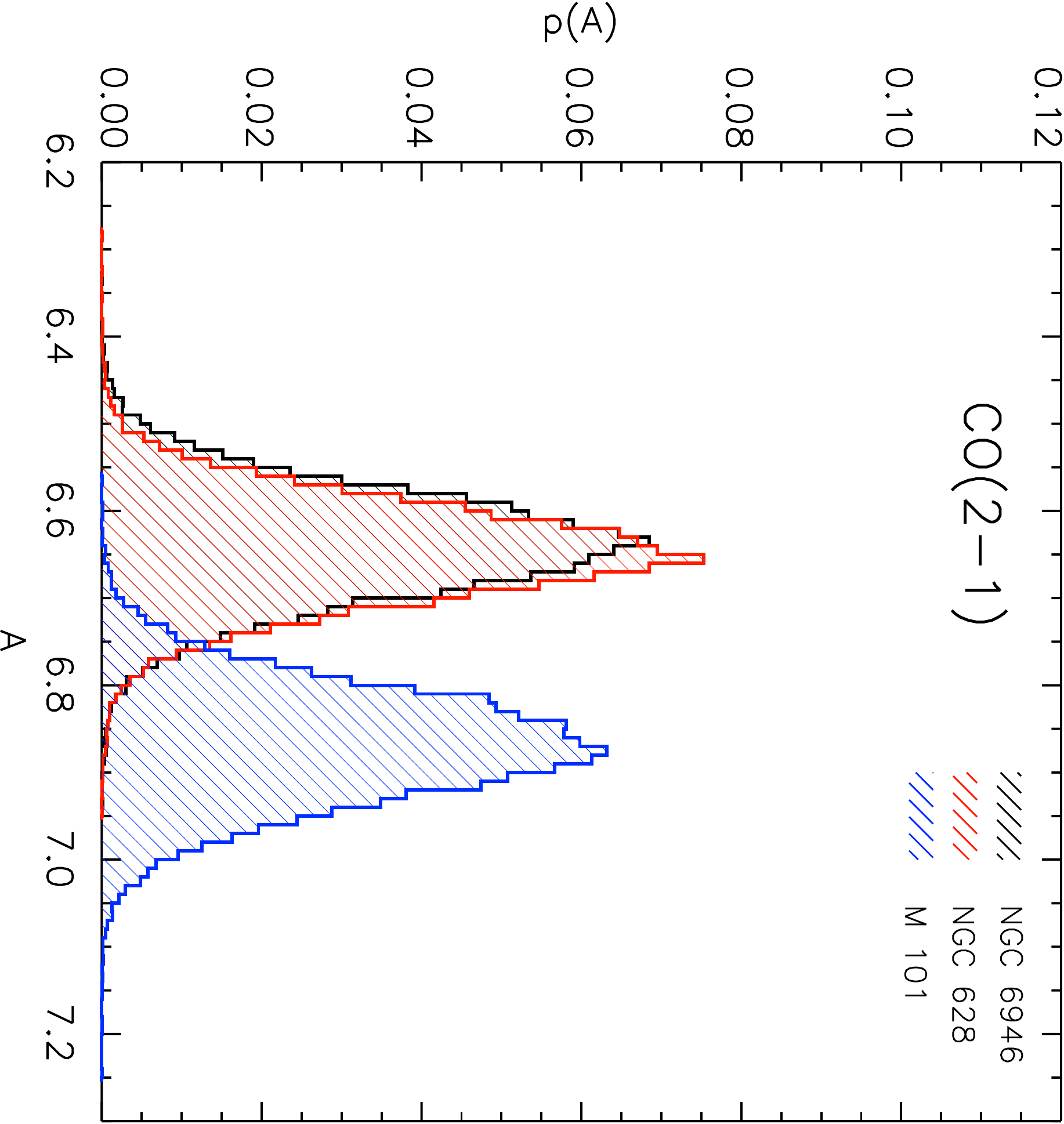,width=0.3\linewidth,angle=90}
\epsfig{file=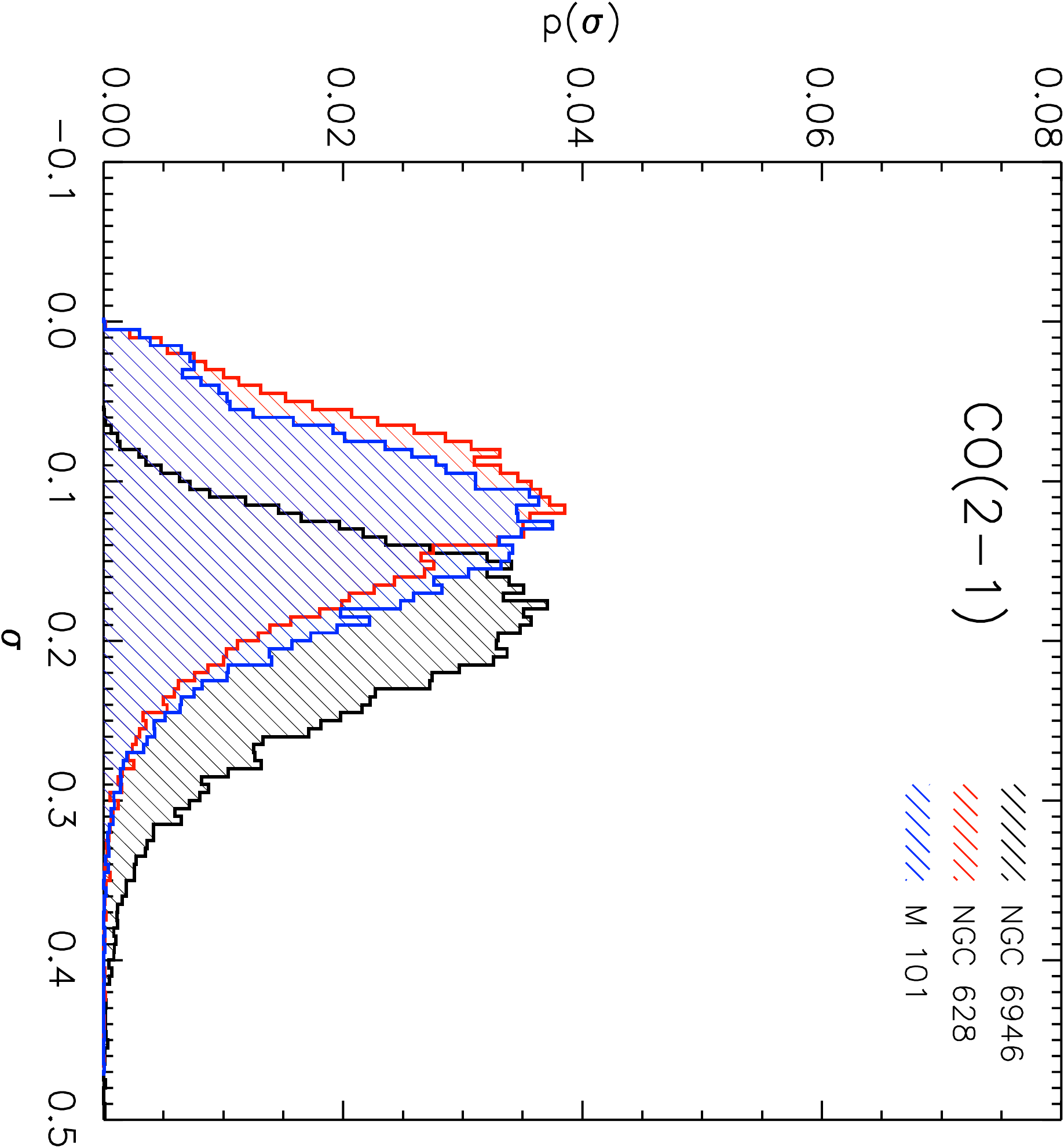,width=0.3\linewidth,angle=90}
\end{tabular}
\caption{The Virial mass-Luminosity relation for clouds (top-left).  Solid blue lines show different $X_\mathrm{CO}$ values.  Down arrows illustrates upper limits from unresolved clouds.  The panels showing the marginal distribution of the slope $\alpha$, the intercept coefficient $A$ and the dispersion of the scatter of the relation $\sigma$ from the Bayesian regression analysis are placed in the same way as Figure \ref{fig_mvir_10}.}
\label{fig_mvir_21}
\end{figure*}

\subsubsection{Virial mass-CO Luminosity relation}\label{mvir-lum}

The virial mass-luminosity relations for the complexes located in the three galaxies are shown in Figure \ref{fig_mvir_10}.  We observe that the complexes roughly fall along the relation defined by a constant CO-to-H$_\mathrm{2}$ factor $X_\mathrm{CO}=2\times10^{20}\mathrm{cm}^{-2}(\mathrm{K}\ \kms)^{-1}$, which was adopted in Equation (\ref{Mco-equ}).  Nevertheless, we notice that the complexes in M101 and NGC 628 show slightly smaller virial masses than NGC 6946 for the same CO luminosities.  Assuming virial equilibrium at such large scales, this difference can be interpreted as smaller values of $X_\mathrm{CO}$ in the regions surveyed in NGC 628 and M101 than the bright eastern part of NGC 6946.  This is consistent with the values reported by \citet{2013ApJ...777....5S} using dust maps as the tracer of the mass.  For the same regions we observed in NGC 628 and M101, they found mean values of the CO-to-H$_\mathrm{2}$ conversion factor slightly below ($X_\mathrm{CO}\sim1.4-1.7\times10^{20}\mathrm{cm}^{-2}(\mathrm{K}\ \kms)^{-1}$) the value used here.  

Similarly to the size-line width relation, we have normalized the CO luminosity of the complexes by a representative observed value, chosen to be $\mathrm{10^6\ K\ km\ s^{-1} pc^{2}}$, to find the best fit for the virial mass-luminosity relation .  Then, the relation utilized for the Bayesian regression fit is given by

\begin{equation}\label{mvir_lco10}
\mathrm{log}\left(\frac{M_\mathrm{vir}}{M_{\odot}}\right)=A+\alpha\ \mathrm{log}\left(\frac{L_\mathrm{CO}}{ \mathrm{10^6\ K\ km\ s^{-1} pc^{2}}}\right)+ \epsilon_\mathrm{scat},
\end{equation}

\noindent where $L_\mathrm{CO}$ is the $\co$ luminosity for complexes, and the $\cotwo$ luminosity for clouds.

Figure \ref{fig_mvir_10} shows the distributions of the linear fit parameters using the Bayesian inference method.  Following the procedure detailed by \citet{2007ApJ...665.1489K}, we have incorporated the upper limits of the virial masses in the fitting from spatially unresolved clouds.  According to this procedure, an additional indicator is introduced in the statistical model, which indicates whether a data point in the dependent variable is censored (i.\ e., non-detected) or not.  The fit is close to a linear relation for galaxies NGC 6946 and NGC 628, but with the latter showing more scatter in the slope distribution likely due to the small range in velocity dispersion observed in this galaxy (see Figure \ref{fig_sizew_10}).  M101 shows a slope peak $\alpha_\mathrm{peak} \sim 1.5$, and the $A_\mathrm{peak}$ intercept coefficient is smaller than the other two galaxies.  Moreover, NGC 628 shows a smaller $A$ coefficient than NGC 6946, consistent with the smaller $X_\mathrm{CO}$ inferred for NGC 628 and M101.  This distribution of the intrinsic scatter about the relation is similar for the three galaxies.  We have compared the values found in this paper to the values found by previous works.  The complexes identified in NGC 6946 are located along the $M_\mathrm{vir}-L_\mathrm{CO}$ relation found by B08, but are systematically above the relation found by S87 for Galactic clouds.  On the other hand, the low luminosity complexes identified in NGC 628 and M101 are closer to the Galactic values found by S87, while high luminosity complexes tend to be closer to the extragalactic values found by B08.  

The virial mass-luminosity relation for the $\cotwo$ clouds is shown in Figure \ref{fig_mvir_21}.  The clouds are roughly consistent with the CO-to-H$_\mathrm{2}$ conversion factor used here, and consistent with the relation found by B08 for extragalactic clouds.  As for complexes, here we have incorporated the upper limits in the fitting from unresolved clouds.  The slope distributions are slightly different for the three galaxies, with M101 showing the steepest relation ($\alpha_\mathrm{peak}$ $\sim$ 1.5) and NGC 628 showing the most shallow one ($\alpha_\mathrm{peak}$ $\sim$ 0.6).

\subsection{The Star Formation Law}
In this section we show the relation between the molecular gas and star formation surface density in the regions we observed in NGC 6946, NGC 628 and M101.  We begin by showing the resulting relations using the boundaries of the $\co$ complexes and $\cotwo$ clouds to define the area over which we calculate the quantities.  Then, we compare these results with the K-S relation calculated by using a uniform grid across the regions observed in the galaxies in our sample.

\subsubsection{$\Sigma_\mathrm{H2}$ vs.\ $\Sigma_\mathrm{SFR}$ relation based on identified structures}\label{k-s_law}
\begin{figure*}
\centering
\begin{tabular}{cc}
\epsfig{file=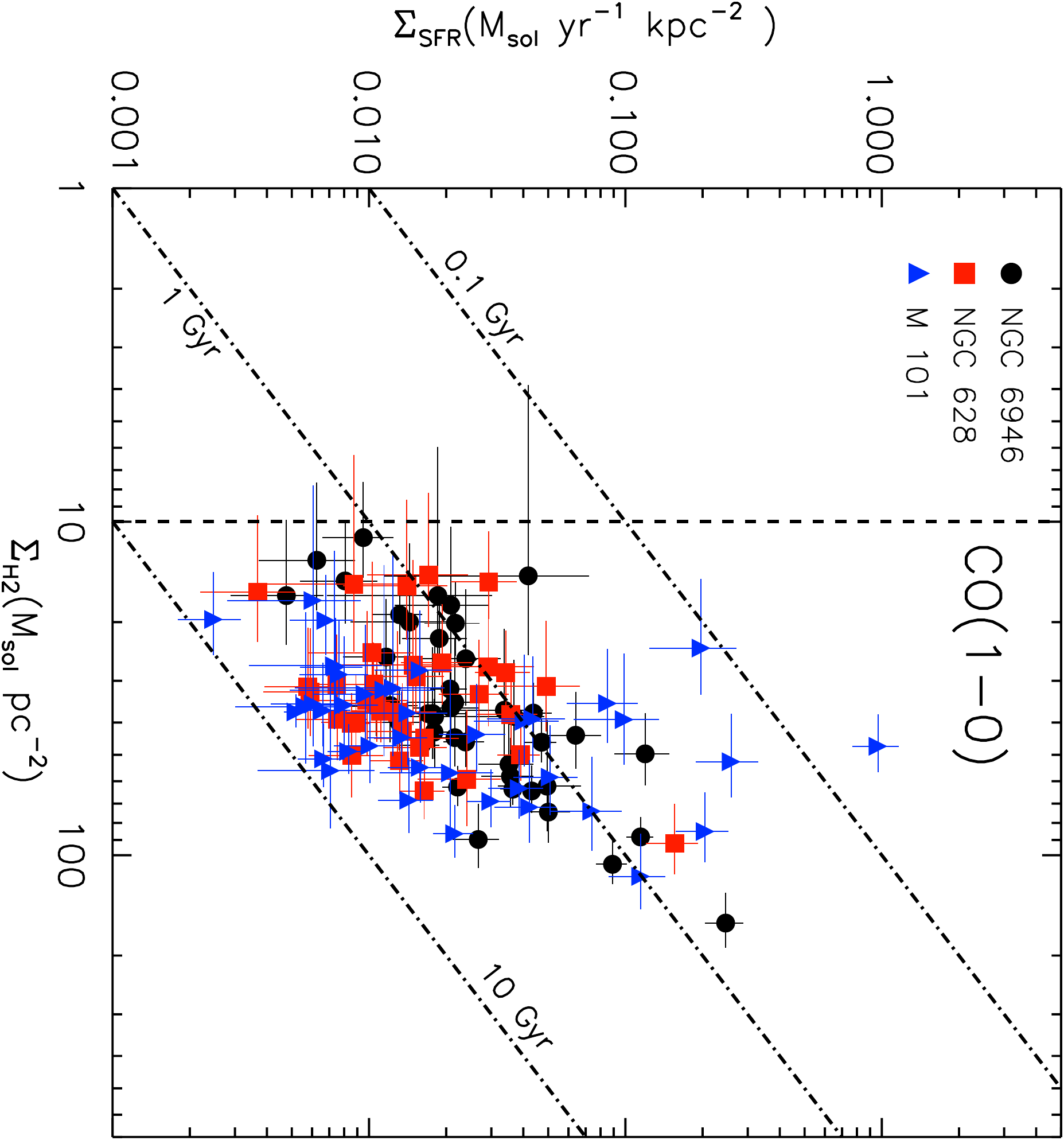,width=0.3\linewidth,angle=90}
\epsfig{file=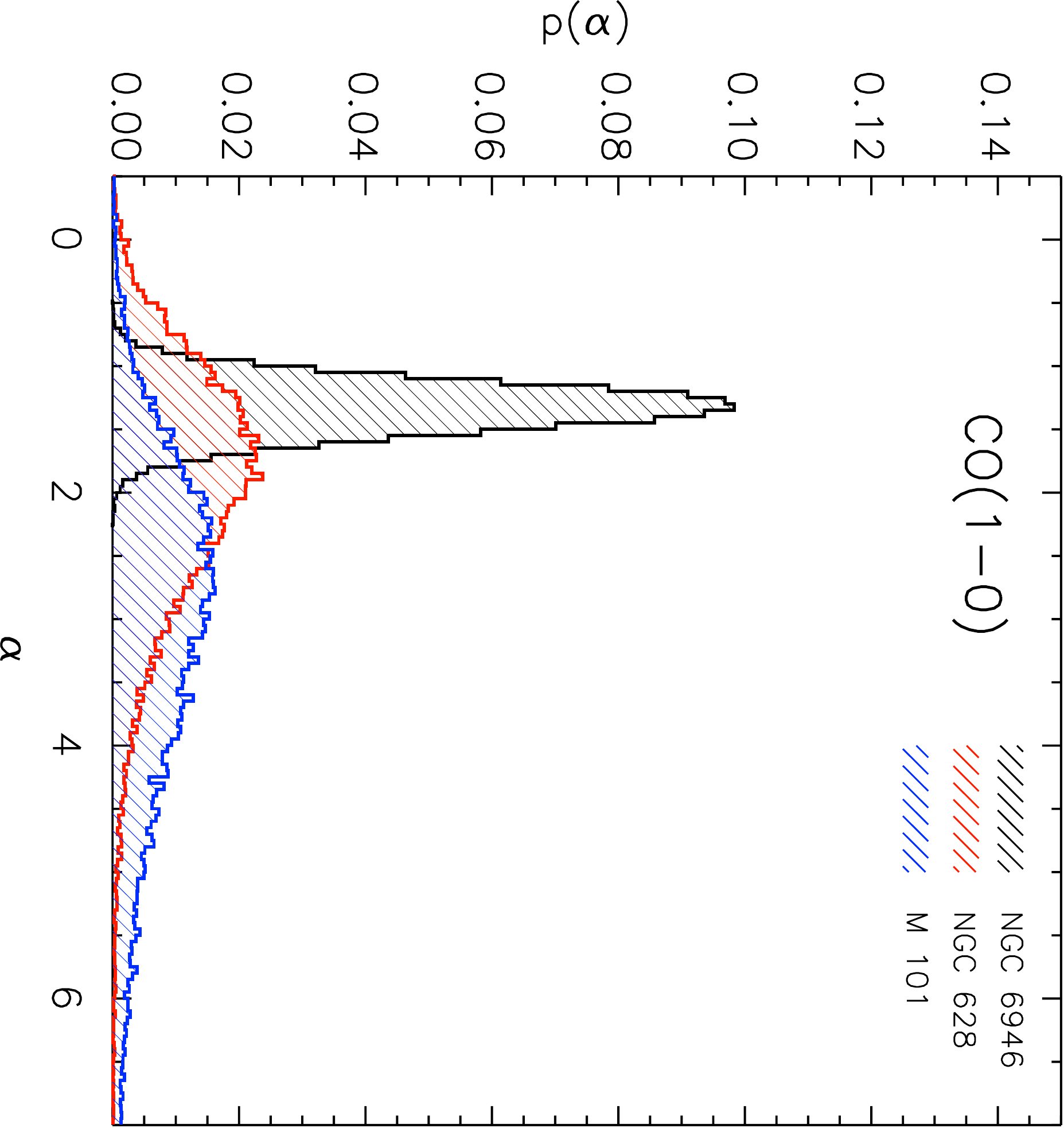,width=0.3\linewidth,angle=90}\\
\epsfig{file=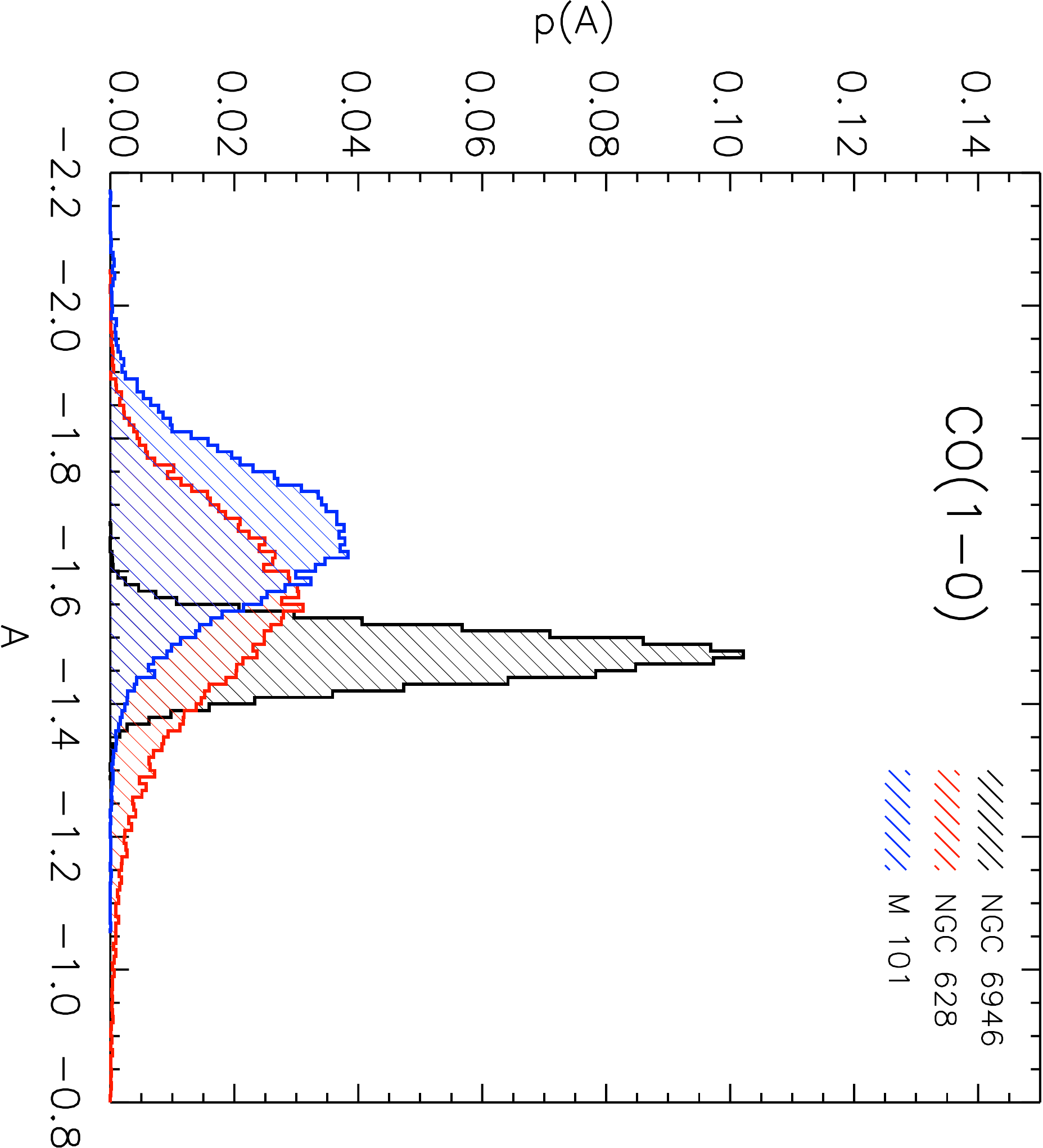,width=0.3\linewidth,angle=90}
\epsfig{file=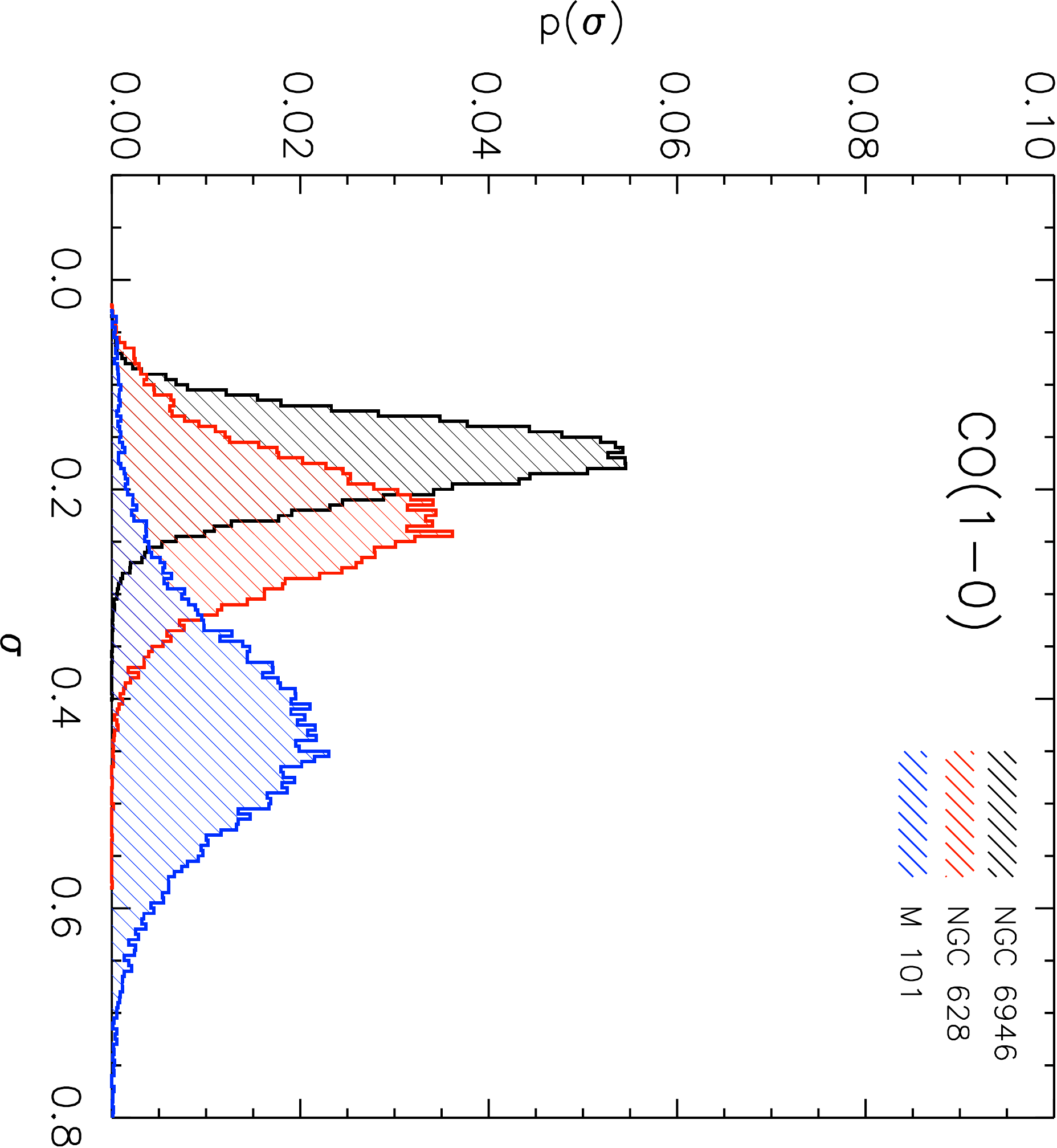,width=0.3\linewidth,angle=90}
\end{tabular}
\caption{Star formation rate vs.\ molecular gas surface density relation for $\co$ complexes (top-left).  The SFR is calculated from the FUV + 24$\mu$m maps.  Black dots represent the complexes identified in NGC 6946, red squares correspond to $\co$ complexes detected in NGC 628 and blue triangles represent complexes identified in M101.  The vertical dashed line illustrates the sensitivity limit of our maps, $\sim$10 $\Msun$ pc$^{2}$.  Black dot-dashed lines show constant molecular gas depletion times (SFE$^{-1}$) of 0.1, 1 and 10 Gyr.  The marginal distribution of the slope $\alpha$ from the Bayesian regression analysis is shown in the top-right panel, the intercept coefficient $A$ is shown in the bottom-left panel, and the dispersion of the scatter $\sigma$ is shown in the bottom-right panel, with log$(\frac{\Sigma_{\mathrm{SFR}}}{M_{\odot}\ \mathrm{yr}^{-1}\ \mathrm{kpc}^{-2}})=A+\alpha$ log$(\frac{\Sigma_{\mathrm{H2}}}{50\ M_{\odot}\ \mathrm{pc}^{-2}})+ \epsilon_\mathrm{scat}$.}
\label{figure_sfr10}
\end{figure*}

The left panel of Figure \ref{figure_sfr10} shows the $\Sigma_\mathrm{H2}$ vs.\ $\Sigma_\mathrm{SFR}$ relation for the $\co$ complexes identified in the three galaxies.  We observe that the complexes are, in general, located in the same area of the $\Sigma_\mathrm{H2}$ vs.\ $\Sigma_\mathrm{SFR}$ diagram.  However, complexes located in NGC 6946 seem to show a stronger $\Sigma_\mathrm{H2}$ vs.\ $\Sigma_\mathrm{SFR}$ correlation than the structures identified in the other two galaxies.  Additionally, there is a subset of complexes in M101 with SFR several times higher than the SFR values found for similar molecular gas surface densities found in NGC 6946 and NGC 628.

The Bayesian regression fitting to the $\Sigma_\mathrm{H2}$ vs.\ $\Sigma_\mathrm{SFR}$ relation is performed similarly to the fit performed on the scaling relations presented above.  In this case, we have normalized the molecular gas surface density by the representative value of $50\ M_{\odot}\ \mathrm{pc}^{-2}$.  Thus, the relation used for the fitting is given by

\begin{equation}\label{sfr_h2}
\mathrm{log}\left(\frac{\Sigma_{\mathrm{SFR}}}{M_{\odot}\ \mathrm{yr}^{-1}\ \mathrm{kpc}^{-2}}\right)=A+\alpha\ \mathrm{log}\left(\frac{\Sigma_{\mathrm{H2}}}{50\ M_{\odot}\ \mathrm{pc}^{-2}}\right)+ \epsilon_\mathrm{scat}.
\end{equation}

\begin{table*}
\caption{Bayesian regression parameters for both $\co$ complexes and $\cotwo$ clouds $\Sigma_\mathrm{H2}-\Sigma_\mathrm{SFR}$ relation.\label{table-bayes-sfr}}
\centering
\begin{tabular}{ccccccccc}
\hline\hline
Relation & $\alpha$ & 90\% HDI & & $A$ &  90\% HDI & & $\sigma$ &  90\% HDI \\
\hline
 & & & & \multicolumn{1}{c}{NGC 6946} &  &  & & \\
\hline
$\Sigma_\mathrm{H2}-\Sigma_\mathrm{SFR}$ $\co$ complexes & 1.33 & [0.97, 1.65] & & -1.48  & [-1.54, -1.41] & & 0.17  & [0.11, 0.23]   \\
$\Sigma_\mathrm{H2}-\Sigma_\mathrm{SFR}$ $\cotwo$ clouds &  1.57 & [1.06, 2.02]  & & -1.51 & [ -1.63, -1.39] & &  0.38  & [0.31, 0.47]\\
\hline
 & & & & \multicolumn{1}{c}{NGC 628}  & & & &\\
\hline
$\Sigma_\mathrm{H2}-\Sigma_\mathrm{SFR}$ $\co$ complexes & 1.38  & [0.43, 3.95]  & & -1.59 & [-1.78, -1.26] & & 0.24  & [0.13, 0.34] \\
$\Sigma_\mathrm{H2}-\Sigma_\mathrm{SFR}$ $\cotwo$ clouds & 1.98  & [0.96, 3.09]  & & -1.54  & [-1.75, -1.37] & & 0.51  & [0.39, 0.70] \\
\hline
 & & & & \multicolumn{1}{c}{M101}  & & & & \\
\hline
$\Sigma_\mathrm{H2}-\Sigma_\mathrm{SFR}$ $\co$ complexes & 2.48 & [1.09, 6.77] & & -1.65 & [-1.84, -1.47] & & 0.45 & [0.19, 0.59]  \\
$\Sigma_\mathrm{H2}-\Sigma_\mathrm{SFR}$ $\cotwo$ clouds &  2.58 & [1.81, 3.71] & &   -1.13   & [-1.34, -0.94] & & 0.58   & [0.44, 0.73]\\
\hline
\end{tabular}
\end{table*}

The distributions of the slope $\alpha$, the intercept coefficient $A$ and the dispersion of the intrinsic scatter $\sigma$ of the regression fit for $\co$ complexes are displayed in Figure \ref{figure_sfr10}.   Table \ref{table-bayes-sfr} shows the values of the regression parameters for complexes.  For NGC 6946, the distribution of the slope has a maximum at 1.33.  On the other hand, the peaks of the slope distributions are  $\sim 1.4$ and 2.5 for NGC 628 and M101 respectively.  In the case of the NGC 6946, the HDI of the distribution of $\alpha$ covers a range of $\sim$ 0.7, extending from 0.97 to 1.65.  For NGC 628, the HDI covers a range of $\sim$ 3.5, extending from 0.43 to 3.95.  Finally, the distribution of the slope of the K-S relation for complexes in M101 shows the largest range extending from 1.09 to 6.77.  Thus, the distribution of the slope is consistent with a super-linear relation for M101.  In the case of NGC 6946 and NGC 628, although the largest area of the distribution of the slope is located above 1, the HDI extends to sub-linear values, especially for NGC 628.  By comparing the distributions of the intrinsic scatter about the $\Sigma_\mathrm{H2}$ vs.\ $\Sigma_\mathrm{SFR}$ relation for the three galaxies, we observe that NGC 6946 offers the tightest relation among the three galaxies in the sample (peak of $\sigma \sim$ 0.17), although significant overlap with the distribution of the intrinsic scatter in NGC 628 is seen.  On the other hand, the intrinsic scatter is larger in M101 (peak of $\sigma$ $\sim 0.45$) due to the large range of SFR values observed in $\co$ complexes, in contrast to the narrow dynamic range in molecular gas surface density values.

\begin{figure*}
\centering
\begin{tabular}{ccc}
\epsfig{file=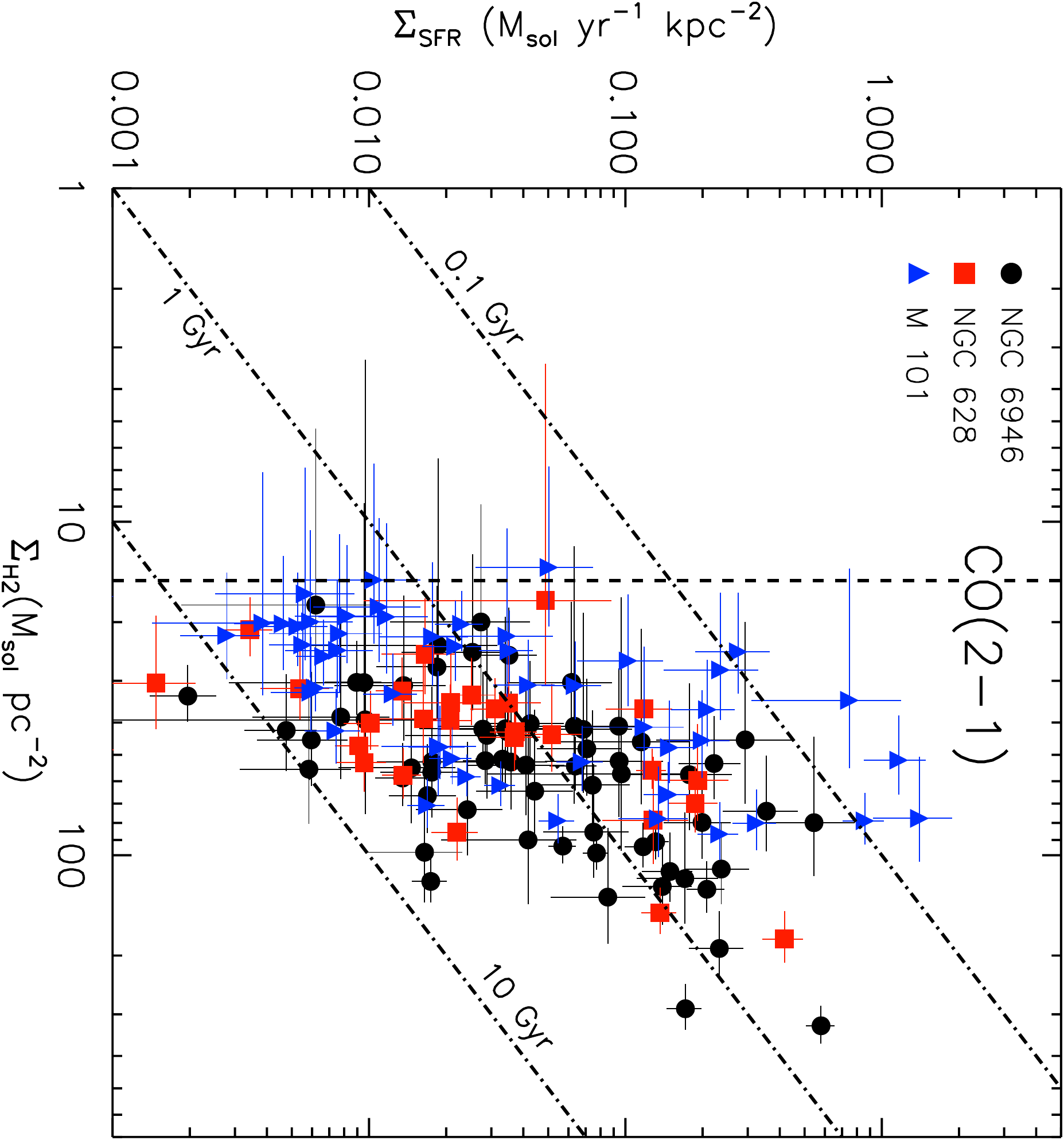,width=0.3\linewidth,angle=90}
\epsfig{file=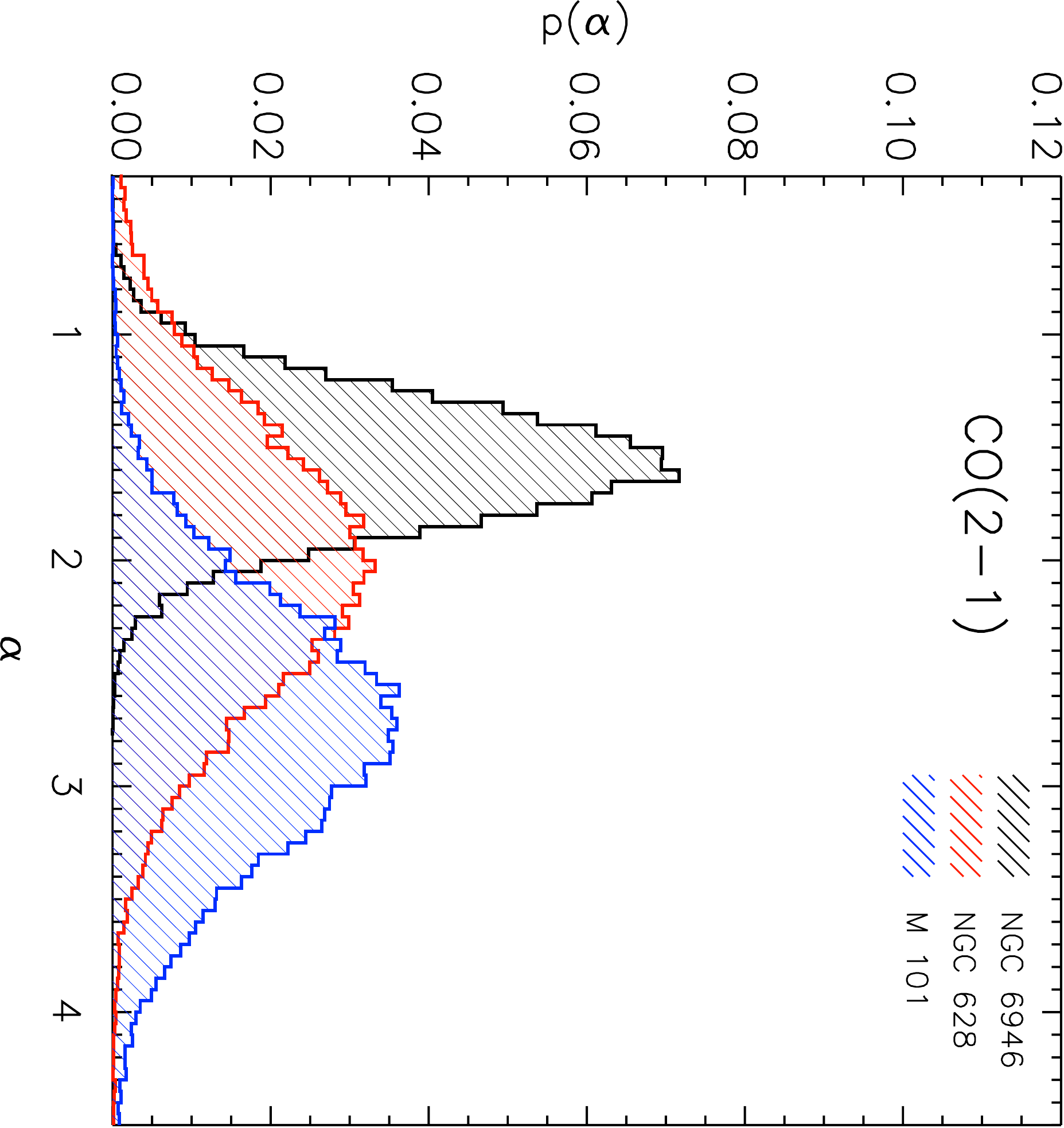,width=0.3\linewidth,angle=90}\\
\epsfig{file=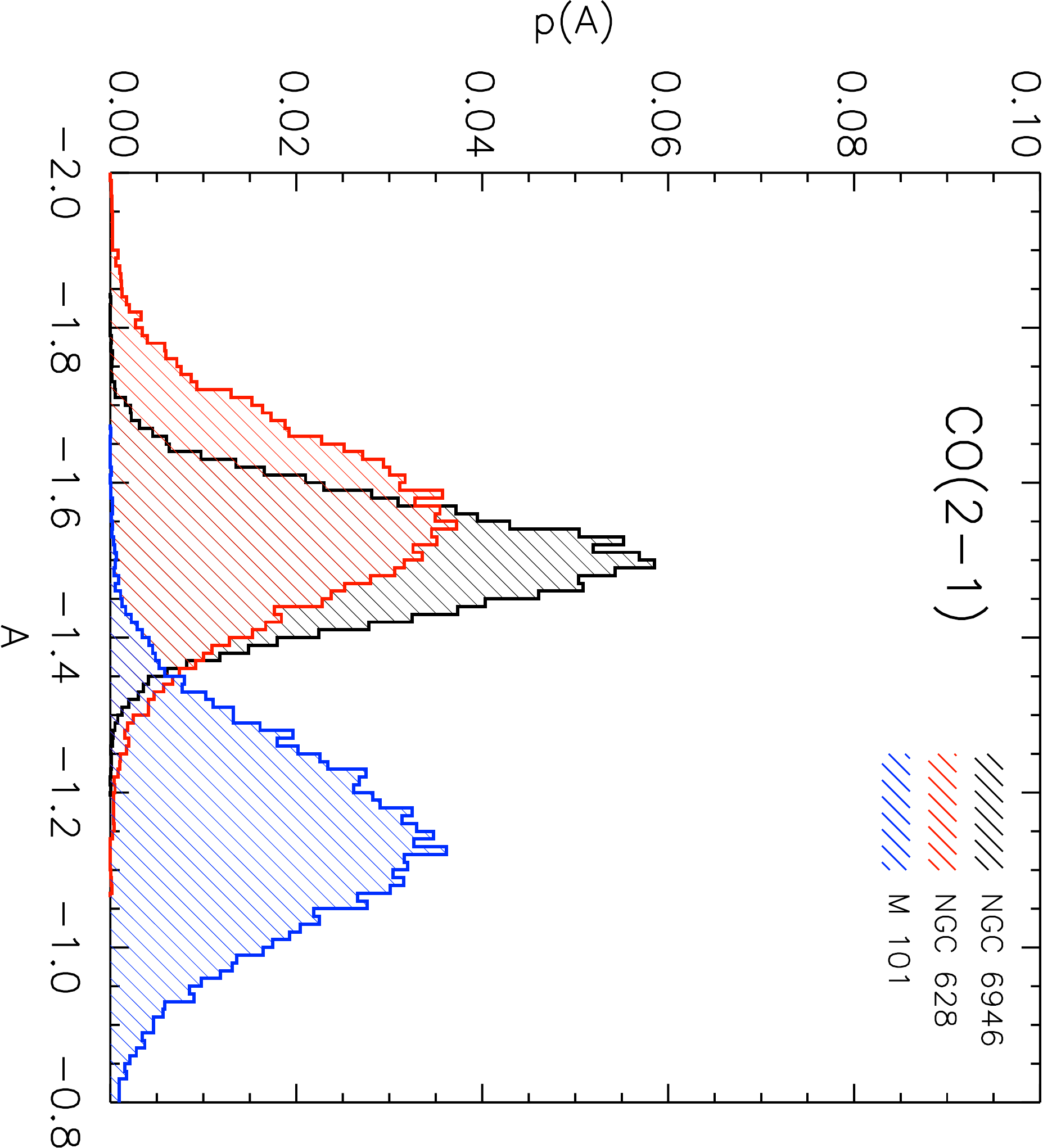,width=0.3\linewidth,angle=90}
\epsfig{file=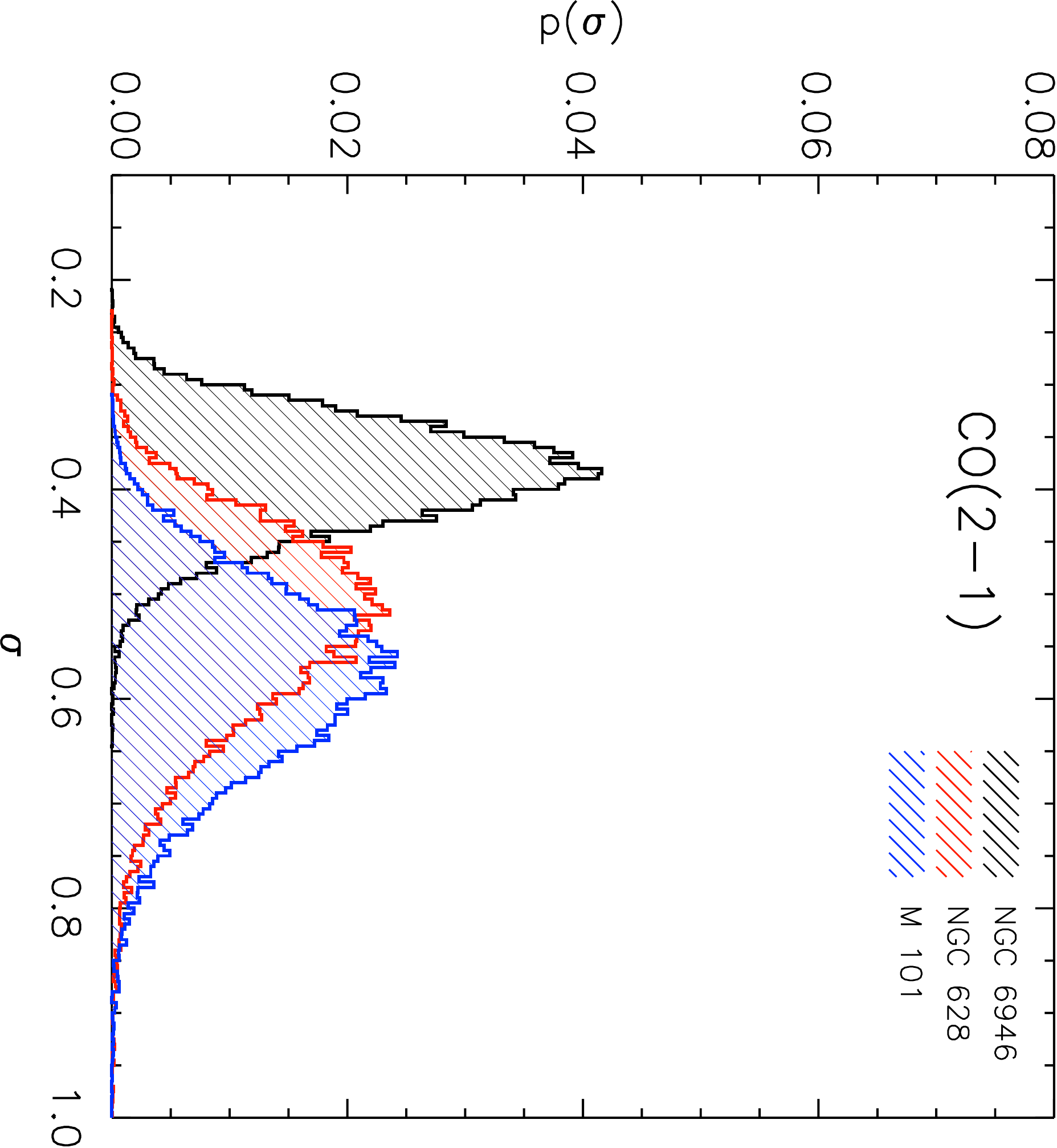,width=0.3\linewidth,angle=90}
\end{tabular}
\caption{Star formation rate vs.\ molecular gas surface density relation for $\cotwo$ clouds (top-left). Symbols and lines are the same as in Figure \ref{figure_sfr10}, except the sensitivity limit is $\sim$15 $\Msun$ pc$^{2}$.  The marginal distribution of the slope $\alpha$, the intercept coefficient $A$, and the dispersion of the scatter $\sigma$ from the linear Bayesian regression analysis are shown in the same panel configuration as Figure \ref{figure_sfr10}, with log$(\frac{\Sigma_{\mathrm{SFR}}}{M_{\odot}\ \mathrm{yr}^{-1}\ \mathrm{kpc}^{-2}})=A+\alpha$ log$(\frac{\Sigma_{\mathrm{H2}}}{50\ M_{\odot}\ \mathrm{pc}^{-2}})+ \epsilon_\mathrm{scat}$.}
\label{figure_sfr21}
\end{figure*}


Figure \ref{figure_sfr21} shows the relation of $\Sigma_\mathrm{H2}$ vs.\ $\Sigma_\mathrm{SFR}$ for $\cotwo$ clouds.  In Table \ref{table-bayes-sfr} we show the values of the regression parameters for clouds.  The general trend is similar to that observed for complexes: a tighter correlation holds for the clouds identified in NGC 6946.  The distributions of the slope show peaks at 1.57, 1.98 and 2.58 for NGC 6946, NGC 628 and M101 respectively.  Overall, clouds show similar $\Sigma_\mathrm{H2}$ vs.\ $\Sigma_\mathrm{SFR}$ relations to those found for large structures identified as complexes.  In the case of NGC 6946 and M101, the HDI of the probability distribution of slopes is consistent with super-linear relations.  For NGC 628, the HDI extends to values below 1.  By comparing the intrinsic scatter for complexes and clouds shown in Figure \ref{figure_sfr10} and \ref{figure_sfr21} respectively, we observe that the scatter about the $\Sigma_\mathrm{H2}$ vs.\ $\Sigma_\mathrm{SFR}$ relation increases for $\cotwo$ clouds.  More scatter could have been introduced by assuming a constant H$\alpha$ extinction factor in our SFR calculations for $\cotwo$ clouds as was mentioned in Section \ref{sfr-co21}.   Additionally, the higher scatter in the K-S relation for $\cotwo$ clouds may be the result of a relatively narrow range of observed $\cotwo$ brightness, added to the intrinsic scatter introduced to the relation by local variations of the molecular gas surface density and star formation at smaller scales than those traced by $\co$ observations.  Given the small regions sampled by our $\co$ and $\cotwo$ observations (with a bias to regions of bright CO emission), it is not possible to generalize the $\Sigma_\mathrm{H2}$ vs.\ $\Sigma_\mathrm{SFR}$ relation found for complexes and clouds as a global trend.  In the next section we will compare the relations that we have found based on $\co$ complex boundaries with the relation derived by using a uniform grid covering the $\co$ and SFR maps, and we investigate the effect of non-detections in the x-axis of the K-S relation on the derived linear relations.

\begin{figure*}
\centering
\begin{tabular}{c}
\epsfig{file=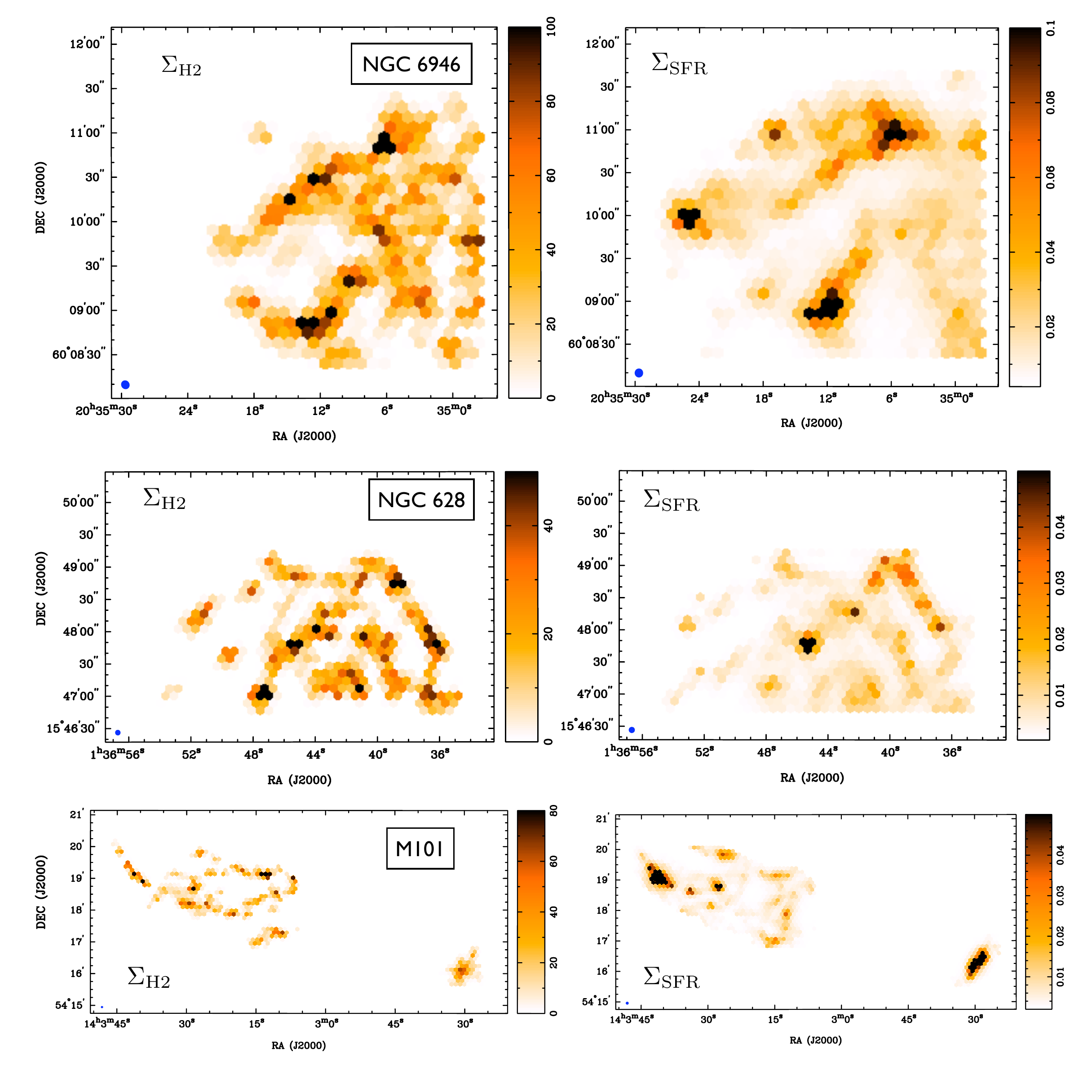,width=0.9\linewidth,angle=0}
\end{tabular}
\caption{Left: Molecular gas surface density for $\co$ map using a uniform grid across the the areas observed in NGC 6946 (top), NGC 628 (middle) and M101 (bottom).  Color bar is in units of $\Msun$ pc$^{2}$.  Right:  Star formation surface density using the same uniform grid approach as in the right panel for NGC 6946 (top), NGC 628 (middle) and M101 (bottom).  Color bar is in units of $M_{\odot}\ \mathrm{yr}^{-1}\ \mathrm{kpc}^{-2}$.}
\label{figure_maps_grid}
\end{figure*}

\begin{figure*}
\centering
\begin{tabular}{ccc}
\epsfig{file=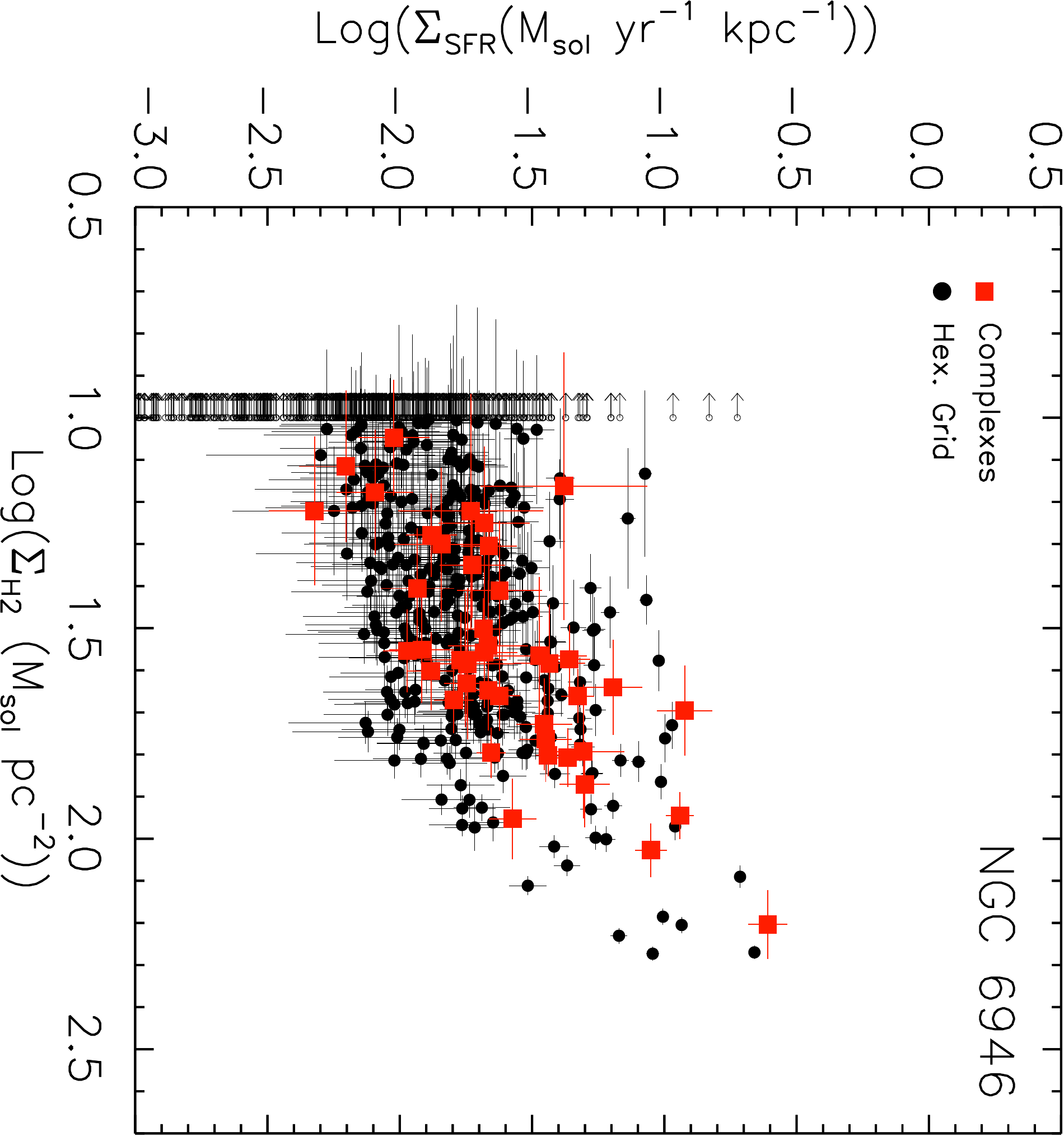,width=0.3\linewidth,angle=90} 
\epsfig{file=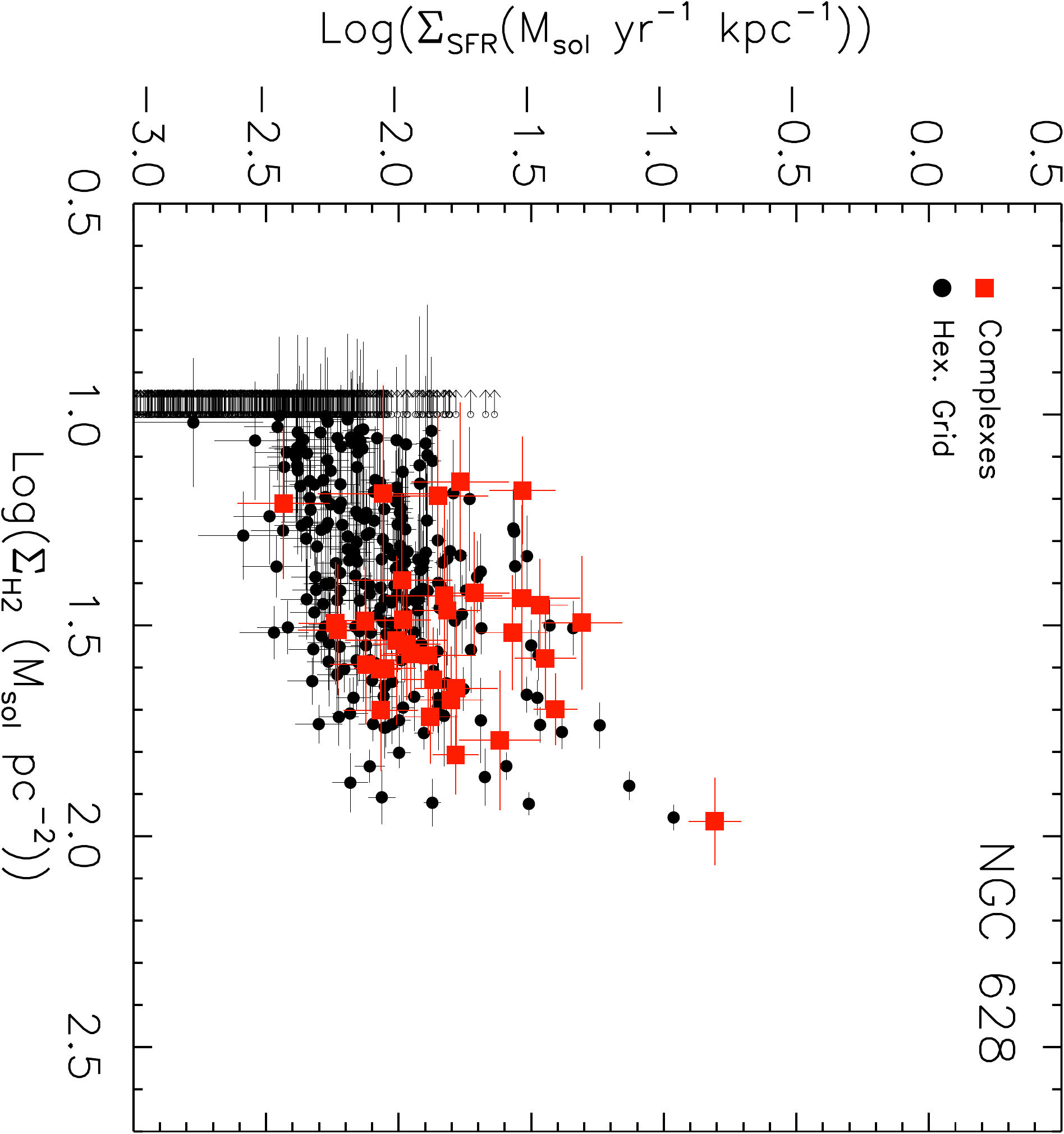,width=0.3\linewidth,angle=90} 
\epsfig{file=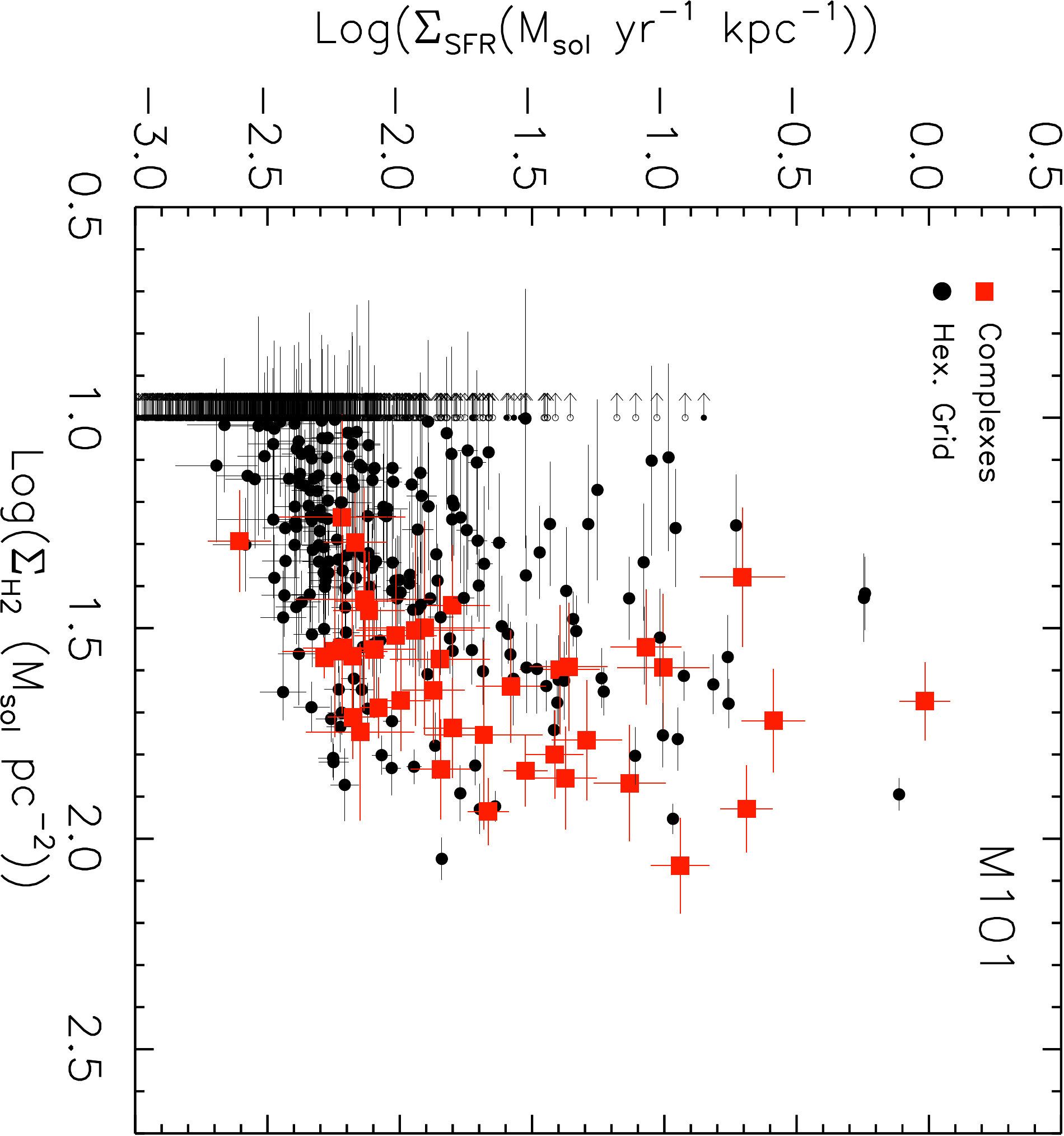,width=0.3\linewidth,angle=90} 
\end{tabular}
\caption{$\Sigma_\mathrm{H2}$ vs.\ $\Sigma_\mathrm{SFR}$ relations using a uniform grid (black dots), and using the $\co$ complexes (red squares) for the three galaxies in our sample NGC 6946 (left), NGC 628 (middle) and M101 (right).  Black arrows illustrate the $\Sigma_\mathrm{H2}$ non-detections at the sensitivity limit of $10\ \Msun\ \mathrm{pc}^{-2}$.}
\label{figure_sfr_h2_grid}
\end{figure*}

\subsubsection{$\Sigma_\mathrm{H2}$ vs.\ $\Sigma_\mathrm{SFR}$ using a uniform grid}\label{k-s_grid}
So far, we have used the sizes of the identified structures (complexes and clouds) to define the area over which to calculate the properties.  That approach is desired (and needed) when the properties require the knowledge of the size and line width of the region under study.  Nevertheless, comparison of star formation and molecular gas surface density require some other considerations.  For instance, the $\Sigma_\mathrm{H2}$ vs.\ $\Sigma_\mathrm{SFR}$ relation is affected by sensitivity bias in both axes.  In our $\co$ maps, our $\Sigma_\mathrm{H2}$ detection limit is $\sim$ 10 $\Msun$ pc$^{-2}$, so in our survey we may be missing low-brightness molecular clouds, more likely to be located in the inter-arm regions.  Several statistical approaches have been implemented to correct for such bias (\citealt{2013AJ....146...19L}; \citealt{2009ApJ...704..842B}).  Additionally, including the full coverage of the region with observations is an important factor in the observed $\Sigma_\mathrm{H2}$ vs.\ $\Sigma_\mathrm{SFR}$ relation.  Defining sample regions using the structure boundaries (as we did in Section \ref{k-s_law}) may lead us to miss of some CO emission that is not attributed to any identified complex or cloud.  An additional bias could be introduced by selecting young clouds with incipient star formation and abundant molecular gas, but excluding more evolved star-forming regions that have destroyed their natal clouds and are weak in CO emission.

\begin{figure*}
\centering
\begin{tabular}{cc}
\epsfig{file=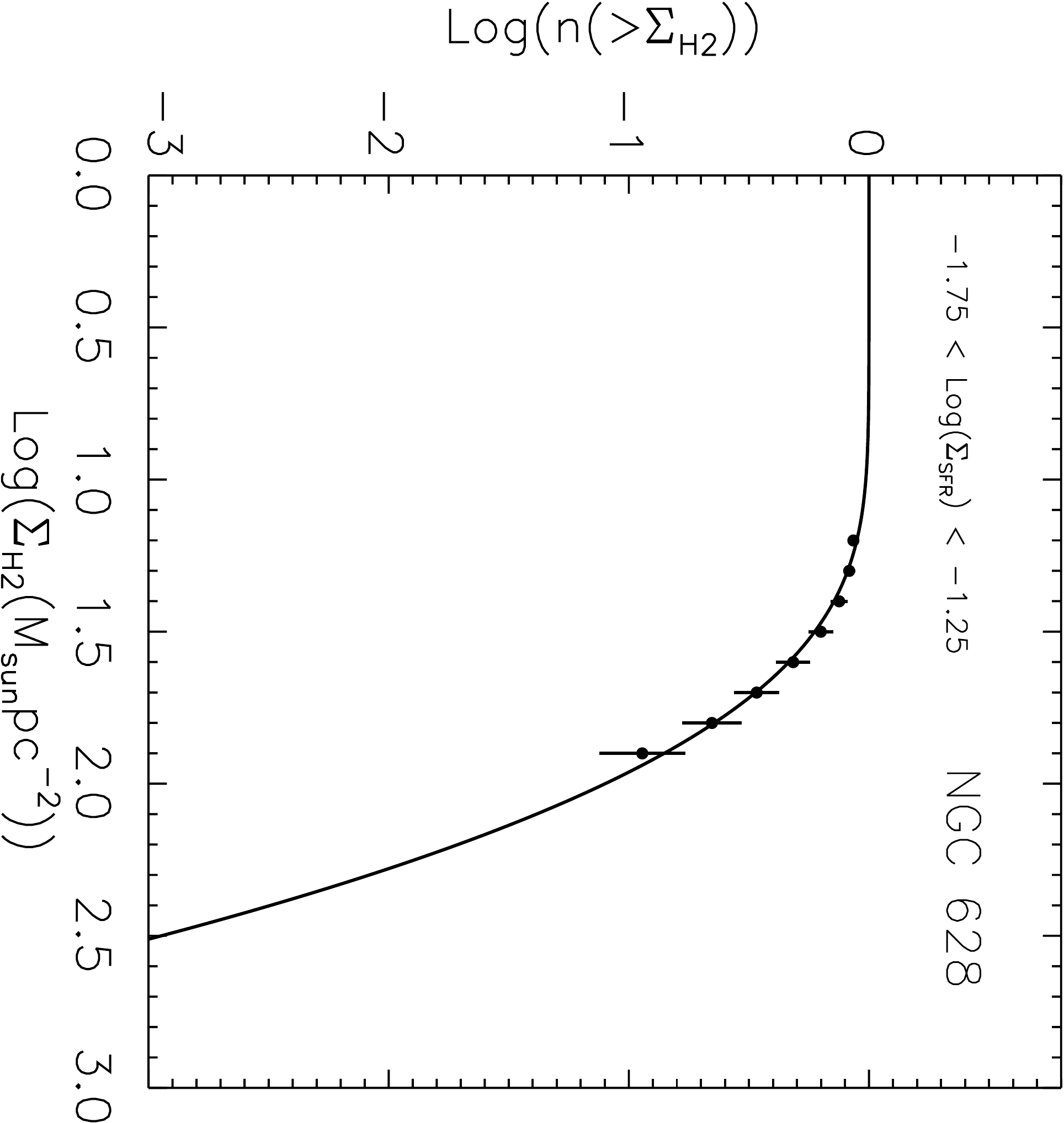,width=0.2\linewidth,angle=90} 
\epsfig{file=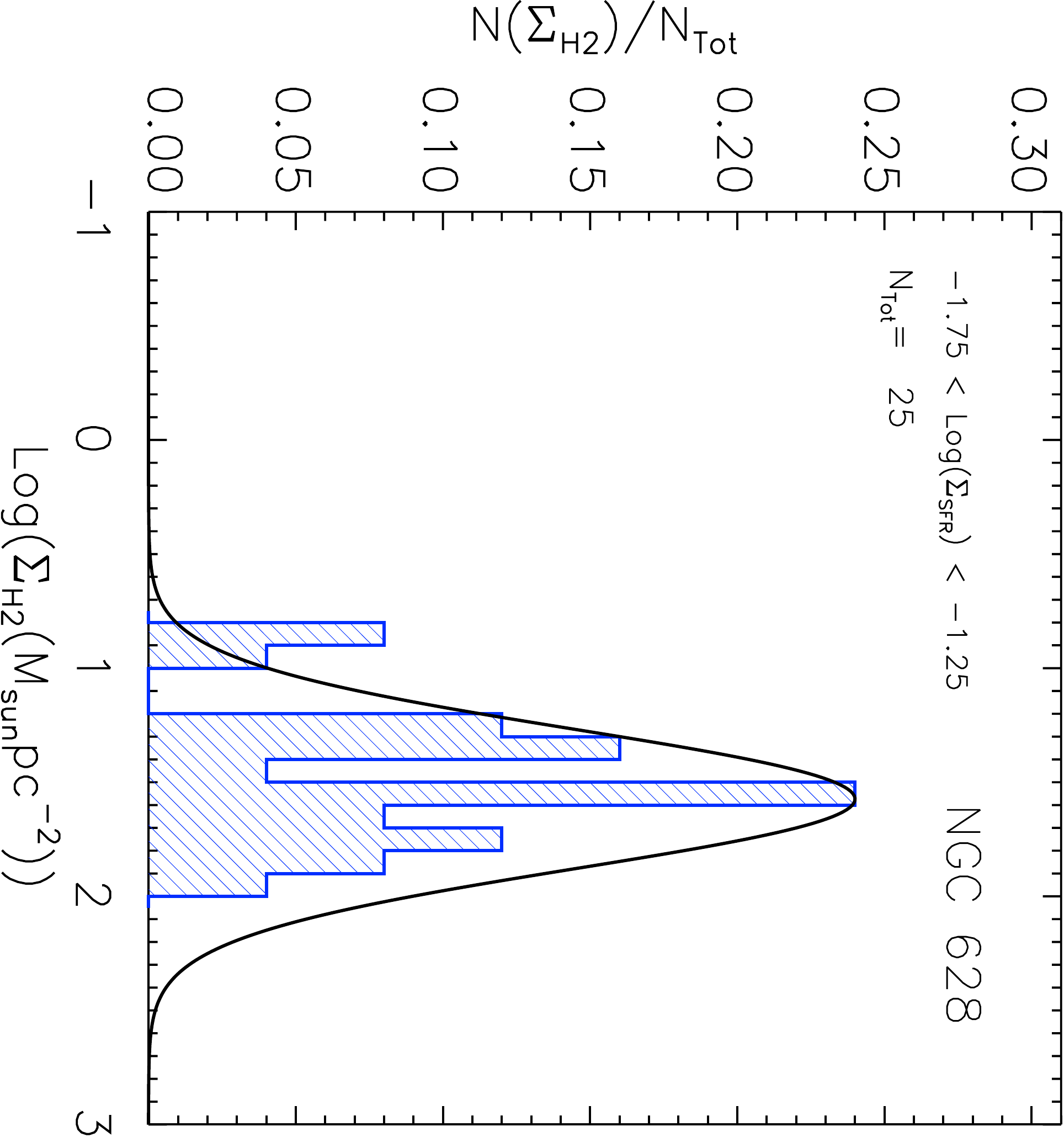,width=0.2\linewidth,angle=90} \\
\epsfig{file=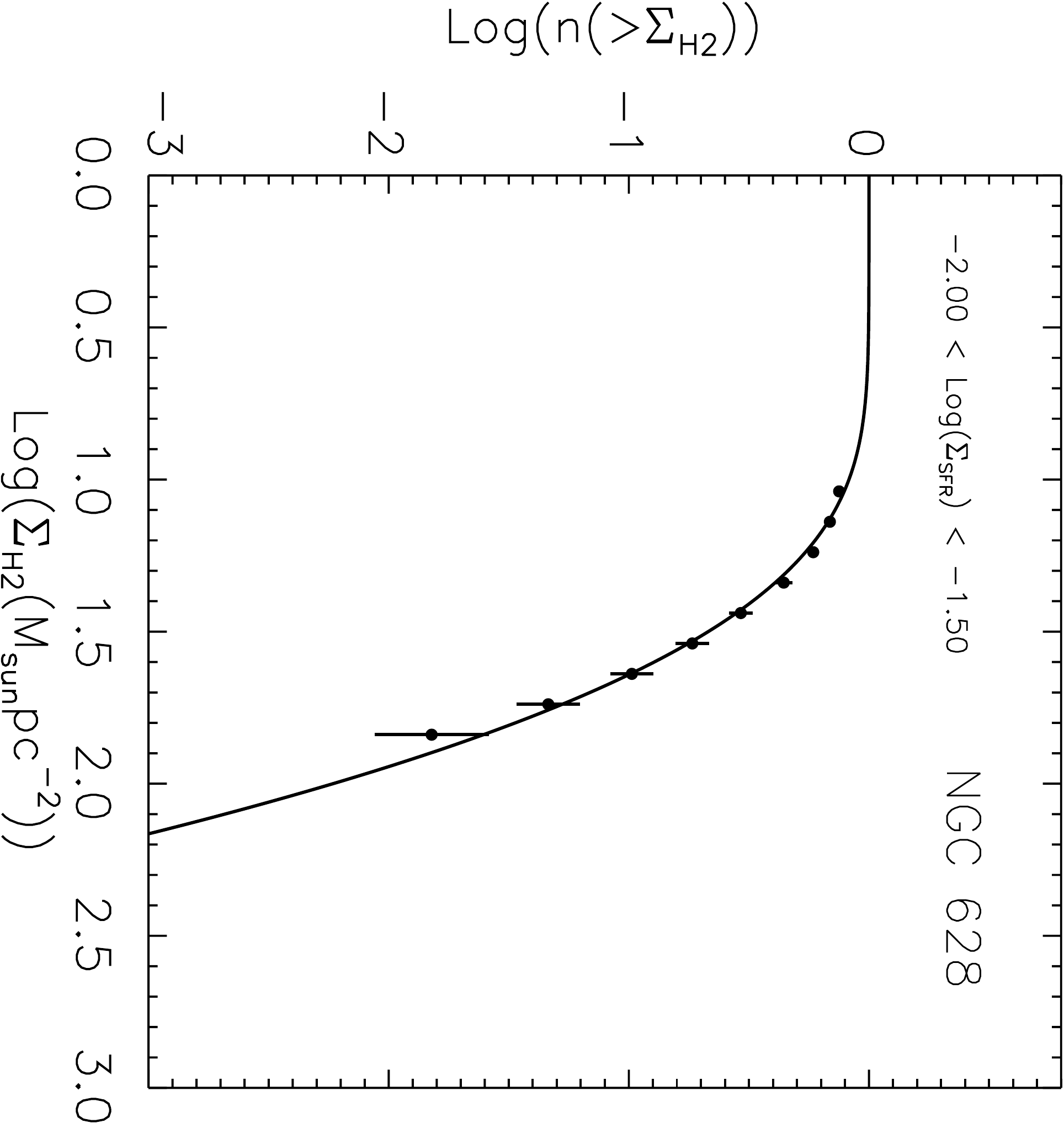,width=0.2\linewidth,angle=90} 
\epsfig{file=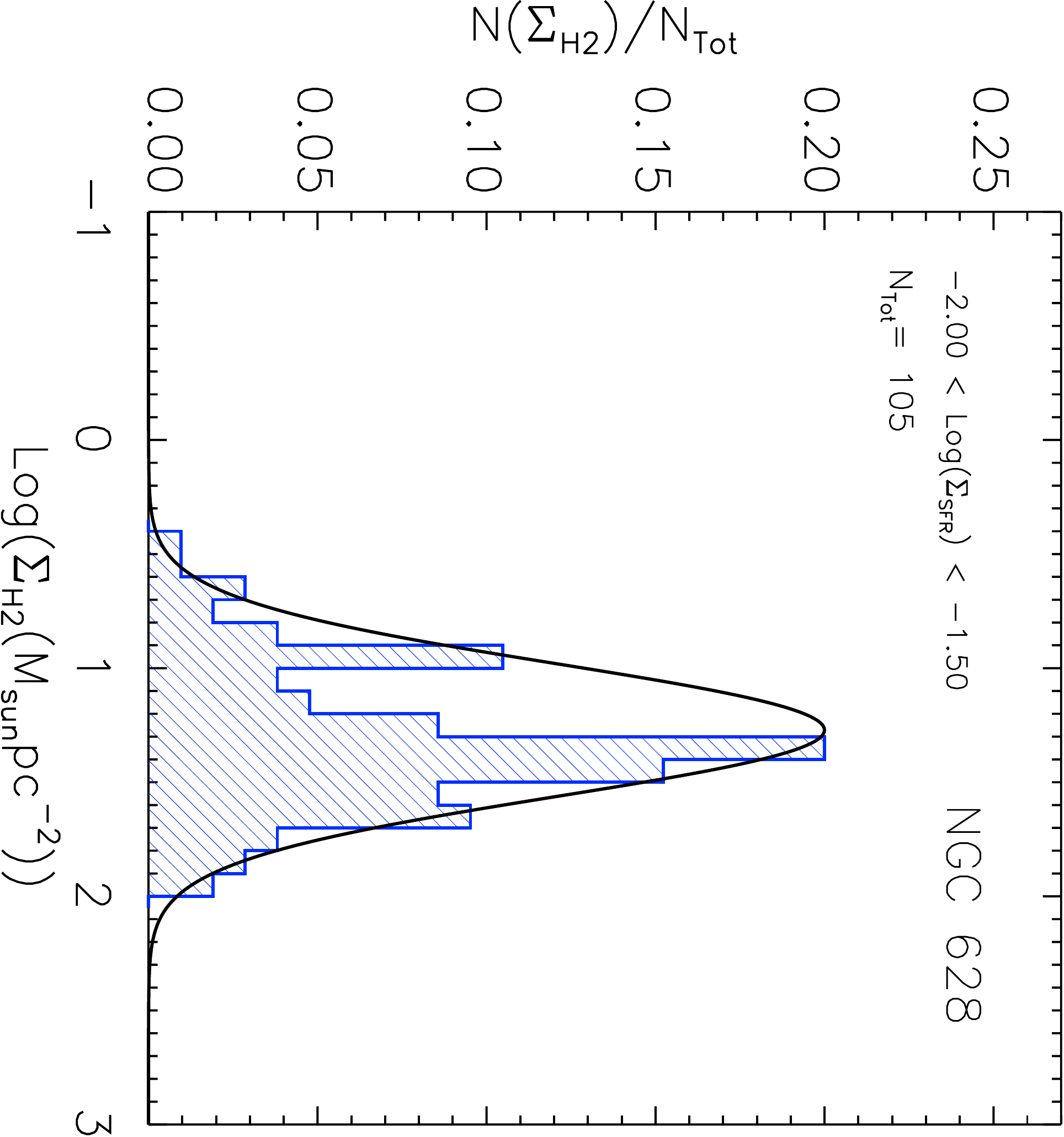,width=0.2\linewidth,angle=90} \\
\epsfig{file=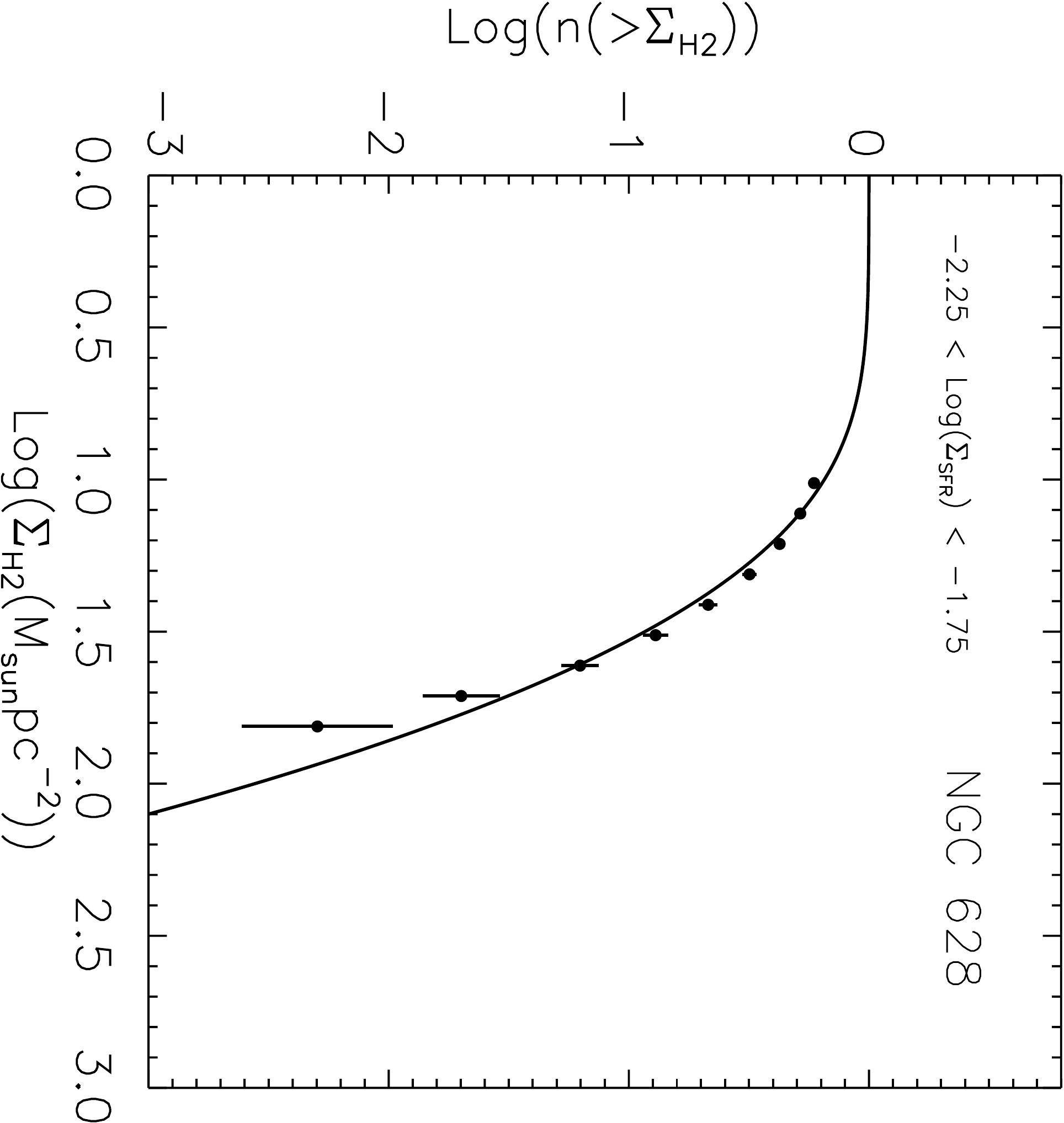,width=0.2\linewidth,angle=90} 
\epsfig{file=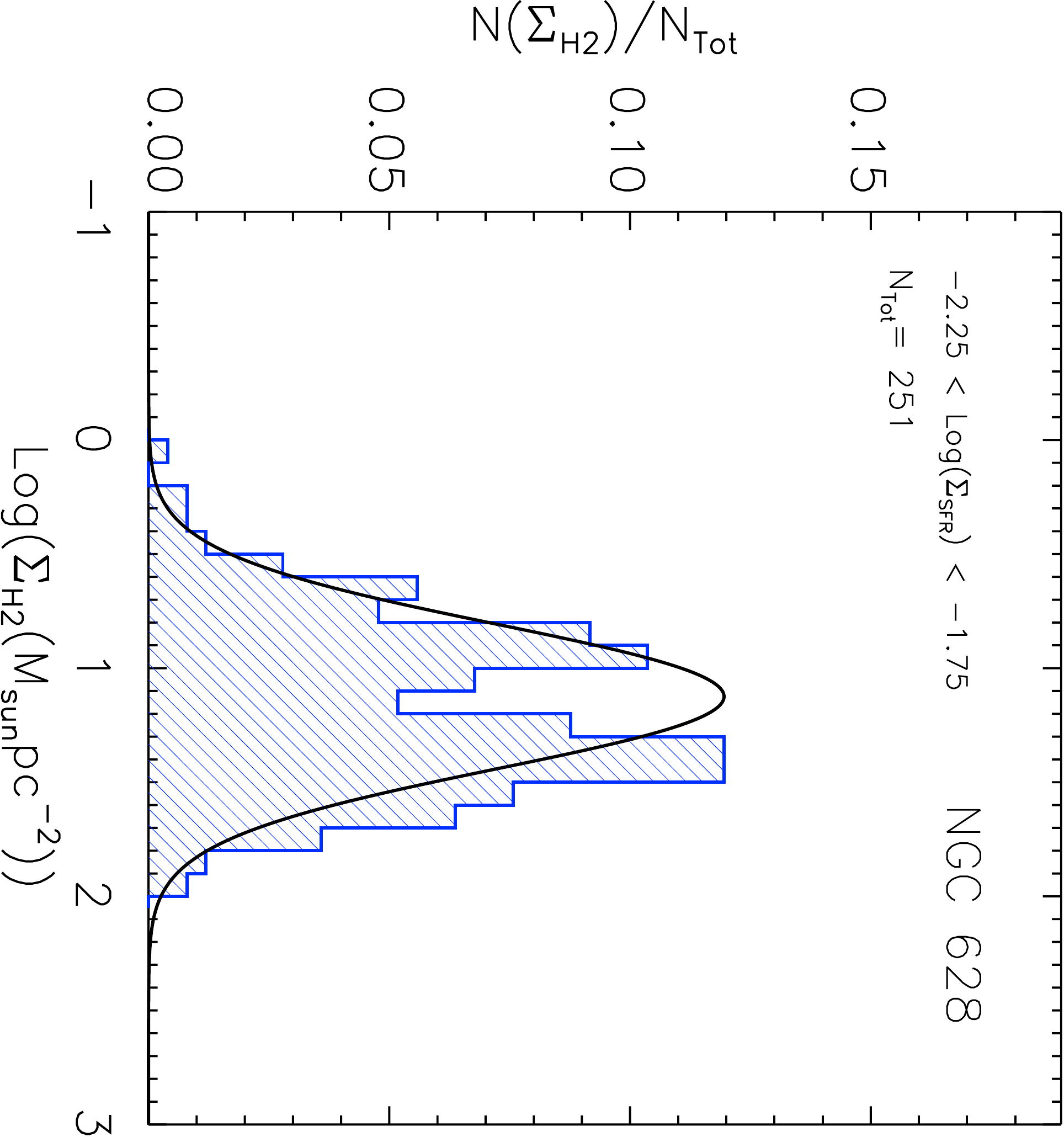,width=0.2\linewidth,angle=90} \\
\epsfig{file=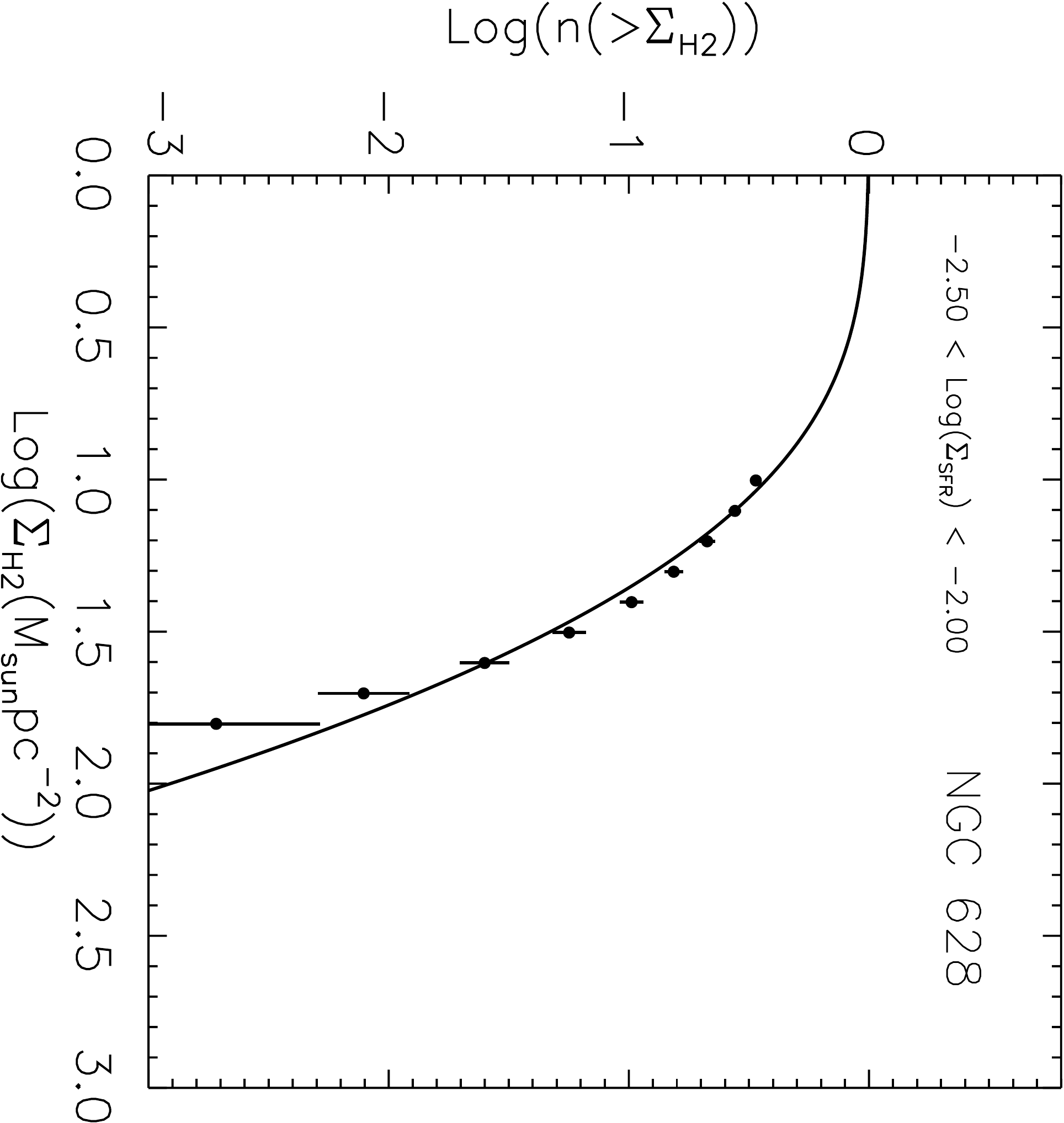,width=0.2\linewidth,angle=90} 
\epsfig{file=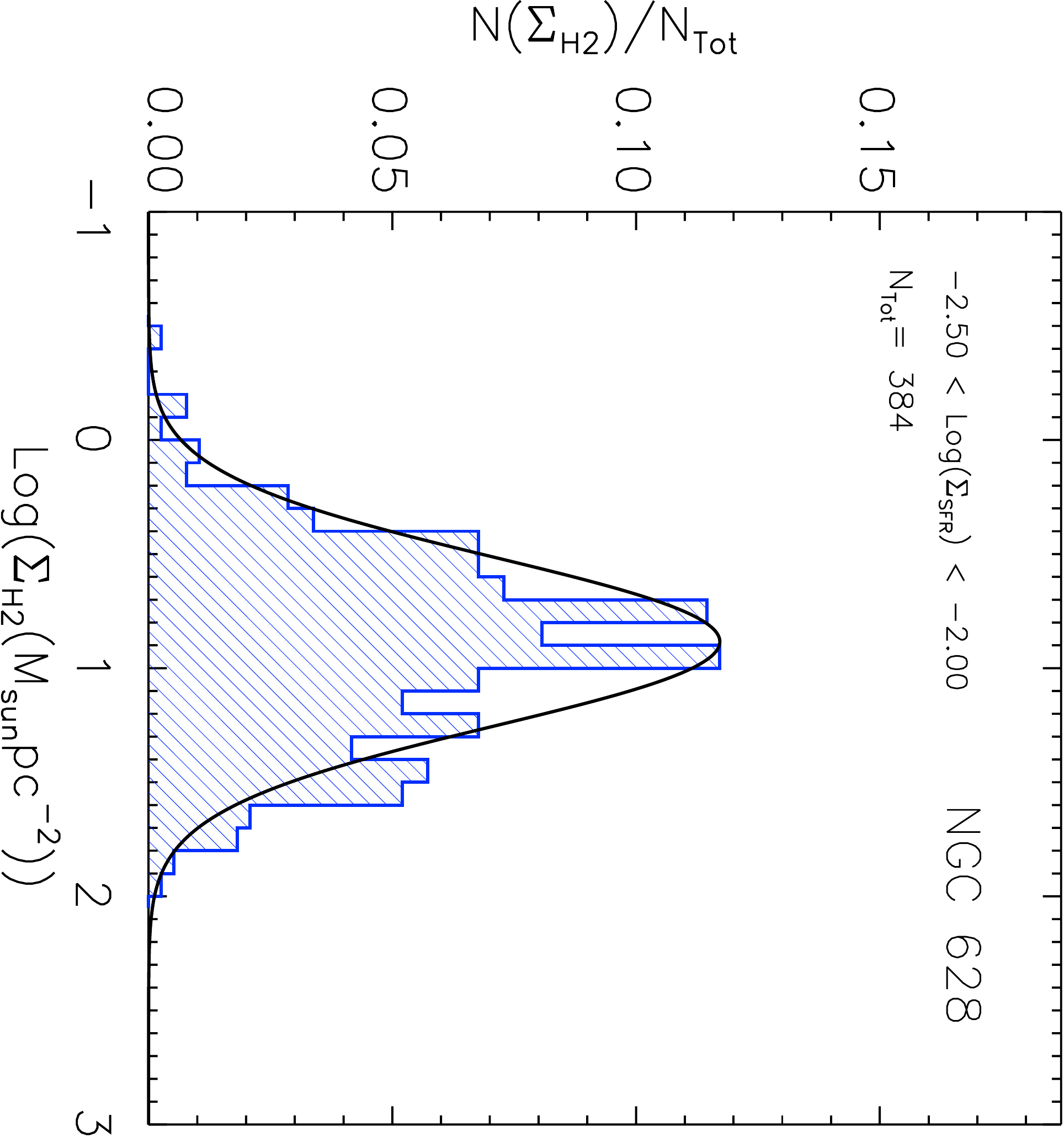,width=0.2\linewidth,angle=90} \\
\epsfig{file=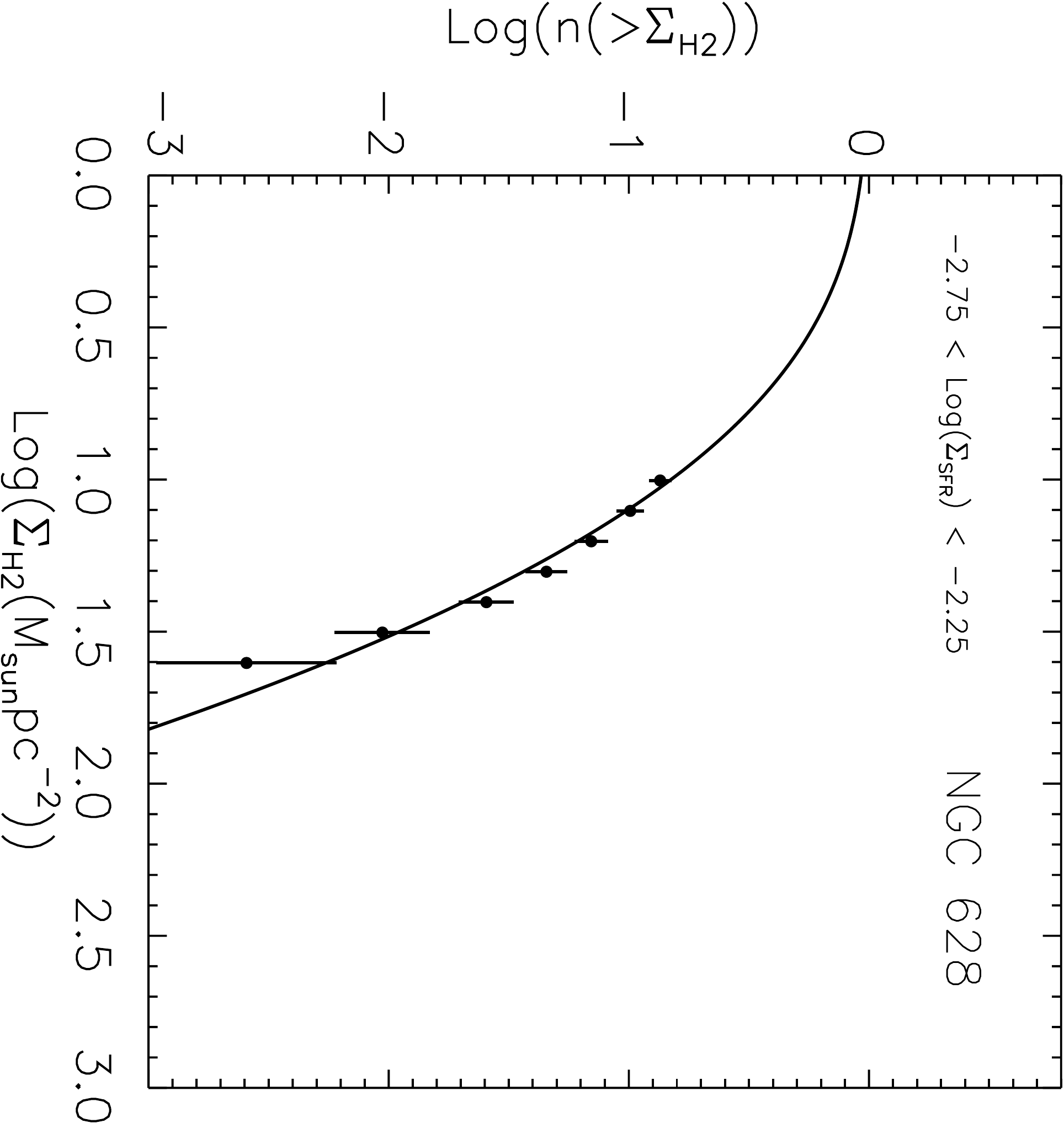,width=0.2\linewidth,angle=90} 
\epsfig{file=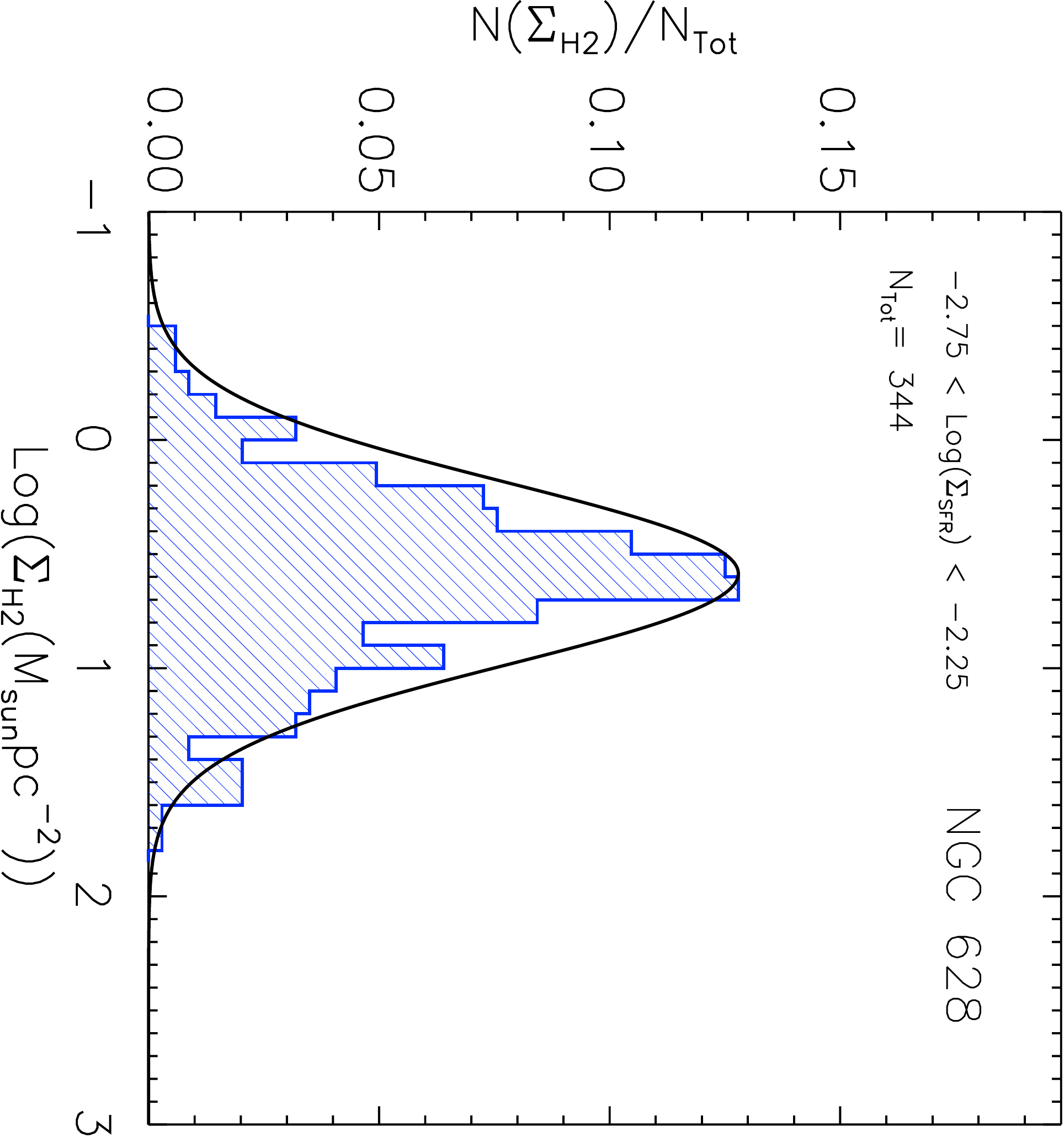,width=0.2\linewidth,angle=90} 
\end{tabular}
\caption{Left column: Cumulative distributions, $f(> \log(\Sigma_\mathrm{H2}))$, for each $\log(\Sigma_\mathrm{SFR})$ bin for galaxy NGC 628.  Black solid lines illustrate the best fitted {\it erfc} function to the distributions.  Right column:  Normalized distribution of $\log(\Sigma_\mathrm{H2})$ including both detections and non-detections.  Non-detections have been randomly distributed using a normal distribution with mean and dispersion parameters provided by the fitted {\it erfc} function.  To help comparison, Gaussian functions with the best fitted mean and dispersion are shown by black solid lines.}
\label{figure_cumulative}
\end{figure*}

The common approach to study the $\Sigma_\mathrm{H2}$ vs.\ $\Sigma_\mathrm{SFR}$ relation in galaxies has been the use of a uniform sampling grid across the region observed (\citealt{2008AJ....136.2846B}; \citealt{2013AJ....146...19L}; \citealt{2009ApJ...704..842B}).  This approach has the advantage of covering uniformly the region under study, allowing the inclusion of CO non-detections into the analysis.  In order to compare the $\Sigma_\mathrm{H2}$ vs.\ $\Sigma_\mathrm{SFR}$ relation for complexes with the relation obtained by using a uniform grid across the regions observed, we have generated gridded maps of SFR and molecular gas surface density.  The $\Sigma_\mathrm{SFR}$ map is created using Equation (\ref{sfr}), i.\ e., we use the combination of FUV and 24$\mu$m.  On the other hand, the $\Sigma_\mathrm{H2}$ map is derived from the $\co$ integrated intensity map.  We have sampled the maps using a hexagonal grid such that each surface element considered in the analysis is a hexagon with edge size of 8$\arcsec$, so each grid elements contains $\sim 1.5$ - 2 times the beam size of the maps.  This edge size corresponds to 213 pc for NGC 6946,  283 pc for NGC 628 and 287 pc for M101.  Every individual grid point in the sample is the average of the pixel values inside the corresponding hexagon.  The re-gridded maps of  $\Sigma_\mathrm{H2}$ and $\Sigma_\mathrm{SFR}$ for NGC 6946, NGC 628 and the two regions observed in M101 are shown in Figure \ref{figure_maps_grid}.  Figure \ref{figure_sfr_h2_grid} shows the $\Sigma_\mathrm{H2}$ vs.\ $\Sigma_\mathrm{SFR}$ scatter plot for the molecular complexes and for the uniformly sampled grid.  We observe that the complexes follow the distribution of the grid values for the three galaxies at high values of $\Sigma_\mathrm{H2}$.  The uniform grid has introduced a group of data points with $\Sigma_\mathrm{H2}$ in the range $\sim$10-50 $\Msun\ \mathrm{pc}^{-2}$, but with a narrow range in $\Sigma_\mathrm{SFR}$ likely to be associated with the inter-arm zone.  In Section \ref{in-on-arm} we discuss the differences in the star formation properties of the clouds in the context of on-arm and inter-arm regions.  Figure \ref{figure_sfr_h2_grid} also shows the distribution in the SFR axis of $\Sigma_\mathrm{H2}$ non-detections for each galaxy.  Assuming a line width window of $\sim 7\ \kms$ to recover the entire flux, we have estimated our $\Sigma_\mathrm{H2}$ non-detection threshold to be $10\ \Msun$ pc$^{-2}$.  We have used the same value for all the galaxies in our sample given that the sensitivity is similar across our maps ($\sim$ 0.4 K).  We notice that while in NGC 628 the $\Sigma_\mathrm{H2}$ non-detections show SFR values below $\sim 0.03\ \Msun\ \mathrm{yr}^{-1}\ \mathrm{kpc}^{-2}$, $\Sigma_\mathrm{H2}$ non-detections can reach higher SFR values of up to $\sim 0.2\ \Msun\ \mathrm{yr}^{-1}\ \mathrm{kpc}^{-2}$ for the galaxies NGC 6946 and M101.  In the next section, we investigate the effect of non-detections on our Bayesian regression analysis of the K-S relationship.

\begin{figure*}
\centering
\begin{tabular}{ccc}
\epsfig{file=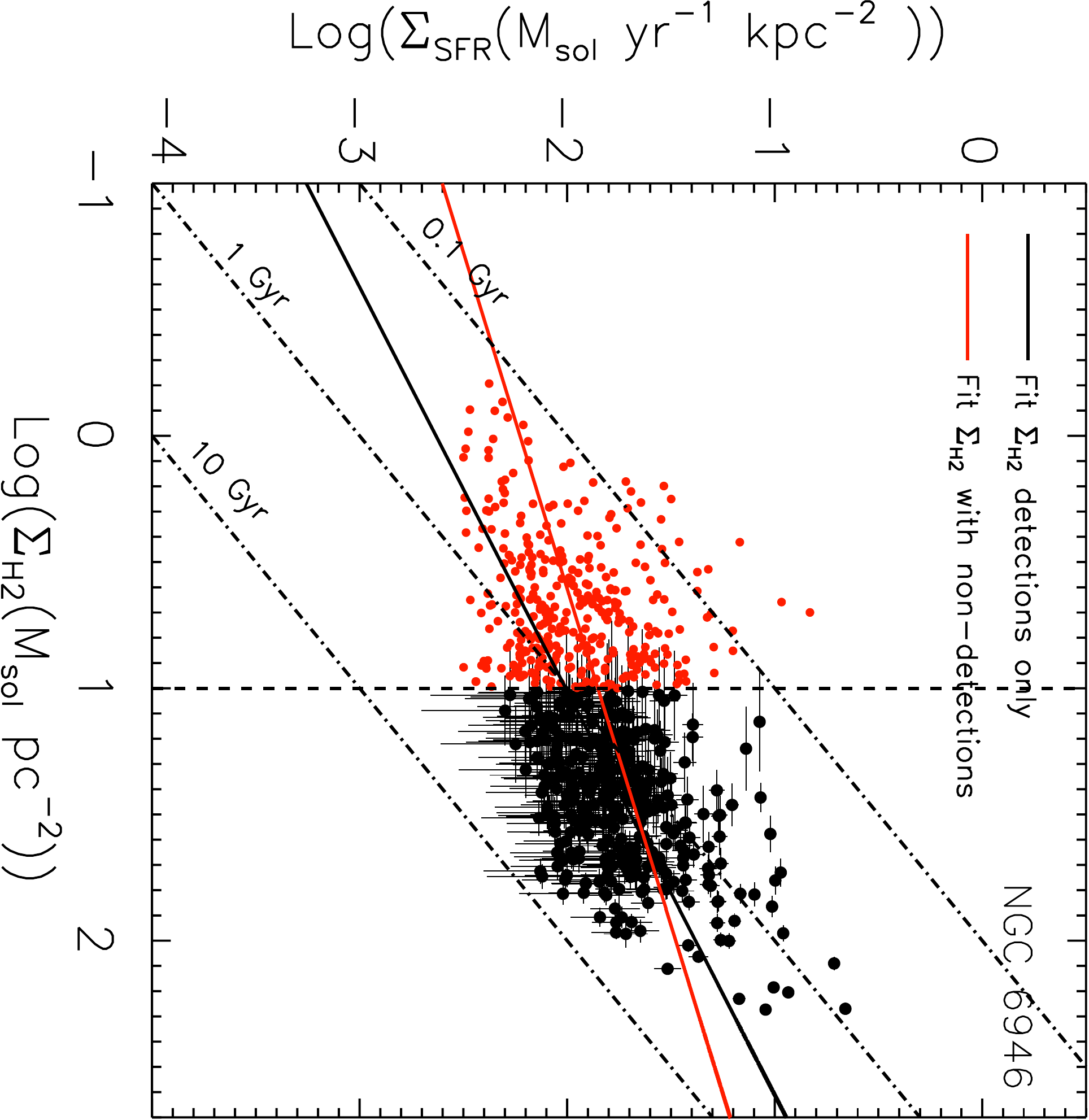,width=0.32\linewidth,angle=90} 
\epsfig{file=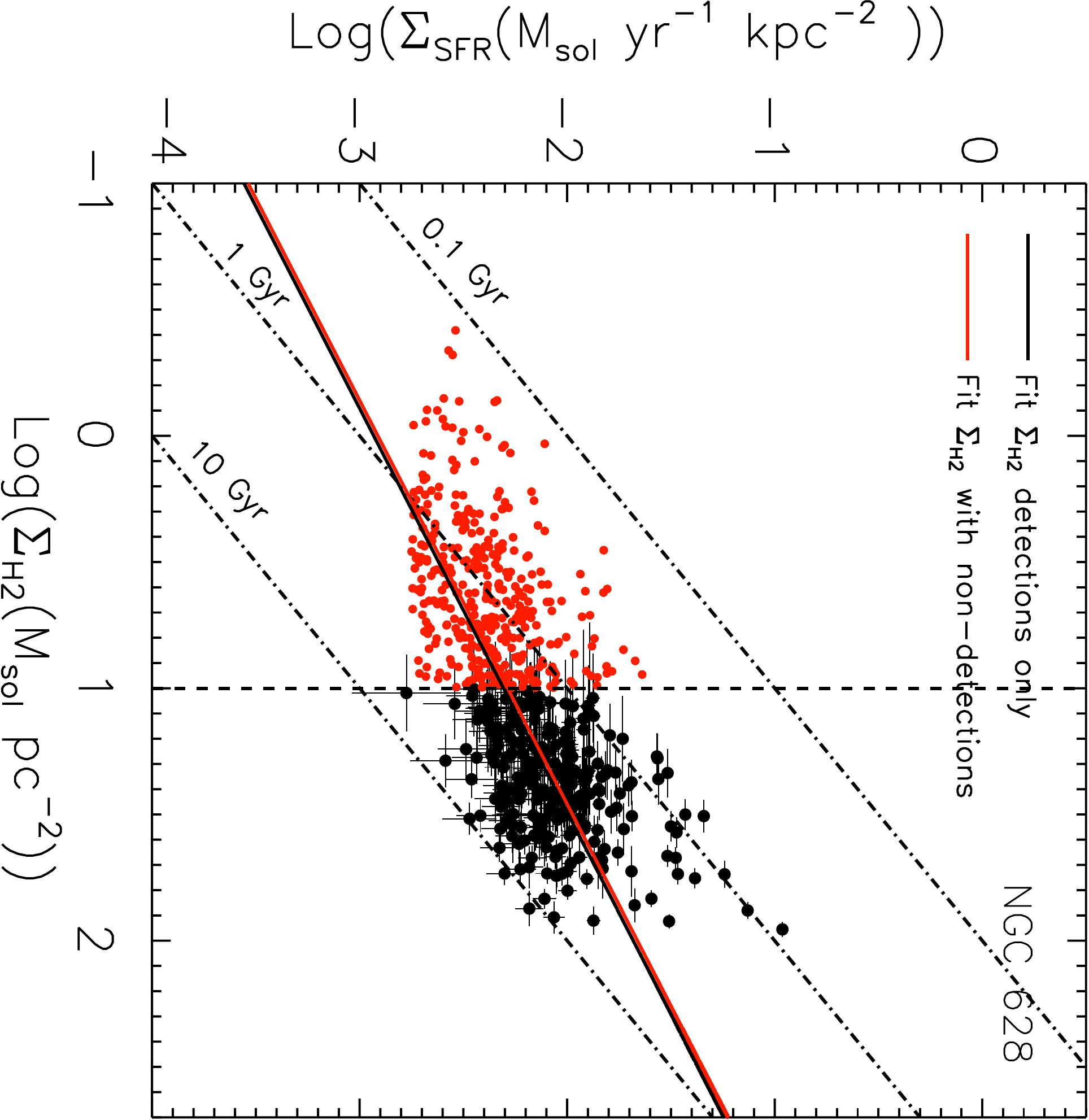,width=0.32\linewidth,angle=90} 
\epsfig{file=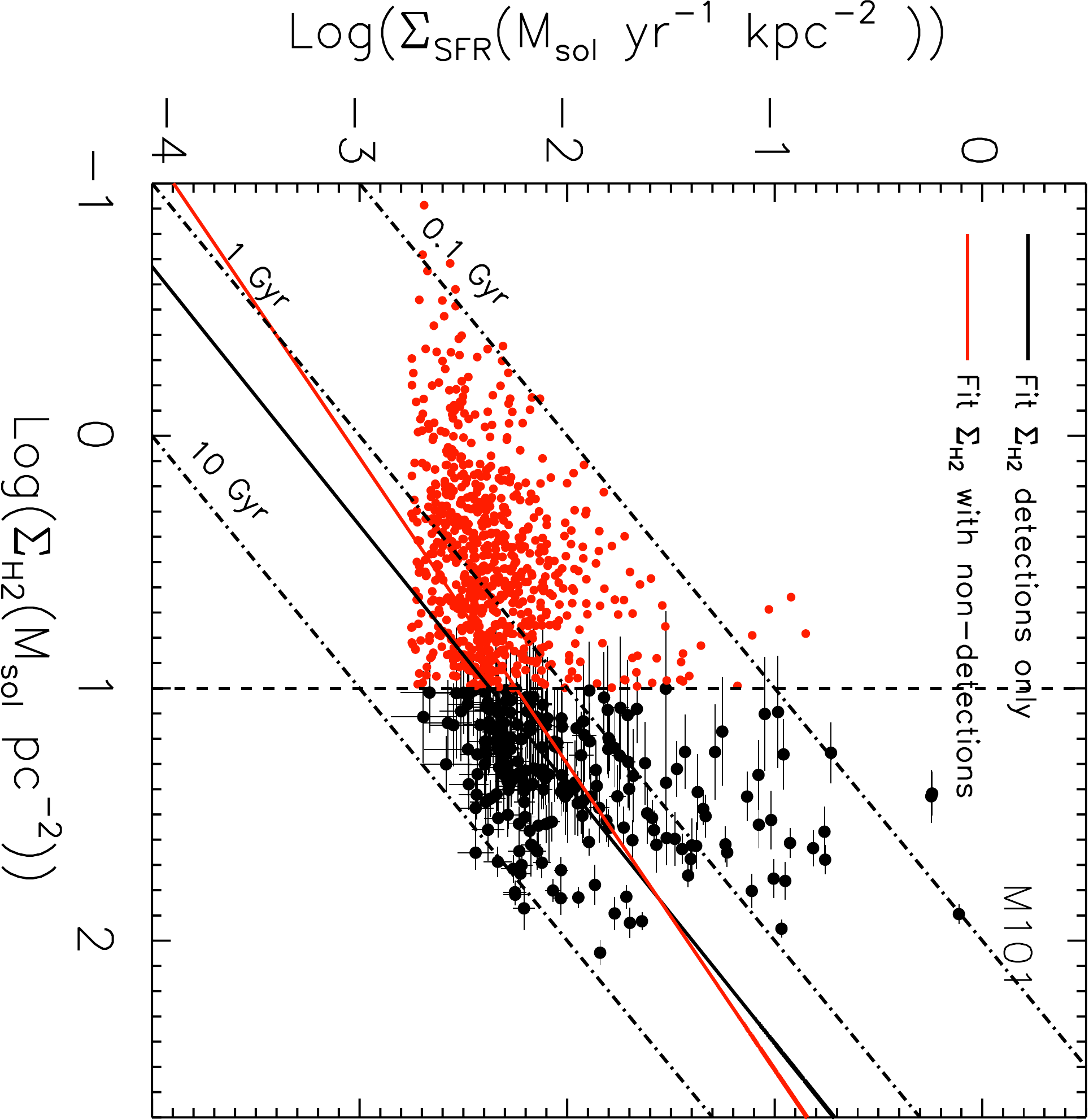,width=0.32\linewidth,angle=90} 
\end{tabular}
\caption{Effect of including non-detections in the Bayesian fitting of $\Sigma_\mathrm{H2}$ vs.\ $\Sigma_\mathrm{SFR}$ relations for the grid sample.  As in Figure \ref{figure_sfr_h2_grid}, NGC 6946 is in the left panel, NGC 628 in the middle and M101 in the right.  Black dots represents only detections, i.\ e., grid points with $\Sigma_\mathrm{H2} > 10\ \Msun \mathrm{pc}^{-2}$.  Red dots show non-detection grid points.  Solid lines are built by selecting the peak values of each parameter distribution of the relation log$(\frac{\Sigma_{\mathrm{SFR}}}{M_{\odot}\ \mathrm{yr}^{-1}\ \mathrm{kpc}^{-2}})=A+\alpha$ log$(\frac{\Sigma_{\mathrm{H2}}}{50\ M_{\odot}\ \mathrm{pc}^{-2}})+ \epsilon_\mathrm{scat}$ (see Figure \ref{figure_bayes_grid}).  The black solid line illustrate the fit to detections only, the red solid line shows the fit to the grid points with non-detections included.}
\label{figure_sfr_h2_grid_nondet}
\end{figure*}

\begin{figure*}
\centering
\begin{tabular}{ccc}
\epsfig{file=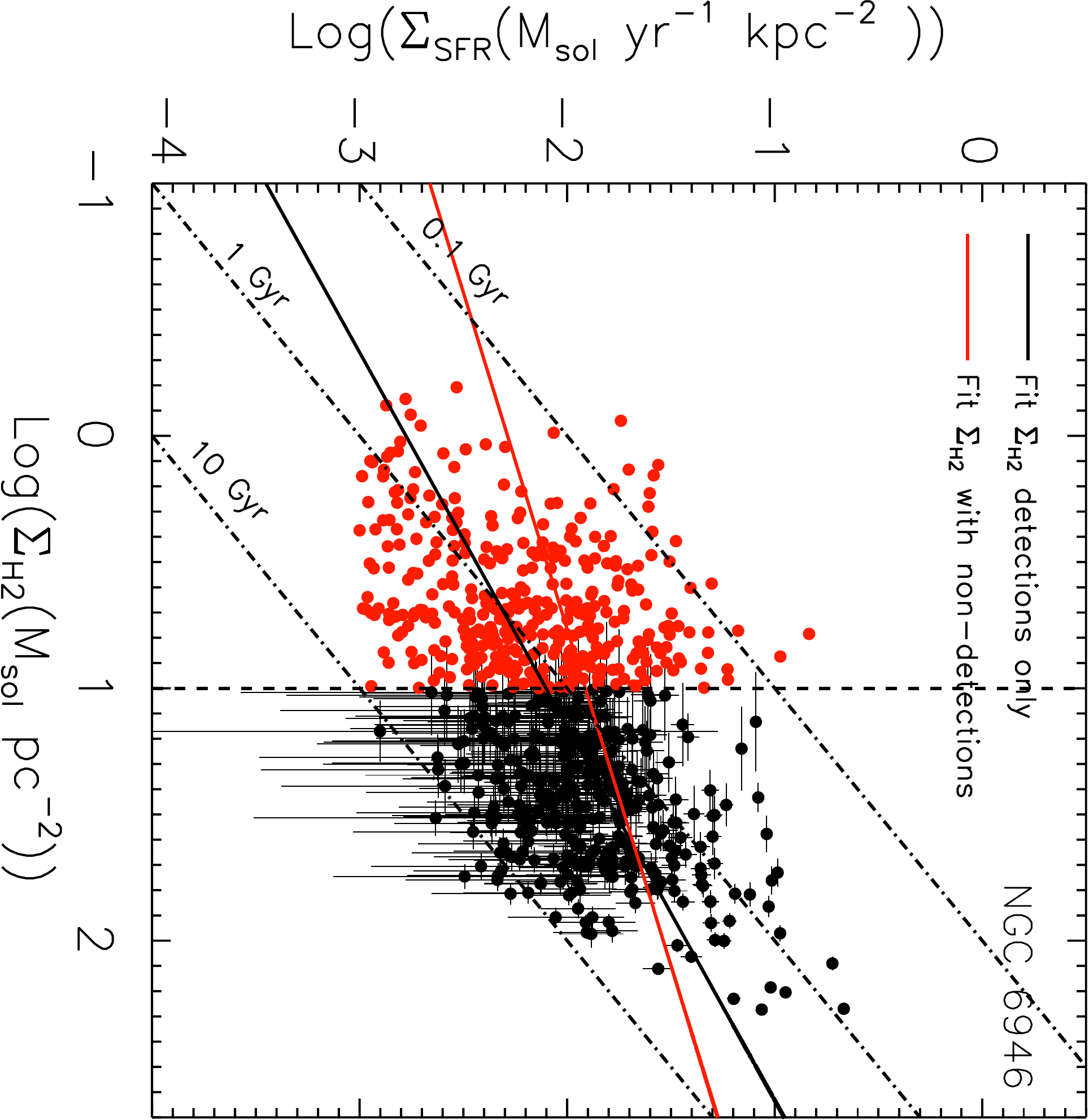,width=0.32\linewidth,angle=90} 
\epsfig{file=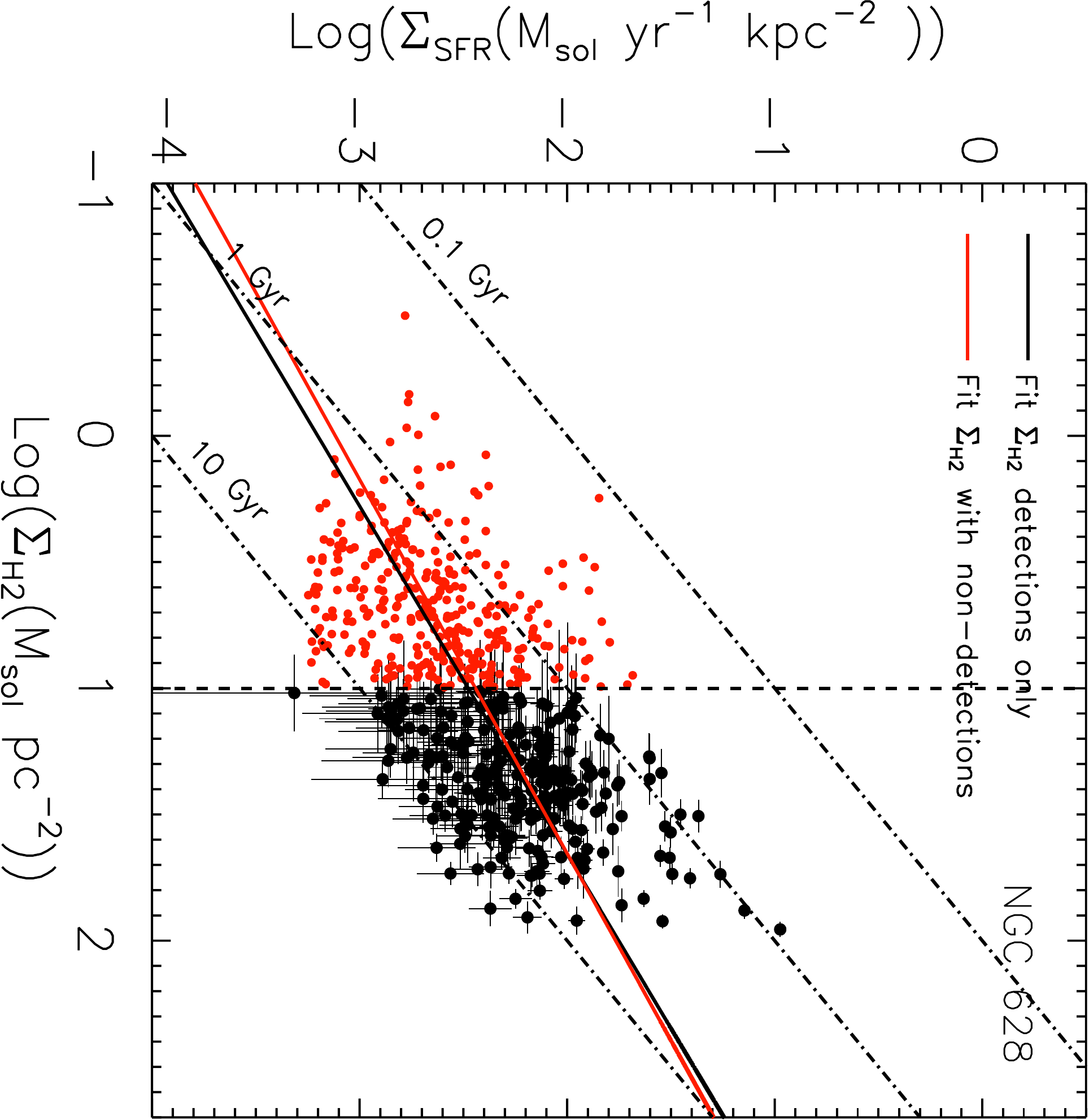,width=0.32\linewidth,angle=90} 
\epsfig{file=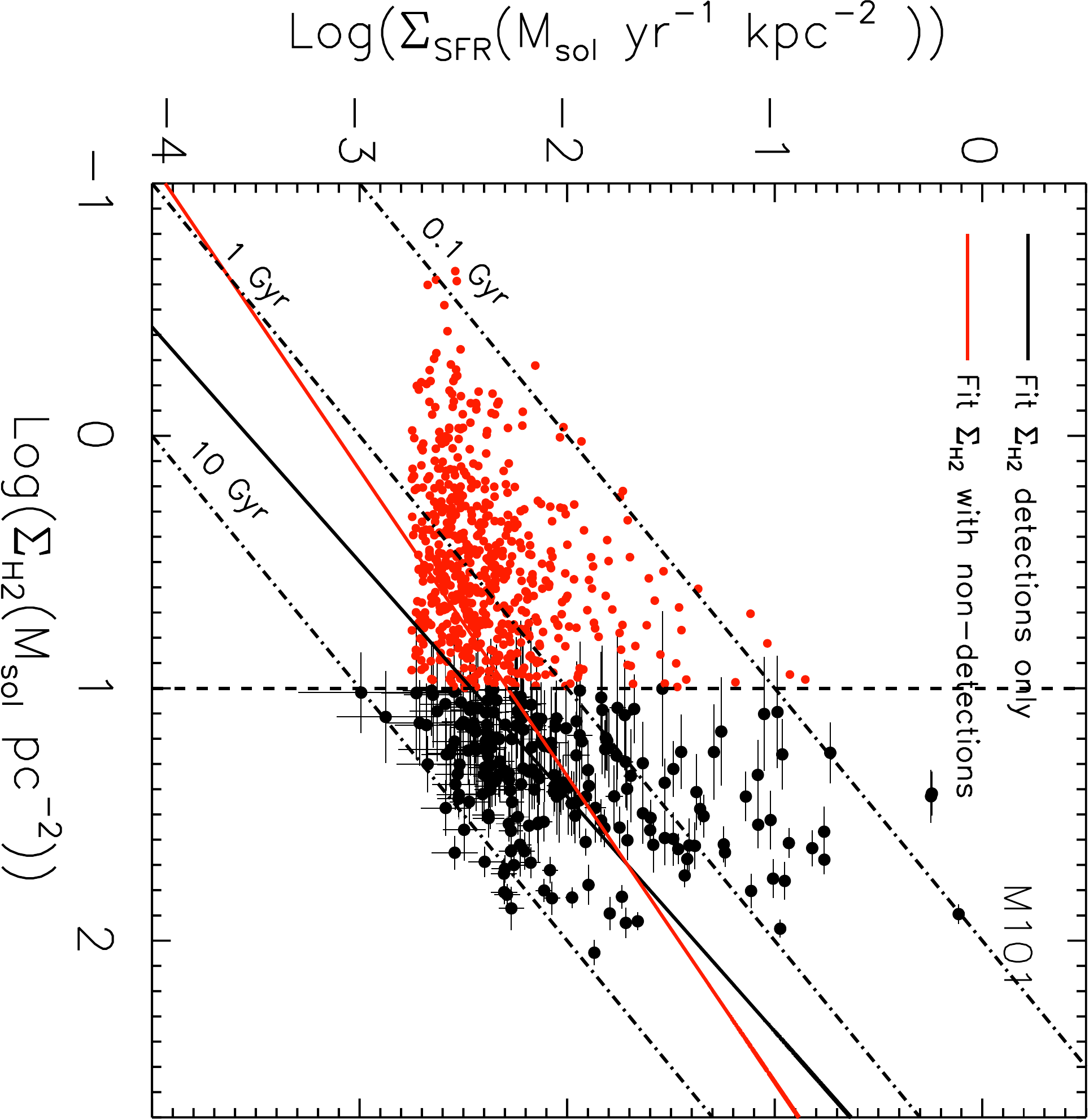,width=0.32\linewidth,angle=90} 
\end{tabular}
\caption{Same as Figure \ref{figure_sfr_h2_grid_nondet}, but $\Sigma_\mathrm{SFR}$ measurements have been corrected by a cirrus component in the 24$\mu$m map.}
\label{figure_sfr_h2_grid_nondet_bg}
\end{figure*}

\subsubsection{$\Sigma_\mathrm{H2}$ non-detections}\label{k-s_grid_h2_nondetect}

The common approach to dealing with non-detections in previous studies of the K-S relationship has been to clip the data below a given sensitivity threshold in the CO flux.  However, as such clipping can lead to a significant bias in fitting a line to the $\log(\Sigma_\mathrm{H2})-\log(\Sigma_\mathrm{SFR})$ relationship, recent studies have proposed different ways to include the non-detections in the K-S analysis (for example, see \citealt{2013AJ....146...19L}; \citealt{2009ApJ...704..842B}).

In this paper, we have included the molecular gas surface density non-detections in the Bayesian analysis by assuming that $\Sigma_\mathrm{H2}$ grid values for each galaxy follow a log-normal distribution at a given $\Sigma_\mathrm{SFR}$.  We adopt this model largely for its simplicity, but note that log-normal density distributions arise naturally in simulations of interstellar turbulence (e.\ g., \citealt{2001ApJ...547..172W}), although the width of the log-normal distribution may vary with the Mach number (e.\ g., \citealt{1998PhRvE..58.4501P}) and/or star formation rate (\citealt{2003ApJ...590L...1K}).  Because we do not know the $\Sigma_\mathrm{H2}$ values for non-detections, we can only use the distribution of detected $\Sigma_\mathrm{H2}$ values to infer the underlying log-normal distribution.   We divide the distribution of observed $\Sigma_\mathrm{SFR}$ into a series of bins in $\log(\Sigma_\mathrm{SFR})$.  In order to assure a significant number of $\Sigma_\mathrm{H2}$ detections in each bin, we have selected a width of $\Delta \log(\Sigma_\mathrm{SFR})=0.5$.  We allow a 50\% overlapping between bins to have a smooth transition of the $\Sigma_\mathrm{H2}$ log-normal distribution parameters across the range of $\Sigma_\mathrm{SFR}$ values.  Our approach is only applied to $\log(\Sigma_\mathrm{SFR})$ bins for which the fraction of $\Sigma_\mathrm{H2}$ detections is larger than 10\%.

For each $\log(\Sigma_\mathrm{SFR})$ bin, we fit the {\it erfc} function to the fractional number of grids with detected molecular gas surface density larger than a given $\log(\Sigma_\mathrm{H2})$ value,

\begin{equation}\label{inv-cumulative}
f(> \log(\Sigma_\mathrm{H2}))\equiv \frac{N(> \log(\Sigma_\mathrm{H2}))}{N_\mathrm{tot}}
\end{equation}

\noindent where $N(> \log(\Sigma_\mathrm{H2}))$ is the cumulative number of grid points with molecular surface density larger than $\log(\Sigma_\mathrm{H2})$, and $N_\mathrm{tot}$ is the total number of grid values in the sample.  The values of the fractional distribution $f(> \log(\Sigma_\mathrm{H2}))$ and their uncertainties are estimated in two steps.  The first task is to estimate the adopted value and its uncertainty for each bin in the differential $N(\log(\Sigma_\mathrm{H2}))$ distribution.  By assuming that the uncertainties in the $\Sigma_\mathrm{H2}$ detections are Gaussian distributed, we generate a new set of observed values $\widetilde{\Sigma}_\mathrm{H2}$ using 

\begin{equation}\label{new_sh2}
\widetilde{\Sigma}_\mathrm{H2}=\Sigma_\mathrm{H2}+\delta \Sigma_\mathrm{H2}
\end{equation}

\noindent where $\Sigma_\mathrm{H2}$ is the observed value, and $\delta \Sigma_\mathrm{H2}$ is drawn from a normal distribution with mean 0 and dispersion equal to the error of the observed $\Sigma_\mathrm{H2}$.  For each generated sample of $\widetilde{\Sigma}_\mathrm{H2}$ values, a new histogram $N(\log(\widetilde{\Sigma}_\mathrm{H2}))$ is built with a fixed bin size of 0.1.  The adopted $N(\log(\Sigma_\mathrm{H2}))$ value for each bin and its uncertainty is calculated from the mean and standard deviation respectively, over 1000 different realisations of $N(\log(\widetilde{\Sigma}_\mathrm{H2}))$.

\begin{table*}
\caption{Bayesian regression parameters for $\Sigma_\mathrm{H2}-\Sigma_\mathrm{SFR}$ relation using an hexagonal grid.\label{table-bayes-grid}}
\centering
\begin{tabular}{lcccccccc}
\hline\hline
Grid elements considered in the fit & $\alpha$ & 90\% HDI & & $A$ &  90\% HDI & & $\sigma$ &  90\% HDI\\
\hline
 & & & & \multicolumn{1}{c}{NGC 6946} &  & & & \\
\hline
$\Sigma_\mathrm{H2} $ detections only & 0.63 & [0.48, 0.68] & & -1.57 & [-1.61, -1.54] & & 0.22 & [0.19, 0.23]   \\
$\Sigma_\mathrm{H2} $ non-detections included  & 0.38  & [0.28, 0.41]  & & -1.59 & [-1.63, -1.56] & & 0.23 & [0.21, 0.24]   \\
$\Sigma_\mathrm{H2} $ detections only 24$\mu$m background subtracted  & 0.68  & [0.53, 0.78]  & & -1.63 & [-1.68, -1.60] & & 0.25 & [0.21, 0.26]  \\
$\Sigma_\mathrm{H2} $ non-detections included 24$\mu$m background subtracted &  0.38  & [0.26, 0.41]  & & -1.65 & [-1.70, -1.62] & & 0.26 & [0.24, 0.27]\\
On-arm sample&  0.63  & [0.50, 0.74]  & & -1.45 & [-1.50, -1.43] & & 0.21 & [0.18, 0.22] \\
Inter-arm sample&  0.23  & [0.08, 0.33]  & & -1.75 & [-1.81, -1.73] & & 0.10 & [0.06, 0.11]\\
On-arm sample 24$\mu$m background subtracted&  0.73  & [0.55, 0.82]  & & -1.51 & [-1.56, -1.48] & & 0.23 & [0.19, 0.24] \\
Inter-arm sample 24$\mu$m background subtracted&  0.28 & [0.06, 0.42]  & &  -1.87 & [-1.94, -1.84] & & 0.12 & [0.08, 0.15]\\
\hline
 & & & & \multicolumn{1}{c}{NGC 628}  & &  & &\\
\hline
$\Sigma_\mathrm{H2} $ detections only & 0.63 & [0.44, 0.74] & & -1.87 & [-1.92, -1.83] & & 0.23 & [0.20, 0.24]   \\
$\Sigma_\mathrm{H2} $ non-detections included   &  0.63  & [0.52, 0.65]  & & -1.85 & [-1.91, -1.83] & & 0.20 & [0.18, 0.21] \\
$\Sigma_\mathrm{H2} $ detections only 24$\mu$m background subtracted  & 0.73  & [0.53, 0.91]  & & -1.97 & [-2.03, -1.91] & & 0.28 & [0.25, 0.30] \\
$\Sigma_\mathrm{H2} $ non-detections included 24$\mu$m background subtracted  &  0.68  & [0.58, 0.75]  & & -1.97 & [-2.02, -1.93] & & 0.26 & [0.23, 0.27]\\
On-arm sample& 0.68  & [0.43, 0.85]  & & -1.79 & [-1.86, -1.74]  & & 0.24 & [0.20, 0.26] \\
Inter-arm sample&  0.38  & [0.16, 0.52]  & & -2.03  & [-2.10, -1.97]  & & 0.17 & [0.14, 0.19] \\
On-arm sample 24$\mu$m background subtracted & 0.78  & [0.49, 1.00]  & & -1.89 & [-1.97, -1.82]  & & 0.28 & [0.24, 0.31] \\
Inter-arm sample 24$\mu$m background subtracted &  0.48  & [0.19, 0.71]  & & -2.15  & [-2.23, -2.07]  & & 0.22 & [0.18, 0.25] \\
\hline
 & & & & \multicolumn{1}{c}{M101}  & & & & \\
\hline
$\Sigma_\mathrm{H2} $ detections only & 0.98 & [0.64, 1.23] & & -1.69 & [-1.79, -1.59] & & 0.44 & [0.40, 0.47]   \\
$\Sigma_\mathrm{H2} $ non-detections included   &   0.83  & [0.69, 0.86]  & & -1.67 & [-1.73, -1.61] & & 0.30 & [0.27, 0.30] \\
$\Sigma_\mathrm{H2} $ detections only 24$\mu$m background subtracted  &  1.03  & [0.70, 1.33]  & & -1.71 & [-1.82, -1.61] & & 0.47 & [0.42, 0.50] \\
$\Sigma_\mathrm{H2} $ non-detections included 24$\mu$m background subtracted   &  0.83  & [0.69, 0.88]  & & -1.71 & [-1.78, -1.65] & & 0.33 & [0.30, 0.33]\\
On-arm sample&  0.88 & [0.46, 1.20] & &   -1.53   & [-1.65, -1.41] & & 0.48  & [0.43, 0.54] \\
Inter-arm sample&  0.48 & [ 0.18, 0.68]  & & -2.09 & [ -2.18, -2.00] & & 0.19 & [0.16, 0.21] \\
On-arm sample 24$\mu$m background subtracted&  0.93 & [0.50, 1.27] & &  -1.55   & [-1.68, -1.42] & & 0.51  & [0.45, 0.56] \\
Inter-arm sample 24$\mu$m background subtracted&  0.53 & [0.23, 0.78]  & & -2.11 & [ -2.23, -2.04] & & 0.21 & [0.17, 0.23] \\
\hline
\end{tabular}
\end{table*}

\begin{figure*}
\centering
\begin{tabular}{c}
\epsfig{file=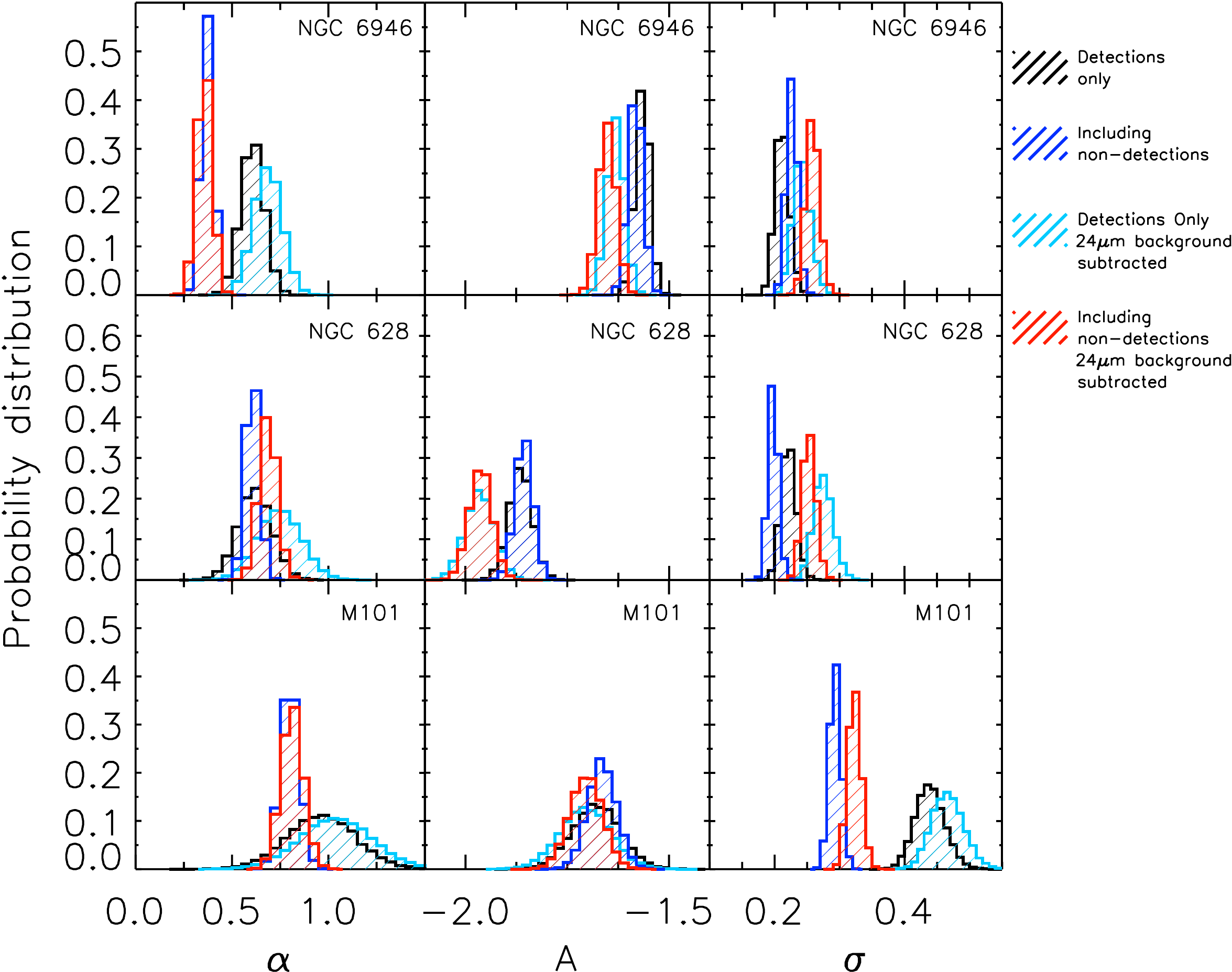,width=0.7\linewidth,angle=90} 
\end{tabular}
\caption{Probability distributions of the parameters $\alpha$, $A$ and $\sigma$ from the Bayesian regression fit to the relation log$(\frac{\Sigma_{\mathrm{SFR}}}{M_{\odot}\ \mathrm{yr}^{-1}\ \mathrm{kpc}^{-2}})=A+\alpha$ log$(\frac{\Sigma_{\mathrm{H2}}}{50\ M_{\odot}\ \mathrm{pc}^{-2}})+ \epsilon_\mathrm{scat}$.  Top row panels show the parameter distributions for NGC 6946, middle row panels show the distributions for NGC 628, and bottom row panels show the distributions for M101.  Left column panels show the distributions of the slope $\alpha$, middle column panels show the distributions of the intercept coefficient $A$, and the right column panels show the distribution of the dispersion of the scatter, $\sigma$.  Black histograms show the resulting distributions considering only $\Sigma_{\mathrm{H2}}$ detections in the Bayesian regression fit.  Blue histograms illustrate the resulting distributions with non-detections included.  Light blue histograms show the distributions considering only detections, but with $\Sigma_{\mathrm{SFR}}$ values corrected by the 24$\mu$m cirrus component.  Finally, red histograms show the distributions considering points with $\Sigma_{\mathrm{H2}}$ non-detections included and $\Sigma_\mathrm{SFR}$ values corrected by 24$\mu$m cirrus component.}
\label{figure_bayes_grid}
\end{figure*}

The second step is to estimate the values and uncertainties of each bin in the fractional cumulative distribution $f(> \log(\Sigma_\mathrm{H2}))$.  They are calculated by generating 1000 different realisations of $f(> \log(\Sigma_\mathrm{H2}))$.  In this case, each realisation of $f(> \log(\Sigma_\mathrm{H2}))$ is the result of applying Equation \ref{new_sh2} to the adopted value of $N(\log(\Sigma_\mathrm{H2}))$ and derived uncertainty.  

Finally, we find the best fit {\it erfc} function to the $f(> \log(\Sigma_\mathrm{H2}))$ distribution for each $\log(\Sigma_\mathrm{SFR})$ bin.  To perform the fit, we have used the MPFIT fitting package given by \citet{2009ASPC..411..251M}.  The non-detections are randomly distributed using a log-normal distribution in $\Sigma_\mathrm{H2}$ with mean and dispersion consistent with the {\it erfc} fit for each $\Sigma_\mathrm{SFR}$ bin.  As an example of our methodology, Figure \ref{figure_cumulative} shows the cumulative functions and complete distribution of both $\Sigma_\mathrm{H2}$ non-detections and detections for each $\log(\Sigma_\mathrm{SFR})$ bin in the galaxy NGC 628.  In Figure \ref{figure_sfr_h2_grid_nondet} we show the $\Sigma_\mathrm{H2}$ vs. $\Sigma_\mathrm{SFR}$ scatter plot for the three galaxies including both detections and non-detections of $\Sigma_\mathrm{H2}$.

Once we have assigned values to $\Sigma_\mathrm{H2}$ non-detections, we can include them in the Bayesian analysis of the K-S relationship.  As in Section \ref{k-s_law}, we have used $2\times 10^4$ random draws to derive the posterior probability distribution of the linear fit parameters in the Bayesian approach.  This procedure was performed for the following cases: considering only $\Sigma_\mathrm{H2}$ detections and including detections $+$ non-detections.  In Figure \ref{figure_bayes_grid} we show the probability distributions of the parameters $\alpha$, $A$ and $\sigma$ from the Bayesian regression fit for each case and for each galaxy in our sample.  Table \ref{table-bayes-grid} shows the peak and 90 \% HDI of the probability distributions for each fitting parameter.  Representative relations built by using the peak values of each parameter distribution are shown in Figure \ref{figure_sfr_h2_grid_nondet}. 

In the case of NGC 6946, the slope $\alpha$ distribution peaks at 0.63 when only $\Sigma_\mathrm{H2}$ detections are considered in the fit.  On the other hand, if all the non-detections are included in the fit, the slope distribution peaks at 0.38.  The distributions of the slope for both cases do not show a significant overlapping.  On the other hand, the distributions of the coefficient $A$ and the dispersion of the intrinsic scatter $\sigma$ when only detections are considered and  when both detections and non-detections are considered are consistent.

In contrast to NGC 6946, the distribution of the slope $\alpha$ for NGC 628 does not show a significant change when the $\Sigma_\mathrm{H2}$ non-detections are included or not to the fitting of the K-S relation.  In both cases, the peak of the distribution of the slope is 0.63, and the 90 \% HDI is slightly larger in the case when only detections are considered.  The distribution of the coefficient $A$ is consistent between the two cases.

For M101, the distribution of the slope peaks at 0.98 when only $\Sigma_\mathrm{H2}$ detections are considered in the fit, while when the non-detections are included the peak is 0.83.  However, both distributions overlap significantly, mainly due to the broad distribution of slopes in the detection-only case. Further evidence of the poor correlation in the K-S relation for M101 is the large scatter $\sigma$ yielded by our fitting approach when only detections are considered in the analysis.

\subsubsection{Cirrus component of the $\Sigma_\mathrm{SFR}$}\label{cirrus_sfr}
So far, our analysis has not considered the systematic uncertainties in the $\Sigma_\mathrm{SFR}$ measurements.  In their analysis of of the IRAM HERACLES CO survey, \citet{2012AJ....144....3L} discuss the bias inherent to each component of the commonly used ``hybrid''  tracers FUV+24$\mu$m and H$\alpha$+ 24$\mu$m.  Additionally, they estimate the contamination of the 24 $\mu$m band by emission not associated with current star formation by using physically motivated dust models.  At 1 kpc resolution, they find that systemic differences among these different tracers are less than a factor of two for $\Sigma_\mathrm{SFR} \gtrsim 10^{-3} \Msun\ \mathrm{yr}^{-1}\ \mathrm{kpc}^{-2}$.

Recent works on the extragalactic K-S law have determined that the index in the power relation between molecular gas and SF surface density is affected by the diffuse emission in the tracers used to estimate the star formation (\citealt{2011ApJ...735...63L}; \citealt{2011ApJ...730...72R}; \citealt{2012ApJ...752...98C}; \citealt{2013ApJ...772L..13M}).  In their study of the spatially resolved star formation law in the galaxies NGC 3521 and M51, \citet{2011ApJ...735...63L} suggest that the difference observed in the exponents of the $\Sigma_\mathrm{H2}$-$\Sigma_\mathrm{SFR}$ power law relation in different works (namely, linear vs.\  super-linear relation) is the result of removing or including the local diffuse background.  

In order to investigate the effect of the fraction of $\Sigma_\mathrm{SFR}$ not associated with current star formation on our Bayesian fitting technique, we apply a simple approach to subtract a cirrus component of the 24$\mu$m map.  A detailed description of our methodology is presented in Appendix \ref{24um_cirr}.  Our approach consists basically of generating a smoothed map of the 24$\mu$m image using a gaussian kernel, which is subsequently subtracted from the original 24$\mu$m map.  The background subtracted 24$\mu$m map is then used to estimate the $\Sigma_\mathrm{SFR}$ using Equation \ref{sfr}. 

With the 24$\mu$m background subtracted $\Sigma_\mathrm{SFR}$ measurements in hand, we perform the same approach to distribute the non-detections presented in Section \ref{k-s_grid_h2_nondetect}.  The resulting $\Sigma_\mathrm{H2}$ vs. $\Sigma_\mathrm{SFR}$ scatter plot for the three galaxies including both detections and non-detections are shown in Figure \ref{figure_sfr_h2_grid_nondet_bg}.  Again, we applied our Bayesian analysis of the K-S relationship to the case considering $\Sigma_\mathrm{H2}$ detections only and to the case when we include both detections and non-detections.  Figure \ref{figure_bayes_grid} shows the probability distributions of the parameters $\alpha$, $A$ and $\sigma$ from the Bayesian regression fit for each case and for each galaxy in our sample.  Table \ref{table-bayes-grid} shows the peak and 90 \% HDI of the probability distributions for each fitting parameter.  The relations for each galaxy based on using the peak values of each parameter distributions are shown as lines in Figure \ref{figure_sfr_h2_grid_nondet_bg}. 




We do not find a significant change in the distribution of the slope between the cases when a cirrus component of the 24$\mu$m image is subtracted from the $\Sigma_\mathrm{SFR}$ measurements (this Section) and when no correction has been made (Section \ref{k-s_grid_h2_nondetect}).  Although the slope distributions are slightly shifted to larger values when the cirrus correction is applied, the overlapping region of the distributions is significant between the two cases.  On the other hand, the coefficient $A$ is shifted to smaller values when the 24$\mu$m cirrus component is subtracted from the $\Sigma_\mathrm{SFR}$ measurements for the galaxies NGC 6946 and NGC 628.  Thus, the subtraction of the cirrus component of the $\Sigma_\mathrm{SFR}$ values has little or no impact on the distribution of the slopes, but it can move the coefficient $A$ to smaller values due to the migration of grid elements to lower values of $\Sigma_\mathrm{SFR}$.


\section{DISCUSSION}\label{discuss}

\subsection{Molecular cloud properties}

Based on the broad parameter distributions generated by the Bayesian regression fit shown in Figure \ref{fig_sizew_10}, we do not find strong evidence for a $R-\sigma_v$ relation for the $\co$ complexes in the three galaxies studied here.  Higher resolution observations do not show evidence for a $R-\sigma_v$ relation either, as the same conclusion can be derived for $\cotwo$ clouds from Figure \ref{fig_sizew_21}.  In their comparative study of GMCs in the galaxies M51, M33 and the Large Magellanic Cloud, \citet{2013ApJ...779...46H} arrive at the same conclusion.  They find that, when the CO emission is decomposed into GMC-like structures, the correlation between size and line width is weak or almost nonexistent in individual galaxies, or even in the aggregate sample smoothed to a common resolution.

Studies of the properties of resolved structures have provided some insights on the nature of the turbulence in the ISM and its role in the star formation process both in the Milky Way (\citealt{2009ApJ...699.1092H}; \citealt{1987ApJ...319..730S}; \citealt{2012MNRAS.425..720S}) and in nearby galaxies (\citealt{2008ApJ...686..948B}; \citealt{2008ApJ...675..330S}; \citealt{2012ApJ...757..155R}).  Several simulation works have proposed a scenario in which the turbulence observed in molecular clouds originates on the largest scales in the ISM (\citealt{2004ApJ...615L..45H}; \citealt{2009A&A...504..883B}).  According to this framework, the velocity dispersion will remain constant for scales larger than the size scale driving the turbulence in the ISM (\citealt{2002A&A...390..307O}).  In their study of the dense ISM in the Central Molecular Zone (CMZ) of the Milky Way, \citet{2012MNRAS.425..720S} find that the trend observed in the $\sigma_v-R$ relation is independent of the presence of dense structures in the region probed by the observations.  They interpret this result as evidence in favor of the turbulence being injected at larger scales in the CMZ.  Over the size scales probed by their observations ($R \sim 2-40$ pc), they do not find a clear flattening of the $\sigma_v-R$ relation, leading them to suggest that the turbulent velocities in the CMZ are driven on scales larger than 30 pc. 

The $\sigma_v-R$ scatter plots of the identified structures in the three galaxies in our sample may provide some insights on the scales at which turbulence is driven in the ISM in spiral galaxies.  Among the three galaxies in our sample, NGC 628 presents the least inclined disk, offering the best place to estimate the cloud velocity dispersions without the influence of cloud blending or crowding.  The $\sigma_v-R$ scatter plots for complexes (Figure \ref{fig_sizew_10}) and for clouds (Figure \ref{fig_sizew_21}) show that, for the scales probed by our observations in NGC 628 ($\sim 70-200$ pc), the velocity dispersions reach a maximum value $\sim 7\ \kms$.  A similar trend is found for complexes and clouds for M101.  However, given the large uncertainties in the slope due to the limited dynamic range in size, as shown in Tables \ref{table-bayes} and \ref{table-bayes21}, it is difficult to draw any conclusion from the observed $\sigma_v-R$ relation about the nature of the turbulence in the ISM in NGC 628 and M101. 

In the case of NGC 6946, we observe some complexes and clouds with velocity dispersions $\sim 10\ \kms$.  In Paper I, our observations of the eastern part of the disk of NGC 6946 showed that the regions with the most active star formation are the regions with the largest clouds and complexes, and with the largest velocity dispersions as well.  Moreover, in one of the active star forming regions of NGC 6946, we find a large velocity gradient in the CO emission map.  Thus, the difference in the star formation rate in those regions, along with other kinematic features in the gas, may be influencing the relation between $\sigma_v$ and size with respect the other regions of the disk.  In fact, in the center of NGC 6946 \citet{2012ApJ...744...42D} find a set of GMCs with velocity dispersions larger than other clouds with similar sizes.  These clouds may be dynamically affected by the presence of a nuclear bar in NGC 6946, resulting in larger line widths compared to complexes or clouds with similar sizes located in other regions of the disk.  A complete census of resolved GMCs in several environments such as the center of a barred galaxy, regions located in a strong spiral arm, and regions with high and low star formation activities is needed to draw a robust picture of the influence of those environments on the turbulent nature of the molecular clouds. 

\subsection{Kennicutt-Schmidt relation}\label{k-s-discuss}

As has been discussed recently (\citealt{2011ApJ...735...63L}; \citealt{2012ApJ...752...98C}; \citealt{2013ApJ...772L..13M}), deriving the the K-S relation and its intrinsic uncertainty is a non-trivial task.  It depends on several factors such as the selection of the tracers used to estimate the gas content and the SFR activity over a particular region, the statistical tools to find the best fit to the power law, or the physical scale probed by the observations, among others.  Analyses that combine different approaches to deal with the effects of these factors may drive the estimated fit parameters and uncertainties to a wide range of values.

While our analysis is subject to some of the same limitations, our Bayesian approach provides us with more realistic estimates for uncertainties of the regression fit parameters.  Among our sample of galaxies, NGC 6946 shows the most constrained K-S relationship at the scale probed by $\co$ complex boundaries.  On the other hand, the parameter distributions for NGC 628 and M101 are poorly constrained, particularly in the distribution of the slope.  The limited dynamic range present in the CO observations of NGC 628 on the one hand, and the large variation of the SFR observed in M101 on the other, makes the derivation of a power law unstable for these two galaxies.  The increased resolution offered by the $\cotwo$ clouds is translated into parameter distributions consistent with super-linear relations for NGC 6946 and M101, but with increased uncertainties.  Additionally, the dispersion of the intrinsic scatter, $\sigma$, shows a factor of $\sim$ 2 higher values compared to the $\co$ complexes for the three galaxies.  This is consistent with the scenario proposed by \citet{2012ApJ...752...98C}, where the slope and the scatter of simulated K-S relations depend on the size scale probed by the region considered in the analysis.

\begin{figure*}
\centering
\begin{tabular}{ccc}
\epsfig{file=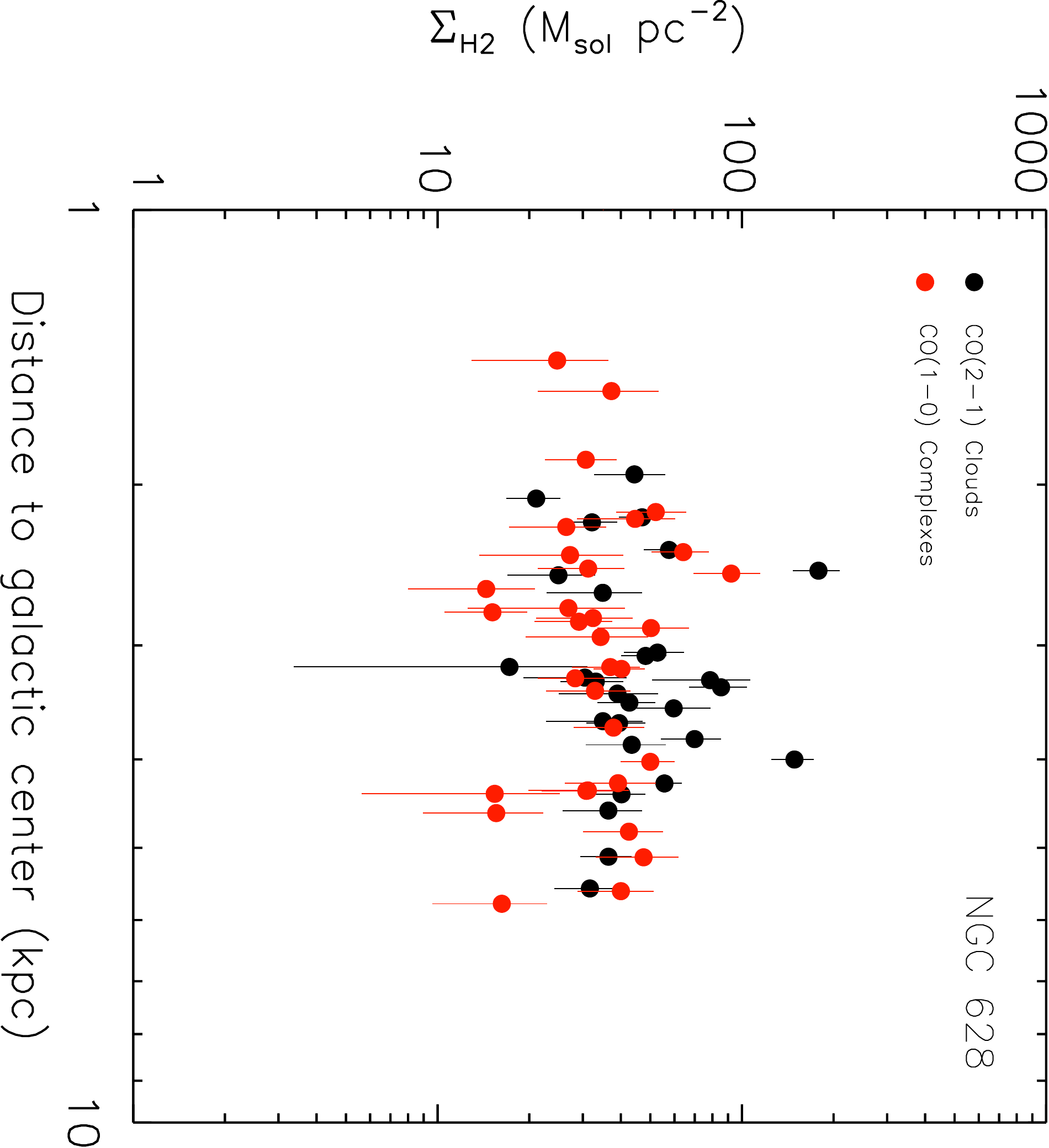,width=0.3\linewidth,angle=90} 
\epsfig{file=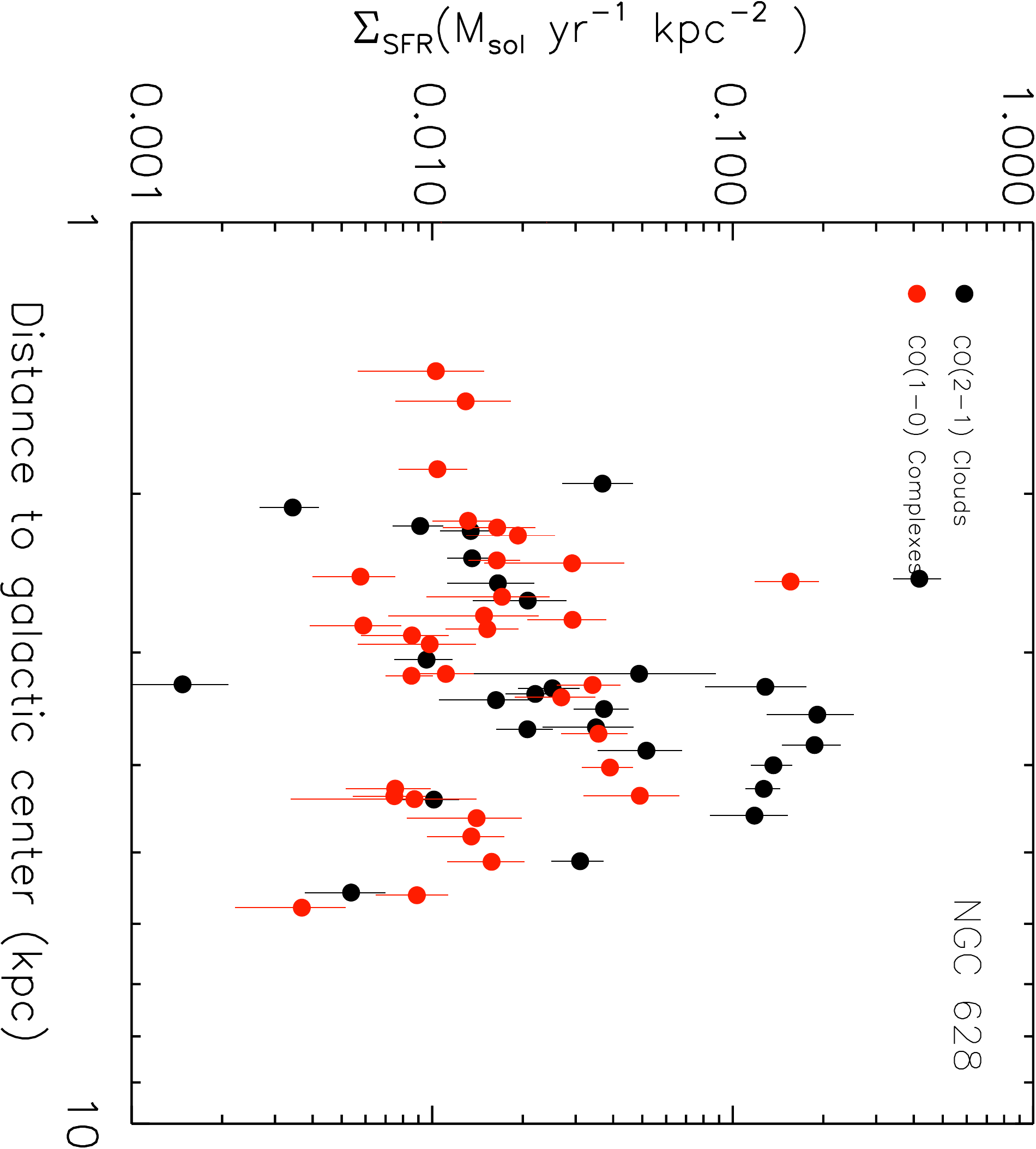,width=0.3\linewidth,angle=90} 
\epsfig{file=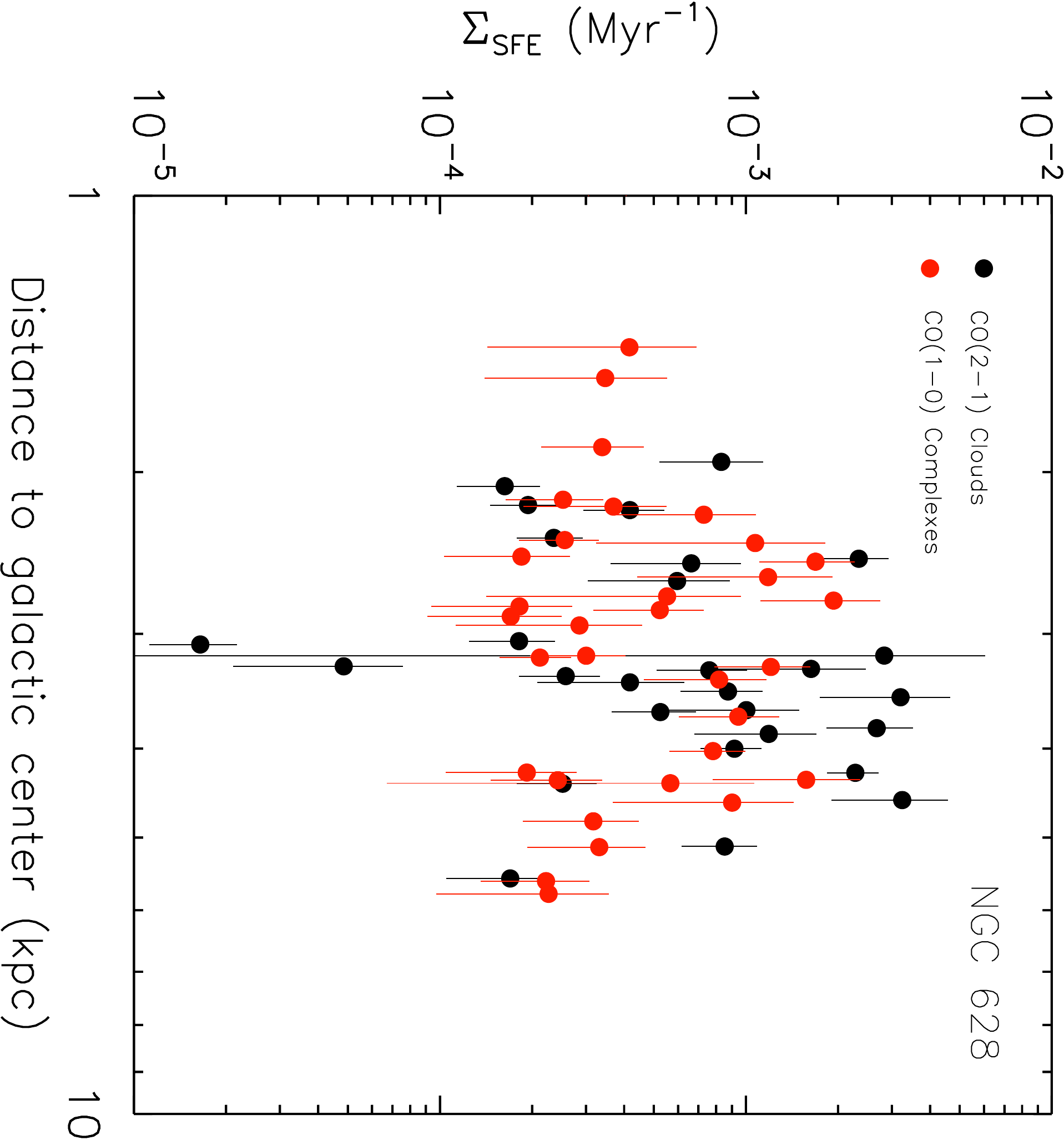,width=0.3\linewidth,angle=90} 
\end{tabular}
\caption{Radial distributions of $\Sigma_\mathrm{H2}$ {\bf (left)}, $\Sigma_\mathrm{SFR}$ {\bf (center)} and $\Sigma_\mathrm{SFE}$ {\bf (right)} of the regions observed in NGC 628, including both $\co$ complexes and $\cotwo$ clouds.  Red dots represent complexes, while black dots illustrate the clouds.  While the molecular gas surface density $\Sigma_\mathrm{H2}$ shows a nearly flat distribution over the disk, the SFR surface density $\Sigma_\mathrm{SFR}$ expands a range of several order of magnitudes. Thus, the star formation efficiency $\Sigma_\mathrm{SFE}$ shows a range of values from 5 $\times 10^{-5}$ to 0.006 Myr$^{-1}$, but with no clear peaks as we observed in NGC 6946 in Paper I.}
\label{figure_rad_sfr_ngc628}
\end{figure*}

\begin{figure*}
\centering
\begin{tabular}{ccc}
\epsfig{file=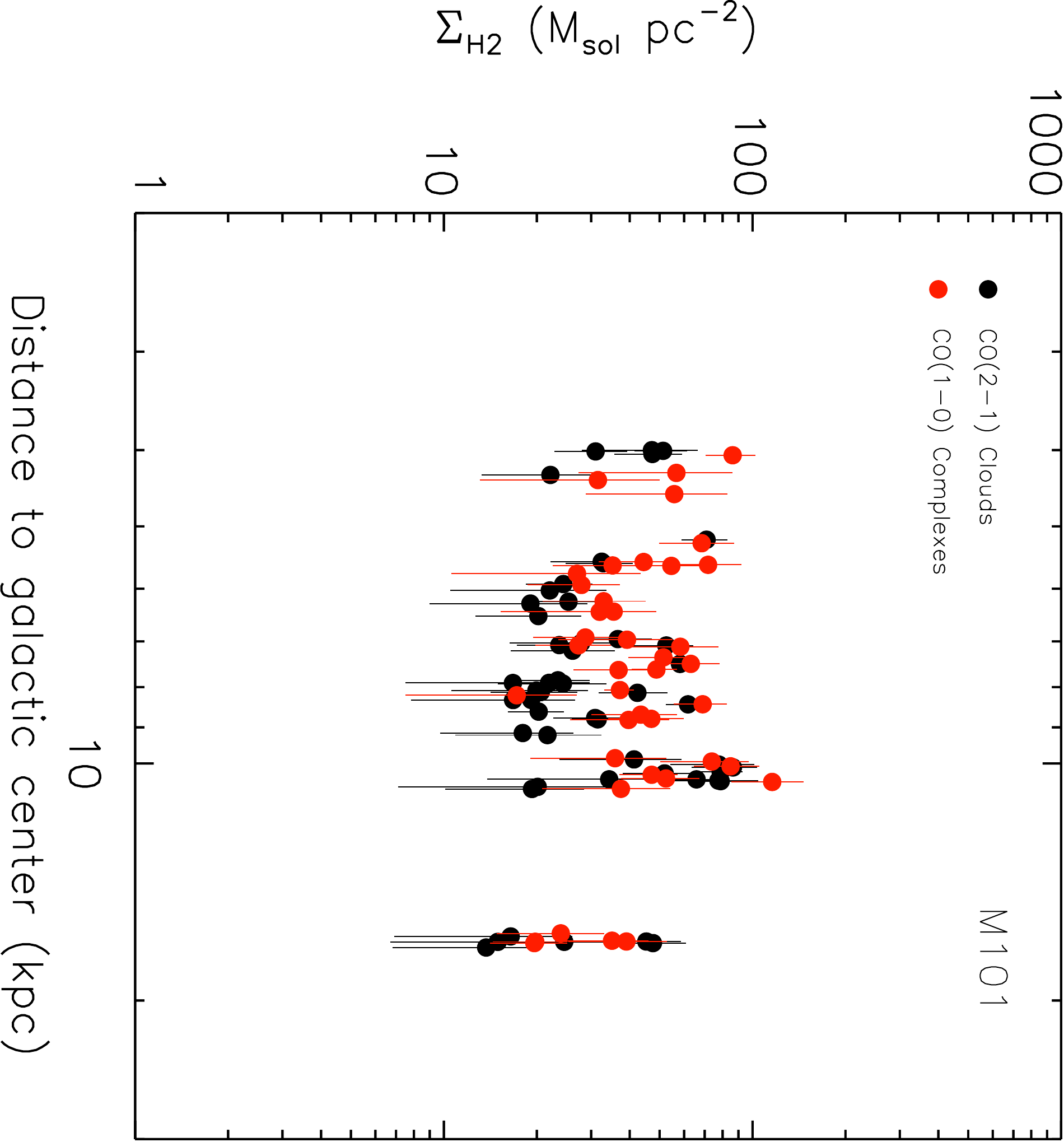,width=0.3\linewidth,angle=90} 
\epsfig{file=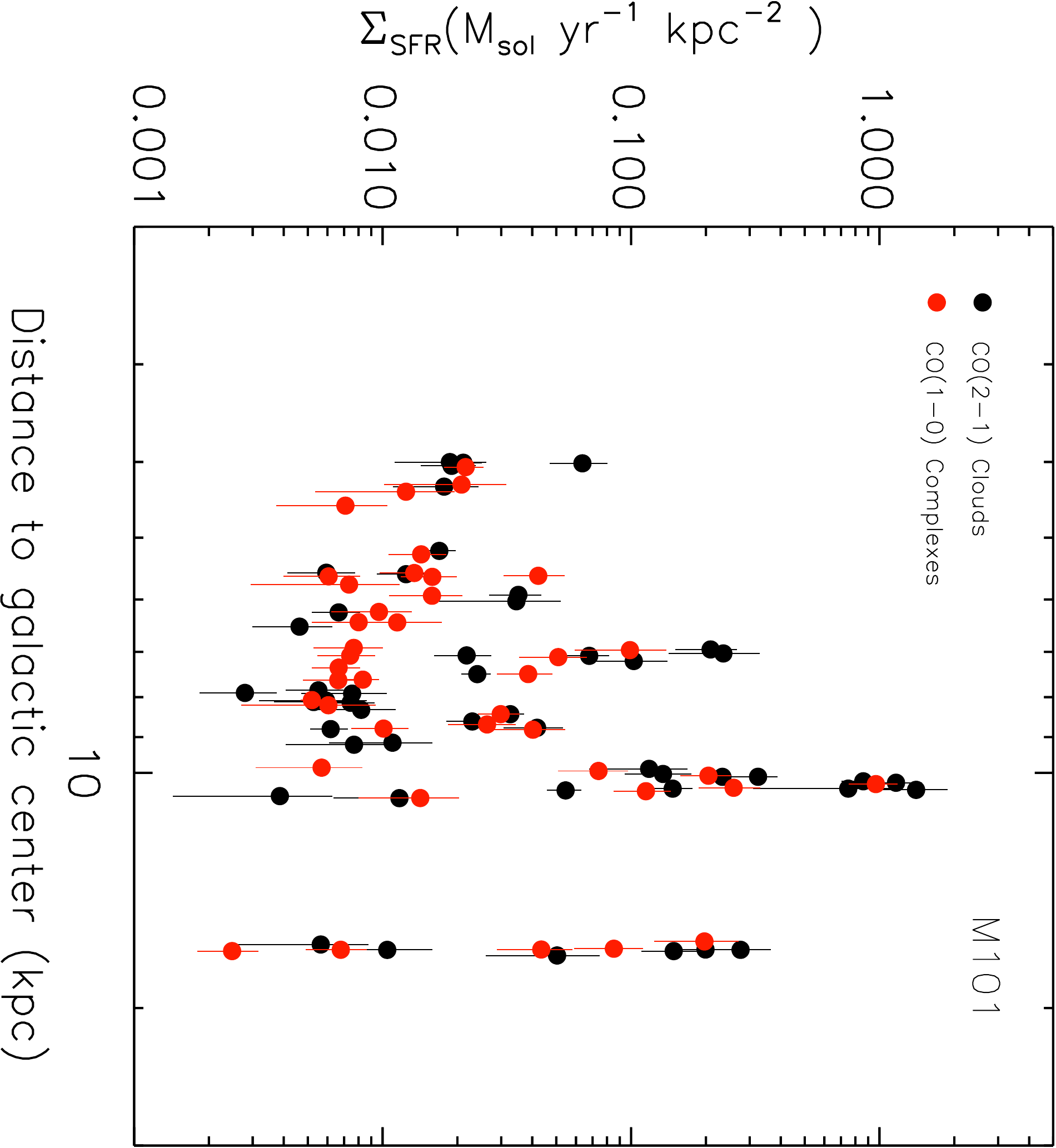,width=0.3\linewidth,angle=90} 
\epsfig{file=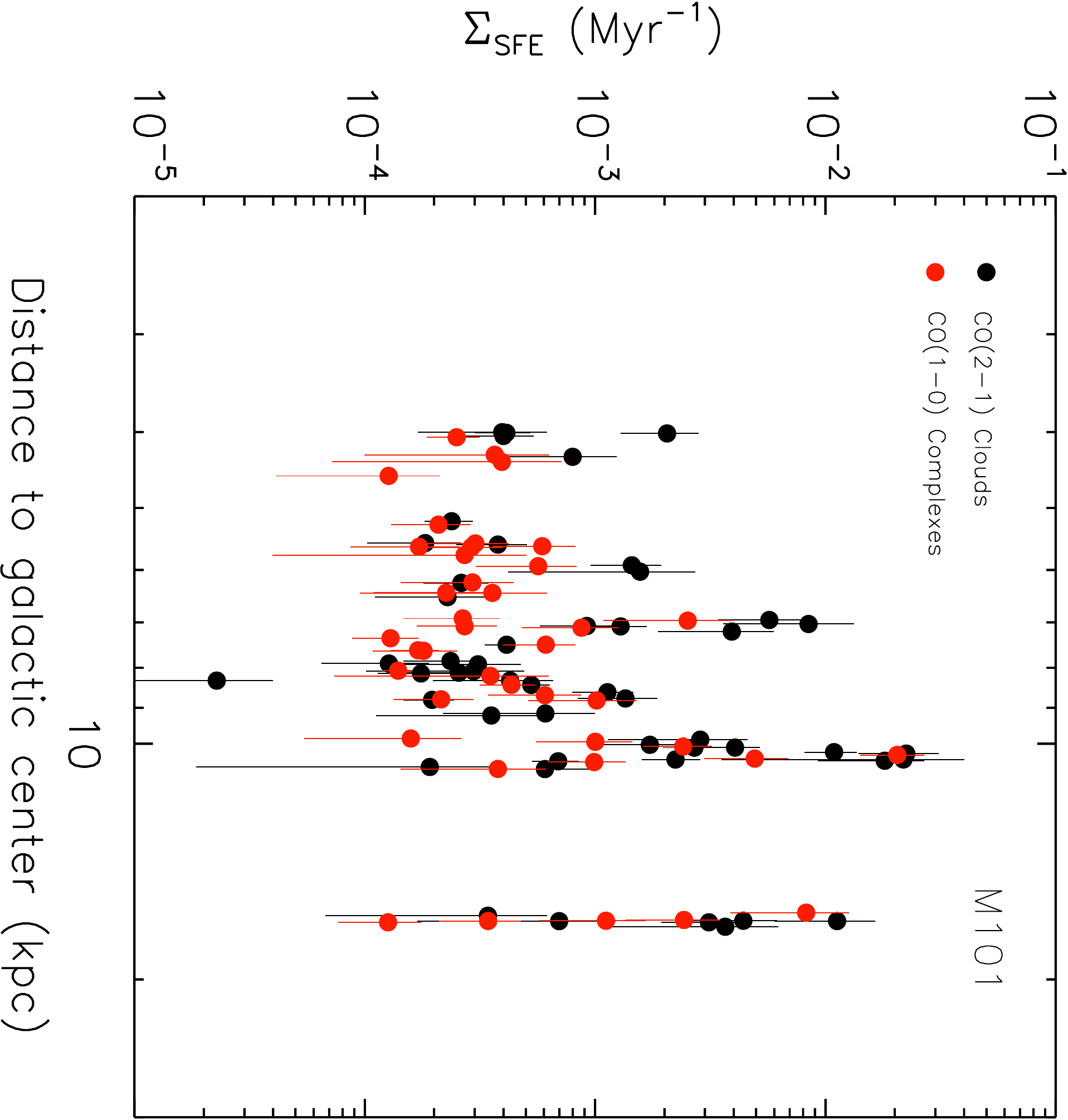,width=0.3\linewidth,angle=90} 
\end{tabular}
\caption{Radial distributions of $\Sigma_\mathrm{H2}$ {\bf (left)}, $\Sigma_\mathrm{SFR}$ {\bf (center)} and $\Sigma_\mathrm{SFE}$ {\bf (right)} of the regions observed over the southern arm of M101, including both $\co$ complexes and $\cotwo$ clouds. Red dots represent complexes, while black dots illustrate the clouds. Similarly to what we have found in galaxy NGC 6946, clear peaks of $\Sigma_\mathrm{SFR}$ and $\Sigma_\mathrm{SFE}$ are observed in M 101.}
\label{figure_rad_sfr_m101}
\end{figure*}

When we study the K-S relation using hexagonal grids, on the other hand, we find slope distributions consistent with sub-linear relations for NGC 6946 and NGC 628, and a less constrained slope distribution for M101 that covers sub-linear and super-linear relations.  Given that the grid approach allows us to include $\Sigma_\mathrm{H2}$ non-detections in the K-S analysis, we are able to assess the change in the fitting parameters when non-detections are considered.  Once the non-detections are included in the fit, we find that the slope distributions for all three galaxies are consistent with sub-linear relations.

Including $\Sigma_\mathrm{H2}$ non-detections can clearly have an impact on the power law index of the K-S relationship.  The left column of Figure \ref{figure_bayes_grid} show that in the case of NGC 6946, the distribution of the slope $\alpha$ is shifted to smaller values when $\Sigma_\mathrm{H2}$ non-detections are included.  On the other hand, for M101, although the slope distribution of the K-S relationship considering $\Sigma_\mathrm{H2}$ detections and non-detections is shifted to values smaller than the case with detections only, both distributions significantly overlap.  For NGC 628, both slope distributions are consistent.  The overall distribution on the $\Sigma_\mathrm{SFR}$ axis of the $\Sigma_\mathrm{H2}$ non-detections ultimately drives the effect of including them in the linear fit.  In the case of NGC 6946, $\sim 75\%$ of the $\Sigma_\mathrm{H2}$ non-detections have $\Sigma_\mathrm{SFR} > 0.006\ \Msun\ \mathrm{yr}^{-1}\ \mathrm{kpc}^{-2}$.  On the other hand, this percentage decreases to $34\%$ in the case of NGC 628, and to $29\%$ in the case of M101.  Thus, a large fraction of grid points with undetected CO but high SFR bias the slope of the linear fit to smaller values in the case of NGC 6946, while a large fraction of $\Sigma_\mathrm{H2}$ non-detections with low SFR keeps the fit parameters similar to the trend found when only $\Sigma_\mathrm{H2}$ detections are considered in the case of NGC 628.

The larger fraction of grid elements with high SFR but non-detected CO in NGC 6946 compared to NGC 628 and M101 could be related to several observational effects.  For example, spatial offsets between CO emission and regions of active SFR may be due to cloud evolution.  Thus, maps with enough angular resolution to resolve this spatial offset will be able to populate the low-CO and high-SFR region of the K-S relation.  In our grid analysis, we have used a fixed grid size of 8$\arcsec$ for the three galaxies, which translates into different physical scales due to differences in the galaxy distances.  For NGC 6946, the grid size is 213 pc, while for NGC 628 and M101 the grid size is 283 pc and 287 pc respectively.  Given the similar grid sizes of the three galaxies, we conclude that, if cloud evolution were responsible for the grid elements with high SFR but weak CO in NGC 6946, similar trends would have been detected in NGC 628 and M101 as well.

In contrast to recent studies, we do not observe a significant change in the slope distribution when a 24$\mu$m cirrus component is subtracted from the $\Sigma_\mathrm{SFR}$ measurements.  In Figure \ref{figure_sfr_h2_grid_nondet_bg} we notice that this correction has migrated some of the $\Sigma_\mathrm{SFR}$ values to the region in $\log (\Sigma_\mathrm{SFR}) \lesssim -2.5$, especially for NGC 6946.  For NGC 6946 and NGC 628, the net effect of the 24$\mu$m cirrus component subtraction is to move the intercept coefficient $A$ to smaller values with respect to the case where no correction is applied.  For M101, no detectable change has been observed.

Alternatively, regions of low CO brightness and high SFR may be related to variations of the $X_\mathrm{CO}$ factor.  By assuming a constant value for $X_\mathrm{CO}$ across the surveyed area, we may be underestimating the amount of molecular mass over regions where CO is dissociated by the radiation field, but H$_\mathrm{2}$ is still present.  Direct comparison with similar spatial resolution maps of other mass tracers such as infrared dust emission will provide independent constraints on the amount of total molecular mass in these regions.  

\begin{figure*}
\centering
\begin{tabular}{ccc}
\epsfig{file=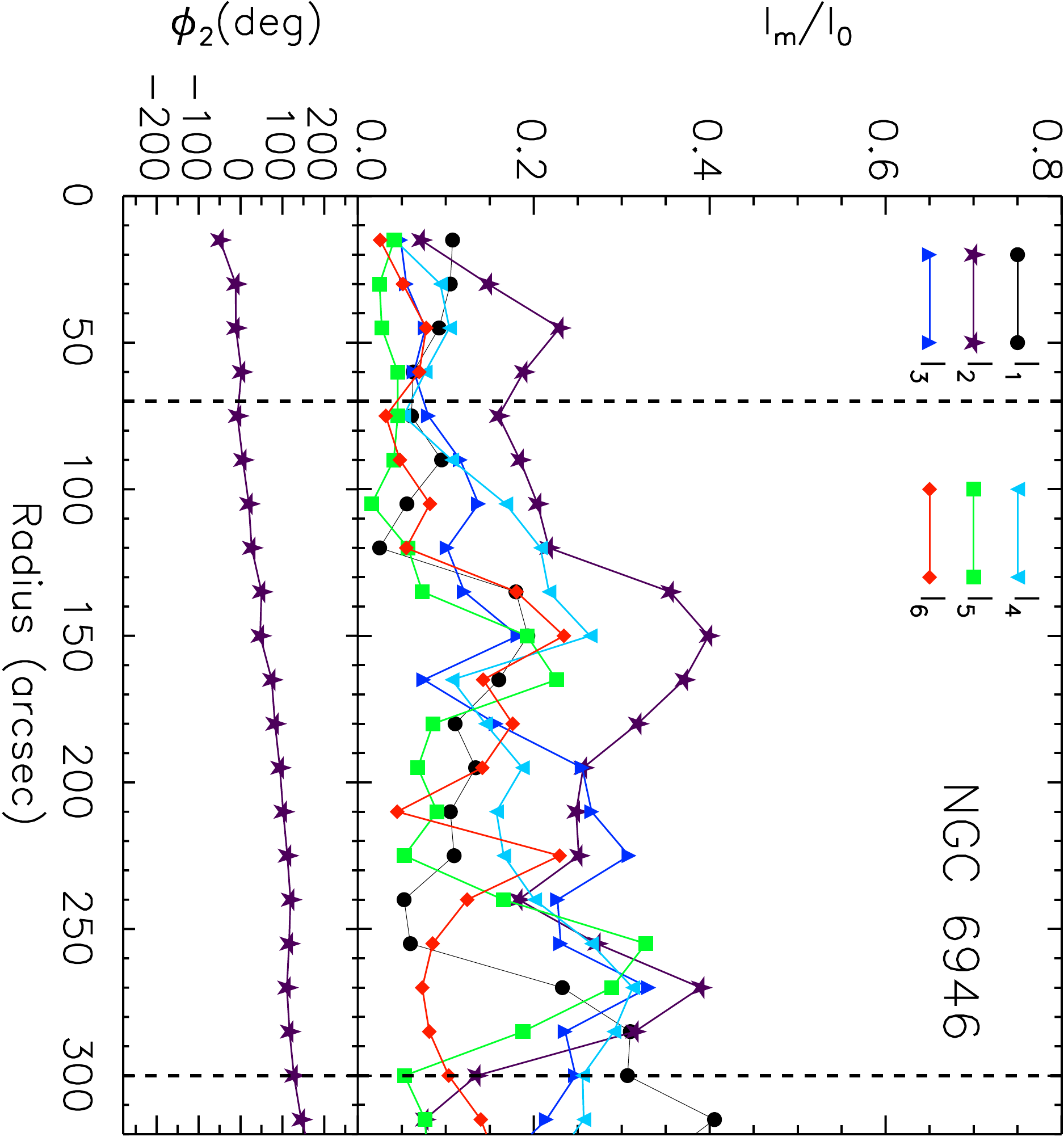,width=0.3\linewidth,angle=90} 
\epsfig{file=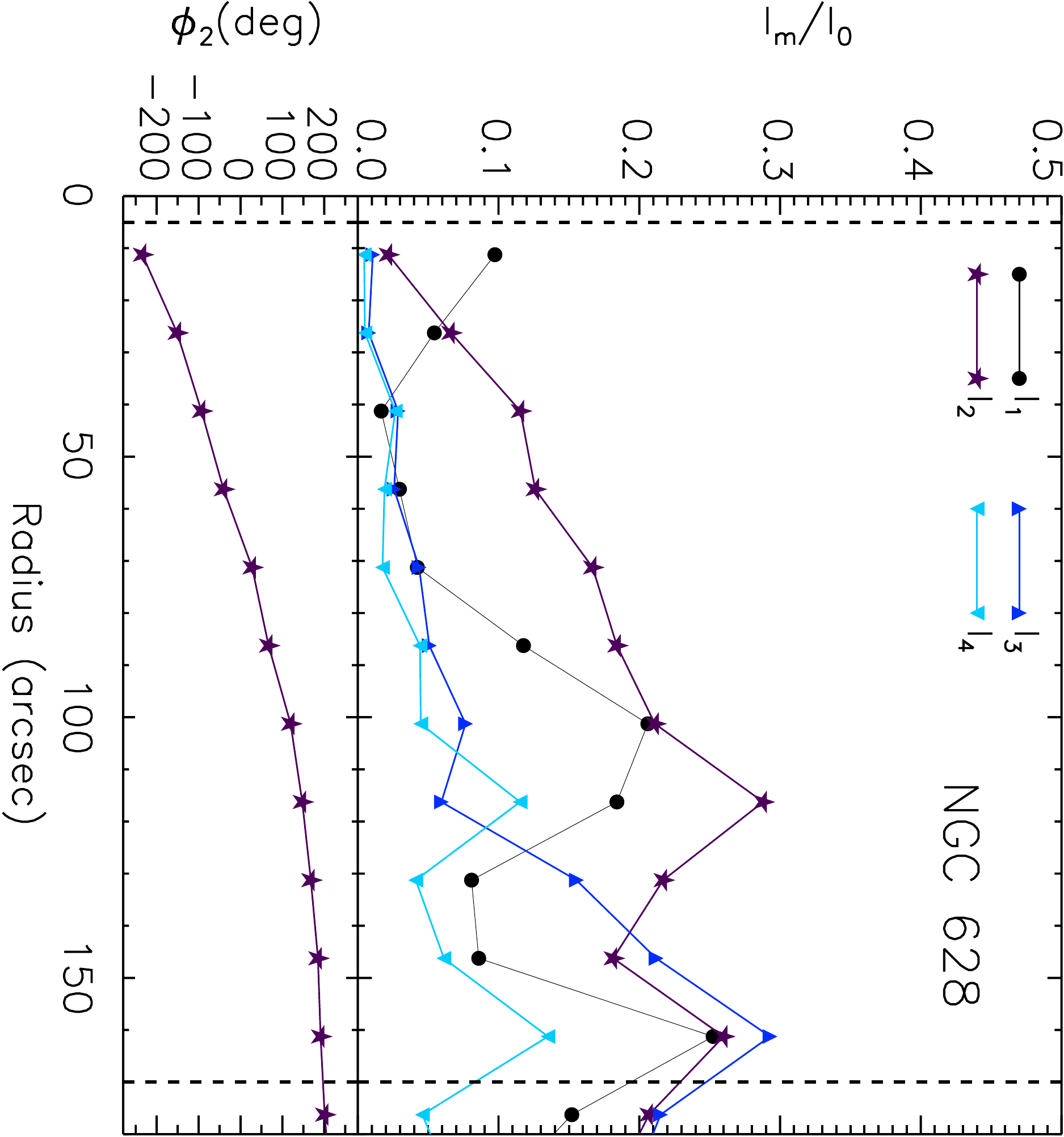,width=0.3\linewidth,angle=90} 
\epsfig{file=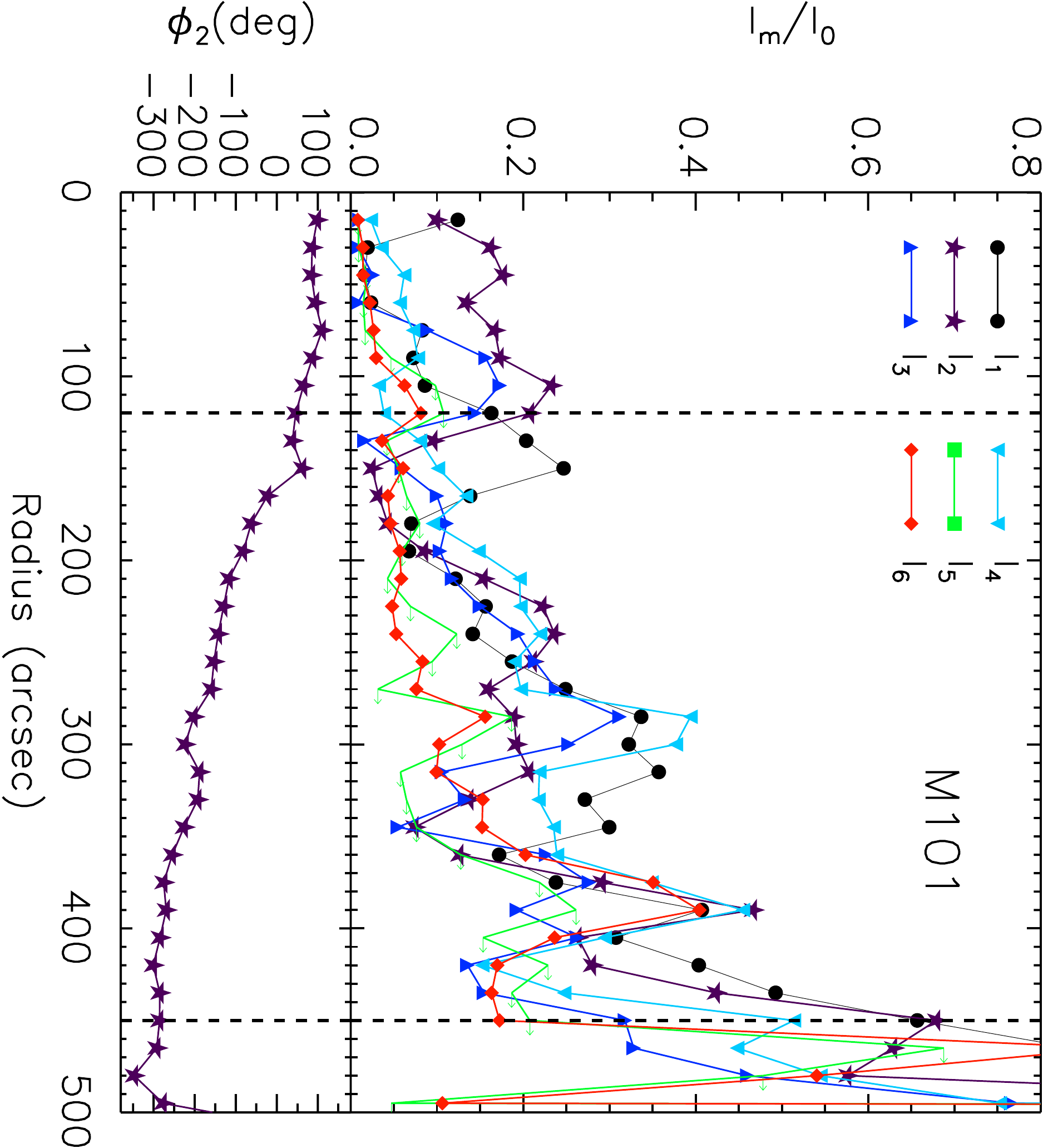,width=0.3\linewidth,angle=90} 
\end{tabular}
\caption{Radial distributions of the Fourier amplitude modes $I_m$ normalized to the $I_0$ mode for galaxies NGC 6946 (left), NGC 628 (center) and M101 (right), using the 3.6$\mu$m images from SINGS.  Vertical dashed lines illustrate the galactocentric radial extent covered by our $\co$ observations.  The radial distribution of the angular phase for $m=2$ ($\phi_2$) is shown in the lower panels.  NGC 628 shows prominent $I_2$ mode in the region we observed the molecular gas, and $\phi_2$ distribution is consistent with a two-arm spiral structure of the disk.  On the other hand, higher Fourier modes are also important for the multi-arm galaxies NGC 6946 and M101.  Although $I_2$ is important in the center of NGC 6946, the flat distribution of $\phi_2$ reflects the presence of a bar structure, which is different from a two-arm structure as seen in NGC 628.}
\label{spiral_structure}
\end{figure*}

\subsection{Distribution of the star formation properties in the disk}\label{rad-prop}
In Figures \ref{figure_rad_sfr_ngc628} and \ref{figure_rad_sfr_m101} we show the distribution of $\Sigma_\mathrm{H2}$, $\Sigma_\mathrm{SFR}$ and $\Sigma_\mathrm{SFE}$ (SFE is the star formation efficiency defined as $\Sigma_\mathrm{SFE} \equiv \Sigma_\mathrm{SFR}/\Sigma_\mathrm{H2}$) with respect to the galactocentric radius for the complexes and clouds identified in galaxies NGC 628 and M101 respectively.  We refer the reader to Paper I (see Figure 14) for NGC 6946 plots.  In the case of NGC 628, we observe a roughly uniform distribution of molecular gas density, with no clear trend with radius.  There are a couple of $\cotwo$ clouds with $\Sigma_\mathrm{H2}> 100\ \Msun$ pc$^{-2}$ which are located on on-arm regions.  The star formation surface density, on the other hand, presents a wider distribution of values ranging from $\sim 10^{-3}$ to $\sim 5\times 10^{-1}\  \Msun\ \mathrm{yr}^{-1}\ \mathrm{kpc}^{-2}$, resulting in a wide distribution of SFE surface densities as well.

In the case of M101, the molecular gas surface density distribution is similar to that observed in NGC 628, i.\ e., there is not a clear trend of $\Sigma_\mathrm{H2}$ with galactocentric radius.  On the other hand, the $\Sigma_\mathrm{SFR}$ shows pronounced peaks at galactocentric radius $\sim 7$, 10  and 17 kpc, which translate to peaks in the SFE for those regions as well.  Those regions have been highlighted in Figure \ref{figure_m101S}.  A similar behavior of the $\Sigma_\mathrm{SFR}$ was observed in NGC 6946 in Paper I, where we identified two specific regions in the eastern part of the disk with high SFE.  

Morphologically speaking, NGC 6946 and M101 seem to be different from NGC 628.  Whereas NGC 6946 and M101 show a multi-arm spiral structure, NGC 628 shows a two-arm spiral pattern in the inner part of the disk (\citealt{2010ApJ...725..534F}).  In order to quantify the significance of grand design spiral structure in the three galaxies in our sample, we use the same approach that we applied in Paper I for the galaxy NGC 6946 to define the spiral arms.  We estimate the arm amplitude by performing a Fourier series expansion of the old stellar population brightness distribution traced by the 3.6 $\mu$m map from SINGS.  Firstly, for each galaxy we have deprojected the 3.6 $\mu$m images using the parameters shown in Table \ref{galaxy-prop}.  Then, we decompose the images into radial bins with a width of 3\farcs75, overlapped by 50\% of the width for smoothness of the decomposition.  For every radial annulus, we fit the function given by

\begin{equation}\label{fourier}
I(r,\phi)=I_0(r)+\sum_{m=1}^{m_\mathrm{max}}  I_m(r) \cos [m\times(\phi-\phi_m(r))] 
\end{equation}

\noindent where $I_m$ is the arm amplitude and $\phi_m$ is the phase for each Fourier component {\it m}, and $m_\mathrm{max}$ is the maximum component considered in the expansion.  We have used the MPFIT package (\citealt{2009ASPC..411..251M}) to determine the best-fit parameters in Equation (\ref{fourier}).  The relative importance of the individual components is ultimately determined by the complexity of the arm structure in the galaxy.  In the case of grand-design galaxies, the spiral arm structure is well recovered by using components up to $m=4$, with a predominance of the mode $m=2$ (\citealt{2010ApJ...725..534F}).  Flocculent and multiarm galaxies, however, usually require higher components to recover their complex arm structure.  In Figure \ref{spiral_structure} we show the resulting $I_m$ amplitudes, normalized to the amplitude of $m=0$, of the three galaxies as a function of the galactocentric radius.  Additionally, Figure \ref{spiral_structure} shows the phase for $m=2$, $\phi_2(r)$, for the three galaxies.  NGC 628 is the galaxy that shows a distribution of amplitudes that resembles the amplitude spectrum expected for a two-arm grand design galaxy, i.\ e, a prominent mode $m=2$ in most of the disk traced by 3.6$\mu$m, and $\phi_2(r)$ steadily increasing with radius.  On the other hand, for galaxies NGC 6946 and M101 the $m=2$ amplitude is not the predominant mode all across the disk, but higher modes are also significant.  Although NGC 6946 presents a prominent $I_2$ in the center of the disk, $\phi_2$ is approximately constant with radius in the inner part.  This behaviour of $\phi_2(r)$ is consistent with the presence of a bar in the center of NGC 6946, and morphologically different from the two-arm spiral structure observed in NGC 628.

\begin{figure*}
\centering
\begin{tabular}{ccc}
\epsfig{file=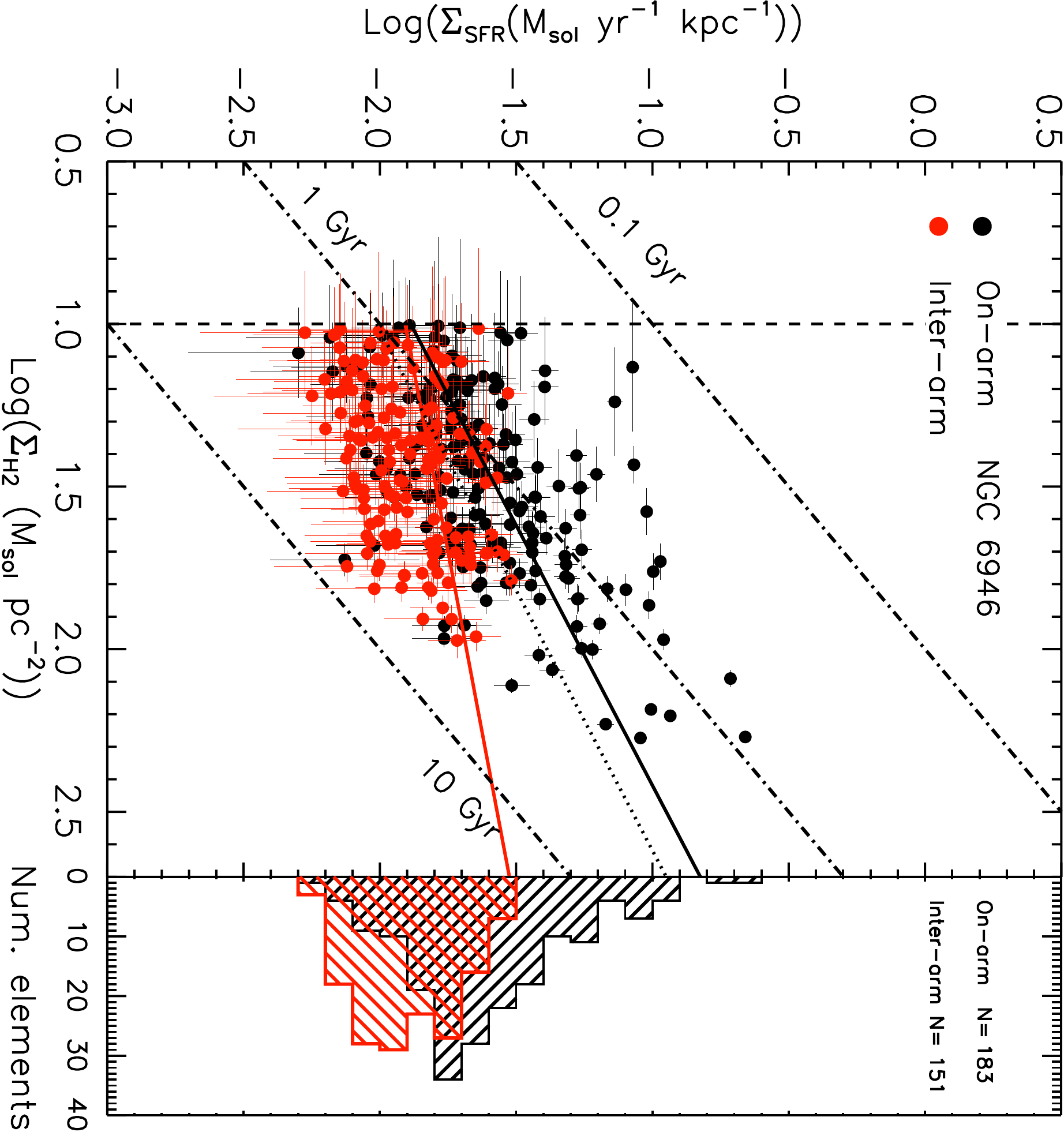,width=0.3\linewidth,angle=90} 
\epsfig{file=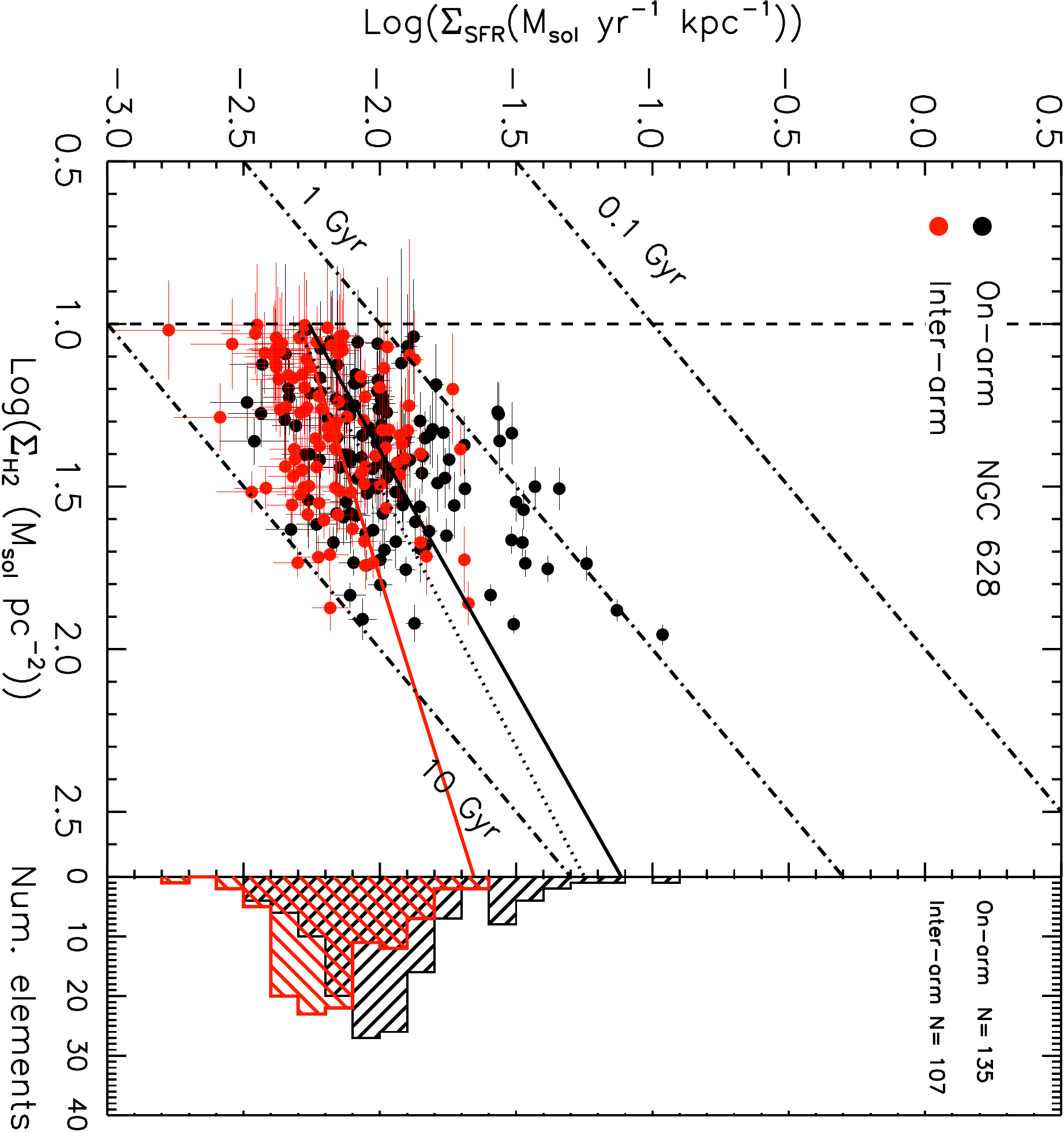,width=0.3\linewidth,angle=90} 
\epsfig{file=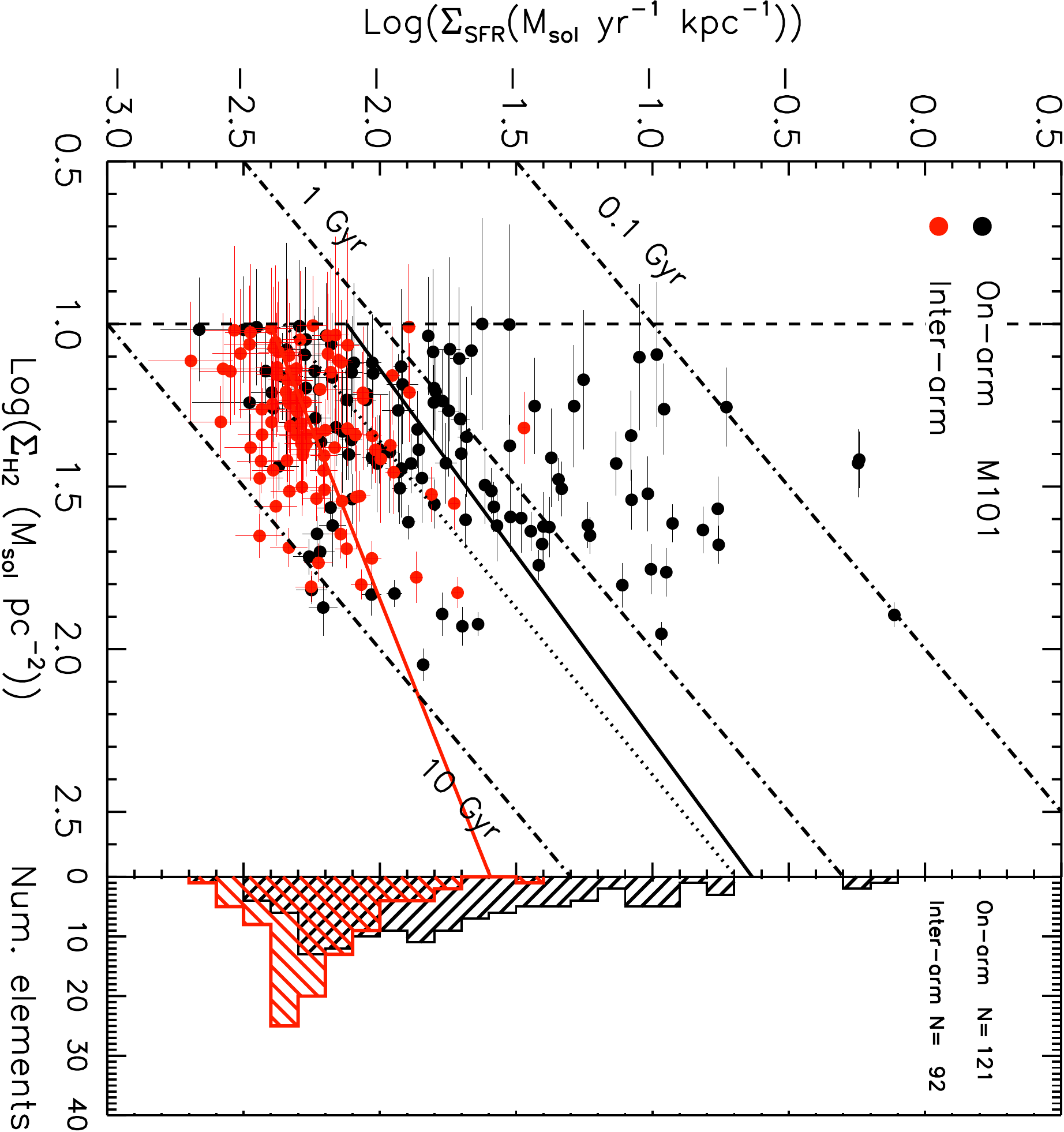,width=0.3\linewidth,angle=90} 
\end{tabular}
\caption{$\Sigma_\mathrm{H2}$ vs.\ $\Sigma_\mathrm{SFR}$ relations using a uniform grid, but in this case we have included the classification of on-arm and inter-arm regions yielded by the Fourier decomposition method.  Black dots represents the grid elements located in on-arm regions.  The red dots correspond to inter-arm regions.  Black solid lines illustrate the regression fit to the on-arm grid elements, while the red solid lines are the regression fit lines for the inter-arm regions.  Black dotted lines represent the regression fit performed to the full sample of grid elements.  We include the histograms of the SFR surface density on the right panels, along with the number of grid elements classified as on-arm or inter-arm regions.}
\label{figure_sfr_h2_grid_on_inter}
\end{figure*}

Why does the distribution of star formation efficiency seem to be different for some regions of flocculent/multi-arm spiral galaxies compared to what is observed in spiral galaxies with symmetric spiral arm structure?  Several authors have used hydrodynamic simulations to model flocculent and multi-arm spiral galaxies by using a ``live'' stellar disk potential (\citealt{2008MNRAS.385.1893D}; \citealt{2011ApJ...735....1W}).  They find that the structures observed in these types of galaxies are driven by mechanisms different from the stellar density waves present in grand design spiral galaxies.  In their simulations, the gas flows into the local potential minima, where it stays and condenses to form the dense gas that triggers the star formation.  Thus, the gas traces the potential minimum, and no clear offset between the gas tracers and spiral potential is observed in the simulations.  Nevertheless, maybe the most remarkable aspect of these simulations is that the spiral arms are not steady on time scales of a few 100 Myr (\citealt{2006MNRAS.371..530C}), with frequent collision and merging of the spiral arms.  Moreover, the simulations show that the largest GMCs are located in the regions where the spiral arms collide.  The collision of spiral arms may induce the formation of gas overdensities, triggering star formation.  In our observations, both NGC 6946 and M101 show spot regions of high star formation on the spiral arm (Figure \ref{figure_rad_sfr_m101} and Figure 14 in Paper I), which may be regions where previous spiral arms have collided.    

For NGC 628, on the other hand, the maximum values of SFE are located across a range of radii in the disk (Figure \ref{figure_rad_sfr_ngc628}).  This is consistent with the picture of the formation of massive molecular clouds attributed to the passing of spiral density waves in grand-design galaxies (\citealt{2002ApJ...570..132K}; \citealt{2004MNRAS.349..270W}).  In this scenario, regions of high star formation activity will be more continuously distributed in radius as they are aligned with the spiral structure, in contrast to the situation observed in NGC 6946 and M101.

\begin{figure*}
\centering
\begin{tabular}{ccc}
\epsfig{file=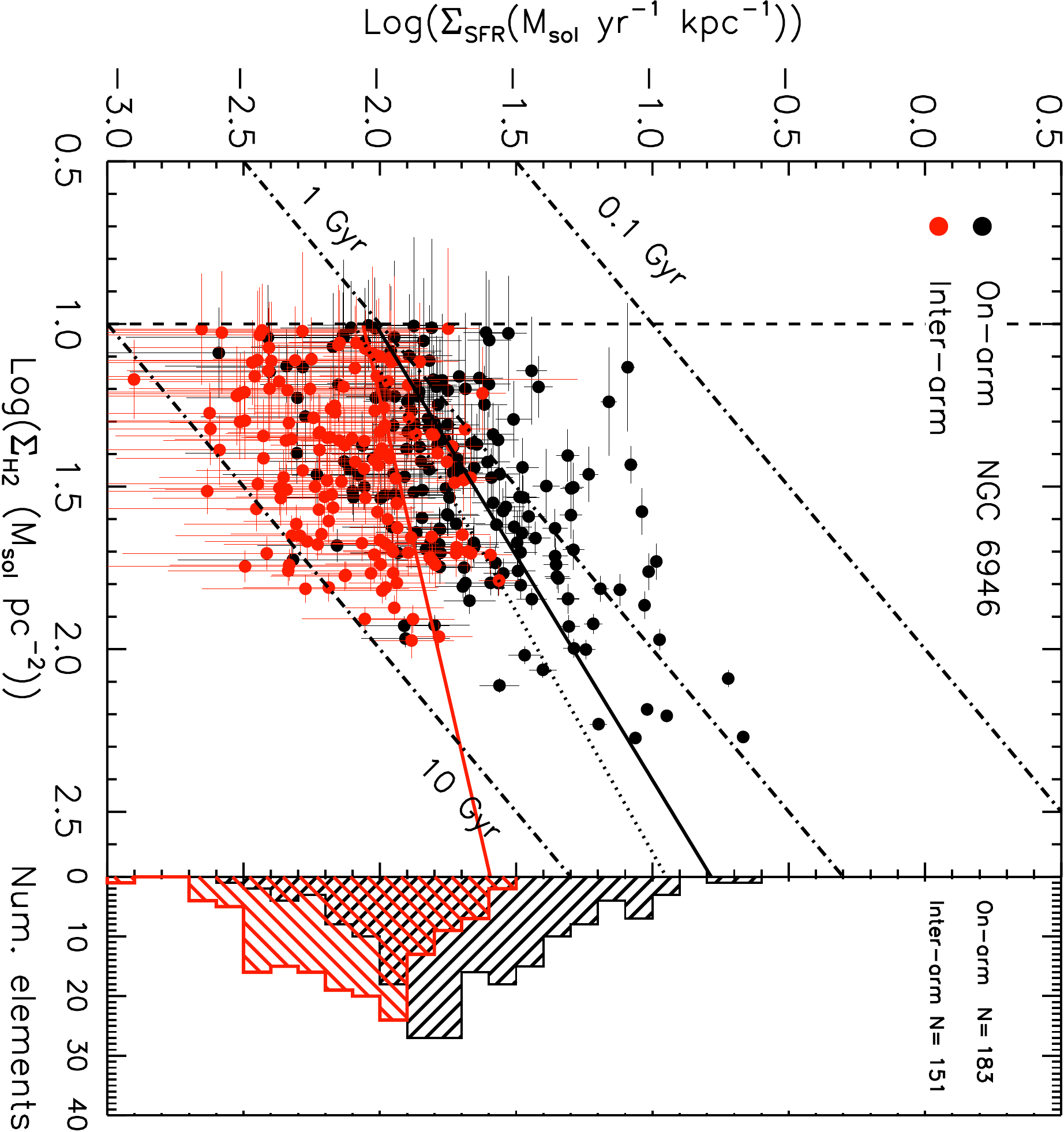,width=0.3\linewidth,angle=90} 
\epsfig{file=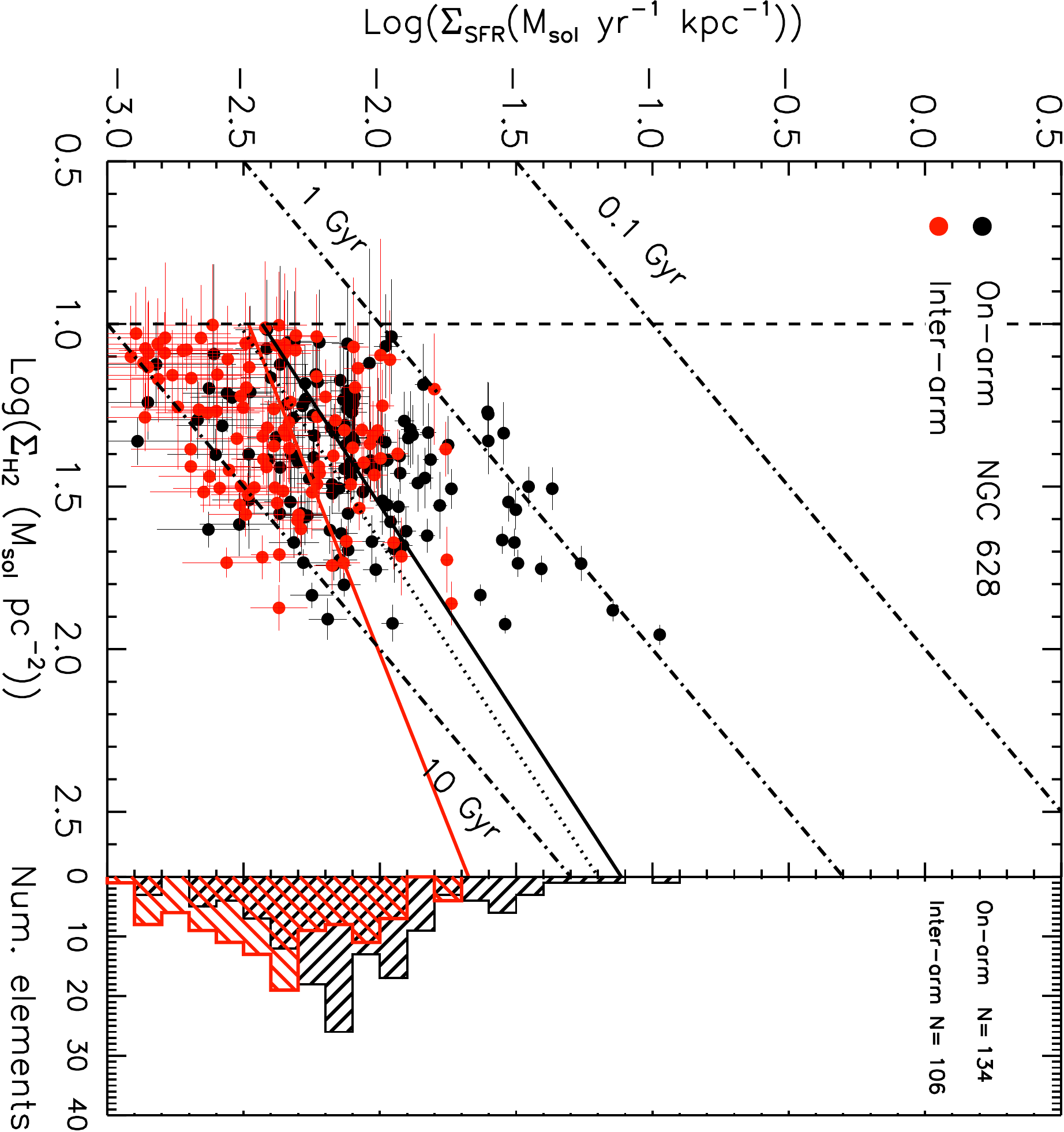,width=0.3\linewidth,angle=90} 
\epsfig{file=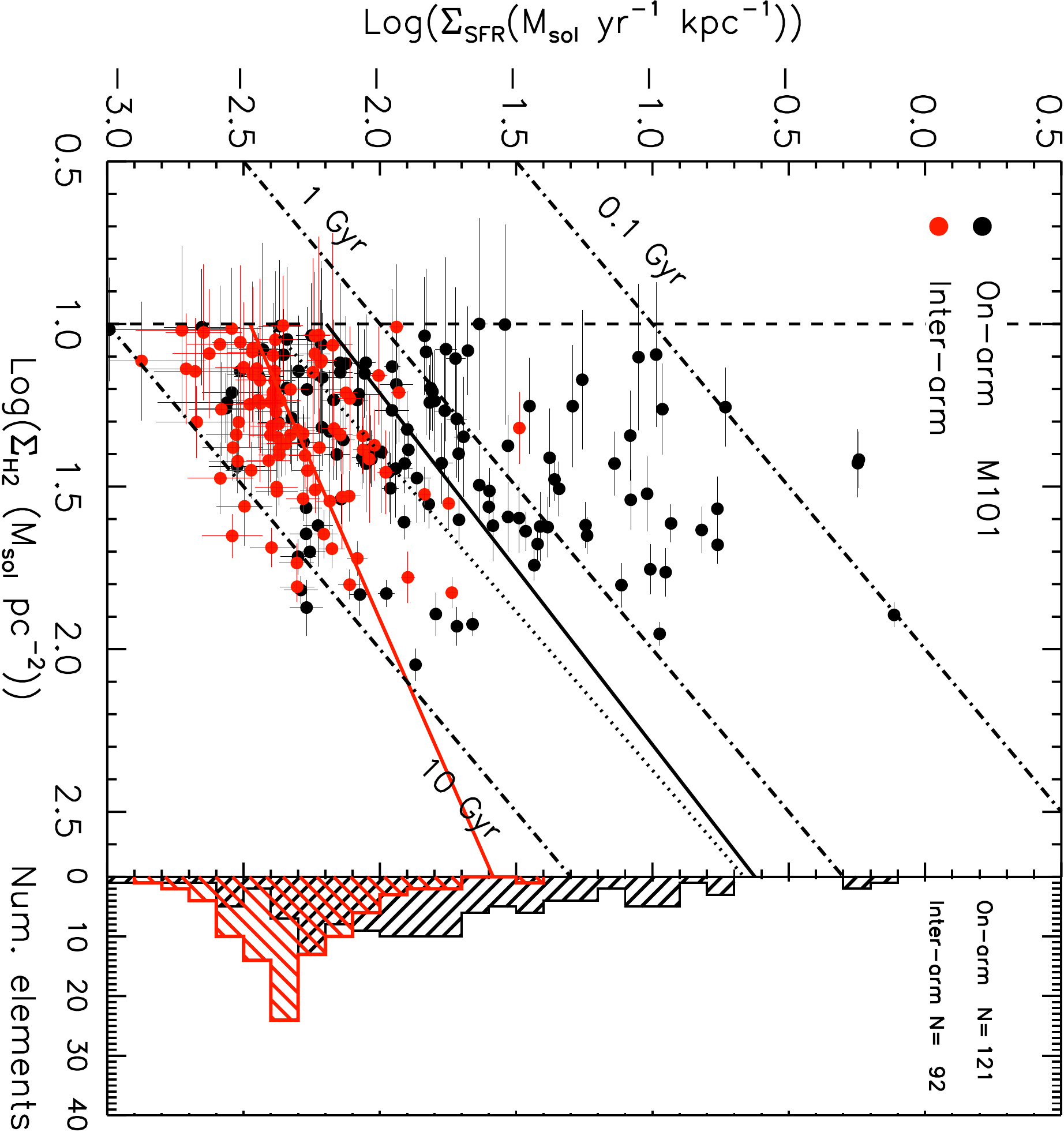,width=0.3\linewidth,angle=90} 
\end{tabular}
\caption{Same as Figure \ref{figure_sfr_h2_grid_on_inter}, but a subtraction of a 24$\mu$m smooth component is applied to $\Sigma_\mathrm{SFR}$.}
\label{figure_sfr_h2_grid_on_inter_sub}
\end{figure*}

\subsection{On-arm vs.\ inter-arm $\Sigma_\mathrm{H2}$-$\Sigma_\mathrm{SFR}$ relation.}\label{in-on-arm}

The Fourier decomposition analysis of the 3.6$\mu$m image allows us to identify the regions associated with the arms and those associated with the inter-arm zone.  We have classified the on-arm and inter-arm regions following the same approach adopted in Paper I.  A reconstructed image of the 3.6$\mu$m is generated from all the modes considered in the Fourier decomposition.  For each radial bin, we have selected the 35\% highest pixels to generate the mask representing the on-arm regions.  That mask is then applied to the grid map in order to assign grid elements to on-arm or inter-arm regions.  For simplicity, we have only considered $\Sigma_\mathrm{H2}$ detections in our on-arm vs. inter-arm analysis.  In Figure \ref{figure_sfr_h2_grid_on_inter}, we show the $\Sigma_\mathrm{H2}$-$\Sigma_\mathrm{SFR}$ relations for the three galaxies in our sample including the on-arm vs.\ inter-arm classification.  For NGC 6946 and M101, it is clear that on-arm regions are distinct from inter-arm regions, consistent with what we found in Paper I for complexes and clouds.  The on-arm regions present higher SFR than inter-arm regions for similar molecular gas surface densities.  In NGC 628, although the main group of points contains both on-arm and inter-arm clouds, we observe a group of on-arm grid elements that deviates toward higher SFR surface density.

In order to investigate whether the K-S relation changes respect to the location in the disk, we have applied the Bayesian regression fitting approach (Equation \ref{sfr_h2}) to the on-arm and inter-arm regions.  Figure \ref{figure_sfr_h2_grid_on_inter} shows the resulting fitting lines for the three galaxies in our sample built from the peak value of each parameter distributions.  In Table \ref{table-bayes-grid} we present the peaks and 90 $\%$ HDI values of the parameter distributions from the Bayesian regression fit.  The overall result obtained for the three galaxies in our sample is roughly similar:  the distributions of slopes of inter-arm regions are shifted to smaller values than those found in K-S relations for on-arm regions.  Also, the distributions of slopes considering on-arm regions are similar to those found when we consider the full sample of grid elements.  Regarding the distribution of the intercept coefficient $A$, we observe that this parameter reaches the smaller values for inter-arm grid regions, followed by the case when all the grid points are considered. We conclude that low level SFR regions mainly present in inter-arm regions does not affect significantly the slope of the K-S relation when all the grid values are considered.  In contrast, the K-S linear relations for the full sample of grid points are biased to smaller values of SFR respect to the on-arm K-S relations, which is evident from the smaller values of the $A$ parameter yielded by the Bayesian regression fit for the full sample respect to the on-arm regions. 

The different behaviours found for on-arm vs.\ inter-arm regions presented above may be reflecting some underlying physical process in the formation of stars in spiral galaxies.  For instance, in their study of the barred spiral galaxy NGC 4303, \citet{2010ApJ...721..383M} found differences in the SFE between the spiral arms, the bar, and the inter-arm regions.  The fact that inter-arm molecular gas has similar surface density compared to on-arm molecular material but with lower SFR, may be an indication of the presence of a higher amount of denser gas inside the molecular clouds located in the spiral arms, especially in the case of M101 where this difference in SFR is more prominent.  The spiral shocks (maybe present on NGC 628), or convergence of gas flows (maybe more related to galaxies NGC 6946 and M101) may lead to an increase in the amount of dense gas in spiral arms.  Several authors have investigated the nature of the extragalactic scaling relation of SFR and dense molecular gas (\citealt{2004ApJ...606..271G}; \citealt{2012ApJ...745..190L}).  \citet{2012ApJ...745..190L} suggest that total star formation rate in a molecular cloud is linearly proportional to the amount of dense gas present in the cloud.  Thus, the steeper $\Sigma_\mathrm{H2}$-$\Sigma_\mathrm{SFR}$ relation observed in Figure \ref{figure_sfr_h2_grid_on_inter} for the on-arm regions, more prominent in NGC 6946, may indicate the presence of denser gas created by processes that depends on the global properties of the dynamics of the gas.  


In our analysis using the uniform grid, we are including regions from the inter-arm zone where the contribution from the diffuse component unrelated to the current star formation may bias the fitted linear relation to smaller slopes.  The effect of the diffuse component is to overestimate the SFR in the lower end of the distribution, flattening the slope of the $\Sigma_\mathrm{H2}$-$\Sigma_\mathrm{SFR}$ linear relation in the logarithmic space.  This may be the cause of the flatter relation we find in our uniform grid analysis compared to the relations found in Section \ref{k-s_law} for complexes and clouds.  As we defined our regions based on $\co$ complexes and $\cotwo$ clouds, we are likely selecting regions less affected by the diffuse contribution.  The small impact of the diffuse emission in the K-S law in regions of high surface density was already reported by \citet{2011ApJ...730...72R} in their study on the central region of the disk of NGC 4254.

Following the approach already introduced in Section \ref{cirrus_sfr} and detailed in Appendix \ref{24um_cirr}, we have subtracted a diffuse component in the 24$\mu$m not associated with current formation from the $\Sigma_\mathrm{SFR}$ measurements.  Figure \ref{figure_sfr_h2_grid_on_inter_sub} shows the resulting K-S relations for the three galaxies in our sample, with the $\Sigma_\mathrm{SFR}$ measurements corrected by the cirrus component in the 24$\mu$m.  We notice that differences between on-arm vs. inter-arm regions is similar to the case when no cirrus correction is applied, with on-arm regions presenting higher $\Sigma_\mathrm{SFR}$ than inter-arm regions, being more evident for NGC 6946 and M101.  Therefore, our approach to subtract a diffuse component in the 24$\mu$m map has not introduced a significant change in the on-arm vs. inter-arm region differentiation.


We apply the Bayesian regression fitting to the K-S relation, but now with the $\Sigma_\mathrm{SFR}$ measurements corrected by the cirrus component in the 24$\mu$m.  Table \ref{table-bayes-grid} shows the resulting peaks and 90 $\%$ HDI values of the parameter distributions.  The overall distribution of the slopes is consistent with the case where no subtraction of the 24$\mu$m smooth component is applied.  In the same direction, the distributions of the intercept coefficient $A$ and the dispersion of the intrinsic scatter $\sigma$ are consistent with the case when $\Sigma_\mathrm{SFR}$ is not corrected for the 24$\mu$m diffuse component.  In the case of NGC 6946, the Bayesian regression fitting to the K-S relation for on-arm regions yields larger slopes than inter-arm regions, which is consistent to the case with no cirrus correction.  On the other hand, for NGC 628 and M101 the difference between on-arm vs. inter-arm region is less prominent.  Again, this is consistent to the K-S relation differences between the on-arm vs. inter-arm regions when no correction of the 24$\mu$m diffuse component is applied to the the $\Sigma_\mathrm{SFR}$ values.

Although our analysis of on-arm and inter-arm regions only includes $\Sigma_\mathrm{H2}$ detections, including the $\Sigma_\mathrm{H2}$ non-detections would not change our main result.  In Section \ref{k-s_grid_h2_nondetect} we see that, in general, the K-S relations present flatter slopes when we include $\Sigma_\mathrm{H2}$ non-detections.  Because the $\Sigma_\mathrm{H2}$ non-detections have a stronger effect on inter-arm grid points, our finding that inter-arm regions have K-S relations with flatter slopes than on-arm regions would be preserved.


\subsection{Caveats and uncertainties of the present work}\label{caveats}
The coverage of our observations have been limited to reduced regions of the molecular gas disk of NGC 6946, NGC 628 and M101.  Therefore, our conclusions can only be applied to the regions covered by our observations and cannot be interpreted as global properties of the galaxies included in our sample.  A full census of the GMCs population across the disks of nearby galaxies is needed in order to perform a detailed study of molecular cloud properties and their relation with the surrounding environment.  Nevertheless, high sensitivity maps with resolutions $\sim 60$ pc that cover a significant part of the molecular disk are observationally challenging, although recent studies have achieved this goal in M51 (\citealt{2013ApJ...779...42S}).  In that direction, we have also successfully completed observations of the full disk of NGC 6946 with CARMA (Rebolledo et al., in preparation).  With the help of facilities such as ALMA, we will be able to obtain high quality images at resolutions close to GMC sizes, allowing us to perform unbiased analyses of the physical properties of the molecular gas in nearby spiral galaxies. 

Equations \ref{sfr} and \ref{sfr_ha} assume a continuous star formation process.  These prescriptions to estimate the SFR may break down for scales close to GMC sizes, as we may be isolating regions with a single stellar population at a specific age, especially with our $\cotwo$ observations.  In order to investigate the age sensitivity and the intrinsic scatter in the SFR estimates by using Equations \ref{sfr} and \ref{sfr_ha} at small scales, \citet{2012AJ....144....3L} simulate the H$\alpha$ and FUV emission after an instantaneous burst of star formation.  They show that H$\alpha$ is mostly emitted in the first 10 Myr.  On the other hand, FUV presents significant emission up to 65 Myr after the burst.  This model predicts a factor of $\sim$2 uncertainty inferring the SFR from H$\alpha$, and a factor $\sim$3-4 uncertainty inferring SFR from FUV.  

Additionally, these equations assume that the ionizing star clusters are massive enough in order to have the high mass tail of the IMF well populated (\citealt{1998ARA&A..36..189K}; \citealt{2007ApJ...671..333K}).  However, this is likely not to be true in regions of low $\Sigma_\mathrm{SFR}$, where the SFR tracer emission could be dominated by low mass stars.  In an effort to understand the different sources of the observed scatter in the K-S relation in spatially resolved studies, \citet{2014MNRAS.439.3239K} present a formalism based on a simple uncertainty principle for SF.  They show that for an idealized galactic environment typical of a disk galaxy, the minimum size required to recover an underlying SF relation is $\sim$ 500 pc, and is determined by the ability to sample independent star-forming regions.  This is a factor of $\sim$3-5 larger than the sizes sampled by our $\co$ observations, a factor $\sim$6-10 larger than the $\cotwo$ cloud sizes, and $\sim$2 larger than the area of the grid elements.  Thus, the high scatter observed in our derived K-S relations, especially for $\cotwo$ clouds, may be the result of our incomplete sampling of independent star forming regions over the areas covered by our observations.  Alternative approaches to measure the SFR at small scales (for instance, via observations of resolved star clusters) can be compared to the SFR traced by FUV or H$\alpha$, allowing us to assess the level of uncertainties in our methodology to estimate $\Sigma_\mathrm{SFR}$.
 
\section{Summary}\label{summary}
We have performed high resolution observations of the molecular gas in the nearby spiral galaxies NGC 628 and M101.  Using CARMA, we have observed the $\co$ emission line over regions of the disk with active star formation and offset from the galactic center.  These observations have supplemented the $\co$ observations of the northeastern region of NGC 6946 reported in Paper I.  Higher resolution observations of $\cotwo$ toward the brightest regions observed in $\co$ have allowed us to resolve some of the largest GMCs.  The results are summarized as follows:

\begin{enumerate}

\item Using the cloud-finding algorithm CPROPS (Section \ref{cl-prop-corr}), we have identified 112 CO emitting complexes with typical sizes of $\sim$150 pc.  The higher resolution observations of $\cotwo$ towards the brightest regions detected in $\co$ allowed us to find 144 structures with sizes $\sim 70$ pc which we have identified as GMCs.  Properties such as size, line width and luminosity were derived using the moment approach implemented in the CPROPS package. 

\item We observed that the size-line width relations for the $\co$ complexes and $\cotwo$ clouds present significant scatter, with some of the regions in NGC 6946 presenting excess velocity dispersions.  On the other hand, line widths in NGC 628 are limited to $\lesssim$ 7 $\kms$.  Additionally, our $M_\mathrm{vir}$-$L_\mathrm{CO}$ scatter plot for both complexes and clouds are consistent with our choice of the CO to H$_2$ conversion factor $X_\mathrm{CO}$=$2\times10^{20}\mathrm{cm}^{-2}(\mathrm{K}\ \kms)^{-1}$.    

\item Linear fitting to the scaling relations between the resolved properties of the identified structures is performed using a Bayesian regression approach.  In agreement with recent studies of GMCs in nearby galaxies, we do not find strong correlations in the size-line width relations when resolution bias correction is applied.  The reduced dynamic range in the sizes of the identified structures does not allow to invoke a physical explanation to the apparent presence of an upper limit in the velocity dispersions observed in NGC 628.

\item The Bayesian linear regression for the $\Sigma_\mathrm{H2}-\Sigma_\mathrm{SFR}$ relation for complexes is consistent with a super-linear K-S relation in the case of M101.  On the other hand, the lower tail of the slope distribution reaches values below 1 for NGC 6946 and NGC 628.  In general, the slopes for $\cotwo$ clouds are similar to those found for $\co$ complexes, but with greater uncertainty.  The distribution of $\alpha$ is consistent with super-linear relations for NGC 6946 and M101, while it reaches values below 1 in the lower end for NGC 628. The higher scatter found in higher resolution maps may be the result of variations in the star formation activity at local scales, which translates to a larger variation of the SFR than the molecular gas surface density from region to region.




\item In contrast to the case when we analyse the K-S relation for structures identified by CPROPS, the Bayesian regression fit using a uniform grid yields sub-linear K-S relations for NGC 6946 and NGC 628.  In the case of M101, the slope probability distribution has a peak at $\sim 1$.  We do not observe a significant change in the slope distributions when a cirrus component in the $\Sigma_\mathrm{SFR}$ is subtracted.  When $\Sigma_\mathrm{H2}$ non-detections are included in the analysis, the slope of the K-S relation may be affected depending on the overall distribution of $\Sigma_\mathrm{SFR}$ along the $y-$axis.

\item Using the 3.6 $\mu$m images of NGC 6946, NGC 628 and M101 as stellar mass density tracers, we have implemented a Fourier decomposition approach to identify the spiral arm structure present in these galaxies.  On-arm and inter-arm regions have been defined based on the 3.6 $\mu$m Fourier reconstructed images.  We observe that, over the regions observed in the galaxies in our sample, the most prominent Fourier mode for NGC 628 is $m=2$, similar to the expected value for a two-arm grand design galaxy.  On the other hand, NGC 6946 and M101 lack a prominent $m=2$ mode, showing significant higher order modes instead, giving evidence for a multi-arm or flocculent structure present in these galaxies.


\item This difference in the distribution of regions with maximum SFR and SFE may be related to the underlying dynamical process that drives the arm structure observed in spiral galaxies (Section \ref{rad-prop}).  In this scenario, in galaxies with nearly symmetric arm shape (e.\ g., NGC 628), the spiral shocks are triggering the star formation along the arms, giving a more uniform distribution of SFE peaks across the disk.  On other hand, galaxies with flocculent or multi-arm spiral structure (e.\ g., NGC 6946 and M101) show regions of high star formation efficiency on specific regions of the spiral arms (Figure \ref{figure_rad_sfr_m101} in this paper, and Figure 14 in Paper I), which may be interpreted as the result of gas flow convergence or regions where previous spiral arms may have collided.  High fidelity observations of the molecular gas and star formation covering the full disk of nearby galaxies with different morphology will be required to test the scenario proposed here.

\item The Bayesian regression fit to the $\Sigma_\mathrm{H2}-\Sigma_\mathrm{SFR}$ relation in logarithmic scale for on-arm and inter-arm regions reveals that the distribution of the slopes are shifted to higher values for on-arm grid points for the three galaxies in our sample.  Same picture is obtained when the $\Sigma_\mathrm{SFR}$ values are corrected by the cirrus component.  Higher star formation activity found in on-arm regions may indicate the presence of denser gas, which may be induced by processes that depend on the global dynamics of the gas.

\end{enumerate}

The authors thanks the anonymous referee for the comments and suggestions that have improved significantly the presentation and the discussion of the paper.  We gratefully acknowledge the efforts by the SINGS, LVL and GALEX NGS teams to make their data public.   We thank the CARMA staff the support in the observations presented in this work.  The construction of CARMA was supported by the Gordon and Betty Moore Foundation, the Kenneth T. and Eileen L. Norris Foundation, the James S. McDonnell Foundation, the Associates of the California Institute of Technology, the University of Chicago, the states of California, Illinois, and Maryland, and the National Science Foundation.  The CARMA development and operations are supported by the National Science Foundation and the CARMA partner universities under a cooperative agreement.  DR acknowledges support from the Australian Research Council Discovery Project Grant DP130100338.  JK is supported by the NSF through grant AST-1211680.  The National Radio Astronomy Observatory is a facility of the National Science Foundation operated under cooperative agreement by Associated Universities, Inc.

\appendix

\section{Systematic uncertainties in SFR tracers}\label{24um_cirr}

The effect of a diffuse component in the $\Sigma_\mathrm{SFR}$ measurements used in our grid analysis is quantified by using a simple approach.  Our approach assumes that the diffuse component not associated to SF is mainly dominated by a cirrus component in the 24$\mu$m map.  Additionally, our approach assumes that the diffuse emission is negligible in regions of high SF, so the $\Sigma_\mathrm{SFR}$ measurements from 24$\mu$m+FUV is a good estimate for the true $\Sigma_\mathrm{SFR}$ in this regime.  On the other hand, it assumes that the $\Sigma_\mathrm{SFR}$ measurements from H$\alpha$ are less affected by a cirrus component at low SF, so any deviation from a linear relation between $\Sigma_\mathrm{SFR}$ from 24$\mu$m+FUV and $\Sigma_\mathrm{SFR}$ from H$\alpha$ is assumed to be due to a cirrus component in the 24$\mu$m map.

Our approach is as follows.  First, we smooth the 24$\mu$m map using a gaussian kernel of a given size.  Then, the smoothed map is multiplied by a factor that accounts for the fraction of the emission associated to the diffuse component.  Finally, this scaled and smoothed map is subtracted from the original 24$\mu$m map.  Thus, our approach relies on two parameters:  the size of the gaussian kernel, $\theta_\mathrm{ker}$, and the assumed fraction of diffuse emission in the smoothed 24$\mu$m map, $f_\mathrm{DE}$.  We have chosen the size of the gaussian kernel based on the unsharp masking approach used by \citet{2011ApJ...730...72R} to estimate the diffuse emission on the SF tracer maps in their study of the galaxy NGC 4254.  They suggest that using kernel sizes below $\sim 6$ kpc overestimates the diffuse component fraction.  Thus, as a first approximation, we have adopted the 6 kpc as the size of gaussian kernel in our approach.  This physical size corresponds to $\sim 170\arcsec$ for NGC 628 and M101, and 225$\arcsec$ for NGC 6946.  The next step is to multiply the smoothed 24$\mu$m map by the $f_\mathrm{DE}$ factor, and subtract this map from the original 24$\mu$m map.  We select the value of $f_\mathrm{DE}$ based on the best linear relation between the $\Sigma_\mathrm{SFR}$ from H$\alpha$ and the cirrus subtracted $\Sigma_\mathrm{SFR}$ from 24$\mu$m+FUV.  We assess the level of linear correlation between these two quantities using a Pearson correlation coefficient, $r_\mathrm{P}$.  We calculate $r_\mathrm{P}$ for different values of $f_\mathrm{DE}=0.2, 0.3, 0.4$, and 0.6, and we select the value that yields the highest $r_\mathrm{P}$.  Table \ref{table-pearson} shows the values for $r_\mathrm{P}$ for each value of $f_\mathrm{DE}$.  Additionally, we have included the value of $r_\mathrm{P}$ when no smooth background subtraction of 24$\mu$m is applied.  We see that in the case of NGC 6946, $f_\mathrm{DE}$=0.3 produces the highest value of $r_\mathrm{P}$=0.929.  For NGC 628, the highest value of $r_\mathrm{P}$=0.888 is obtained with $f_\mathrm{DE}$=0.6.  Finally, $f_\mathrm{DE}$=0.2 gives the highest value of $r_\mathrm{P}$=0.932 for M101.  

\begin{figure*}
\centering
\begin{tabular}{cc}
\epsfig{file=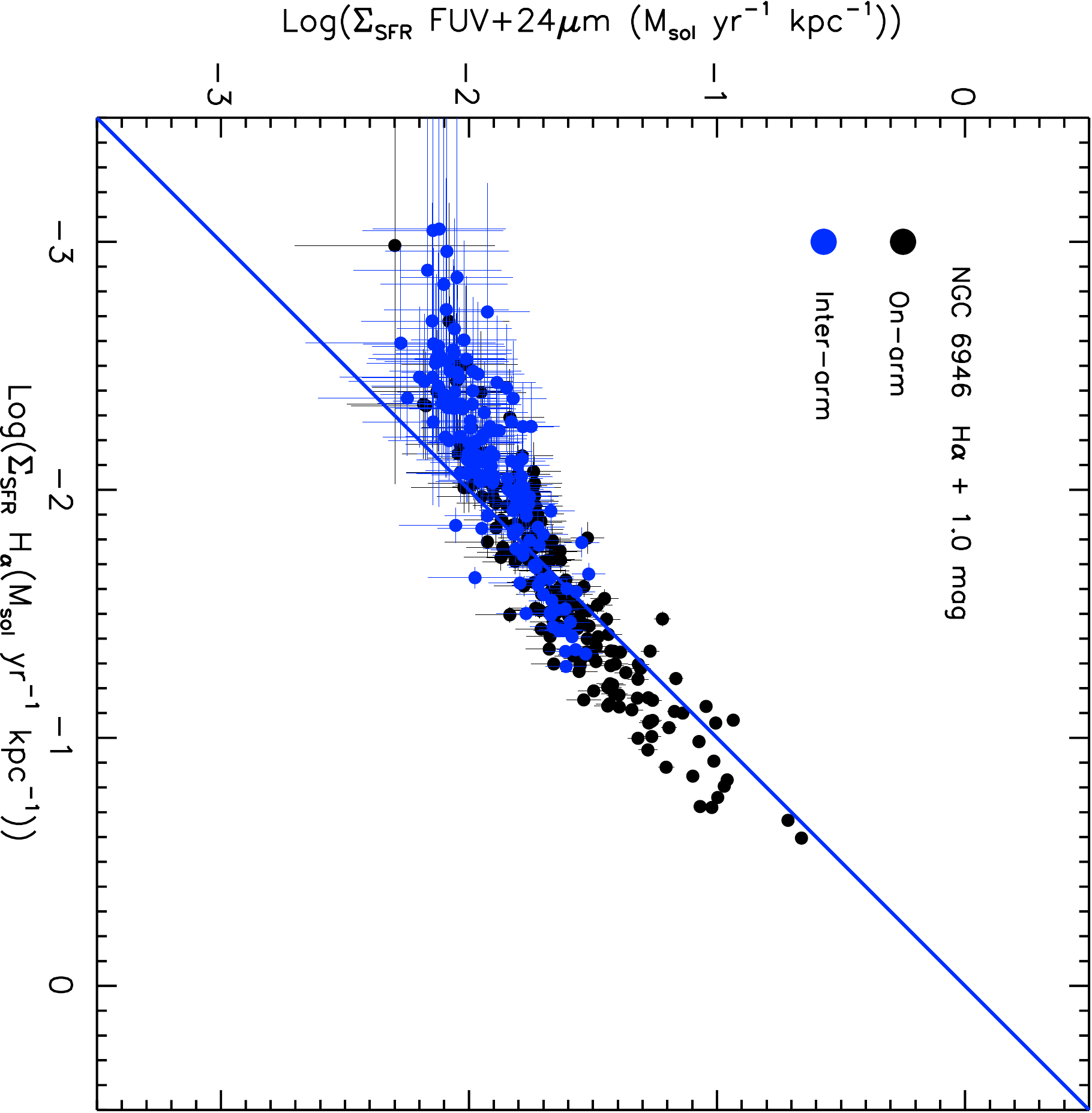,width=0.33\linewidth,angle=90} 
\epsfig{file=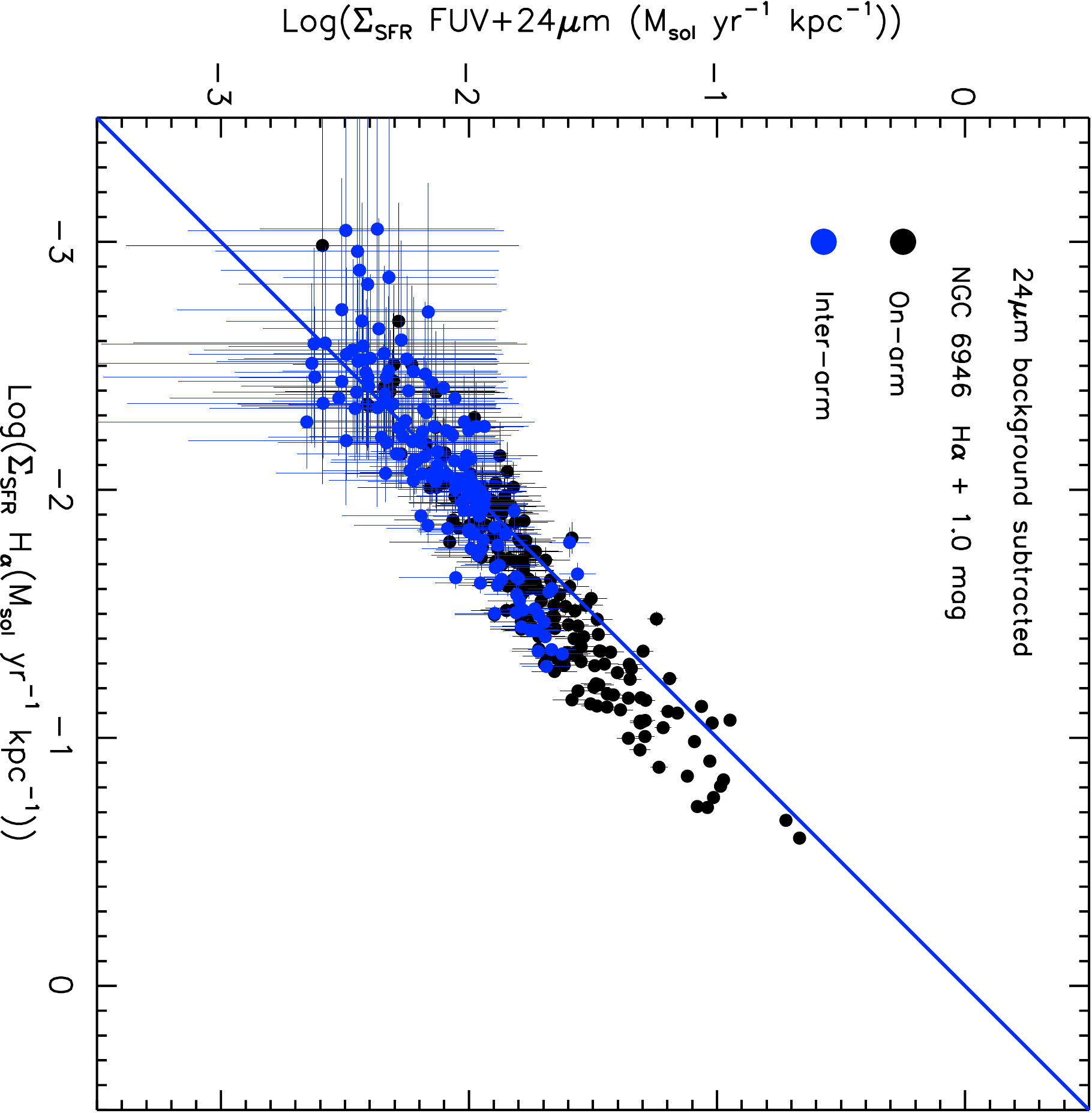,width=0.33\linewidth,angle=90} 
\end{tabular}
\caption{Left: Comparison between $\Sigma_\mathrm{SFR}$ using H$\alpha$ and the $\Sigma_\mathrm{SFR}$ using FUV + 24$\mu$m for NGC 6946 (same as Figure \ref{fig_ha_fuv} but for the uniform grid). Black dots represent on-arm regions, while blue dots correspond to inter-arm regions.  Right:  Same as left panel but now a subtraction of a 24$\mu$m smooth component is applied to $\Sigma_\mathrm{SFR}$ using FUV + 24$\mu$m.  As in Figure \ref{fig_ha_fuv}, we use A$_\mathrm{H\alpha}=1.0$ magnitude extinction for regions in NGC 6946.  After the 24$\mu$m smooth component subtraction is applied, the linear relation between the two $\Sigma_\mathrm{SFR}$ approaches is better recovered.}
\label{fig_ha_fuv_grid}
\end{figure*}

In Figure \ref{fig_ha_fuv_grid} we illustrate the effect of applying the 24$\mu$m cirrus correction to the $\Sigma_\mathrm{SFR}$ measurements for NGC 6946.  In the left panel, we observe that at $\Sigma_\mathrm{SFR} \sim 0.01\  \Msun \mathrm{yr}^{-1}\ \mathrm{kpc}^{-2}$, the measurements of $\Sigma_\mathrm{SFR}$ from 24$\mu$m+FUV overestimates the $\Sigma_\mathrm{SFR}$ from H$\alpha$.  After a cirrus component of 24$\mu$m is subtracted using $f_\mathrm{DE}$=0.3 and $\theta_\mathrm{ker}$=225$\arcsec$, both approaches to estimate $\Sigma_\mathrm{SFR}$ present a tighter correlation for $\Sigma_\mathrm{SFR} \lesssim 0.01\  \Msun \mathrm{yr}^{-1}\ \mathrm{kpc}^{-2}$.

\begin{table}
\caption{Pearson correlation coefficient for $\Sigma_\mathrm{SFR, H\alpha}$ vs. $\Sigma_\mathrm{SFR, 24\mu m+FUV}$ linear correlations.\label{table-pearson}}
\centering
\begin{tabular}{cccccccc}
\hline\hline

$\theta_\mathrm{ker}$ & & No diffuse & \multicolumn{5}{c}{$f_\mathrm{DE}$}  \\
 \cline{4-8}
& &  correction & 0.2 & 0.3 & 0.4 & 0.5 & 0.6   \\
\hline
& & & \multicolumn{2}{c}{NGC 6946} &  &  &   \\
\hline
 $225\arcsec$ & $r_\mathrm{P}$ & 0.912 &  0.926 & 0.929  & 0.920 & 0.864 & 0.907  \\
\hline
& & & \multicolumn{2}{c}{NGC 628} &  & & \\
\hline
$170\arcsec$ & $r_\mathrm{P}$ & 0.849 &  0.862 & 0.869  & 0.876 & 0.882 & 0.888  \\
\hline
& & & \multicolumn{2}{c}{M101} &  & &   \\
 \hline
$170\arcsec$  & $r_\mathrm{P}$ & 0.928 &  0.932 & 0.930  & 0.927 & 0.916 & 0.885  \\

\hline
\end{tabular}
\end{table}

\bibliography{biblio}

\end{document}